\documentclass[12pt,a4paper]{book}
\usepackage{doublespace,subfigure,boxedminipage,latexsym,amssymb,multirow,rotating,fancyhdr}
\usepackage{aran} 
\newcommand{\gev}  {\mbox{\,GeV}}
\newcommand{\tev}  {\mbox{\,TeV}}
\newcommand{\kev}  {\mbox{\,keV}}
\newcommand{\gevc} {\mbox{\,GeV$/c$}}
\newcommand{\evcc} {\mbox{\,eV$/c^2$}}
\newcommand{\kevcc}{\mbox{\,keV$/c^2$}}
\newcommand{\gevcc}{\mbox{\,GeV$/c^2$}}
\newcommand{\tevcc}{\mbox{\,TeV$/c^2$}}
\newcommand{\invpb}{\mbox{\,pb$^{-1}$}}
\newcommand{\gt}   {\ensuremath{>}}
\newcommand{\lt}   {\ensuremath{<}}
\newcommand{\degs} {\ensuremath{^{\circ}}}
\newcommand{\m}    {\mbox{\,m}}
\newcommand{\cm}   {\mbox{\,cm}}
\newcommand{\mm}   {\mbox{\,mm}}
\newcommand{\um}   {\mbox{\,$\mu$m}}
\newcommand{\s}    {\mbox{\,s}}
\newcommand{\us}   {\mbox{\,$\mu$s}}
\newcommand{\ns}   {\mbox{\,ns}}
\newcommand{\ee}   {\mbox{e$^{+}$e$^{-}$}}
\newcommand{\pos}  {\ensuremath{\mathrm{e^+}}}
\newcommand{\ele}  {\ensuremath{\mathrm{e^-}}}
\newcommand{\nue}  {\ensuremath{\nu_{\mathrm{e}}}}
\newcommand{\num}  {\ensuremath{\nu_{\mu}}}
\newcommand{\nut}  {\ensuremath{\nu_{\tau}}}
\newcommand{\anue} {\ensuremath{\bar{\nu}_{\mathrm{e}}}}
\newcommand{\anum} {\ensuremath{\bar{\nu}_{\mu}}}
\newcommand{\anut} {\ensuremath{\bar{\nu}_{\tau}}}
\newcommand{\fbar} {\ensuremath{\bar{\mathrm{f}}}}

\newcommand{\Z}    {\ensuremath{\mathrm{Z}}}
\newcommand{\W}    {\ensuremath{\mathrm{W}}}
\newcommand{\suc}  {\ensuremath{SU(3)_{\rm C}}}
\newcommand{\sul}  {\ensuremath{SU(2)_{\rm L}}}
\newcommand{\uy}   {\ensuremath{U(1)_{\rm Y}}}
\newcommand{\uem}  {\ensuremath{U(1)_{\rm em}}}
\newcommand{\grav} {\ensuremath{\mathrm{\tilde{G}}}}
\newcommand{\cha}  {\ensuremath{\chi^\pm}}
\newcommand{\chaOp}{\ensuremath{\chi^{+}_1}}
\newcommand{\chaOm}{\ensuremath{\chi^{-}_1}}
\newcommand{\chaO} {\ensuremath{\chi^\pm_1}}
\newcommand{\neu}  {\ensuremath{\chi}}

\newcommand{\slep} {\ensuremath{\tilde{\ell}}}
\newcommand{\slepR}{\ensuremath{\tilde{\ell}_{\mathrm{R}}}}
\newcommand{\slR}  {\ensuremath{\tilde{l}_{\mathrm{R}}}}
\newcommand{\sll}  {\ensuremath{\tilde{l}}}
\newcommand{\sel}  {\ensuremath{\mathrm{\tilde{e}}}}
\newcommand{\selR} {\ensuremath{\mathrm{\tilde{e}_R}}}
\newcommand{\smu}  {\ensuremath{\tilde{\mu}}}
\newcommand{\smuR} {\ensuremath{\mathrm{\tilde{\mu}_R}}}
\newcommand{\stau} {\ensuremath{\tilde{\tau}}}
\newcommand{\stauO}{\ensuremath{\tilde{\tau}_1}}
\newcommand{\stauR}{\ensuremath{\mathrm{\tilde{\tau}_R}}}

\newcommand{\snu}  {\ensuremath{\tilde{\nu}}}

\newcommand{\roots} {\ensuremath{\sqrt{s}}}
\newcommand{\Ebeam} {\ensuremath{E_{\mathrm{beam}}}}
\newcommand{\Lum}   {\ensuremath{\mathcal{L}}}
\newcommand{\p}     {\ensuremath{\vec{p}}}
\newcommand{\Mmess} {\ensuremath{\mathrm{M_{m}}}}

\newcommand{\tanb}  {\ensuremath{\tan\beta}}
\newcommand{\N}     {\ensuremath{\mathrm{N_{5}}}}
\newcommand{\Fm}    {\ensuremath{\mathrm{F_{m}}}}

\newcommand{\rootF} {\ensuremath{\mathrm{\sqrt{\Fm}}}}
\newcommand{\rootFo}{\ensuremath{\mathrm{\sqrt{F_0}}}}
\newcommand{\half}  {\ensuremath{\frac{1}{2}}}

\newcommand{\thrhlf}{\ensuremath{\frac{3}{2}}}
\newcommand{\dm}    {\ensuremath{\Delta m}}
\newcommand{\eff}   {\ensuremath{\varepsilon}}
\newcommand{\MP}    {\ensuremath{\mathrm{M_{P}}}}
\newcommand{\GN}    {G_{\mathrm{N}}}
\newcommand{\MZ}    {\ensuremath{m_{\rm Z}}}
\newcommand{\MW}    {\ensuremath{m_{\rm W}}}
\newcommand{\QU}    {\ensuremath{\mathrm{Q_{U}}}}
\newcommand{\ldet}  {\mbox{$\ell_{\mathrm{det}}$}}
\def \nch  {N_{\mathrm{ch}}}
\def \mvis {M_{\mathrm{vis}}}
\def \pt   {p_{\mathrm{t}}}
\def \ww   {\mathrm{WW}}
\def \zz   {\mathrm{ZZ}}
\def \qq   {\mathrm{q\bar{q}}}
\def \bb   {b\bar{b}}
\def \gaga {\gamma\gamma}
\def \ngood{N_{\mathrm{good}}}
\def \tt   {\tau\tau}
\def \ptmiss{\not\!\!\pt}
\def \Emiss {\not\!\!E}
\def \Evis  {E_{\rm{vis}}}
\newcommand{\nNF} {\ensuremath{N_{95}}}
\newcommand{\nbNF}{\ensuremath{\bar{N}_{95}}}
\newcommand{\sNF}{\ensuremath{\sigma}_{95}}
\newcommand{\sbNF}{\ensuremath{\bar{\sigma}_{95}}}
\newcommand{\cls} {\ensuremath{\mathrm{CL}_s}}
\newcommand{\clsb}{\ensuremath{\mathrm{CL}_{s+b}}}
\newcommand{\clb} {\ensuremath{\mathrm{CL}_b}}

\begin{document}
\pagenumbering{roman}      
\pagestyle{empty}
\begin{singlespace}
\begin{titlepage}
\begin{center}
\vspace{2.5cm}
\begin{spacing}{2.5}
{\LARGE\bf\mdseries\scshape
 Searches for Gauge Mediated \linebreak
 Supersymmetry Breaking at ALEPH \linebreak
 with Centre--of--Mass Energies \linebreak 
 up to 209 GeV} \\
\end{spacing}
\vspace{2.0cm}
\Large
by \\
\vspace{0.5cm}
Ar\'an Garc\'\i a-Bellido \\
\vspace{3cm}
\large
\includegraphics[width=40mm]{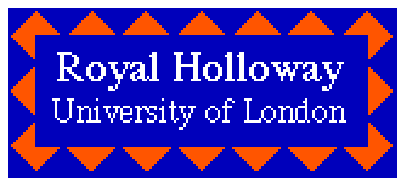}\\
\vspace{2cm}
Department of Physics \\
Royal Holloway \\
University of London \\
\vspace{3.5cm}
\normalsize
A thesis submitted to the University of London \\
for the degree of Doctor of Philosophy \\ 
September, 2002
\end{center}
\end{titlepage}

\newpage
\thispagestyle{empty}
\begin{center}
\large {\bf 
Searches for gauge mediated supersymmetry breaking \\
at ALEPH with centre--of--mass energies up to 209 GeV} \\
\vspace{1cm}
Ar\'an Garc\'\i a-Bellido \\
\vspace{0.1cm}
Royal Holloway University of London
\end{center}
\vspace{1cm}
\begin{spacing}{1.5}
A total of 628\,pb$^{-1}$ of data collected with the ALEPH detector 
at centre-of-mass energies from 189 to 209\,GeV is analysed in the
search for gauge mediated supersymmetry breaking (GMSB) topologies. In this
framework, a novel search for six-lepton final states when the stau is the
next-to-lightest supersymmetric particle (NLSP) and has negligible lifetime
is performed. 
Other possible signatures at LEP are studied and the ALEPH final
results described for two acoplanar photons, non-pointing single photons,
acoplanar leptons, large impact parameter leptons, detached slepton
decay vertices, heavy stable charged sleptons and multi-leptons plus
missing energy final states. 
No evidence is found for new phenomena, and lower limits on the
masses of the relevant supersymmetric particles are derived.
A scan of a minimal GMSB parameter space is performed and 
lower limits are set for the NLSP mass at 54\,GeV$/c^2$ and for the mass
scale parameter $\Lambda$ at 10\,TeV$/c^2$, independent of the NLSP lifetime.
Including the results from the neutral Higgs boson searches,
a NLSP mass limit of 77\,GeV$/c^2$ is obtained and values of $\Lambda$ up
to 16\,TeV$/c^2$ are excluded.
\end{spacing}

\newpage
\thispagestyle{empty}
\begin{center}
{\large \bf 
Acknowledgements}
\end{center}
\vspace{0.3cm}
\begin{spacing}{1.42}
I am grateful to the following people for their help and support during the
last three years. 

It would have been impossible for me to come to Royal Holloway without the
effort and perseverance of Mike Green who has also been a great advisor.
 
I cannot be thankful enough to Grahame Blair
for his supervision throughout my PhD. I treasure the conversations I have
had with him whether related to work, physics or my future. He has been a
source of knowledge, guidance, motivation and encouragement. 
I just hope one day I can become that good. 

I would like to thank Fabio Cerutti for his help and dedication with the
scan, based on his previous work. 
The people involved in GMSB searches at ALEPH,
Gary Taylor, Barbara Clerbaux, Chris Booth, Fabio Cerutti, Paolo
Azzurri, Eva Bouhova, Luke Jones and Jason Nielsen were all of immense help to
explain and provide their results. Needless to say, this thesis could not
have been done without them. A special thank you to Christoph Rembser, always joyful and
supportive, for pointing me towards the six-lepton topology and for doing
such a fantastic job in the GMSB LEP combination.

Grahame Blair, Mike Green and Glen Cowan have read all or parts of this
work and their comments and suggestions have been very helpful. Glen was
always there to sort out my endless questions about statistics. 

I acknowledge the support from CLRC and CERN who funded my stays in
Geneva. Special thanks to Dieter Schlatter, Roberto Tenchini, Fabiola
Gianotti and Fabio Cerutti for making the second one possible.  
The Physics Department also funded my attendance to schools and
conferences. 

Thanks to everyone in the HEP group at RHUL for being such a great crowd. 
David, Graham, Onuora, Omar, Fernando, Ra\'ul, Jordi and Ana made life really
enjoyable at CERN. Thanks for pulling me out of the office and teaching me
the intricacies of rock climbing and kitchen-cloth squeezing.

My aunt Paloma in Oxford and my cousins Juan and Ester
in Geneva have provided the warmth and affection of the family
abroad. In Geneva, I had the good advice, hospitality, 
and a car from my cousins and I could even babysit from time to time the
adorable Sara; in Oxford, I was the one babysitted. 

Finally, my parents. I owe them everything. This thesis is dedicated to
them. 
\end{spacing}
\flushright Egham, September 2002.
\end{singlespace}

\pagestyle{plain}
\tableofcontents
\newpage
\pagestyle{thesisheadings}
\pagenumbering{arabic}     
\addcontentsline{toc}{chapter}{Outline}
\chapter*{Outline}
\label{intro}

\begin{spacing}{1.5}
This thesis describes the final searches performed by the ALEPH
collaboration at 
LEP for supersymmetry (SUSY) when a light gravitino is the lightest supersymmetric
particle. The first chapter is devoted to symmetries in particle physics: 
those of the Standard Model and supersymmetry. 
The limitations of the Standard Model are reviewed
and the solutions given by supersymmetry are analysed. The fact that SUSY
is a broken symmetry, and a very difficult one to break, will lead to a
variety of possible models. Gravity mediation and gauge mediation,
the subject of this thesis, will be presented and the 
production and decay of the predicted new particles will be introduced. 

In Chapter~\ref{detector} the ALEPH detector and its analysis tools are
explained. Chapter~\ref{sixl} documents the search for selectron or smuon 
production developed by the author to cover the case of heavy neutralinos
and light staus with zero lifetime. Chapter~\ref{gmsbphen} summarises all
other searches carried out by other collaborators within ALEPH in the context
of gauge mediated SUSY breaking (GMSB) scenarios, including neutral Higgs boson searches. 
Together with the six-lepton topology described in Chapter~\ref{sixl}, 
these results are interpreted in terms of excluded areas in the minimal
GMSB parameter space.
Chapter~\ref{scan} describes the scan
performed by the author over these parameters and the combined results on
mass limits, cross sections, and the input parameters of the theory. 

Chapter~\ref{concl} will compare ALEPH results with other collaborations
and overview the status of GMSB models after LEP. A brief look ahead is also
given. 
\end{spacing}
\thispagestyle{empty}
\chapter{The Standard Model and beyond: Supersymmetry}
\label{theory}
\begin{center}
\begin{spacing}{1.5}
\setlength{\fboxsep}{5mm}
\begin{boxedminipage}[tb]{0.9\linewidth} \small
The Standard Model of fundamental interactions is probably one of the most 
accurately tested theories in physics. Its foundations will be briefly
reviewed here, stressing the importance of symmetries as the organising
principle behind it. Symmetries provide an important route to advance
the understanding and development of a theory. But sometimes as well as
discovering a fundamental symmetry it is as important to know how to break it. The
concept of spontaneous symmetry breaking is also central to the Standard
Model, and it leads to the main untested sector of the theory: the
Higgs sector. Hereafter, the problems of the Standard Model are
outlined and a possible solution in terms of a new (ultimate) symmetry will
emerge. 

Supersymmetry is regarded as the most likely incarnation of physics beyond
the Standard Model to be accessible at present experiments. 
The strong points of the theory are studied in this chapter and special 
attention is given to the consequences of the breaking of this symmetry.
The different phenomenological
consequences for current particle colliders will be described paying special
attention to the framework of this thesis: models where the electroweak and
strong interactions (and not gravity) mediate the breaking.
\end{boxedminipage}
\end{spacing}
\end{center}
\newpage

\section{Symmetries in physics: the Standard Model}
The Standard Model (SM) of particle physics describes the electromagnetic, weak
and strong interactions in terms of the internal symmetries they exhibit. 
The discovery of a symmetry in Nature provides the means not only to
describe observations, but to \emph{construct} theories and make powerful
predictions based on calculations from conserved quantities. 

In the Standard Model~\cite{Glashow:1961tr,Weinberg:1967tq,salam}, 
the forces between the fundamental particles are
mediated by the exchange of spin-1 bosons. 
Thus the electromagnetic
force is `carried' by the photon which couples to charged particles, the
weak force couples all fermions by the W and Z bosons, 
and the strong force couples quarks via the gluons and accounts for the
stability of nucleons.  
Table~\ref{tab:interactions} lists all four known interactions and
the corresponding force carriers.
The constituents of matter are spin-$\half$ fermions, the leptons
and quarks, which come in three families (or generations), as listed in
Tab.~\ref{tab:fermions}. 
\begin{table}[htb]
\begin{center}
\begin{tabular}{|llc|lcc|} \cline{4-6}
\multicolumn{3}{c}{} & \multicolumn{3}{|c|}{Carrier particles} \\ \hline
Interaction & Strength & Range & Name & Mass ($\gevcc$) & Spin \\ \hline \hline
Strong          & $\alpha_{3} \sim 0.121$  & $10^{-15}$m & Gluons (g) & 0 & 1 \\
Electromagnetic & $\alpha_{1} \sim 1/128$  & $\infty$    & Photon ($\gamma$) & 0 & 1 \\
Weak            & $\alpha_{2} \sim 10^{-6}$ & $10^{-17}$m &Z$^0$,W$^{\pm}$ & 91.2, 80.4 & 1 \\ 
Gravitation     & $\GN \sim 10^{-39}$ & $\infty$ & Graviton (G) & 0 & 2 \\ \hline
\end{tabular}
\end{center}
\caption[Interactions and carrier particles]{\label{tab:interactions} \small The
four interactions in Nature, with the corresponding boson carriers.
The values of the couplings $\alpha$ are measured at the Z mass; 
they evolve with energy. The graviton has not yet been observed. 
The gravitational constant $\GN$ has units of (GeV/$c^2$)$^{-2}$. Adapted
from Ref.~\cite{h&m}.}
\end{table}
\begin{table}[htb]
\begin{center}
\begin{tabular}{|c|c|llc||llc|} \cline{3-8}
 \multicolumn{2}{c|}{} & \multicolumn{3}{c||}{Leptons} &  \multicolumn{3}{|c|}{Quarks} \\ \cline{3-8}
 \multicolumn{2}{c|}{} & Name & $~~~~~~Q$ & Mass (MeV/$c^2$) & Name & $~~~~~~~Q$ & Mass (MeV/$c^2$)\\ \cline{3-8} \hline \hline
 \multirow{6}{0.3cm}{\rotatebox{90}{Families}}  
 &\multirow{2}{0.4cm}{1$^{\rm{st}}$} & electron  & e$~~-1$ & 0.511 & up & $u~~+2/3$ & 1--5 \\
 &                         & e-neutrino&$\nue~~~~\:0$& $< 3\times10^{-6}$ & down & $d~~-1/3$ & 3--9 \\ \cline{2-8}
 &\multirow{2}{0.4cm}{2$^{\rm{nd}}$} & muon  & $\mu~~-1$ & 106 & charm & $c~~+2/3$ & 1150--1350 \\
 &                         & $\mu$-neutrino & $\num~~~~\:0$ & $< 0.19$ & strange & $s~~-1/3$ & 75--170 \\ \cline{2-8}
 &\multirow{2}{0.4cm}{3$^{\rm{rd}}$} & tau  & $\tau~~-1$ & 1780 & top & $t~~+2/3$ & 174$\pm$5$\times$$10^3$ \\
 &                         & $\tau$-neutrino & $\nut~~~~\:0$ & $< 18.2$ & bottom & $b~~-1/3$ & 4000--4400 \\ \hline
\end{tabular}
\end{center}
\caption[Fermion summary table]{\label{tab:fermions} \small  The fermions
in the SM are grouped into three families (or generations). 
Leptons feel the electroweak force whilst quarks also feel the strong
force. Masses are given in MeV/$c^2$~\cite{pdg}. Only upper limits exist on
the mass of neutrinos. The quark masses cannot be precisely determined
since QCD forbids isolated quarks. For each fermion $f$ there is an
antifermion $\bar{f}$ with opposite electric charge $-Q$.}
\end{table}

By imposing certain symmetries, one can write down the Lagrangian and thus
derive the equations of motion for all particles described above. This is
the beauty of the theory: specify a global symmetry group and the
nature of the interaction is fully determined. So by building on
experiment, one learns the following symmetries need to be fulfilled by the
Standard Model Lagrangian: 
\begin{itemize}
\item Poincar\'e invariance~\cite{h&m}. The theory must be invariant under continuous
spacetime transformations such as translations, rotations and Lorentz
transformations, i.e. the SM obeys the laws of Special Relativity. These
symmetries imply the conservation of linear momentum, angular momentum and
4-momentum, respectively. 
\item The CPT symmetry, which swaps the charge,
parity and time flow of the process, must be respected exactly. This
implies the conservation of the corresponding discrete quantum numbers.  
\item And finally, the gauge (phase) symmetries.
The Standard Model is based on the local symmetry group
$\suc\otimes\sul\otimes\uy$. This structure uniquely determines the
form of the interactions. The $\suc$
symmetry is that of the strong interaction between quarks, which can
transform one quark to another by means of eight gauge bosons (the gluons)
conserving the colour charge. 
Similarly, the gauge invariance under $\sul$ requires three gauge fields
(eventually the $\W^+,\W^-$ and $\Z^0$) and conserves the weak isospin
$I_L$. This symmetry allows an up-type fermion to interact with a down-type
fermion via the weak force. 
And the group $U(1)$ is associated with one boson field and conserves the
hypercharge Y, giving rise to the electromagnetic force. 
Actually, only left-handed fermions which are
weak isospin doublets like $(^{\nue}_{\rm\,e})_{\rm L}$ 
will couple to the $\sul$ gauge bosons. Right-handed
fermions are isospin singlets like $\rm{u_{\rm R}}$, and can only interact if
they carry non-zero hypercharge or colour quantum numbers.
\end{itemize}

At this stage the theory is gauge invariant, i.e. transforming the fermion
fields in the theory by a local phase shift requires the introduction of
gauge bosons to leave the Lagrangian unchanged. But all the particles in
the theory are massless. Indeed, fermion masses should appear in terms like 
$m\bar{\psi}\psi=m(\bar{\psi}_{\rm L}\psi_{\rm R}+
\bar{\psi}_{\rm R}\psi_{\rm L})$, which
explicitly breaks the $\sul\otimes\uy$ symmetry in which left-handed fields transform
independently of right-handed fields. And similarly for vector boson mass
terms like $\half m^2\W_{\mu}\W^{\mu}$.
Nevertheless, by inspecting Tables~\ref{tab:interactions}
and~\ref{tab:fermions}, it is obvious that fermions have mass and the weak
bosons are (very) massive.
The Z and W bosons \emph{have} to be massive to give the weak interactions
a very short range. Thus some mechanism to
break the $\sul\otimes\uy$ symmetry into $\uem$ is needed if the theory is
to describe massive W and Z bosons and a massless photon. 

\subsection{Electroweak symmetry breaking}
An elegant way of introducing massive gauge bosons for the weak
interactions without explicitly breaking the $\sul\otimes\uy$ gauge
invariance of the Lagrangian, 
was proposed by Higgs and others~\cite{Higgs:1964ia,Englert:1964et} 
using the concept of \emph{spontaneous symmetry breaking}. 
Briefly, the introduction of a complex spin-0 (scalar)
field $\phi$ does the job if it is a doublet in $\sul$ and its potential
respects the most general $\sul\otimes\uy$ gauge invariant
form: 
\begin{equation}
\label{higgspot}
V(\phi) = \mu^2\phi^{\dagger}\phi + \lambda(\phi^{\dagger}\phi)^2 ~~;~~ 
\phi=\sqrt{\half}\pmatrix{\phi_1+i\phi_2 \cr \phi_3+i\phi_4}
\end{equation}
where $\mu$ is the mixing parameter and $\lambda$ is the quartic
scalar self-coupling.
For $\mu^2<0$ and $\lambda>0$, the above potential has an infinite number
of non-zero solutions for which only the field's norm is known:
$|\phi|^2=\half(\phi_1^2+\phi_2^2+\phi_3^2+\phi_4^2)=-\mu^2/2\lambda\equiv
v^2/2$, where $v$ is the energy scale at which the electroweak symmetry
will be broken. 
\begin{figure}[tb]
\begin{center}
\vspace{-0.5cm}
\includegraphics[width=0.79\linewidth]{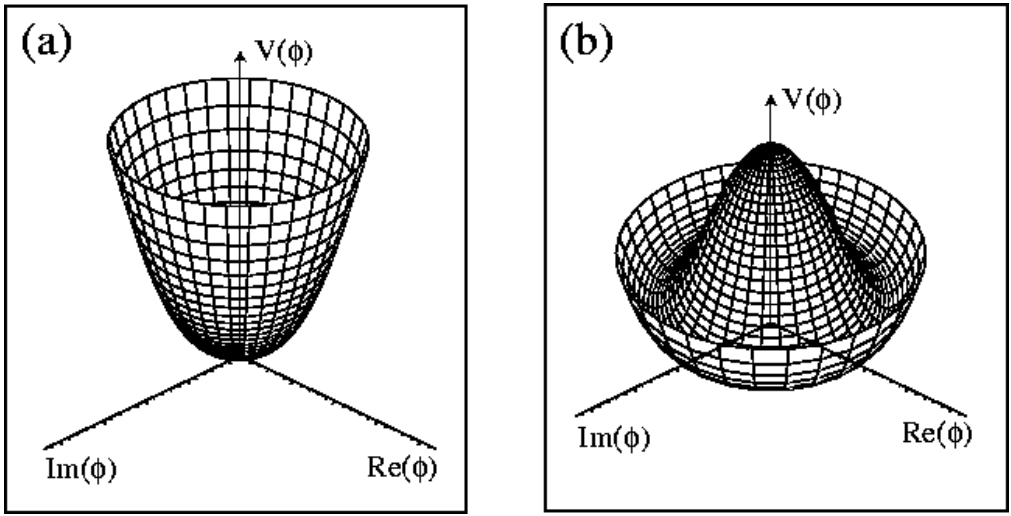}
\caption[The Higgs potential]
{\label{fig:higgspot}{\small The scalar potential $V(\phi) =
\mu^2\phi^{\dagger}\phi + \lambda(\phi^{\dagger}\phi)^2$ for a
single complex field $\phi=\phi_1+i\phi_2$, with (a) $\mu^2>0$
and (b) $\mu^2<0$. In a the minimum (ground state) is at
$\rm{Re}\phi=\rm{Im}\phi=0$, while for b the minimum of the potential is at
$|\phi|=v/\sqrt{2}=174\gev$, which describes a circle in the
(Re$\phi$,Im$\phi$) plane. 
Any point in the circle is equally
likely, thus the spontaneous symmetry breaking occurs when only one vacuum
state gets chosen, e.g. $\phi_1=v ; \phi_2=0$. From Ref.~\cite{stevea}.}}
\end{center}
\end{figure}
The fundamental state is thus degenerate (the circle of vacuum states at
$V(\phi)=0$ in Fig~\ref{fig:higgspot}b and $\sul$ is broken when only one of the
infinite points in the circle is chosen by the physical vacuum. 
Such a choice, for example $\phi_3=v;\phi_{1,2,4}=0$, constitutes a
spontaneous breaking of the invariance around the circle, but leaves $\uem$
unbroken.  
This is the essence of spontaneous symmetry breaking in the Higgs mechanism:
the physical vacuum adopts a specific ground state with non-zero
expectation value $v/\sqrt{2}$ (thus naturally breaking the symmetry) 
but the Lagrangian remains invariant.

By spontaneously breaking a continuous symmetry, massless particles
called Goldstone bosons must be produced~\cite{Goldstone:1962es}. 
In this case the electroweak gauge symmetry
had four scalar degrees of freedom $\phi_{1,2,3,4}$, but only $SU(2)$ is broken
thus only three massless Goldstone bosons are
expected. These three degrees of freedom are interpreted as the
longitudinal components of the three now massive vector bosons 
$\W^+,\W^-$ and $\Z^0$. The fourth degree
of freedom is predicted to be the Higgs boson $H^0$, the only particle in
the Standard Model yet to be discovered.  

In summary, the Higgs mechanism has allowed the massless weak bosons to
become massive by eating up the Goldstone bosons arising from the $SU(2)$
breaking, while keeping the photon massless. Furthermore, it predicts the
existence of a scalar boson with mass $m_H=\sqrt{-2\mu^2}$, which is a
free parameter and must be experimentally measured. The fermion
masses can then be generated by Yukawa couplings $g_f$ between the fermion
fields and the Higgs field of the form:
$g_f[\bar{\psi}_{\rm L}\phi\psi_{\rm R}+\bar{\psi}_{\rm R}\phi^{\dagger}\psi_{\rm L}]$. 
The inner products of Higgs and left-handed
fermion doublets of $\sul$ yield gauge invariant singlets, so these terms
can now be added to the Lagrangian without destroying its invariance. 

\subsection{Defects of the SM}
The Standard Model is not only an elegant description of the symmetries
behind matter and interactions. It has been tested in detail by experiments
and agreement has always been found between the measured and predicted
values from the theory~\cite{pdg}. However, despite its experimental
success, there remain some theoretical issues which make it impossible for
it to be a complete description of Nature. They are reviewed here.
\subsubsection{Gravity}
The Standard Model is an accurate description of the electromagnetic, weak
and strong interactions, but does not include the fourth known force:
gravity. Newton's constant has dimensions of the inverse of a mass
squared ($\GN=1/\MP^2$, where $\MP\sim 10^{19}\gev$ is the Planck mass 
where quantum effects become important) whereas the other interactions have
dimensionless coupling constants. 
In General Relativity, gravity, as a force, manifests itself through
deformations in spacetime. In the Standard Model the spacetime is assumed to
be flat and forces to arise from the exchange of quanta. This is probably
one of the remaining great issues to be resolved in physics: how to make
the Standard Model and General Relativity compatible. 
\subsubsection{Number of free parameters}
Ideally, one would expect that the three coupling constants ($\alpha_1$,
$\alpha_2$, $\alpha_3$ for each symmetry group in
Tab.~\ref{tab:interactions})  
would be the only free parameters of the theory. Assigning the
matter particles to a specific representation of the gauge group and then
measuring the couplings should be enough to have a complete theory of
Nature. Conversely, the SM has at least 28 arbitrary parameters to account
for: three gauge couplings; two parameters in the Higgs potential
($\mu$ and $\lambda$); 12 fermion masses (including the neutrinos); three
mixing angles and one phase in the quark sector and another three mixing
angles and three phases in the lepton sector for massive neutrinos; and
finally one CP violation phase in the strong interaction.
\subsubsection{Fermion masses}
Looking back at Tab.~\ref{tab:fermions} the diversity of
observed masses for the fermions is striking. 
As was mentioned in the previous section,
the Higgs mechanism incorporates massive fermions into the theory while 
keeping the Lagrangian gauge invariant. 
The couplings between the Higgs and the
fermion fields are proportional to the fermion masses: $g_f=\sqrt{2}m_f/v$. If
$v/\sqrt{2}=174\gev$, why is the top-quark Yukawa coupling so close to one and
the electron's five orders of magnitude weaker? The SM does not explain
this mass hierarchy.

The recent experimental evidence for neutrino
oscillations~\cite{nu,Ahmad:2002jz},  
only possible if they are massive, strongly suggests physics beyond the
SM. The extreme smallness of their masses ($m_{\nu}\lesssim 0.2\evcc\sim
10^{-6}m_{\rm e}$) indicates that their origin cannot
be the Higgs mechanism. Such small masses may only arise if the existence
of right-handed neutrinos is postulated (Dirac masses, e.g. 
$\overline{\nu_{\rm R}}m_D\nu_{\rm L}$) 
or if a Majorana nature of the neutrinos is invoked
(if the neutrino is its own antiparticle: 
$\nu_{\rm L}^Tm\nu_{\rm L}$)~\cite{Altarelli:1999wi}.   
In both cases, the origin of neutrino masses must be linked with an energy
high enough to violate lepton number L. 
\subsubsection{Dark matter and the baryon asymmetry}
Conventional matter, made of baryons like the proton and the neutron,
represents $\lesssim 5\%$ of the total energy content of the Universe
according to Big-Bang nucleosynthesis~\cite{Olive:1999ij} 
and cosmic microwave background (CMB) radiation~\cite{Lange:2000iq}
analyses.  
Studies of large-scale structure, rotation curves in
spiral galaxies, luminous matter and CMB, suggest that the
total amount of dark matter in the Universe corresponds to about 30$\%$
of the total energy density
$\rho_c\sim5\times10^{-6}\gev/\cm^3$~\cite{rubakov}.  
If the SM seems to describe only $5\%$ of $\rho_c$, what kind of matter
is this remaining 30\%? 

Furthermore, why is the Universe made of matter and not antimatter? 
The SM equations are the same for matter and antimatter, 
but then there is no evidence for antimatter galaxies or stars in the
Universe, only matter. Three conditions
are required to generate a baryon asymmetry~\cite{Sakharov:1967dj}: 
the baryonic quantum
number B cannot be conserved; the C and CP symmetries are violated; and a
period of thermal inequilibrium is necessary for the asymmetry not to be washed
away. Although the SM could accommodate in non-perturbative processes a
violation of the baryon number, and CP violation arises naturally in a three
generation model, it seems the amount of CP
violation predicted in the SM is not sufficient~\cite{Farrar:1994hn}. 
Today there are $\sim$10$^{9}$ photons for each baryon in the Universe, so
at some point there had to be roughly one extra quark per billion $\qq$ pairs. The
origin of this asymmetry requires the existence of physics beyond the SM
with a stronger CP violation. 
\subsubsection{Hierarchy or naturalness problem}
\begin{figure}[hb]
\begin{center}
\vspace{-1cm}
\subfigure[\scriptsize W and Z boson contribution]{
\includegraphics[width=0.32\linewidth]{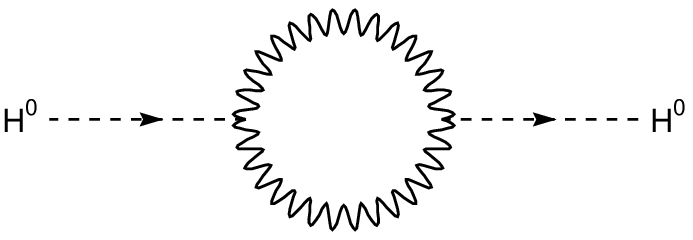}}
\subfigure[\scriptsize Fermion contribution]{
\includegraphics[width=0.32\linewidth]{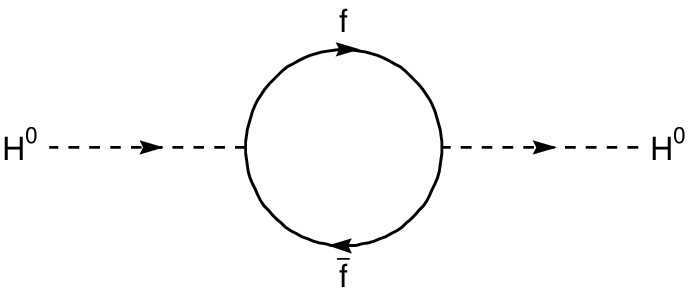}}
\subfigure[\scriptsize Scalar contribution]{
\includegraphics[width=0.32\linewidth]{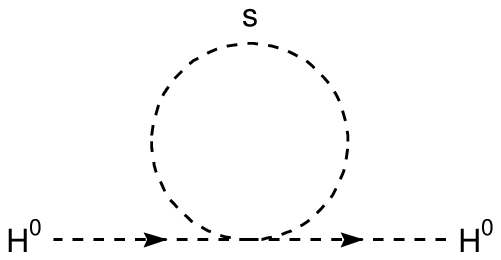}}
\caption[One-loop contributions to the Higgs mass]
{\label{fig:loops}{\small One-loop quantum corrections to the Higgs mass.}}
\end{center}
\end{figure}

Paradoxically, it is a particle whose existence is not yet proven that
poses the most serious threat to the Standard Model. The Higgs boson mass
is a free parameter in the Standard Model, but for the theory to be valid
it must be bound from below and, specially, from above. 
However, the Higgs mass in the Standard Model gets contributions from loops
with the gauge bosons, fermions and with itself, as seen in
Fig.~\ref{fig:loops}. 
The problem arises when considering radiative corrections from scalar
particles, as in Fig.~\ref{fig:loops}c. Gauge boson and fermion masses are protected
against divergences, so their loops do not pose a problem.
 
Scalar particles tend to get masses of the order
of the largest mass scale in the theory. And if the SM is to hold unscathed
up to the energy where gravity becomes strong, $\MP$, then the quantum
corrections to the Higgs mass diverge. 
Normally one introduces a cutoff energy
scale $\Lambda$ between the W mass and $\MP$, above which the SM is
invalid and some new physics must appear. Thus the Higgs mass will be
given by some `bare' mass term $m_0^2$ (where no assumption on new physics
is made) and the corrections that involve the new energy scale: $m_H^2
\simeq m_0^2 + \delta m_H^2 + \dots$, with: $\delta
m_H^2=\mathcal{O}\left(\frac{\alpha}{\pi}\right)\Lambda^2$.
Then, if on the one hand the Higgs mass involves
radiative terms of order $\Lambda^2\sim 10^{38}\gev^2$, 
but on the other hand it is known
to be of the order of the electroweak symmetry breaking scale
($m_{H}^2\sim-v^2$), how is it
possible that the bare mass is so finely tuned as to exactly absorb the
enormous radiative contributions and be left with the correct physical
value? This unnatural fine tuning (of up to 34 orders of magnitude) 
is referred to as the \emph{hierarchy} or \emph{naturalness problem}. 
It can also be expressed as: why is the W mass
much smaller than the Planck mass? Or equivalently, why is the Coulomb
potential for a typical particle mass $m$ in the atom so much greater than
the Newtonian potential: e$^2\gg \GN m^2=m^2/\MP^2$?

All these reasons lead to the idea that the SM is only an effective theory
at the present energy scale $\sim$100$\gev$ and that there must be a new
more fundamental theory just around the corner, at 
$\Lambda\lesssim \mathcal{O}(1)\tev$. Supersymmetry is widely accepted as
the most promising extension of the SM to explain some of the theoretical
problems outlined here.

\section{The last symmetry: Supersymmetry}
The symmetries of the SM involve scalar charges, such as colour or hypercharge,
that link particles with the same spin: $Q|J\rangle = |J\rangle$. 
The basic idea in supersymmetry (SUSY) is to have fermionic charges that
relate  fermions and bosons, so that:
\begin{equation}
\hat{Q}|{\rm fermion}\rangle = |{\rm boson}\rangle {\rm ~~~~and~~~~} 
\hat{Q}|{\rm boson}\rangle   = |{\rm fermion}\rangle.
\end{equation}
Supersymmetry is therefore a transformation that relates states of
different spin. Particles like bosons and fermions that in
principle have a completely different nature may be linked and form part of
the same representation. 

Perhaps the most compelling argument for supersymmetry is that it is the last
undiscovered symmetry allowed in scattering processes. Considering possible
extensions of the Poincar\'e group, Coleman and Mandula~\cite{Coleman:1967ad}
proved that the addition of any new operator which transforms as a boson
leads to a trivial S-matrix, i.e. in particle scattering experiments the
only allowed outcomes would be completely forward or backward scattering,
which of course completely disagrees with observation.
But this very strict `no-go' theorem can be avoided if the operator is
fermionic and as was demonstrated later~\cite{Haag:1975qh}, supersymmetry
is actually the only extension of the Poincar\'e group which does not lead
to a trivial S-matrix. So this argument should be persuasive enough: `in
Nature something not illegal is compulsory'. 

Supersymmetry is not only attractive aesthetically as an underlying theory
for the SM, it is also a step towards unification with
gravity\footnote{Making supersymmetry a local (spacetime dependent)
symmetry necessarily involves gravity. Local SUSY transformations are
equivalent to a local coordinate transformation; therefore supersymmetry is
entangled with a new approach to spacetime, which must be present in any
viable `Theory of Everything'~\cite{Ellis:2002mx}.}. 
Furthermore, it cures several of the SM shortcomings. 

The supersymmetric algebra derived from the above conditions requires the
existence of new particles with the same mass, gauge quantum
numbers and couplings as the SM ones. Thus for each fermion in the SM there
is a bosonic partner and {\it vice versa}, structured in \emph{supermultiplets}
like:
\begin{equation}
\pmatrix{{\rm fermion} (J=\half) \cr {\rm sfermion} (J=0)} ~~~;~~~
\pmatrix{{\rm g.~boson} (J=1) \cr {\rm gaugino} (J=\half)} ~~~;~~~
\pmatrix{{\rm graviton} (J=2) \cr {\rm gravitino} (J=\thrhlf)}
\end{equation}
which describe matter and Higgses, gauge fields and gravity, respectively. 
The convention in naming the superpartners (or sparticles) is to add
a prefix `s' (for scalar) to each fermion and a suffix `ino' for each boson. 
Could any of the known particles be one of these superpartners? No, it is
impossible to pair together any of the known fermions with the known bosons,
for the simple reason that their internal quantum numbers do not
match. For example, leptons have non-zero lepton number ${\rm L}=1$, but bosons
have ${\rm L}=0$, thus they cannot form part of the same supermultiplet.

By predicting this plethora of new particles supersymmetry solves neatly
the hierarchy problem. The radiative corrections from fermions and scalars
(diagrams b and c in Fig.~\ref{fig:loops}) now cancel each other naturally at
all levels of loops~\cite{Wess:1974kz}. 
For each positive scalar loop there is now a negative
fermion loop with the same couplings. 
The radiative contribution to the Higgs mass becomes~\cite{Ellis:2002mx}:  
\begin{eqnarray}
\label{radcorr}
\delta m_H^2 &=& 
-\frac{g_f^2}{16\pi^2}\left(\Lambda^2+m_f^2\right)
+\frac{g_s^2}{16\pi^2}\left(\Lambda^2+m_s^2\right)+\dots
\nonumber \\
             &\simeq& \mathcal{O}\left(\frac{\alpha}{\pi}\right)(m_s^2-m_f^2)
\end{eqnarray}
where the quadratic divergences have disappeared and the residual term is
much smaller than $m_H^2$ if the masses of the SUSY bosons and fermions are
similar: 
\begin{equation}
\label{lowenergy}
|m_s^2-m_f^2| \lesssim 1\tev^2
\end{equation}
This means that a light Higgs boson, such as electroweak data
constraints~\cite{lepew} and recent direct searches seem to
suggest~\cite{leph}, might be only possible if SUSY exists at a relatively
low energy.  

\begin{figure}[tb]
\begin{center}
\includegraphics[width=0.7\linewidth]{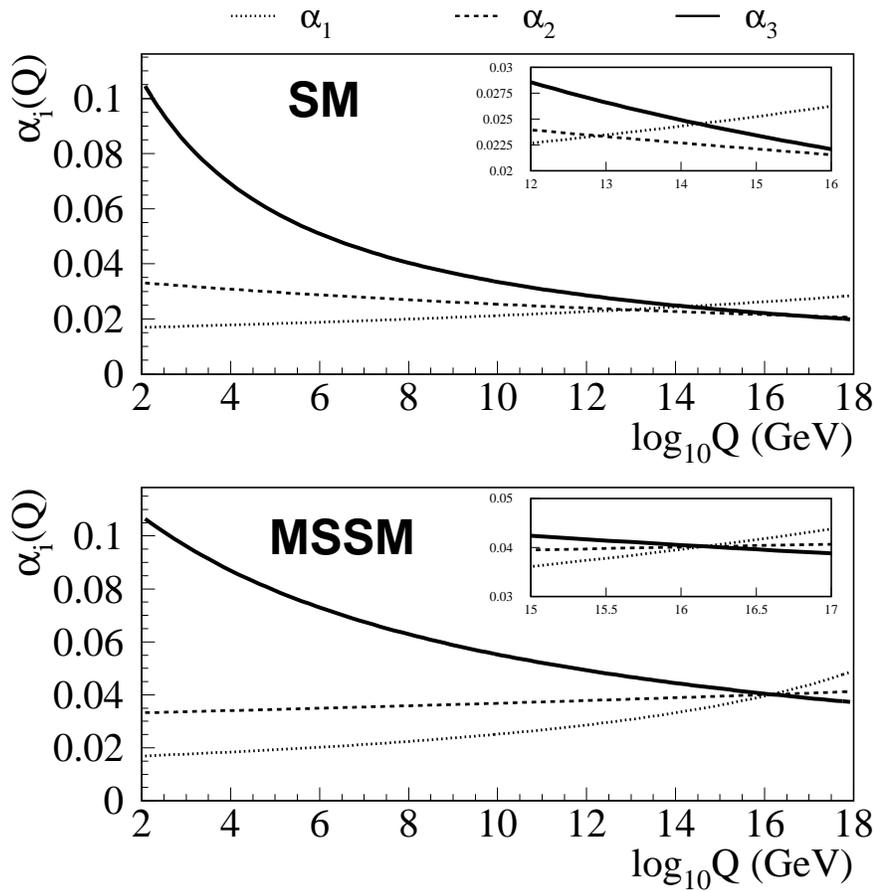}
\caption[Gauge couplings evolution with energy in the SM and the MSSM]
{\label{fig:running}{\small Evolution of the gauge couplings
$\alpha_i$ as a function of the energy Q, with the particle content of the
Standard Model (above) and with the minimal supersymmetric extension of the
SM (below). The energy scale where SUSY appears has been assumed to be 
$10^2\gev$.}}
\end{center}
\end{figure}

Another appealing feature of SUSY is that
it permits the unification of the three forces at a very high energy
scale, which cannot be achieved within the SM.
The three coupling constants
($\alpha_1,~\alpha_2,~\alpha_3$) are not constants at all,
but depend on the energy. Every particle with mass $m_X$ below a scale Q will
contribute to the running of the couplings via loop corrections to the force
carriers propagators and vertices~\cite{Amaldi:1991cn}:  
\begin{equation}
\label{running}
\frac{1}{\alpha_i({\rm Q})} = \frac{1}{\alpha_i(m_X)}+8\pi\:b_i\:\ln
\left(\frac{{\rm Q}}{m_X}\right) \hspace{1cm}i=1,2,3
\end{equation}
where the coefficients $b_i$ depend on the number of colours and the number
of active flavours (particles whose mass threshold is below Q), including
Higgs bosons. 
These coefficients are different in the SM and SUSY, as can be seen in
Fig.~\ref{fig:running}. In the SM the three lines do not intersect at a
same energy, even allowing for experimental error bands on the measurements at
present energies. But SUSY models do in fact permit a complete unification
of the couplings at around $\QU=10^{16}\gev$. Although this is a nice feature
of supersymmetry, it does not take into account anything else happening
during some 14 orders of magnitude from the scale where SUSY is turned on
up to the GUT (Grand Unified Theory) scale. 

There are two more characteristics of SUSY models worth discussing. 
The first is that two Higgs doublets are needed to give mass to up-type
fermions and down-type fermions separately. In the SM only one doublet $\phi$
and its complex conjugate $\phi^{\dagger}$ was sufficient, but in SUSY
models $\phi^{\dagger}\phi$ terms are not allowed since they do not
transform appropriately under supersymmetric transformations. Then, two
Higgs doublets $\phi_1=(\phi_1^+,\phi_1^0)$ and $\phi_2=(\phi_2^+,\phi_2^0)$
with opposite hypercharge ($\mp1$) are needed in order to give masses 
to the SM fermions. 
Both will acquire a non-zero vacuum expectation value, fixed like before:
$v_1^2+v_2^2=(246\gev)^2$, but their ratio is a free parameter,
called $\tanb$:
\begin{equation}
\label{tanb}
\tan\beta \equiv
\frac{\langle\phi_2\rangle}{\langle\phi_1\rangle}=\frac{v_2}{v_1}  
\end{equation}
This makes phenomenology somewhat more interesting since there are now eight
degrees of freedom and only three of them get `eaten' to give mass to the
weak gauge bosons, leaving five physical massive particles: the lighter
CP-even neutral $h$, the heavier CP-even neutral $H$, the CP-odd neutral
$A$ and two charged $H^{\pm}$.
Once the two Higgs supermultiplets have been introduced, they can couple to
each other in terms like $\mu\phi_1\phi_2$, where
$\mu$ is the mass mixing parameter, analogous to the Higgs mass $m_H$ in the
SM. 

The other very important phenomenological consequence of SUSY models is
that new terms which violate lepton- and baryon-number are now allowed in the
Lagrangian and still are consistent with gauge symmetries.
Couplings which violate both B and L are strongly constrained from nuclear
physics and specially from the proton lifetime~\cite{Dreiner:1997uz}. 
In the SM, B$-$L conservation was never imposed by hand but rather was an
`accidental' outcome of the theory. 
Since the strength of these couplings is severely limited, one can impose
the conservation of a new multiplicative quantum number: R-parity,
defined as:
\begin{equation}
\label{Rp}
R_P = (-1)^{3B-L+2S}
\end{equation}
where $S$ is the spin of the particle. Standard Model particles will have
$R_P=+1$ and SUSY particles $R_P=-1$. To assume R-parity conservation, which
is to some extent to impose B$-$L conservation by hand, has three fundamental
consequences: 
\begin{itemize}
\item sparticles are always produced in pairs,
\item heavier sparticles decay to lighter ones, and
\item the lightest sparticle is stable because it has no allowed decay mode.
\end{itemize}
The last point is perhaps the most attractive, since it implies the
existence of heavy stable particles. If these are neutral they could very
well constitute that 30\% of the energy density of the Universe that must
be in the form of cold dark matter~\cite{rubakov}. 
However, R-parity conservation is not obligatory and there may exist terms,
with maybe significant strength, that only violate B or L. Nonetheless,
the work throughout this thesis assumes that R-parity is strictly conserved. 

\subsection{Supersymmetry breaking}
The major problem with SUSY is that it must be a broken symmetry.
There is not a single piece of experimental evidence for SUSY in
particle physics so far. 
If a supersymmetric companion of the electron existed with its same
mass and spin-0 it would have been discovered long ago. The problem then is
to write a SUSY Lagrangian with terms that make sparticles heavier than
their SM counterparts, but do not introduce quadratic divergences to the
Higgs mass (Eq.~\ref{lowenergy}). 
These terms are thus said to induce a `soft SUSY breaking'. The origin of
such terms is not known and several possibilities exist~\cite{Witten:1981nf}. 
So in general one
writes all possible terms~\cite{Girardello:1982wz} and does not make any
assumption about \emph{how exactly} SUSY is broken. 
But to parametrise our ignorance on the origin of these terms, at least 105
new free parameters must be introduced. They account for the sparticle
masses, their mixing angles, 40 new CP violating phases, etc\dots
Thus a theory that was simple and extraordinarily predictive becomes
almost unmanageable, plagued with many more free parameters than before.
The hope is that these parameters will be eventually explained by GUT-scale
physics in terms of a few fundamental ones. Making some assumptions on how
SUSY breaking is transmitted and imposing experimental constraints will
lead to more predictive models as will be seen shortly.

The minimal supersymmetric extension of the Standard Model, or MSSM for
short, can then be constructed in the most general way, allowing for
R-parity violating terms and soft-SUSY breaking terms. The properties of
the mass spectrum are now reviewed.  
\begin{table}[tb]
\begin{center}
\begin{tabular}{|c|c|c|c|c|c|} \hline
\multicolumn{2}{|c|}{Standard Model Particles} & \multicolumn{4}{c|}{Supersymmetric Particles}\\ \cline{3-6}
\multicolumn{2}{|c|}{} & \multicolumn{3}{c|}{Weak Eigenstates} & Mass Eigenstates \\ \cline{1-5}
particle & spin & \multicolumn{2}{c|}{particle} & spin & \\ \hline \hline
$q=u,d,c,s,t,b$ & \half & $\tilde{q}_{\rm L},\tilde{q}_{\rm R}$ & squarks & 0 & $\tilde{q}_1,\tilde{q}_2$ \\ \hline
$\ell={\rm e},\mu,\tau$ & \half & $\slep_{\rm L},\slep_{\rm R}$ & sleptons & 0 & $\slep_1,\slep_2$ \\ \hline
$\nu=\nue,\num,\nut$ & \half & $\tilde{\nu}$ & sneutrinos & 0 & $\tilde{\nu}$ \\ \hline
g & 1 & $\tilde{{\rm g}}$ & gluino & \half & $\tilde{{\rm g}}$ \\ \hline
$\gamma$ & 1 & $\tilde{\gamma}$ & photino & \half & \multirow{3}{2.1cm}{~~~$\chi^0_{1,2,3,4}$ neutralinos}\\
$\Z$ & 1 & $\tilde{\Z}$ & zino & \half &  \\ 
$h,H,A$ & 0 & $\tilde{H}^{0}_{1,2}$ & neutral Higgsinos & \half &  \\ \hline
$\W^{\pm}$ & 1 & $\tilde{\W}^{\pm}$ & wino & \half & $\chi^{\pm}_{1,2}$ \\ 
$H^{\pm}$ & 0 & $\tilde{H}^{\pm}$ & charged Higgsinos & \half & charginos \\ \hline  
G         & 2 & $\grav$       & gravitino         &
\ensuremath{\frac{3}{2}} &  $\grav$ \\ \hline 
\end{tabular}
\caption[The MSSM particle zoo]
{\label{tab:susypart}{\small The MSSM particle zoo.}}
\end{center}
\end{table}

Table~\ref{tab:susypart} lists the complete particle content of the MSSM. 
For each left- or right- handed fermion $f_{\rm L,R}$ there exist two different
sfermions $\tilde f_{\rm L,R}$ with spin-0 and weak isospin $\half$ and 0,
respectively. Of course the sfermions being scalars do not have a defined
helicity, but they do have different weak isospin depending on the helicity of
their SM partners. Once electroweak symmetry is broken these different weak
isospin states are allowed to mix.  
Thus for each flavour a $2\times 2$ mixing matrix exists which will give
the physical mass states from the weak eigenstates $\tilde f_{\rm L,R}$. It
takes the following general form~\cite{Ellis:2002mx}: 
\begin{equation}
\label{sfmix1}
\pmatrix{\tilde f_{\rm L} & \tilde f_{\rm R}}
\pmatrix{m^2_f+m^2_{\tilde f_{\rm L}}+m^2_D & a_fm_f \cr
         a_fm_f & m^2_f+m^2_{\tilde f_{\rm R}}+m^2_D }
\pmatrix{\tilde f_{\rm L} \cr \tilde f_{\rm R}}
\end{equation}
with 
\begin{equation}
\begin{array}{rl} 
\label{sfmix2}
m^2_D & \equiv \MZ^2\cos2\beta\left( I_3+Q\sin^2\theta_W \right) \\ 
a_f & \equiv \left( A_f - \mu^{\tan\beta}_{\cot\beta}\right)~~{\rm for}~~
f~=~^{e,\mu,\tau,d,s,b}_{u,c,t} 
\end{array}
\end{equation}
where $I_3$ is the third component of weak isospin, $\sin^2\theta_W=0.23$ and
$A_f$ are the trilinear Higgs-sfermion-sfermion couplings arising
from soft SUSY breaking. 
The off-diagonal terms depend on $m_f$, so it is clear that there will be more
mixing for $\tilde t_{\rm L,R}$, and if $\tanb$ is large also for $\tilde
b_{\rm L,R}$ and $\stau_{\rm L,R}$. 
The mass eigenstates can be obtained by diagonalisation. 
\begin{equation}
\label{sfdiag}
\pmatrix{\tilde f_1 \cr \tilde f_2} =
\pmatrix{ \cos\theta_{\tilde f} & \sin\theta_{\tilde f} \cr
	  \sin\theta_{\tilde f} & \cos\theta_{\tilde f} }
\pmatrix{\tilde f_{\rm L} \cr \tilde f_{\rm R}}
\end{equation}
where $\theta_{\tilde f}$ is the mixing angle. For example, in the case of
stau mixing, Eqs.~\ref{sfmix1} and~\ref{sfdiag} are then related by: 
\begin{equation}
\label{taumix}
\frac{m_{\tilde{\tau}_{\rm R}}^2-m_{\tilde{\tau}_{\rm L}}^2}{m_{\tau}}=
\frac{2(A_{\tau}-\mu\tanb)}{\tan2\theta_{\stau}}
\end{equation}
In general, the lighter mass eigenstate $\stauO$ will correspond to the
right handed $\stauR$ state
if there is no mixing ($\theta_{\stau}=0$
and $A_{\tau}=\mu\tanb$).
As the mixing increases (i.e. for increasing values of
the mixing angle) the lighter $\stauO$ becomes an admixture of right- and
left-handed staus, until $\theta_{\stau}=45\degs$ which corresponds to
maximal mixing. In this case the right- and left-handed components of the
lightest stau are equal. 
Mixing is therefore more relevant in the third family: the lightest stau
can be lighter than selectrons and smuons, 
and the lightest stop will generally be the lightest squark.
Mixing will also occur between the massive fermionic companions of the 
electroweak gauge bosons. The photino $\tilde{\gamma}$, zino $\tilde{\Z}$ 
and neutral higgsinos $\tilde{H}^0_{1,2}$ will mix to form the 
`neutralinos': $\neu^0_{1,2,3,4}$, where the lower the index the lighter the
particle. Similarly, the charged wino $\tilde{\W}^{\pm}$ and higgsino
$\tilde{H}^{\pm}$ give four charged mass eigenstates: the
`charginos' $\neu^{\pm}_{1,2}$. The corresponding mixing matrices for
neutralinos and charginos can be found in Ref.~\cite{Martin:1997ns}. 

\subsection{Super-Higgs mechanism and the gravitino}
Returning to the problem of SUSY breaking, it turns out to be impossible to
break SUSY in a phenomenological acceptable way if the only particles and
interactions are those of the MSSM. In the SM, the Higgs vacuum
expectation value determines the scale of electroweak breaking. But the
specific masses of the bosons and fermions are dictated by the coupling of
the forces that communicate the information of electroweak breaking: gauge
and Yukawa couplings, respectively. 
In the MSSM, to give masses to the gauginos, for example, soft terms in the
Lagrangian should contain tree level interactions of the type
scalar-gaugino-gaugino, but these are not allowed in
supersymmetry~\cite{Martin:1997ns}.  
Similarly, squarks would be unacceptably light if their mass was generated
at tree level as is the case in the SM~\cite{Ellis:2002mx}.

Thus the soft mass terms must arise indirectly or radiatively: 
from the coupling of new particles to the particles in the MSSM. 
Supersymmetry breaking must therefore occur in a `hidden' sector at a large
energy scale. There the spontaneous breaking of local supersymmetry occurs
and as a result some field F condensates, acquiring a non-zero expectation
value F$_0$.
Similarly to the Higgs mechanism, a Goldstone degree of freedom must now
appear, but in this case it is a fermion since SUSY is a fermionic
symmetry. This massless particle is the so-called `goldstino', and its two
polarisation states are `eaten' by the massless gravitino, giving it a total
of four polarisation states to become a massive spin-$\thrhlf$
particle. The end result is then a massive gravitino and a massless
graviton: local supersymmetry is thus successfully broken. This 
\emph{super-Higgs mechanism} is actually the only consistent way of
breaking SUSY, as the Higgs mechanism was the only consistent way of
breaking gauge symmetry~\cite{Ellis:2002mx}. 
The gravitino mass is then given by:
\begin{equation}
\label{mgrav}
m_{\grav} = \frac{{\rm F_0}}{\sqrt{3}\MP}
\end{equation}
where F$_0$ is the energy scale at which SUSY breaking occurs in the hidden
sector (like $v^2$ in the Higgs mechanism) and $\MP$ is the reduced Planck
mass: $\MP=(8\pi \GN)^{-1/2}=2.4\times 10^{18}\gevcc$. By dimensional
analysis, it is intuitively clear that the gravitino should be massless in
unbroken SUSY (F$_0 \to 0$) or if gravity becomes irrelevant
($\MP\to\infty$ or $\GN\to 0$). 

\begin{figure}[tb]
\begin{center}
\includegraphics[width=0.66\linewidth]{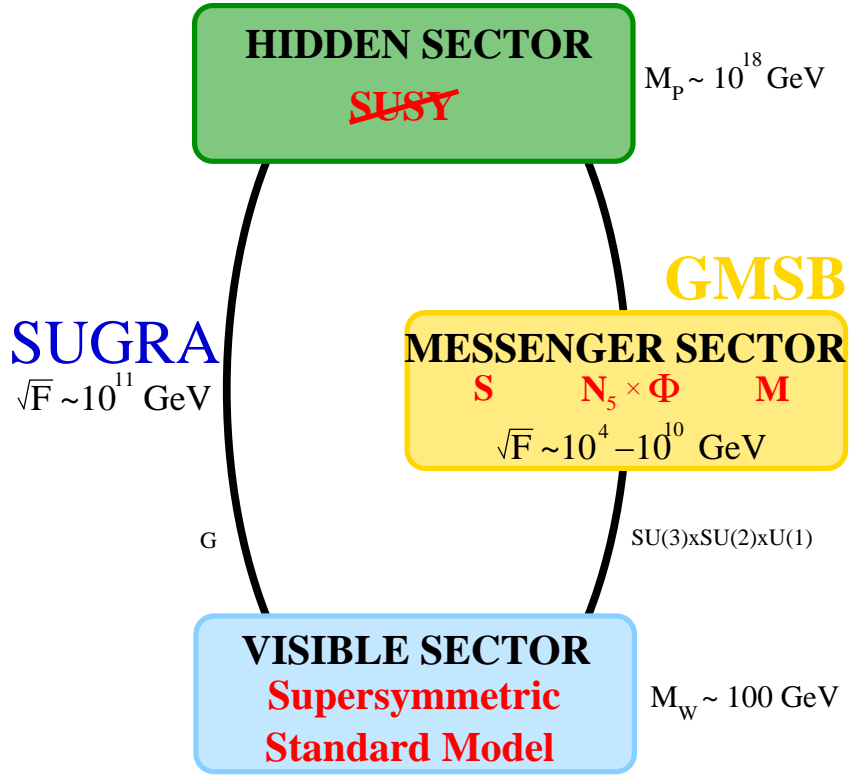}
\caption[Diagram of SUSY mediation and energy scales]
{\label{fig:susymed}{\small Supersymmetry breaking and its
mediation. SUSY is broken (spontaneously) in a high-energy `hidden' sector
and its effects are transmitted radiatively to the `visible' sector where
all MSSM particles reside. The mediation can occur via gravity or gauge
interactions.}}
\end{center}
\end{figure}

Now MSSM particles still have to be informed of the SUSY breaking. This can be
achieved in different ways. Since all massive particles feel gravity, a
natural candidate to communicate the hidden sector with the `visible'
sector where all MSSM particles sit, is the gravitational force. 
Thus \emph{gravity mediated SUSY breaking} or supergravity (SUGRA) models,
postulate the coupling between Planck-scale physics and the MSSM
particles via gravity (exchanging gravitons). These weak couplings (of the
order $1/\MP$) generate acceptable soft mass terms and a definite MSSM
spectrum. Another possibility is that the SUSY breaking is transmitted to
the MSSM particles by the usual gauge $\suc\otimes\sul\otimes\uy$
interactions. This case is referred to as \emph{gauge mediated SUSY
breaking} (GMSB) models and they imply the existence of an intermediate
energy sector of particles, called `messengers' $\Phi$ which will directly feel
the SUSY breaking and then transmit it to the MSSM particles by radiative
corrections, giving rise to a given MSSM mass spectrum. This last type
of mediation is the framework of this thesis. A schematic drawing of the
structure of SUSY breaking and mediation for these two scenarios is
presented in Fig.~\ref{fig:susymed}. Both alternatives are discussed in the next
sections.

A very nice feature of the super-Higgs mechanism is that it naturally induces
electroweak symmetry breaking. If one assumes, as the unification of
couplings suggests, that all fermion and boson masses unify at the
GUT-scale with values $m_0$ and $m_{1/2}$ respectively, one can calculate all
MSSM masses in terms of these GUT-scale input values. It turns out that $m_0$
evolves with energy in such a way that drives the squared mass of 
the up-type Higgs negative due to the large (negative) radiative
corrections from the top-quark Yukawa coupling and the stop mass. 
Indeed, electroweak symmetry breaking is generated radiatively
and a link between the SUSY breaking Higgs masses and the electroweak scale
$\MZ$ can be obtained~\cite{Giudice:1998bp}:
\begin{equation}
\label{muEWB}
\mu^2 = -\frac{\MZ}{2} + \frac{m_{H_1}^2-m_{H_2}^2 \tan^2\beta}
{\tan^2\beta-1}
\end{equation}
This again expresses the hierarchy problem in a different format: the
$\mu$ parameter on the left can be as big as the highest energy in the
theory (it does not break any SM symmetry), but both $\MZ$ and the Higgs
masses on the right have to be of the order the electroweak scale. What
makes $\mu$ then lie in a physically acceptable region?

\section{Gravity mediated SUSY breaking}
If gravitational interactions are responsible for the soft mass terms,
Planck-scale physics is the messenger of SUSY breaking~\cite{Nilles:1984ge}. 
A simplifying assumption, although not inevitable, is to consider the
\emph{universality} of masses and couplings. First, since the gauge
couplings seem to unify at $\QU=10^{16}\gev$ (see Fig.~\ref{fig:running}),
the gaugino masses $M_1$, $M_2$ and $M_3$ for the photino, wino and gluinos
respectively are also assumed to unify with a
common value of: $m_{1/2} \equiv M_i \sim m_{\grav}$~\cite{Ellis:2002mx}.
Secondly, the scalar masses-squared and the trilinear couplings are taken
to be flavour-diagonal and universal at $m_0^2 \sim m_{\grav}^2$ and
$A_f=A_0Y_f\sim m_{\grav}$, where $Y_f$ are the Higgs-fermion Yukawa
coupling matrices. All the above relations are evaluated at the unification
scale $\QU$ or $\MP$, and all of them are of the order F$_0/\MP$ set by
gravitational interactions. 
To calculate the mass of each particle in the visible sector, one runs the
energy down from the known energy dependence of these
parameters~\cite{Martin:1997ns}. 
This set of assumptions lead to a very predictive model, called minimal
supergravity (or mSUGRA), where only five parameters are needed to fully
calculate the mass spectrum in the MSSM:
\begin{equation}
m_{1/2},~m_0,~A_0,~\tanb,~{\rm sign(\mu)}
\end{equation}
where the absolute value of $\mu$ can be calculated with Eq.~\ref{muEWB}
and only its sign remains a free parameter. 

Since the soft masses have to be of the order of $100\gevcc$, and these are
determined by parameters proportional to F$_0/\MP$,
the scale of SUSY breaking $\rootFo$ in these type of models is predicted
to be of the order of $10^{11}\gev$. In this type of models the gravitino
is then expected to be heavy, comparable to the MSSM particles with
$m_{\grav}\sim 100\gevcc$ or so. 

It is then found that the gaugino masses at any scale satisfy the
relation~\cite{Martin:1997ns}:
\begin{equation}
\label{unirel}
M_3=\frac{\alpha_3}{\alpha}\sin^2\theta_W M_2 = 
\frac{3\alpha_3}{5\alpha}\cos^2\theta_W M_1
\end{equation} 
which at the electroweak scale ($\alpha_3=0.118$ and $\alpha=1/128$)
becomes:  $M_3:M_2:M_1 \simeq 7:2:1$.
It can also be derived that in mSUGRA the neutralino $\neu_1^0$ is
usually the lightest supersymmetric particle (LSP)~\cite{Martin:1997ns}. 
If R-parity is conserved, signatures for SUSY in colliders will consist of
missing energy  carried away by the heavy, stable and very weakly
interacting neutralino. 

\section{Gauge mediated SUSY breaking}
In GMSB models~\cite{Giudice:1998bp} 
the splitting of masses in the MSSM is generated not at the
Planck scale, but at some lower energy scale where pairs of very heavy
messenger quarks and leptons exist. 
The gravitational interaction is still responsible for
the communication between the hidden sector and this messenger sector,
but its effects are now negligible in the MSSM. 
Gravity will couple to a singlet S in the
messenger sector which will acquire a vacuum expectation value $\rootF$.
This singlet breaks supersymmetry in the messenger sector and gives mass to
all other messenger superfields $\Phi$ there. In turn, these messengers are
able to communicate with the MSSM via gauge interactions, but now from a
much lower scale than $\MP$. 

The key element in these models is then the structure of the messenger
sector. The heavy messengers can be generally described as supermultiplets
of $SU(5)\supset SU(3)\otimes SU(2)\otimes U(1)$, for which the parameter
$\N$ is introduced and represents the number of families of $\Phi$. 
To a good approximation, once SUSY is broken in the messenger sector, 
the fields $\Phi$ will share a common mass $\Mmess$.  

\begin{figure}[h]
\begin{center}
\renewcommand{\subfigtopskip}{-20pt}
\subfigure[Gauginos]{
\includegraphics[width=0.3\linewidth]{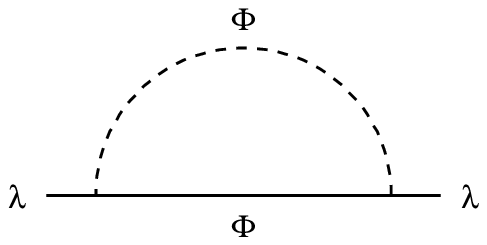}}
\subfigure[Scalars]{
\includegraphics[width=0.6\linewidth]{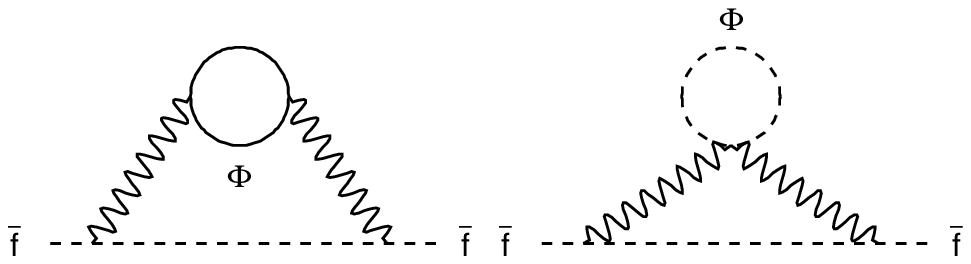}}
\caption[Coupling of messenger particles to gauginos and scalars through
radiative loops]
{\label{fig:gsmasses}{\small In GMSB, the (fermionic and bosonic)
messengers $\Phi$ give supersymmetry breaking masses to (a) gauginos 
$\lambda$ at one loop and to (b) sfermions $\tilde f$ at two loops. The
coupling is via gauge interactions.}} 
\end{center}
\end{figure}

The MSSM masses are then generated at one loop in the case of gauginos
$\lambda_i$ and two loops in the case of scalars $\tilde f$, taking into account that couplings
are in both cases given by gauge $\alpha_i$ strengths. Examples of the
Feynman diagrams contributing to the MSSM masses are shown
in Fig.~\ref{fig:gsmasses}. 

The masses can then be calculated at the messenger mass scale $\Mmess$,
as~\cite{Giudice:1998bp}:
\begin{eqnarray}
\label{gauginomass}
m_{\lambda_i} &=& \frac{\alpha_i}{4\pi}\:\Lambda\N \\
\label{scalarmass}
m^2_{\tilde f} &\simeq& \left(\frac{\alpha_i}{4\pi}\right)^2\:\Lambda^2\N
\end{eqnarray}
where the parameter $\Lambda\equiv\Fm/\Mmess$ represents the universal mass
scale of SUSY particles (like $m_0$ or $m_{1/2}$ in SUGRA).
The most noticeable consequence is the degeneracy between squark and
slepton masses. In SUGRA models the different flavour masses arise directly
from a high energy $\MP$ where flavour symmetry is respected. There
is no control then on the mass difference between squarks and sleptons,
which could be in principle very large. These possible flavour-breaking
soft mass terms are very dangerous, because they lead to flavour changing
neutral currents (FCNC) and large CP violation. Both of which are strongly
constrained by observations~\cite{pdg}. 
Measurements of the $\bar{K}^0$-$K^0$ mass difference
and $\mu\to{\rm e}\gamma$ decays set stringent bounds on the
possible mass difference between squarks and sleptons~\cite{Giudice:1998bp}. 
In GMSB models instead, sfermion masses arise at a lower energy $\Lambda$,
much below the scale at which flavour symmetry is expected to be
valid. Thus the SM Yukawa couplings are already present when the mass
generation takes place, and FCNC are then automatically suppressed. 
Also, the trilinear couplings $A_f$ are negligible at the $\Mmess$ scale,
since they are generated at two loops and are further reduced by a factor
$\alpha_i/4\pi$. They will be in any case very small at the electroweak
scale and are not considered as a relevant parameter. 

Another important characteristic of these models is the gravitino
mass. Since soft masses are proportional to the coupling constants, the LSP
will be the gravitino, the least interacting sparticle. 
By reducing the energy scale at which soft mass terms are generated 
from $\MP$ to $\Mmess$, the `effective' SUSY breaking scale 
(in the messenger sector) $\rootF$ is now much lower. 
The messenger mass could be $\sim 100\tevcc$, 
thus to produce MSSM particles with the right mass 
($m\sim\Fm/\Mmess \sim 100\gevcc$), $\rootF$ can be as low as
$10^{4}\gevcc$. 
From this `effective' SUSY breaking scale the gravitino mass is derived now
as~\cite{Giudice:1998bp}:
\begin{equation}
\label{mgravfm}
m_{\grav} = \frac{\Fm}{k\sqrt{3}\MP} = 
\frac{2.4}{k}\left( \frac{\rootF}{100\tev}\right)^2\evcc
\end{equation}
where $\Fm=k{\rm F_0}$ and $k$ is a model dependent parameter describing
how SUSY breaking is communicated from the hidden sector to the messenger
sector. 

If the gravitino is the LSP all MSSM particles will eventually decay into
it, and since it couples gravitationally one would naively expect those
decays to be extremely slow. But the gravitino contains two longitudinal 
components from the goldstino it absorbed by means of the super-Higgs
mechanism.  And these components with helicity $\pm\half$ and gauge
couplings may lead to decay rates high enough to be of experimental
importance. 
 
Finally, to completely specify the MSSM spectrum and its phenomenology, 
it is only necessary to use six parameters:
$\Mmess,~m_{\grav},~\Lambda,~\N,~\tanb$ and sign($\mu$).
Hence, GMSB models are highly predictive as will be discussed in
Chapter~\ref{scan}. 

\section{Collider signatures}
\label{coll.sign}
\begin{figure}[ptb]
\begin{center}
\renewcommand{\subfigtopskip}{0pt}
\renewcommand{\subfigcapskip}{-10pt}
\subfigure[Sfermion production]{
\includegraphics[width=0.99\linewidth]{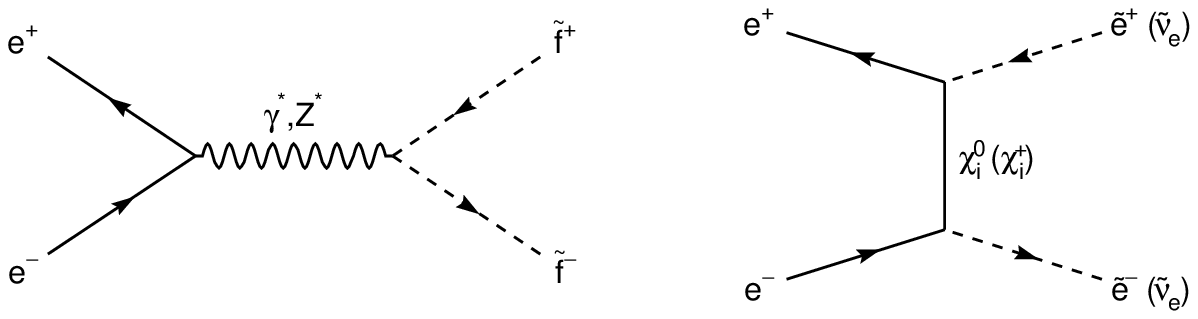}}
\subfigure[Neutralino and chargino production]{
\includegraphics[width=0.99\linewidth]{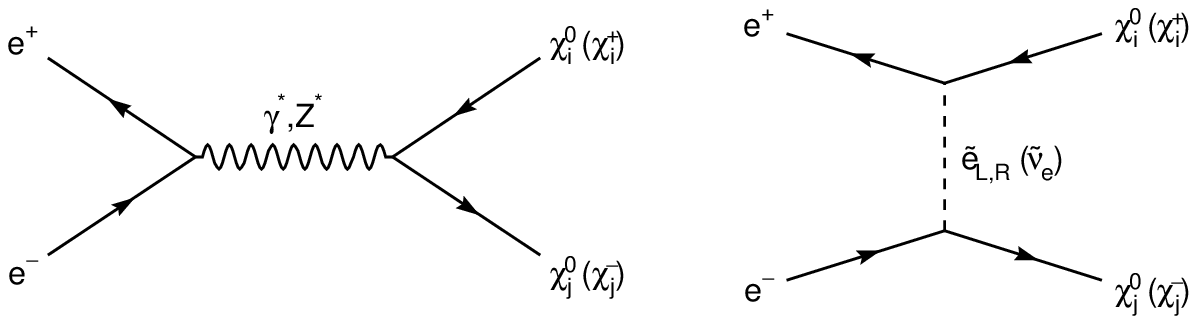}}
\subfigure[Single gravitino production]{
\includegraphics[width=0.99\linewidth]{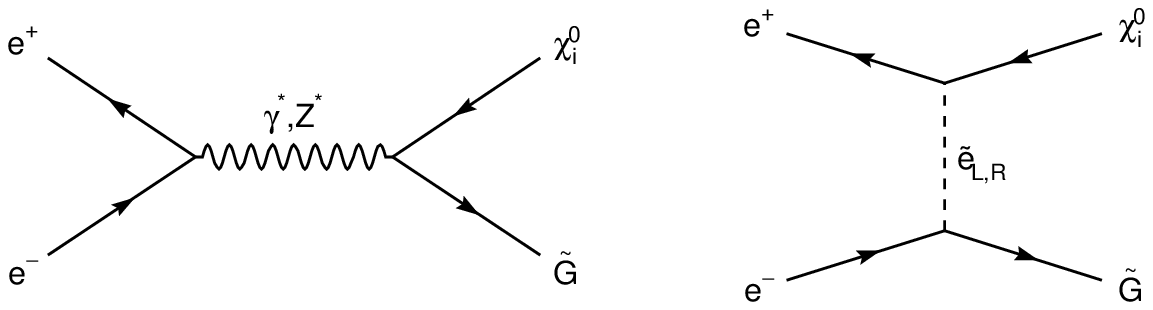}}
\subfigure[Higgsstrahlung and associated production]{
\includegraphics[width=0.99\linewidth]{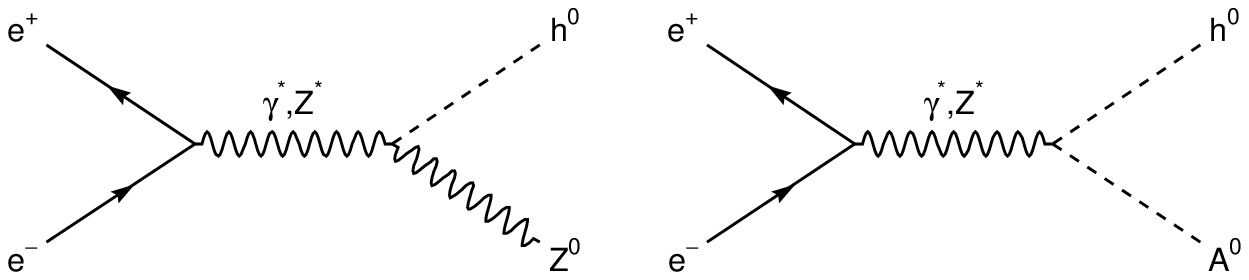}}
\caption[Production of SUSY particles and
MSSM Higgs neutral bosons at LEP when R-parity is conserved]
{{\label{fig:lep2prod}}{\small Production of SUSY particles and
MSSM Higgs neutral bosons at LEP when R-parity is conserved.}}
\end{center}
\end{figure}

The best probe for supersymmetry is naturally direct production of
superpartners at high energy colliders. Figure~\ref{fig:lep2prod} shows the
Feynman diagrams for the production of all relevant SUSY particles at
LEP, where electrons and positrons collide head-on. 

\begin{figure}[tb]
\begin{center}
\renewcommand{\subfigtopskip}{-10pt}
\subfigure[Contours of the NSLP decay length (Eq.~\ref{nlsplifetime} 
with $\kappa_{\gamma}=1$) as a function of the NLSP mass and the gravitino
mass, for pair-produced particles at $E=104\gev$.]{
\includegraphics[width=0.48\linewidth]{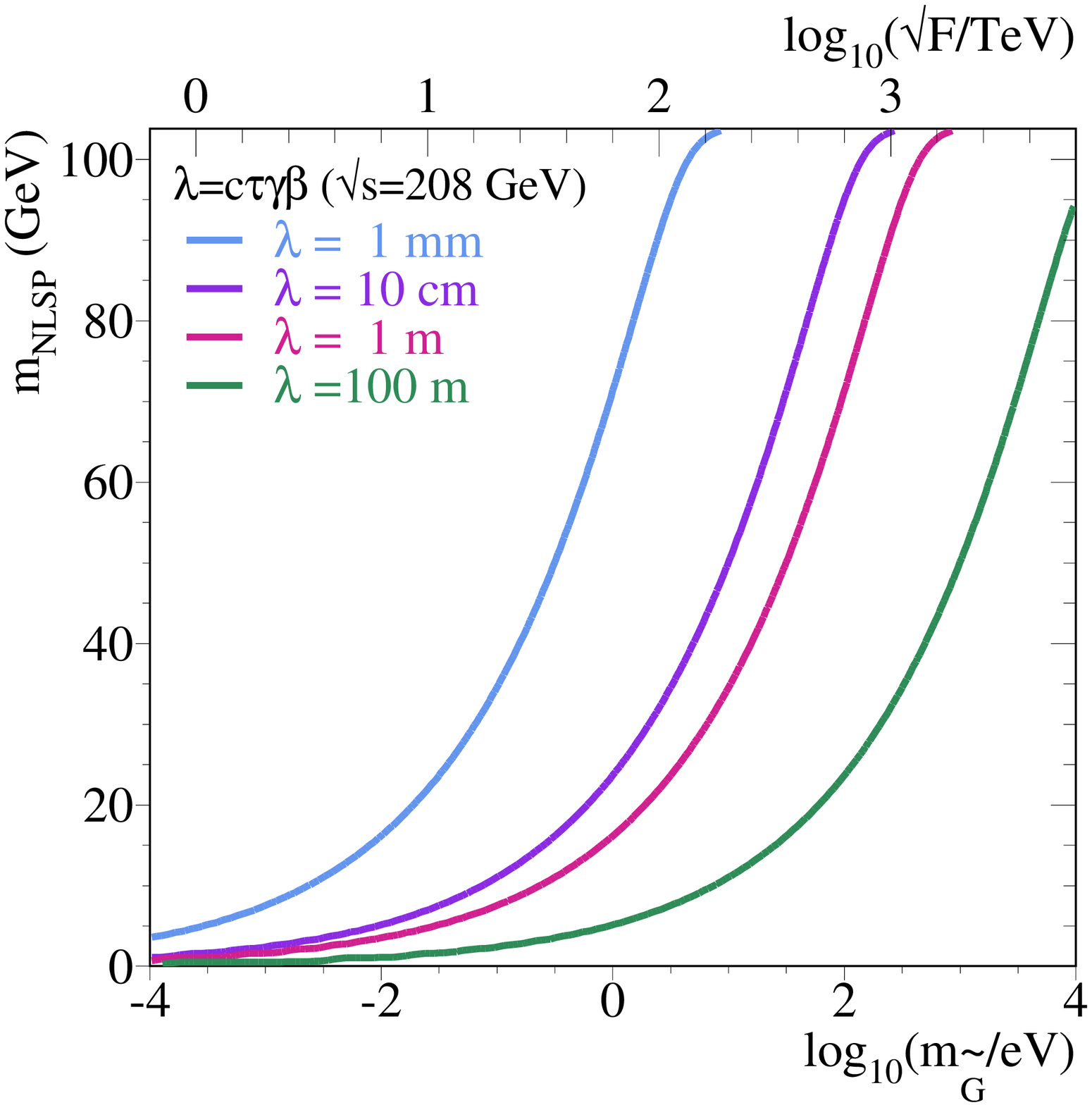}}
\subfigure[Domains of stau NLSP and lightest neutralino NLSP as a function of the
number of messenger families and the messenger mass. The solid lines
represent different values of $\tanb=3,10,30$. From Ref.~\cite{Feng:1998zr}]{
\includegraphics[width=0.49\linewidth]{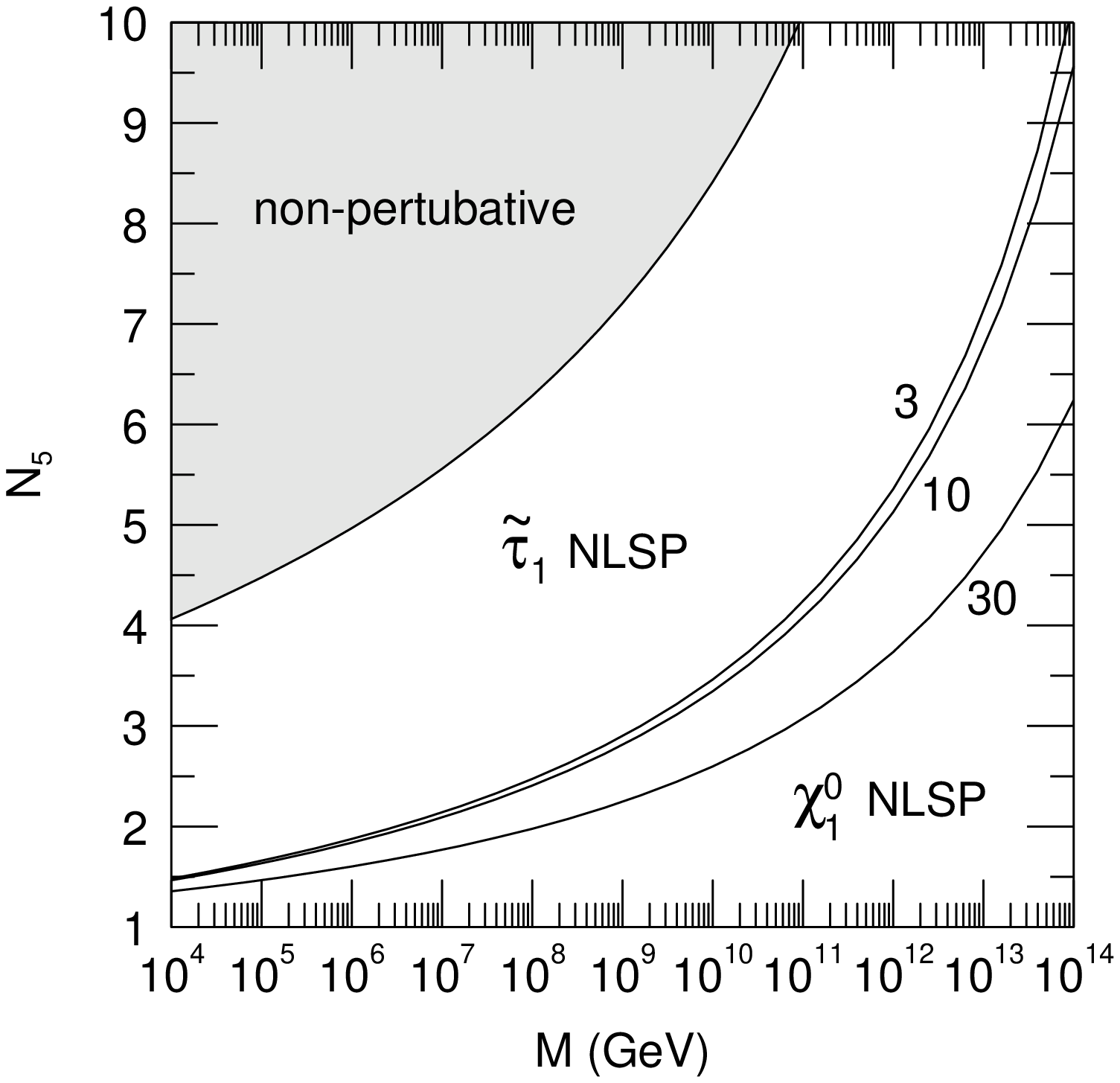}}
\caption[The NLSP nature as a function of parameters]
{\label{fig:NLSPnature}{\small The NLSP nature as a function of parameters.}}
\end{center}
\end{figure}

In Gauge Mediated SUSY Breaking
models, experimental signatures differ significantly from models with a
neutralino LSP. If R-parity is conserved, as is assumed throughout this
work, all SUSY decay reactions will terminate in the next-to-lightest
supersymmetric particle (or NLSP) decaying to its SM partner and the
gravitino LSP.  The identity of the NLSP, therefore, is central to the
phenomenology. 
Furthermore, the NLSP may be long lived and thus very
striking signatures are possible. The NLSP decay length is controlled by 
the SUSY breaking scale or equivalently (Eq.~\ref{mgrav}) by the 
gravitino mass:
\begin{equation}
\label{nlsplifetime}
\lambda_{\rm NLSP} = c\tau_{\rm NLSP}\gamma\beta = 
\frac{0.01}{\kappa_{\gamma}}\left( \frac{100\gevcc}{m_{\rm NLSP}}\right)^{-5}
\left( \frac{m_{\grav}}{2.4\evcc} \right)^2
\sqrt{\frac{E^2}{m_{\rm NLSP}^2}-1}\cm
\end{equation}
where $\kappa_{\gamma}$ is the photino component of the neutralino NLSP, or
one in all other cases (see Fig.~\ref{fig:NLSPnature}a).
If $m_{\grav}$ is large enough and the NLSP decay
length is measured as a vertex displacement, it provides a unique method to
probe the value of $\rootFo$, the scale of SUSY breaking, as opposed to
other measurements which in general are only sensitive to $\Lambda$.
Searches for NLSP's decaying in the middle of the detector also benefit
from lower background rates, hence the interest in studying these signatures.

As regards the NLSP type, depending on the GMSB parameters it will
generally be either the neutralino or the lightest stau (see
Fig.~\ref{fig:NLSPnature}b.
This can be deduced from Eqs.~\ref{gauginomass} and~\ref{scalarmass} where 
sparticles with only $\uy$ interactions will be lighter than the rest since 
 $\alpha_1<\alpha_2\ll\alpha_3$. The tiny selectron and smuon mixing
factors (Eq.~\ref{sfmix2}) allows one to generally treat them as degenerate, 
unmixed mass eigenstates, thus $m_{\sel}=m_{\smu}$ will be assumed for the
rest of this work.   
In some cases, specially for low $\tanb$, 
the three  right handed sleptons are degenerate in mass and act as
co-NLSP. Also possible in
a much more restricted corner of parameter space is the neutralino-stau
co-NLSP where the mass difference between the neutralino and the sleptons is
less than the tau mass. Table~\ref{tab:nlspcond} lists the conditions
in the mass hierarchy for each type of NLSP scenario and its
decay~\cite{Ambrosanio:1997rv}. A neutralino NLSP will predominantly decay
to $\gamma\grav$ and not Z$\grav$ for $m_{\neu} \lesssim
100\gevcc$~\cite{Dimopoulos:1997yq}. From now on, references to the slepton
NLSP case or scenario, should be understood as corresponding to the stau
NLSP region, including also the more restricted case of sleptons as co-NLSP. 
\begin{table}[tb]
\begin{center}
\vspace{-0.6cm}
\begin{tabular}{|rll|}\hline
Case & Condition & Final decay \\ \hline \hline
neutralino NLSP & $m_{\neu}<m_{\stauO}-m_{\tau}$ & $\neu\to\gamma\grav$ \\ 
stau NLSP       & $m_{\stauO}<\min(m_{\neu},m_{\slR})-m_{\tau}$ & 
                  $\stauO\to\tau\grav$ \\
slepton co-NLSP & $m_{\slR}<\min(m_{\neu},m_{\stauO}+m_{\tau})$ & 
                  $\slR\to l\grav ; \stauO\to\tau\grav$ \\
neutralino-stau co-NLSP & $|m_{\stauO}-m_{\neu}|<m_{\tau} ; m_{\neu}<m_{\slR}$ 
                        & $\neu\to\gamma\grav ; \stauO\to\tau\grav$ \\
\hline
\end{tabular}
\caption[The four possible NLSP scenarios]
{\label{tab:nlspcond}{\small The four possible NLSP scenarios,
with the corresponding NLSP decay, neglecting the electron and muon masses.}}
\end{center}
\end{table}

\begin{figure}[h]
\begin{center}
\includegraphics[width=0.6\linewidth]{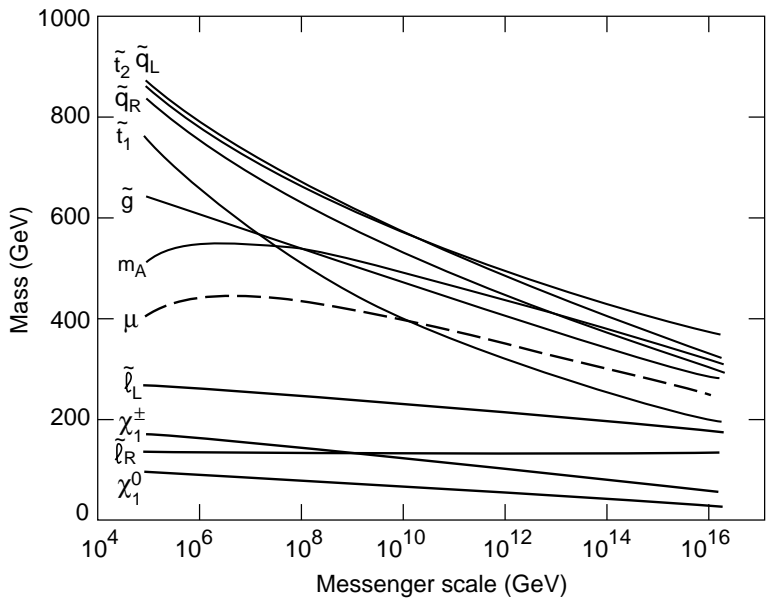}
\caption[Evolution of sparticle masses with $\Mmess$]
{\label{fig:mass_spec}{\small Sparticle masses dependence with $\Mmess$ in
a GMSB model with $\N=1$, $\tanb=3$ and a messenger scale bino mass of
$115\gevcc$. From Ref.~\cite{Dimopoulos:1997yq}.}}
\end{center}
\end{figure}

The superpartner mass spectrum in GMSB models will ultimately determine which
SUSY particles can be produced at LEP and their subsequent decay chains.  
The strongly interacting states, including the generally lighter stops, are
usually too heavy to be produced at LEP. Only the lightest neutralino
$\neu_1^0$, slepton $\slep_{\rm R}^{\pm}$ or Higgs $h^0$ will in general be
accessible at LEP2. The charginos, heavier neutralinos and left-handed
sleptons are too heavy to be relevant to discovery at these 
energies\footnote{
Even if no other SUSY particle is accessible at LEP, direct production of
gravitinos is still possible. The cross section for
$\ee\to\grav\grav\gamma$ scales with 1/$m_{\grav}^4$, so only ultra light
gravitinos could be observed through direct production. ALEPH has searched
for this process and derives an upper limit on the gravitino
mass of $1.3 \times 10^{-5}\evcc$~\cite{Heister:2002ut}.
}. A typical mass spectrum of GMSB models is shown in
Figure~\ref{fig:mass_spec}. 

In the following, the lightest neutralino $\chi_1^0$ will be abbreviated to
$\chi$ and referred to as `the' neutralino, when no other heavier
neutralino or chargino is present. The symbol $\ell$ will stand for the
three charged leptons, e, $\mu$ or $\tau$; and $l$ will only refer to the
electron or muon. 
\thispagestyle{empty}
\chapter{Experimental Overview}
\label{detector}
\begin{center}
\begin{spacing}{1.5}
\setlength{\fboxsep}{5mm}
\begin{boxedminipage}[tb]{0.9\linewidth} \small
The Large Electron Positron collider (LEP) started operation in 1989 and
ended in the fall of 2000. 
The largest accelerator of its kind in the world, it was built to study 
in detail the Z and W massive vector bosons. From 1989 to 1995 it ran with
a centre-of-mass energy ($\roots$) close to the Z mass, scanning the
electroweak sector and the properties of the Z particle. This phase is
known as LEP1. From 1996 onwards, LEP increased in 
energy through the 161$\gev$ threshold, for the W boson pair-production,
and reached 209$\gev$ in 2000. This phase, referred to as LEP2, continued the
precision measurements, enlarging our understanding of the Standard Model, 
probing the theory to a high level of accuracy.

The LEP programme has thus set a milestone in the experimental
determination of the parameters of the SM and has confirmed the theoretical
predictions with extremely high precision.
Once the detectors and the accelerator have been dismantled, the LEP tunnel
will house the Large Hadron Collider (LHC), a proton-proton accelerator
with $\roots = 14\tev$, which is expected to begin operation in 2007. 
This chapter describes briefly the LEP collider and, in more detail, the 
ALEPH detector and its performance.
\end{boxedminipage}
\end{spacing}
\end{center}
\newpage
\section{The LEP collider}
\begin{figure}[tb]
\begin{center}
\includegraphics[width=9.5cm,angle=-90]{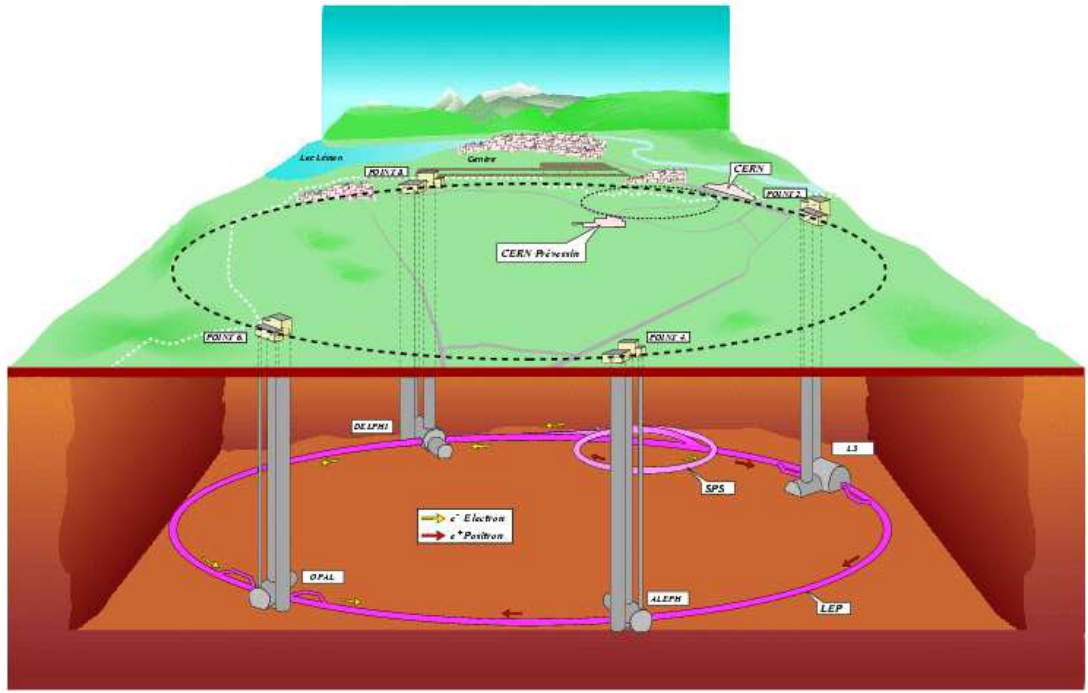}
\caption[An overview of the LEP
accelerator]
{\label{fig:lepview}{\small An overview of the LEP
accelerator. From Ref.~\cite{cernphoto}.}}
\end{center}
\end{figure}

The Large Electron Positron collider (LEP) lay between 100 and 150$\m$
underground on the Swiss-French border, near Geneva. It was a 26.7\,km long 
storage ring designed to accelerate electrons and positrons and produce
collisions at four points where general-purpose detectors were placed to
observe and record the resulting new particles. These detectors were ALEPH,
DELPHI, OPAL and L3.
Figure~\ref{fig:lepview} shows the LEP accelerator complex along with the
detectors. For a detailed description of LEP see
Refs.~\cite{unknown:1984mj,lep2phys}, here only an overview is given.

Beams of counter-rotating electrons and positrons were injected into LEP 
at an energy of 22$\gev$ from the Super Proton Synchrotron (SPS). 
Once the particles were in the highly evacuated beam-pipe, they were 
accelerated by superconducting NbTi radio frequency (RF) cavities and
normal Cu cavities operating at a frequency of 352\,MHz.
To maintain a circular orbit, the beams were bent by 3368 dipole magnets and
focused by 816 quadrupole and sextupole magnets.
When particles follow a circular trajectory,  they loose some of their
energy by emitting synchrotron radiation. Since this effect is inversely 
proportional to the square of the accelerator radius, it is necessary to 
build very large accelerators. Hence the large radius of LEP helped to
minimise the amount of energy that cavities had to give back to the
particles to attain the nominal $\roots$.

The normal mode of operation consisted of four equally spaced \emph{bunches}
of electrons and positrons which crossed each other every 89$\us$ (the
collision frequency was thus $f \sim 11$\,kHz).
With approximately 10$^{11}$ particles per bunch, typical total
beam current values were 6\,mA, or 750\,$\mu$A per bunch. 
At the interaction points (IPs), where
collisions are produced and the dimensions of the bunches have to be as
small as possible to enhance the collision rate, 
each bunch was approximately 1$\cm$ long, 200$\um$ wide horizontally
($\delta_x$) and 8$\um$ vertically ($\delta_y$). 
This yielded a typical LEP2 \emph{instantaneous luminosity} of 
10$^{31}\cm^{-2}\s^{-1}$. The instantaneous luminosity parameter 
contains all the accelerator capabilites and is proportional to the  
rate of interactions (collisions). 
It is defined as:
\begin{equation}
\Lum = \frac{N_{\rm{bunch}}N_{\pos}N_{\ele}f}{4\pi\delta_x\delta_y}
\end{equation}
where $N_{\rm{bunch}}$ is the number of bunches and $N_{\mathrm{e}^{\pm}}$ 
is the number of electrons/positrons per bunch. The luminosity was measured
independently by LEP and the four experiments with dedicated subdetectors. 

\section{The ALEPH detector}
\label{aleph}
The ALEPH (Apparatus for LEp PHysics) detector was designed to cover as 
much solid angle as possible around the point where the beams were made 
to collide, 
with high granularity and hermeticity. This is to ensure
that all particles emerging from the $\ee$ collisions are measured and
identified. In practice, a coverage of $\sim$$3.9$\,srad was achieved, allowing
the detailed study of Standard Model physics at LEP and searches for new 
physics. Its shape was cylindrical, built around the
interaction point, with 12$\m$ in diameter and 12$\m$ in length, weighing
over 3000\,tons and having some 700\,000 readout channels. 

Particles originating at the IP and traveling outwards 
traversed through several subdetectors arranged in cylindrical 
layers. They first passed through a series of three low-density tracking
devices, designed to measure the trajectory of charged particles, and then
encountered high-density calorimeters where all but muons and neutrinos
were completely stopped depositing all their energy. 
Only muons could penetrate to the last subdetector, the muon chambers, 
being therefore tagged as such. 
The tracking subdetectors were immersed in a highly uniform 1.5\,T magnetic
field created by a 6.4$\m$ long and 5.3$\m$ diameter superconducting
solenoid. The trajectories of charged particles are curved in the
strong magnetic field, describing a helix which spirals around the beam
axis. This curvature provides a measurement of their momentum. 

\begin{figure}[tb]
\begin{center}
\vspace{-1cm}
\includegraphics[width=0.84\linewidth]{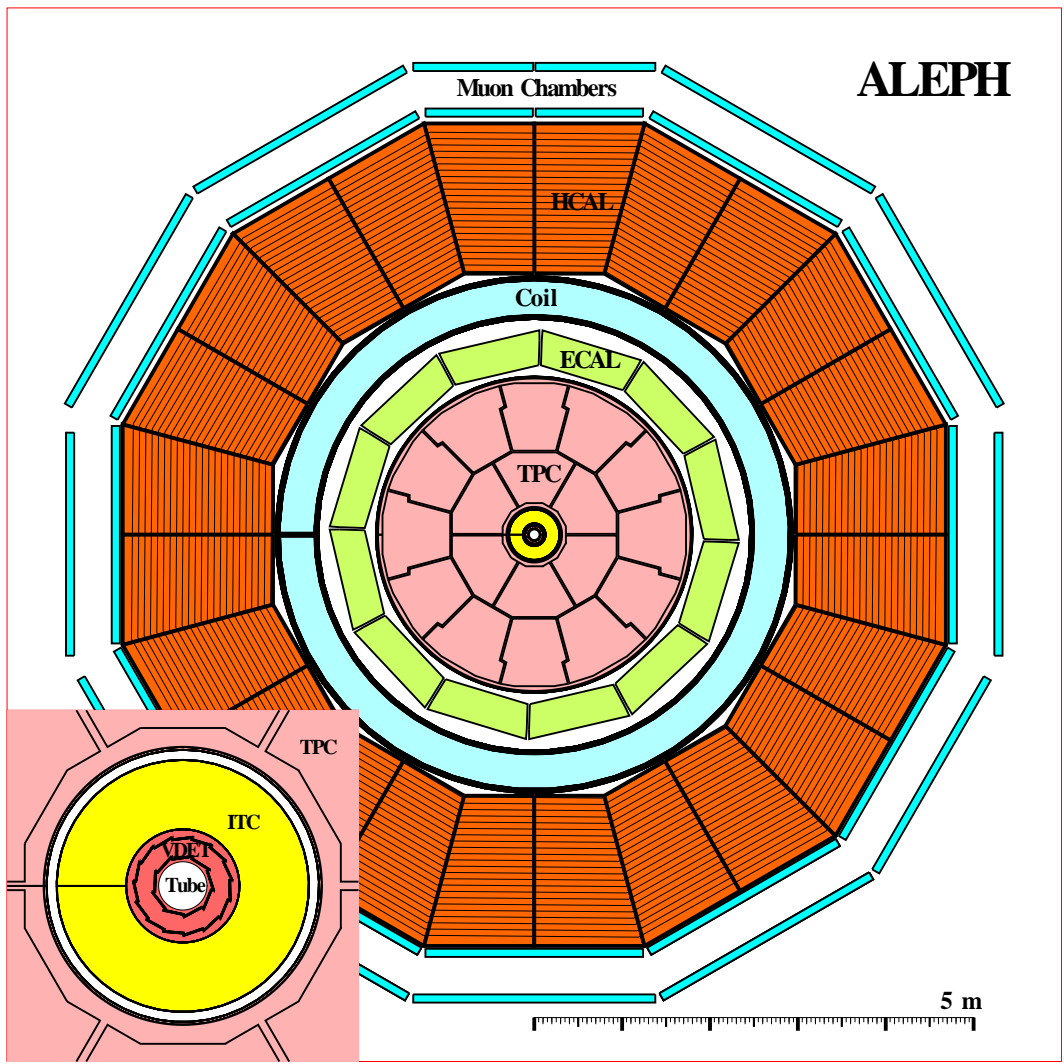}
\caption[End-on view of the ALEPH detector showing
all its subdetectors]
{\label{fig:alephdet}{\small End-on view of the ALEPH detector showing
all its subdetectors~\cite{alephwwwed}.}}
\end{center}
\end{figure}

Figure~\ref{fig:alephdet} shows all the subdetectors forming the ALEPH
detector.
Moving radially away from the beam-pipe, there is the Microvertex Detector
(VDET), the Inner Tracking Chamber (ITC) and the Time Projection Chamber
forming the tracking volume of the detector. They are followed by the
Electromagnetic Calorimeter (ECAL) and then the magnet so as not to degrade
the ECAL energy resolution. The iron return yoke for the magnet was
segmented into layers and instrumented with streamer tubes acting as a
Hadron Calorimeter (HCAL). Finally, surrounding the HCAL, two additional
double layers of streamer tubes formed the muon chambers. 
Another three pairs of detectors LCAL, SiCAL and BCAL provided low angle
coverage and measured Bhabha scattered electrons and positrons.
They were located very close to the beam-pipe, BCAL giving measurements of 
instantaneous luminosity from outside the detector, while LCAL and SiCAL 
measured integrated luminosity. 
The ALEPH detector is described in Refs.~\cite{Decamp:1990jr,alephhb1,alephhb2} 
and its performance is described in Ref.~\cite{Buskulic:1995wz}. 

\begin{figure}[tb]
\begin{center}
\vspace{-0.5cm}
\includegraphics[width=0.6\linewidth]{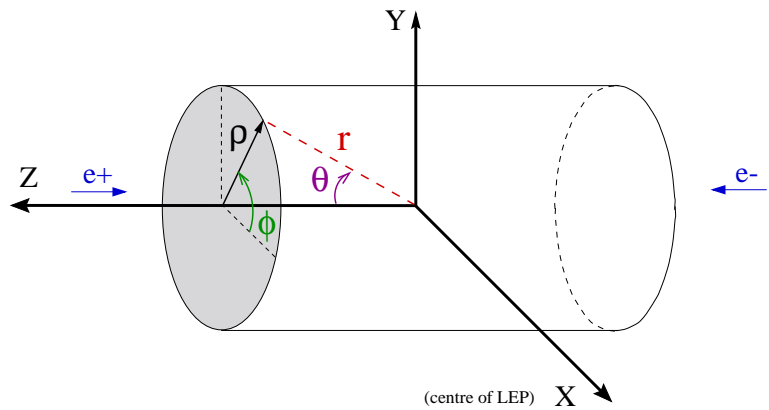}
\caption[ALEPH coordinate system]
{\label{fig:coord}{\small The ALEPH coordinate system.}}
\end{center}
\end{figure}

The coordinate system is defined as follows: the origin is the
geometrical centre of the detector, also the nominal IP; the $x$-axis is
horizontal and points to the centre of the LEP ring; the $z$-axis is along the
electron beam direction and makes an angle of 3.59\,mrad upwards with the local
horizontal; the $y$-axis then makes an angle of 3.59\,mrad with the local
vertical. This small displacement with respect to the local vertical is due
to the fact that the LEP tunnel was tilted by 1.4\% for geological reasons. 
Given the detector and event geometry, it is often more useful to use 
spherical (r,$\theta$,$\phi$) or cylindrical coordinates ($\rho$,$\phi$,z),
defined as:
\begin{eqnarray}
\nonumber x =& r\sin\theta\cos\phi &= \rho\cos\phi \\
          y =& r\sin\theta\sin\phi &= \rho\sin\phi \\
\nonumber z =& r\cos\theta &= z      
\end{eqnarray}
where the polar angle $\theta$ is measured with respect to the
$+z$-axis and extends between 0 and $\pi$.
The azimuthal angle $\phi$ is measured in the ($x,y$) plane
starting at the $x$-axis.
The coordinate system is depicted in Fig.~\ref{fig:coord}.

\section{Charged Particle Tracking}
\subsection{Vertex Detector}
\label{vdet}
The Vertex Detector (VDET)~\cite{Creanza:1998wg} 
was the innermost subdetector of ALEPH. It was very close
to the IP to provide highly accurate measurements of charged particles'
trajectories. It specially served to identify very short-lived particles,
such as $B$-mesons, made of $b$-quarks, which travel a very small distance
before decaying (typically $\sim$$400\um$). Its principal aim was to determine
whether a particle came from the IP or rather originated from a displaced
vertex. It was upgraded in 1995 to prepare for LEP2. With
respect to the original VDET, it had improved radiation hardness,
covered a wider angle and introduced less passive material. 
The main physics purpose of the upgrade was to improve the efficiency of
$b$-tagging. At LEP2 energies, the
Higgs boson is expected to decay mainly to $b$-quarks (BR($H\to b\bar{b}$)
$\gt$ 70\%), hence $b$-tagging is crucial in Higgs boson searches. 

\begin{figure}[h]
\begin{center}
\vspace{-0.5cm}
\subfigure{\includegraphics[width=0.35\textwidth,angle=180]{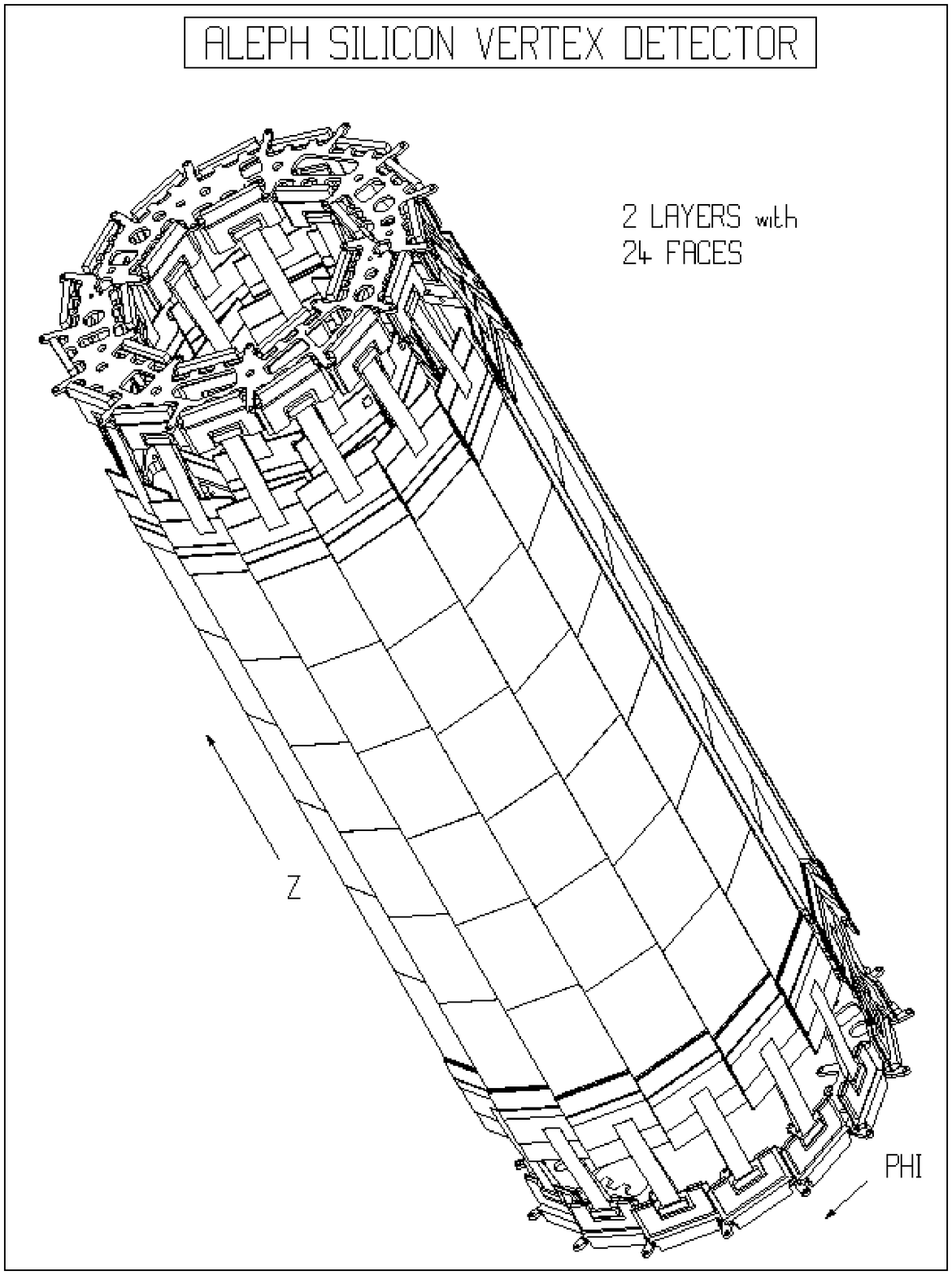}}
\subfigure{\includegraphics[width=0.47\textwidth]{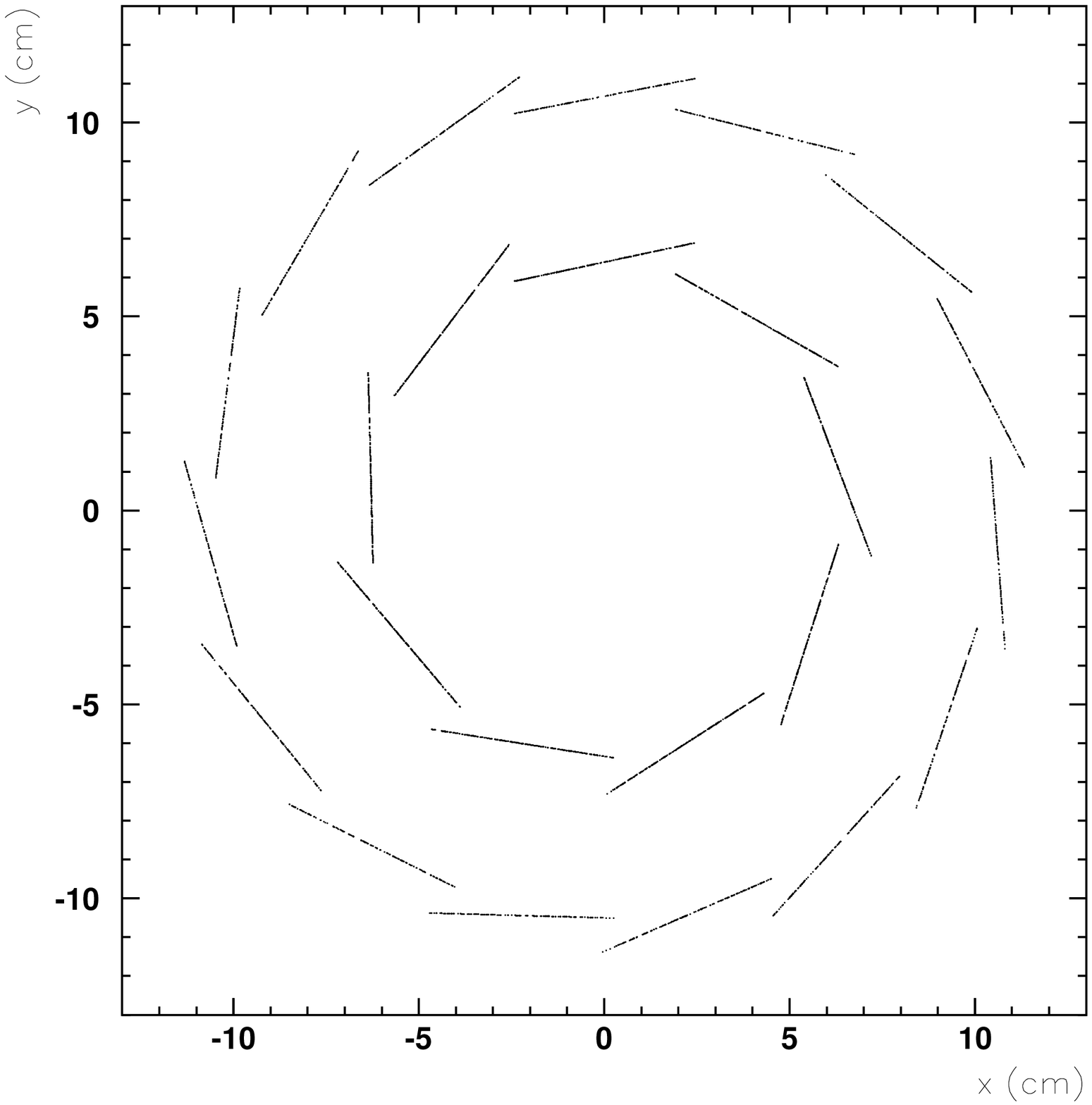}}
\caption[The vertex detector]
{\label{fig:vdet}{\small Left, full view of VDET. Right, end
view showing the position of the faces~\cite{alephhb2}.}}
\end{center}
\end{figure}

The final VDET consisted of two approximately cylindrical layers. The
inner layer, resting on the beam-pipe at 6.3$\cm$ from the beam, was made of
9 `faces', each face measuring 40$\cm$ in length along the $z$-axis 
and formed by six double-sided silicon microstrip detectors, or `wafers'. 
The outer layer, at 11.0$\cm$ from the beam, had 15 faces and
ensured a maximal lever-arm between the beam-pipe and the Inner Tracking
Chamber. Figure~\ref{fig:vdet} shows a full view of the VDET and the
geometry of the faces from an end-on view. 

The two sides of each wafer had readout strips arranged
orthogonally to each other. When a charged particle passed through the
wafer, a VDET hit was formed giving a two dimensional measurement: 
the $r\phi$ side provided a $\phi$ measurement while the $z$ side recorded $z$. 
The readout pitch on the $r\phi$ side was 50$\um$ and 100$\um$ on
$z$. However, the spatial resolution was improved by using the signal
recorded on several strips to determine the position of each hit down to
10$\um$ in $r\phi$ and 15$\um$ in $z$. 
This precision allowed very precise measurements of particle lifetimes, as
short as 300\,fs. 
The angular acceptance was such that
a particle with $|\cos\theta| \lt 0.95$ hit at least one layer of the
VDET. 
\subsection{Inner Tracking Chamber}
\label{itc}
\begin{figure}[h]
\begin{center}
\includegraphics[width=10cm]{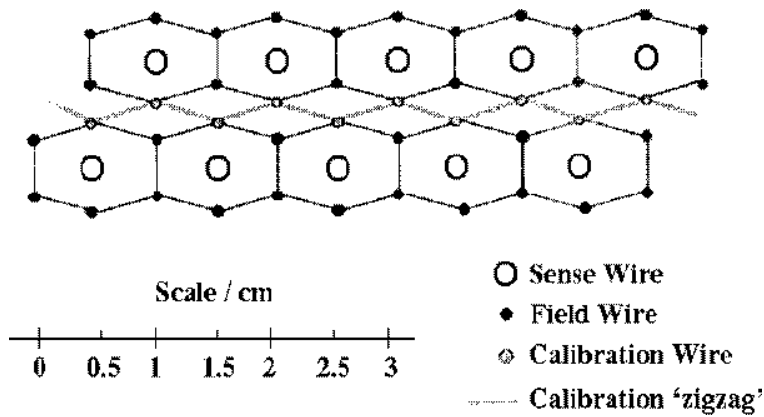}
\caption[Drift cell geometry of the
ITC]
{\label{fig:itccell}{\small Drift cell geometry of the
ITC~\cite{alephhb1}.}} 
\end{center}
\end{figure}

The ITC~\cite{Barber:1988ht} was a 2$\m$ long cylindrical multiwire drift
chamber. There were
eight concentric layers of wires running parallel to the beam ($z$)
direction. Its inner radius was $r=16\cm$, where it supported the outer layer
of the VDET, and its outer radius extended to $r=26\cm$. Its full volume was
filled with a mixture of argon (80\%) and carbon dioxide (20\%) at
atmospheric pressure. Each of the 960 sense wires, arranged in eight layers,
was held at a positive potential of $\sim$2\,kV and was surrounded by five
field wires and a calibration wire all held at ground potential. 
This formed an hexagonal `drift cell'. The cell structure is shown in
Fig.~\ref{fig:itccell}. 

When a charged particle traversed an ITC cell it ionised the gas and
produced electrons which were attracted to the sense wires. The ionisation
charge drifted to the sense wire with an average velocity across the
cell of 50$\um/\mathrm{ns}$. 
This negative current pulse was detected at both ends of 
the sense wire. The $z$ coordinate was thus obtained by the difference in
arrival time of the signal at each end. The $r\phi$ coordinate was deduced
by converting the drift time into a drift distance using a parametrisation
of the non-linear relationship between the two. The spatial resolution was
around 150$\um$ in the $r\phi$ coordinate and only 5$\cm$ in the $z$
coordinate. Each wire could only produce one hit per event, and therefore
there were a maximum of eight hits per track. 

The drift cells were small so that they could provide information extremely
quickly: the cell size determines the maximum drift time.
This was achieved at the cost of having many wires, readout channels 
and the consequent problem of fitting more electronics.  
But then, the ITC was able to provide the only tracking
information for the first-level trigger (see Sec.~\ref{trig}) in less than
3$\us$ after a beam crossing.
\subsection{Time Projection Chamber}
\label{tpc}
\begin{figure}[tbh!]
\begin{center}
\includegraphics[width=10cm,angle=-90]{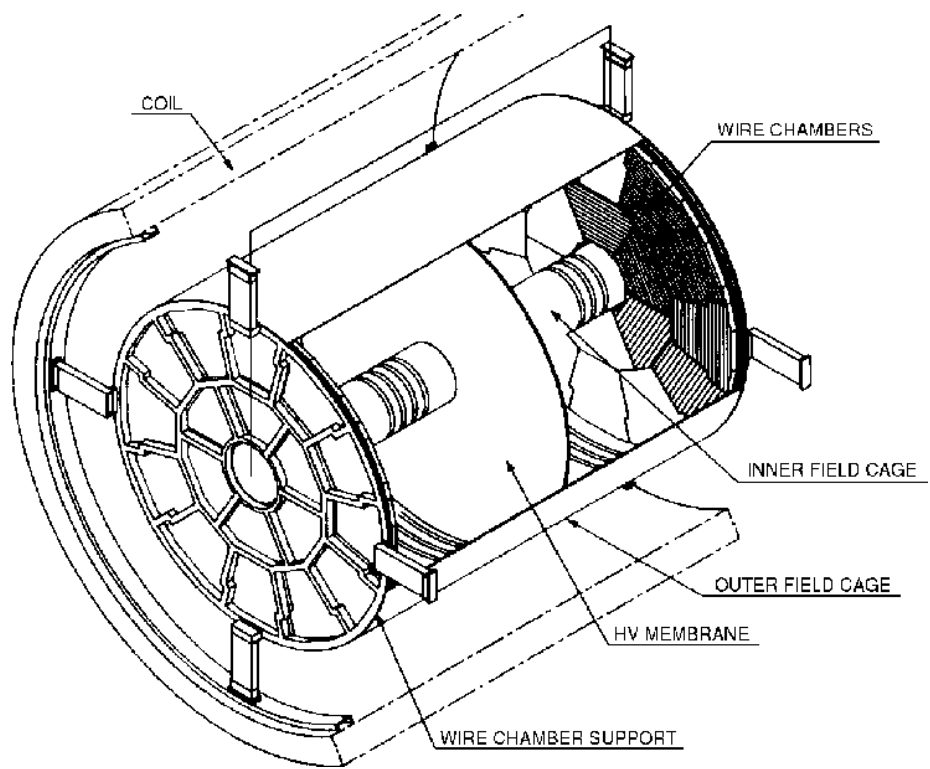}
\caption[TPC overall view]{\label{fig:tpc}{\small TPC overall view. From
Ref.~\cite{alephhb1}.}} 
\end{center}
\end{figure}

The Time Projection Chamber (TPC)~\cite{Atwood:1991bp} was the main tracking
detector in ALEPH 
providing three-dimensional points for track reconstruction. 
It was a 4.4$\m$ long cylindrical drift chamber with its axis parallel to the
beam, an inner radius of 31$\cm$ and an outer radius of 180$\cm$.
There were three main components: the field cage, made up by two
cylinders (inner and outer), two circular end-plates, and eight `feet'
which supported the TPC weight, attaching it to the
magnet. A diagram of the TPC is shown in Fig.~\ref{fig:tpc}.

The chamber was divided into two halves by a central membrane kept at
high potential ($-27$\,kV). This membrane, along with the cathode planes 
of the end-caps (held at a potential near ground) and the inner and
outer field cages, produced a uniform electric field in the $\pm z$
direction. The passage of a charged particle ionised the gas (ArCH$_4$) and
the electron cloud drifted towards the nearest end-cap following the
11\,kV/m electric field. The parallel magnetic field within the tracking
volume ensured that drift electrons did not diffuse radially and had tight
helix trajectories (the clouds were contained in the $r\phi$ plane).

\begin{figure}[tb]
\begin{center}
\vspace{-0.5cm}
\subfigure{\includegraphics[width=0.3\textwidth,angle=-90]{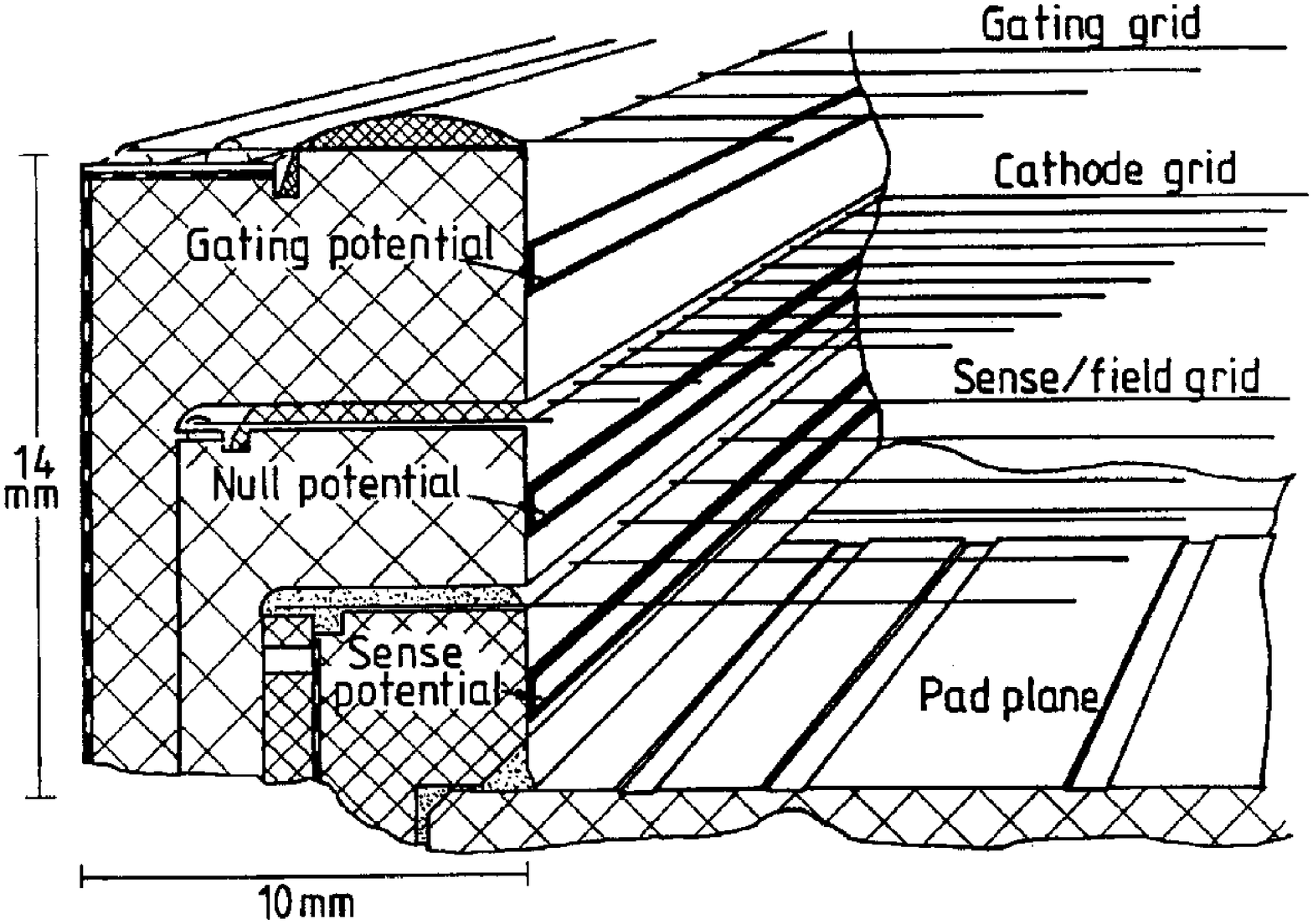}}
\subfigure{\hspace{0.4cm}\includegraphics[width=0.42\textwidth]{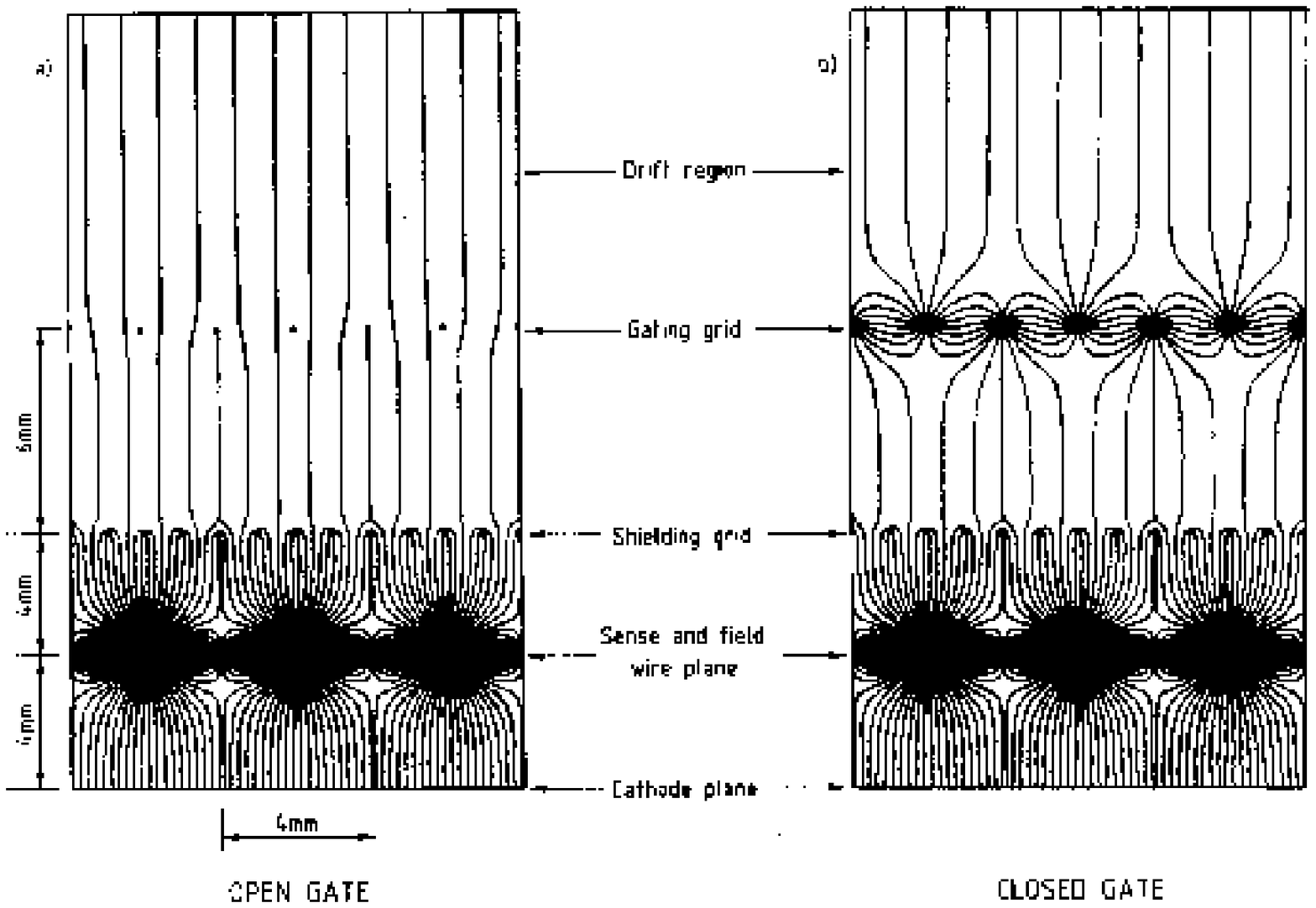}}
\caption[TPC sector edge and gating diagrams]{\label{fig:tpcedge}{
\small Schematic view of a TPC sector edge
(left) showing wire grids, pad plane and potential strips. On the right,
the gating grid of the TPC showing electric field lines in the (a) open and
(b) close gate states. From Ref.~\cite{alephhb1}.}}
\end{center}
\end{figure}
The TPC end-plates were formed by three consecutive wire grids. See
Fig.~\ref{fig:tpcedge} for a diagram of an end-plate edge. 
The first one was the \emph{gating grid} which will be discussed shortly. 
Next was the \emph{cathode grid}, which was grounded and was responsible 
for the drift of the ionisation cloud.
And finally, electrons were collected on the \emph{sense grid}. It consisted of
alternating sense wires, held at ground potential, and field wires, held at a
high positive potential, which formed conventional proportional
multiwire chambers with cathode `pads' to read out the pulses.
There were 21 rows of cathode pads precisely positioned behind the wires and
capacitively coupled to them.
Hence the $z$ coordinate was calculated from the arrival time of the signal 
on the pad and the known drift velocity of electrons in the drift volume. 
The drift velocity was constantly measured by laser calibration. 
The $r$ and $\phi$ coordinates were obtained from the position of the pads 
involved in the measurement. Hits could overlap in $r\phi$, e.g. if two 
particles ionised the same $r\phi$ region at different $z$, 
producing two pulses in the same pad. 
The pad recognised the different pulses by their different
time profile allowing separation of the hits.
Thus the TPC could measure up to 21 three-dimensional
points for a track at polar angle $\theta$ of 90$\degs$ with a resolution
of 180$\um$ in $r\phi$ and 800$\um$ in $z$. 
The average resolution decreased with particle momentum and $\theta$. 
Events with low momentum and low polar angle, like two-photon processes,
have the worst resolution.

If an ionisation avalanche was produced close to the sense wires, positive
ions could drift back to the main TPC volume and distort the local electric 
field leading to track distortion. 
The \emph{gating grid} in the end-caps was used to prevent charge build-up 
in the drift volume of the TPC. 
It was thus located between the drift volume and the 
cathode wires so that if the gating grid was in the `closed' state, positive
ions did not enter the drift volume. In the `open' state, a negative
potential was applied to the gating wires and the gate was transparent to the
passage of drifting charged particles. In the closed state, alternate gate
wires were kept at positive and negative potentials creating dipole fields
which prevented the passage of charged particles. See
Fig.~\ref{fig:tpcedge} for a diagram of the gate electrostatic
configuration in both open and closed states.
The gate was open 2$\us$ before a beam-crossing and stayed open for
45$\us$ if the Level-1 trigger reached a `yes' decision (Sec.~\ref{trig}).
This allowed the ionisation electrons to be
collected. A negative decision from the Level-1 trigger closed the gate
until the next beam-crossing. 

\begin{figure}[ht]
\begin{center}
\subfigure{\includegraphics[width=0.48\linewidth]{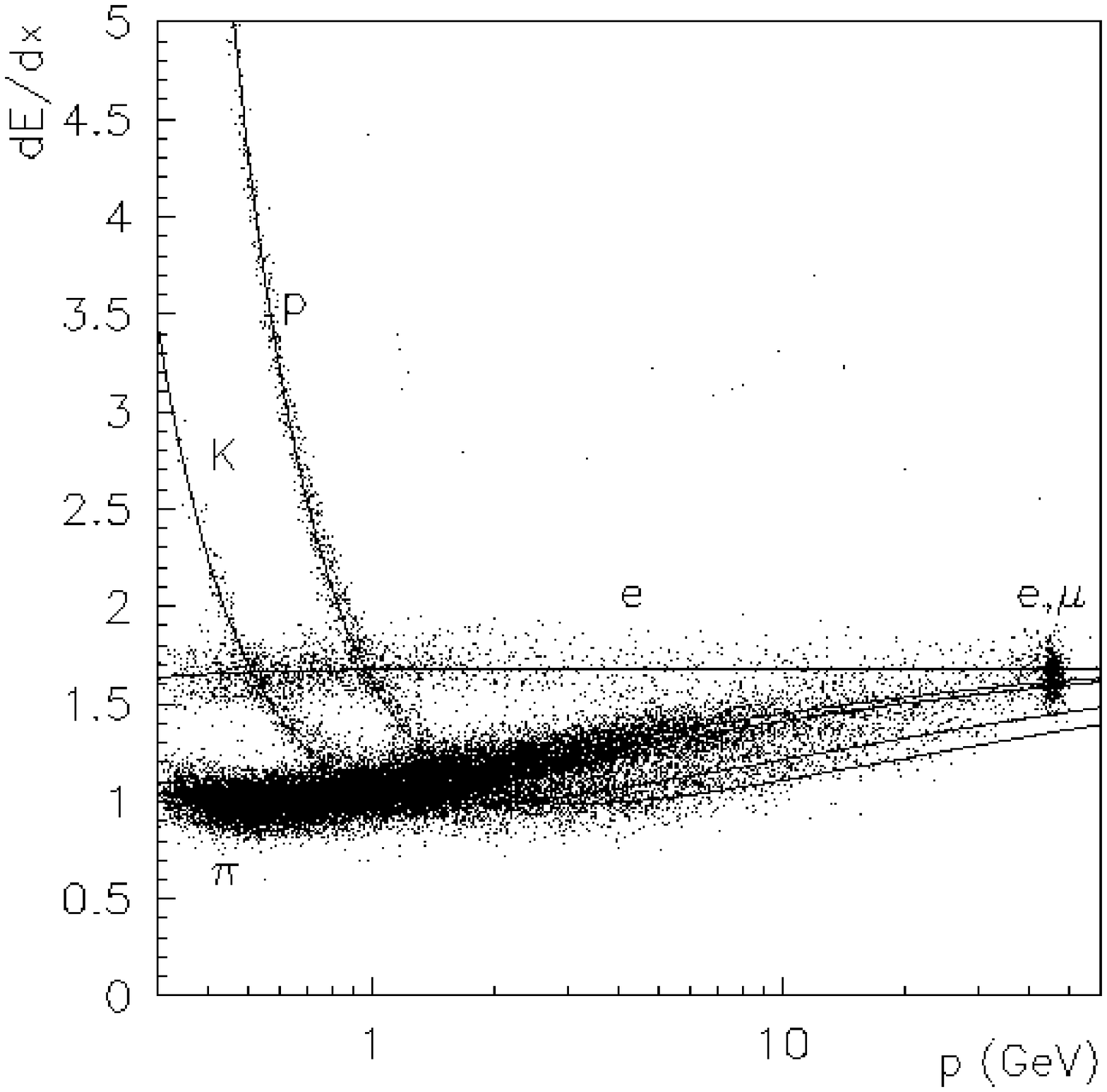}}
\subfigure{\hspace{0.5cm}\includegraphics[width=0.47\linewidth]{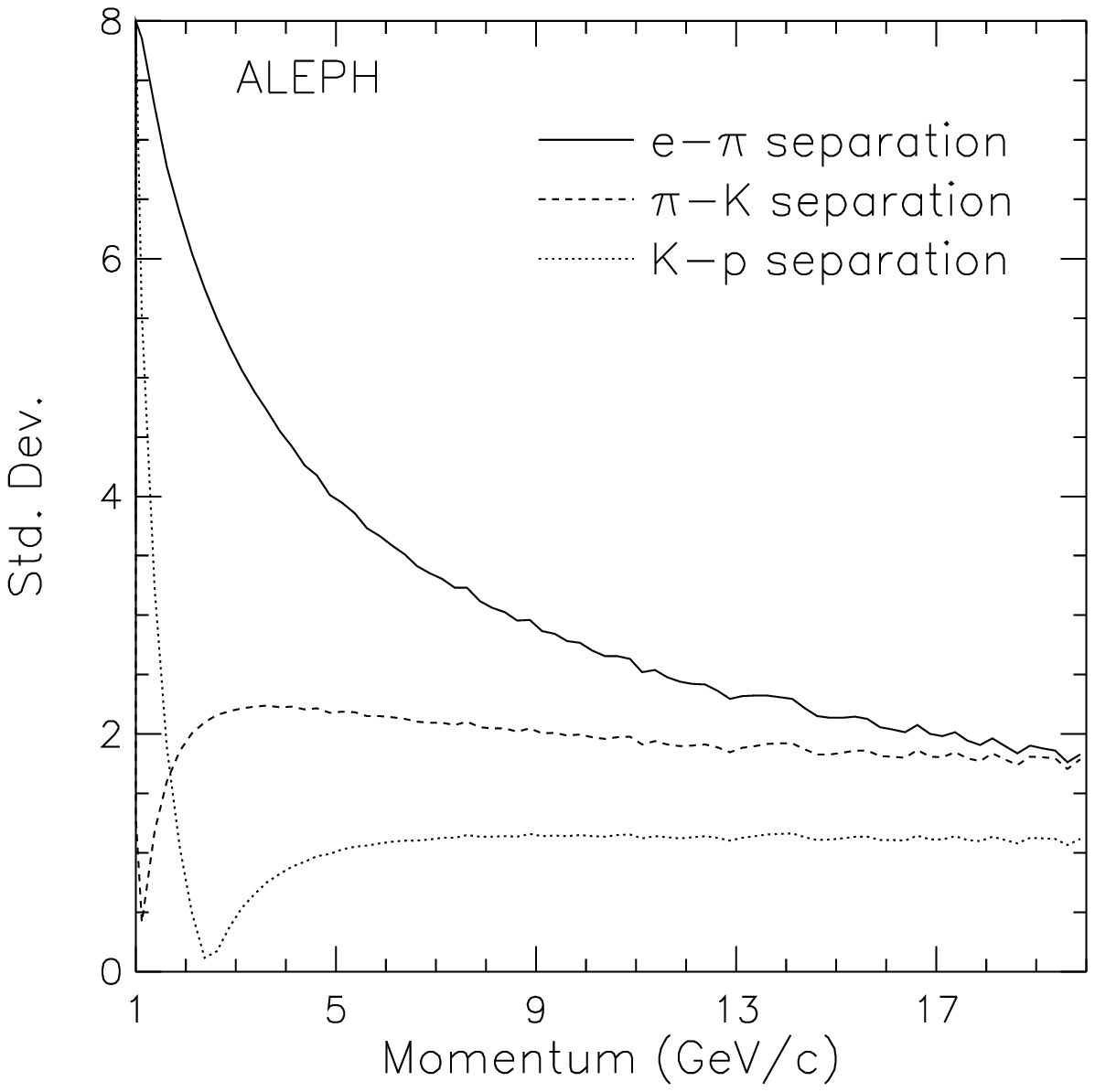}}
\caption[TPC particle identification by d$E$/d$x$ and momentum
measurements] {\label{fig:dedx}{\small On the left, the measured d$E$/d$x$
as a  function of particle momentum for a sample of about 40\,000 tracks.
Lines are the fitted parametrisation to a modified Bethe-Bloch formula
for electrons, muons, pions, kaons and protons. 
On the right, the average separation in standard deviations between
different particle types as a function of momentum. 
It has been computed using all tracks in hadronic Z
decays which have at least 50 d$E$/d$x$ measurements. 
From Ref.~\cite{Buskulic:1995wz}.}}
\end{center}
\end{figure}

Furthermore, the TPC wires were also used to measure the
specific ionisation of charged tracks since the pulse height on
the sense wires is proportional to the specific ionisation
energy loss (d$E$/d$x$).
The coordinates ($r,\phi,z$) along a particle trajectory provide
a measurement of its momentum. The d$E$/d$x$ for each particle depends
only on its velocity in a certain material, so by combining the d$E$/d$x$ and
the momentum measurements the mass of the particle can be inferred and thus
its identity derived. There were a total of 338 possible measurements of
d$E$/d$x$ for tracks within the angular acceptance of the TPC. A typical
resolution of 7.2\% was achieved in hadronic events. Not all tracks
had d$E$/d$x$ information since a minimum of 50 wire samples were required
to obtain a reasonably accurate measurement. Thus particles with polar
angles near 0 or $\pi$ did not have identification information from the
TPC. Figure~\ref{fig:dedx} shows how the d$E$/d$x$ information from the TPC is
used to obtain particle identification. 

\section{Calorimetry and Luminosity Monitors}
\subsection{Electromagnetic Calorimeter}
\label{ecal}
The main purpose of the Electromagnetic Calorimeter (ECAL)~\cite{alephhb1} 
was to measure the energy of photons, electrons and positrons in an event. 
The ECAL covered
polar angles of $|\cos\theta| \lt 0.98$ and was capable of measuring tracks
throughout the polar angle range of the TPC. It was inside the
superconducting solenoid to avoid a degradation of its performance
resulting from particles decaying in the uninstrumented coil. 

The ECAL was composed of a 4.7$\m$ long barrel and two end-caps, all with
similar properties. The calorimeter consisted of 45 layers of lead with
a proportional wire chamber in between each layer. The signals from the wire
chambers were read out by cathode planes arranged into towers pointing to
the interaction point. There were more than 73\,000 such towers, each
subdivided into three `storeys' 
which each yielded an energy measurement. 

An incoming electron, positron or photon interacted in one of the 2$\mm$
lead sheets and generated an electromagnetic shower by 
\emph{bremsstrahlung} in the case of electrons and by 
$\ee$ \emph{pair-production} in the case of photons. When the particles from
the shower traversed the wire chambers, they ionised the gas and the
cathode pads collected the pulses from the wires. This allowed a
position measurement of the shower development to be made from the readout
of the cathode pads. 
The information from the wires with their faster readout was used to make a
trigger decision. The choice of lead as absorber was due to its short
radiation length ($X_0 = 6.4$\,g/cm$^2$~\cite{pdg}). 
The total thickness of lead was about 22$X_0$, or $\sim$$40\cm$, which
ensured that over 98\% of the energy of a 50 GeV electron was contained in
the lead sheets. Hadrons and muons traversed the ECAL and could deposit
some energy inside it since it was about one interaction length thick. 

\begin{figure}[h]
\begin{center}
\includegraphics[width=0.42\linewidth,angle=-90]{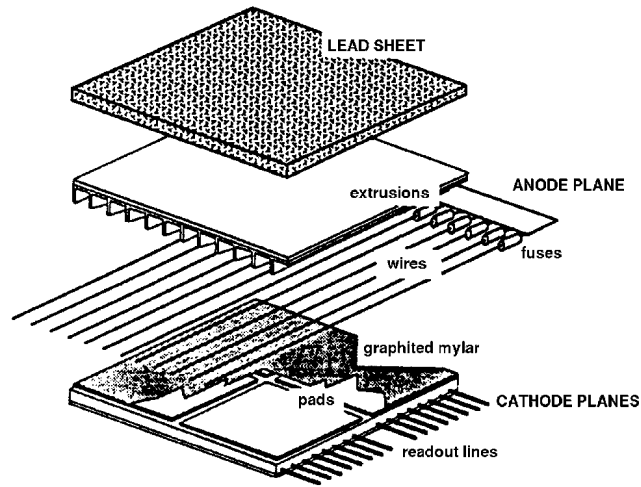}
\caption[Detail of an ECAL layer]{\label{fig:ecal}{\small Detail of an ECAL
layer. There are 45 such layers~\cite{alephhb1}.}}
\end{center}
\end{figure}

The large number of towers provided high granularity which was needed for
spatial separation of the electromagnetic showers and helped to improve
particle identification. Since each tower gave three energy measurements, 
the shower profile could also be studied, adding more information to
discriminate between electromagnetic and hadronic showers. 
An ECAL layer is illustrated in Fig.~\ref{fig:ecal}. 

The ECAL had an angular resolution of~\cite{Buskulic:1995wz}:
\begin{equation}
\sigma_{\phi} = \frac{\sigma_{\theta}}{\sin\theta} = 
 0.25 + \frac{2.5}{\sqrt{E(\gev)}} ~ \mathrm{mrad}
\end{equation}
and the energy resolution was parametrised using Bhabha scattered
electrons as: 
\begin{equation}
\label{eres}
\frac{\sigma_{E}}{E} = 0.009 + \frac{0.18}{\sqrt{E(\gev)}}
\end{equation}

The end-caps met the barrel at polar angles of $\sim$$40\degs$. There were a
number of cables in this region which reduced the total depth in radiation
lengths. The energy resolution was degraded by around 30\% in this region as
a result. 
\subsection{Hadron Calorimeter}
\label{hcal}
\begin{figure}[tb]
\begin{center}
\includegraphics[width=0.76\linewidth,angle=-90]{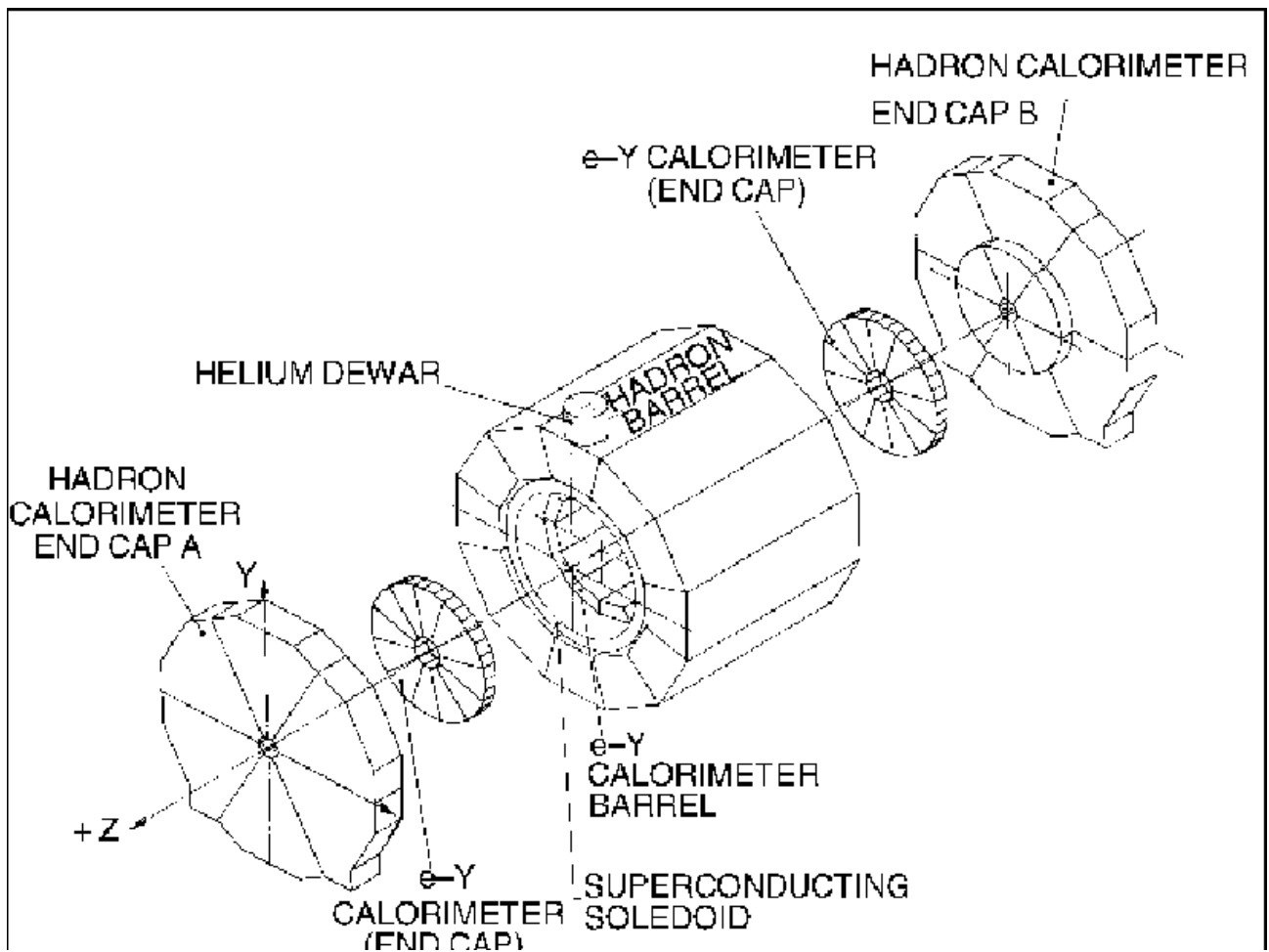}
\caption[Overview of the calorimeters]{\label{fig:hcal}{\small The ALEPH
calorimeters. The HCAL end-caps measure about 9$\m$ in
diameter~\cite{alephhb1}.}} 
\end{center}
\end{figure}

The Hadronic Calorimeter (HCAL)~\cite{alephhb1} was used to finally stop and
measure the energy of hadrons as well as tag the trajectories of muons. 
Its structure was very similar to that of the ECAL, comprised of a barrel 
and two end-caps. However, it was slightly rotated with respect to the ECAL
to avoid an overlap between the inactive `cracks' of both detectors. 
This can be seen in Fig.~\ref{fig:hcal}, which offers a view of both ECAL
and HCAL. 

The HCAL was made of 23 iron layers separated by plastic streamer tubes. 
It weighed in excess of 2500 tons and particles at normal incidence
encountered 1.2$\m$ of iron, equivalent to 7.2 interaction lengths for
hadrons. The iron acted as the return yoke of the magnetic
flux from the solenoid and also as the main mechanical support for the
ALEPH detector. 

A hadron reaching the iron layers interacted through the strong
force with the nucleons of iron and decayed into
a shower of low energy particles. Hadronic showers are broader and deeper
than electromagnetic showers which explains why the HCAL needed to be
thicker than the ECAL. 
The ionisation caused by the shower in the tubes was collected by cathode
pads which were grouped into 4068 projective towers pointing to the IP, like
the ECAL. 
The pad electrodes gave a measurement of the integrated energy from each
tower, but did not provide any information on the $r$ coordinate, the depth
of the shower, which is very important to determine the hadronic shower
profile.  
A digitised cathode strip running along each streamer tube provided the 
necessary $r\phi$ measurements by indicating which plane of the HCAL 
had recorded a hit. This enabled a better discrimination between muons
and hadrons by their different hit pattern (see Sec.~\ref{lepid}). 

For pions at normal incidence the energy resolution is parametrised
by~\cite{Buskulic:1995wz}: 
\begin{equation}
\frac{\sigma_{E}}{E} = \frac{0.85}{\sqrt{E(\gev)}}
\end{equation}

\subsection{Muon Chambers}
\label{mucha}
Muons traversed both calorimeters with a characteristic trail of hits
with no shower development. Surrounding the HCAL there were two further
double-layers of streamer tubes known as muon chambers. Only muons could 
penetrate that far after the HCAL and these chambers acted
as a muon tracking detector. The two layers were separated by more than
40$\cm$ to enable a measurement of the track angle with an accuracy of
$\sim$$15$\,mrad. 

\subsection{Luminosity Monitors}
\label{lumcal}
The luminosity delivered by LEP was determined by using three pairs of
specific detectors. An accurate measurement of the luminosity $\Lum$ is
necessary in order to obtain precise reaction cross sections from the event
count rates. Cross sections, giving the probability of an
interaction to occur, are usually given in $barns$, defined
as 1 b = $10^{-28}\m^{-2}$.

These detectors measured the number of small angle Bhabha events,
coming from the QED process $\ee~\to~\gamma^*~\to~\ee$,
whose cross section is known very accurately, to calculate the
luminosity\footnote{The same process, via exchange of a Z boson, introduces
some corrections in the total cross section which were poorly known at the
start of LEP. By restricting the measurements to low $\theta$, the
contribution from Z production is effectively reduced, and the rate of
events is much larger. This is why these detectors sit very
close to the beam-pipe.}. 

For $N$ events and cross section $\sigma$:
\begin{equation}
\int\Lum dt = \frac{N}{\sigma} \hspace{5mm} \mathrm{with} \hspace{5mm}
\sigma = \frac{1040}{s}\left(
\frac{1}{\theta_{\rm{min}}^2} - \frac{1}{\theta_{\rm{max}}^2} 
\right) \mathrm{nb}
\end{equation}
where $s$ in $\gev$ is the centre-of-mass energy and $\theta_{\mathrm{min/max}}$ 
are the angular bounds which define the acceptance cone subtended 
by the monitors. 

Each luminosity monitor consisted of two units on either side of the
interaction point to detect the outgoing back-to-back $\ele$ and $\pos$.
LCAL and SiCAL were inside ALEPH itself, both at about
2.5$\m$ away from the IP. BCAL, with 12 layers of tungsten and scintillator
counters, was about 7$\m$ away, outside ALEPH. 

SiCAL had 12 annular tungsten-sheets sandwiched between silicon
detectors. The active region covers polar angles from 24 to 58\,mrad. 
LCAL was similar in construction to the ECAL ---lead sheets interspersed
with multiwire proportional chambers---, covering angles just beyond 
SiCAL, from 45 to 190\,mrad. LCAL can be thought of as a low angle
extension of the ECAL. 
LCAL and SiCAL had low event rates so they provided an integrated
luminosity measurement. 
However, for LEP2, additional background shielding was
required close to the beam-pipe which reduced the angular acceptance of
SiCAL to 34\,mrad. So, at high energies, SiCAL lacked enough data to 
provide an accurate measurement of $\int \Lum dt$ and was used only to 
extend the calorimeter hermeticity in ALEPH. LCAL was therefore the only 
source of integrated luminosity. 

BCAL, on the contrary, having a much lower angular acceptance
($\theta_{\mathrm{min}} \sim 7$\,mrad) due to its position, could detect a
larger number of Bhabha events and had high enough statistical precision
to provide an instantaneous luminosity value. BCAL was therefore used as an
online luminosity monitor of the background conditions. 

\section{Trigger System}
\label{trig}
At LEP, there was approximately one beam crossing every 89$\us$ but the
detectors could not possibly record all those events at such an enormous
rate. A decision on whether the event was `interesting' was required to
activate the readout system and record and store the event. 
However, of all the beam crossings, most resulted in `uninteresting' events,
like beam interactions with gas molecules in the accelerator, or off-momentum
beam particles scattered by a collimator close to the IP. Only a small
fraction of the beam crossings actually consisted of electron-positron
interactions. The \emph{trigger} system was designed to accept these events 
and reject the others. It also reduced the dead-time of the
detector that results from readout. 

The trigger system was built in three stages, the first two were hardwired
logic to achieve a very fast response and the third one implemented in off-line
software. Each one based their `yes' or `no' decision on different
subdetectors depending on their capability to produce fast information
about an event. Table~\ref{tab:trigger} summarises the
subdetectors used and the decision time. 
\begin{table}[h]
  \begin{center}
    \begin{tabular}{|c|c|c|c|}
      \hline
      Stage & Decision Time & Max. Rate (Hz) & Information Used \\
      \hline \hline
      \multicolumn{2}{|l|}{LEP beam crossing} & $\sim 11000$ & --- \\ \hline
              &          &           & Hit Patterns in ITC \\
      Level-1 & 5~$\mu$s & $\sim 10$ & Pad/wire readout    \\
              &          &           & from ECAL+HCAL+LCAL \\
      \hline
      Level-2 & 50~$\mu$s & $\sim 5$ & Hit Patterns in TPC \\
      \hline
      Level-3 & 62~ms & 1-3 & All subdetectors \\
      \hline
    \end{tabular}
    \caption[Summary of the ALEPH trigger system.]{Summary of the ALEPH
trigger  system.}
    \label{tab:trigger}
  \end{center}
\end{table}

The first level trigger (Level-1) delivered a positive decision within
5$\us$, much less than the beams crossing, if a good charged track had been
registered by the ITC and/or the calorimeters had an energy deposit. This
positive decision allowed the TPC to collect the drift electrons and
partially reconstruct the tracks. If the Level-1 gave a `no' decision, the
TPC gate grid was closed and all the subdetectors reset for the next beam
crossing, avoiding any dead time. 
The Level-2 needed around 50$\us$ to form a decision based on the TPC
tracks: it required good tracks (exceeding a hit threshold) pointing to the 
interaction region. If the event satisfied the Level-2 conditions, the
whole detector output was read out. 

The Level-1 and Level-2 criteria valid for the 1999 and 2000 run periods is the
following, with only minor modifications on the threshold values for
previous years: 
\begin{itemize}
\item Single Electron/Photon Triggers. Drift chamber (ITC for Level-1, TPC
for Level-2) track segment pointing at an energy deposit in the ECAL of
200\,MeV for electrons or a 2$\gev$ energy deposit for photons (with no track).
\item Total Energy Trigger. ECAL energy deposits should exceed certain
thresholds: $5.5\gev$ in the barrel or $4.5\gev$ in either end-cap or $1.7\gev$
in both end-caps.
\item Back-to-Back Track Trigger. Two ITC (or TPC for Level-2) tracks
pointing in opposite directions from the IP.
\item Single Muon Trigger. ITC or TPC track with at least 4 HCAL hits.
\item Bhabha Luminosity Triggers. Hits in the LCAL with opposite azimuth.
\end{itemize}
In fact, the trigger achieved an efficiency greater than 99\% in accepting
the `interesting' events, such as Z$\,\to\,q\bar{q}$, Z$\,\to\,\ell^+\ell^-$, 
Bhabha scattering, or two-photon events. 

\section{Event reconstruction}
\label{evrec}
Event reconstruction is the process by which raw data from the
different subdetectors is transformed and correlated to form physically
meaningful final data. It mainly consists of \emph{track
reconstruction}, which assigns each hit in the tracking chambers to a
particle track, and \emph{energy flow analysis}, which associates charged
particle tracks with energy deposits in the calorimeters, enabling particle
identification and improving the overall energy resolution. 
Event reconstruction begins just after the data is stored on tape and is
run by JULIA (Job to Understand Lep Interactions in
Aleph)~\cite{julia}. Particle identification, as well as the rest of the user
analysis of the events is performed by the computer program ALPHA (ALeph
PHysics Analysis package)~\cite{alpha}.

\subsection{Track reconstruction}
\label{trkreco}
The information from all three tracking detectors is used to reconstruct a
charged particle trajectory, or track. The reconstruction algorithm starts
in the TPC. It links nearby hits to form segments. Several segments will be
connected to form a track if the hypothesis that a helix is described is
satisfied. This TPC track will then be extrapolated down to the ITC and VDET
where matching hits will be assigned to the track. This solves any
ambiguities that could arise from the two-dimensional ($r\phi$ and
$z$) hits in the VDET. 

A final track fit is then performed taking into account the segment errors
by means of a Kalman filter technique~\cite{Billoir:1984mz} to provide a
fully reconstructed track.  
Table~\ref{tab:trkmomres} shows the
momentum resolution achieved by incorporating the diferent subdetector
segments into the track for a Z$\,\to\,\mu^+\mu^-$ sample of events. 
\begin{table}[h]
\begin{center}
\begin{tabular}{|r|c|c|} \hline
 Tracking Detectors & Max. Hits & $\sigma_{\pt}/\pt^2~(\gevc)^{-1}$
 \\ \hline \hline
 TPC           & 21     & 1.2$\times 10^{-3}$ \\
 TPC+ITC       & 21+8   & 0.8$\times 10^{-3}$ \\
 TPC+ITC+VDET  & 21+8+2 &0.6$\times 10^{-3}$\\ \hline
\end{tabular}
\caption[Momentum resolution in the ALEPH tracking subdetectors]{
\label{tab:trkmomres}\small Transverse momentum resoultion 
$\sigma_{\pt}/\pt^2$ for $\pt$ in
GeV/$c$ when the information from the different subdetectors is
incorporated into the track reconstruction~\cite{Buskulic:1995wz}. The
number of maximum hits available for each subdetector corresponds to a
track with 90$\degs$ polar angle.}
  \end{center}
\end{table}

A `good' track, as applied to the analysis described here, is defined as
follows:  
\begin{itemize}
\item A minimum of four hits in the TPC, which restricts good tracks to the
range in polar angle: $|\cos\theta| \lt 0.96$
\item Transverse impact parameter: $|d_0| \lt 1\cm$
\item Longitudinal impact parameter: $|z_0| \lt 5\cm$
\item Transverse momentum: $p_t \gt 2$\,MeV/$c$
\end{itemize}

\subsection{Energy Flow}
\label{eflow}
The energy flow algorithms are designed to provide an accurate calculation
of the total visible energy and momentum of an event. The energy
resolution is fundamental in SUSY searches since their characteristic
signature is large missing energy or momentum.

The event is first of all `cleaned' by removing possible cosmics (the muons
from cosmic showers initiated in the upper atmosphere) and electronic
noise. Clusters of energy deposition are built in the calorimeters by adding
together `storeys' (in the ECAL) or towers (in the HCAL) which have
energies above 30\,MeV and share a common edge or corner. 
Good tracks, as defined in the previous section, are then extrapolated to
these calorimeter clusters.  
This forms the so-called \emph{energy flow objects},
which are groups of topologically connected tracks and clusters. 
Several such objects are defined: 
\begin{itemize}
\item Electrons. They are identified using the ECAL shower profile. Their
energy is calculated using their track momenta and mass. If the associated
calorimeter cluster energy is greater than 3 times the expected energy,
that extra energy is associated with a photon from bremsstrahlung. 
\item Muons. Identified by 400\,MeV energy deposits per HCAL plane and up
to 1$\gev$ in an ECAL cluster. 
\item Photons. Neutral energy clusters in the ECAL. If the invariant mass
of two photons matches the mass of a $\pi^0$, a neutral pion is assumed.
\item Charged and Neutral Hadrons. The remaining charged tracks are
attributed to charged hadrons, and the remaining energy to neutral
hadrons. 
\end{itemize}

Finally, this analysis provides a list of objects which is expected to be
an accurate representation of the true particles involved. The kinematic
variables calculated from energy flow objects provide the basis for any
further analysis of the event. For example, the visible mass of an
event ($M_{\rm vis}$) is the invariant mass of all energy flow objects in that
event; or the transverse momentum ($p_t$) is the transverse component of
the vector sum of momenta of all energy flow objects.

\begin{figure}[tb]
 \begin{center}
 \vspace{-0.5cm}
 \includegraphics[width=0.54\linewidth]{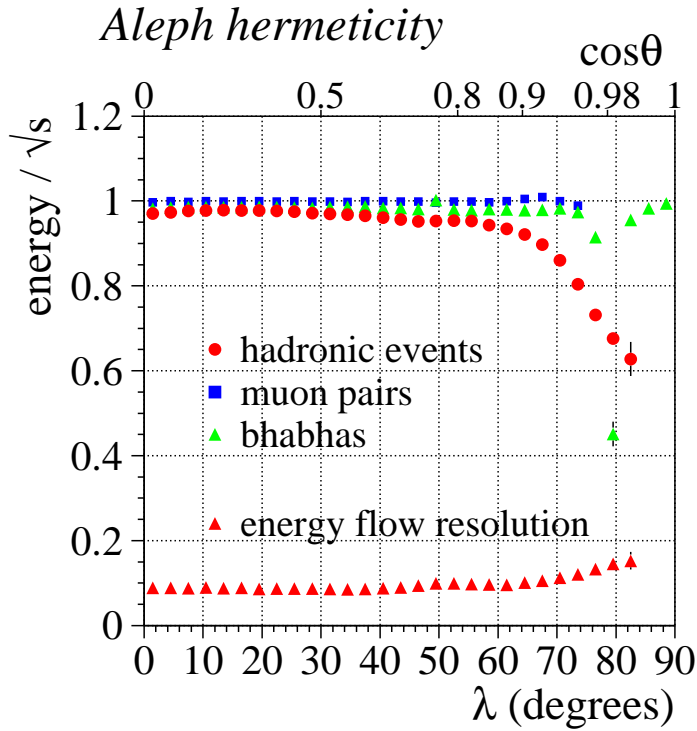}
   \caption[ALEPH hermeticity and energy flow resolution]
    {\label{fig:hermeticity}{\small Energy flow resolution normalised
to one as a function of the polar angle $\theta$ ($\lambda = \pi/2 -
\theta$). An average energy flow resolution of 7\% is achieved. 
The three sets of markers at the top of the plot show the
reconstructed energy as a fraction of the total input energy (\roots).  
The sharp drop at $\cos\theta \sim 0.98$ for Bhabha events is due to the
gap region between the LCAL and ECAL. From Ref.~\cite{Nachtman:1997em}.}}  
 \end{center}
\end{figure}

The energy flow resolution as a function of the polar angle is shown by the
lower set of markers in Fig.~\ref{fig:hermeticity}, calculated for
hadronic events Z$\,\to\,q\bar{q}$. The distribution of the reconstructed
energy is fitted to a Gaussian giving an average relative resolution of
$\sim$$7\%$, reaching a maximum of 15\% for $\cos\theta \gtrsim 0.98$.
The ALEPH hermeticity is also shown in that figure, as a
measurement of the reconstructed energy normalised to $\roots$. 
The gradual decrease with $\cos\theta$ in the measured energy for hadronic
events arises from the high multiplicity of such events. Events with
$\cos\theta \gtrsim 0.9$ are more likely to have objects going down the
beam-pipe, leading to a decrease in the measured energy.
Otherwise, Bhabha and muon-pair events have low multiplicity
and are very well reconstructed. 
However, there is a sudden drop in the measured energy for Bhabha events at  
$\cos\theta \sim 0.98$. This corresponds to the insensitive region where
the LCAL meets the ECAL end-cap. 

The high hermeticity of the detector and the precise measurements of the
energy contained in the events, permit searches for new physics to be
carried out successfully in ALEPH. 

\subsection{Lepton Identification}
\label{lepid}
Lepton identification has a major role in the analysis described here, and
needs to be described in some detail. 
\subsubsection*{Electron Identification}
Electron identification, as described in the previous section, relies on
the energy profile deposited in the ECAL and additionally on the d$E$/d$x$
information from the TPC. Three quantities, called \emph{estimators}, are
defined such that the measured value of a variable ($q$) is compared to the mean
expected value ($\langle q \rangle$), taking the resolution ($\sigma_q$) into 
account. The estimator (Q) will have a standard Gaussian distribution:
$Q\,=\,\frac{q-\langle q \rangle}{\sigma_q}$, centered at zero and with
width unity.  
The estimators are: the transverse shower shape $R_T$, 
based on the compactness of the shower; the longitudinal shower shape  
$R_L$; and $R_I$, based on the ionisation energy loss in the TPC. 

The electron shower in the ECAL
tends to be more compact in the transverse direction than that of a hadron
like the pion. This leads to the definition of the $R_T$ estimator, based
on the compactness of the shower. The momentum $p$ of the electron track is
known from the TPC data. That track is extrapolated to the ECAL and the energy
in the four towers closest to the extrapolated point is summed, giving
$E_4$. The energy fraction $E_4/p$ is the variable used in the $R_T$
estimator:
\begin{equation}
R_T = \frac{E_4/p(\mathrm{meas})-E_4/p(\mathrm{exp})}{\sigma_{E_4/p}}
\end{equation} 

The second estimator $R_L$ is defined to take into account the
longitudinal development of the electron shower, using the inverse of the
mean depth of the energy deposition in the ECAL ($X_L$):
\begin{equation}
X_L = \frac{E_4}{\sum_{i=1}^{4}\sum_{j=1}^{3}E_i^jS_j} 
~~~~\Longrightarrow ~~~~
R_L = \frac{X_L(\mathrm{meas})-X_L(\mathrm{exp})}{\sigma_{X_L}}
\end{equation}
where the energy is summed over $i$ storeys and $j$ segments, $S$ is the
mean depth of energy deposition on that segment. 

The estimator $R_I$ is similarly defined, by comparing the measured
d$E$/d$x$ to what is expected for a particle with the mass and charge of an
electron. 

It is also important not to misidentify electrons produced originally in
the IP with \emph{photon conversions}, 
i.e. photons which interact with the VDET, ITC or TPC material and produce an
$\ee$ pair. This $\ee$ pair does not have to be
symmetric, and sometimes only one particle is energetic enough to be
registered. Since there are many SM backgrounds involving photons, it is
important not to take a conversion for a `genuine' electron coming from the
signal SUSY process. This is achieved by trying to identify pairs of tracks
coming from a single displaced vertex. To tag conversions it is required that:
\begin{itemize}
\item the distance between the two tracks in $xy$ plane at the point where
they are parallel and closest together, is less than 2\cm,
\item the distance between the two tracks in the $z$ direction at that same
point is less than 3\cm,
\item the invariant mass of the two tracks at that point assuming they are
both electrons is less than 40\,MeV.
\end{itemize}

Finally, for an object to be identified as an electron it must pass the
following criteria: 
\begin{itemize}
\item It must be a `good' track and not compatible with a photon conversion
\item $p \geq 2\gevc$ 
\item $R_T \geq -3$
\item $|R_L| \leq 3$
\item $|R_I| \leq 3$
\end{itemize}

\begin{figure}[tb]
\begin{center}
\vspace{-1cm}
\subfigure{\includegraphics[width=0.4\linewidth]{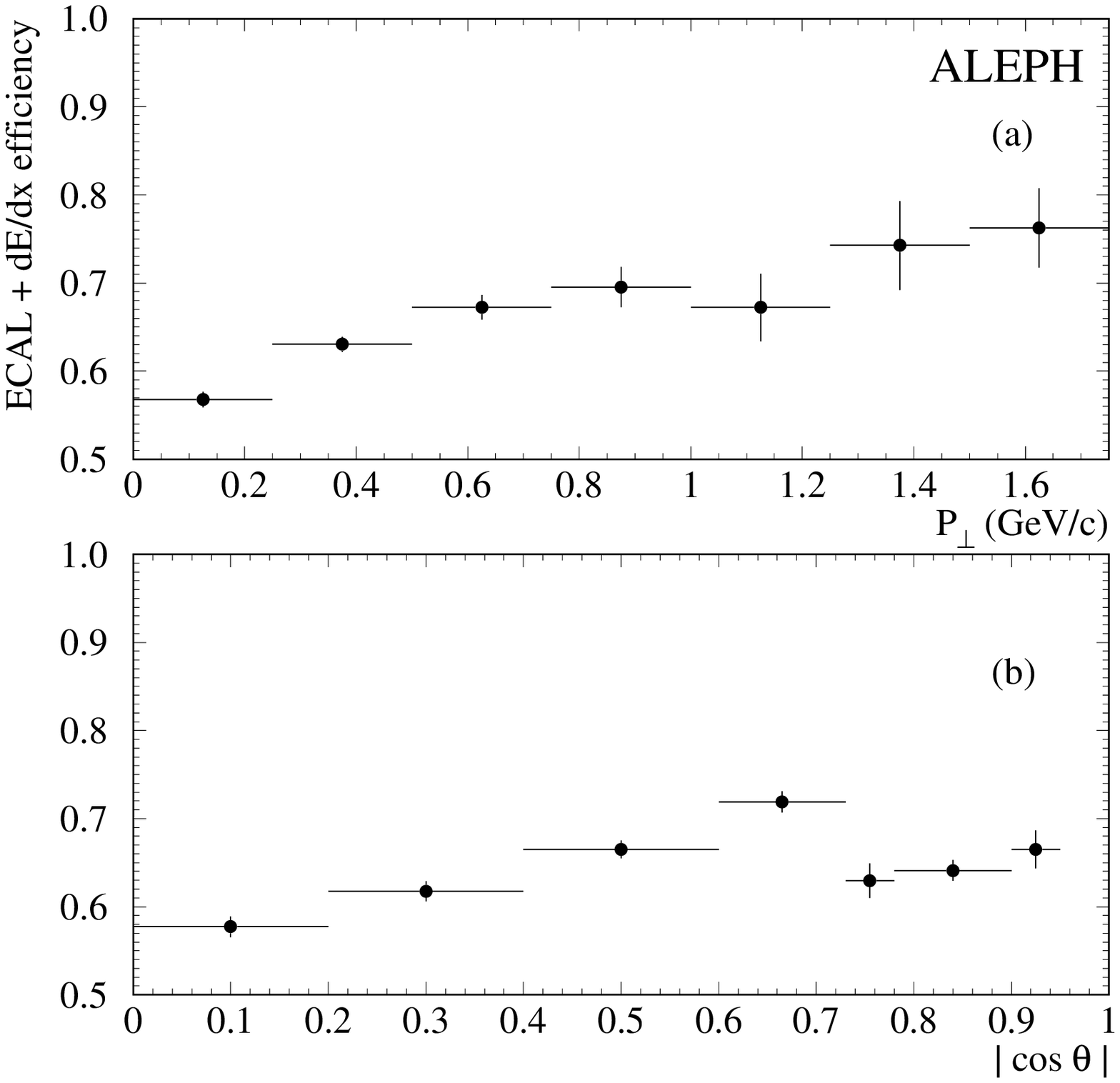}}
\subfigure{\hspace{1cm}\includegraphics[width=0.5\linewidth]{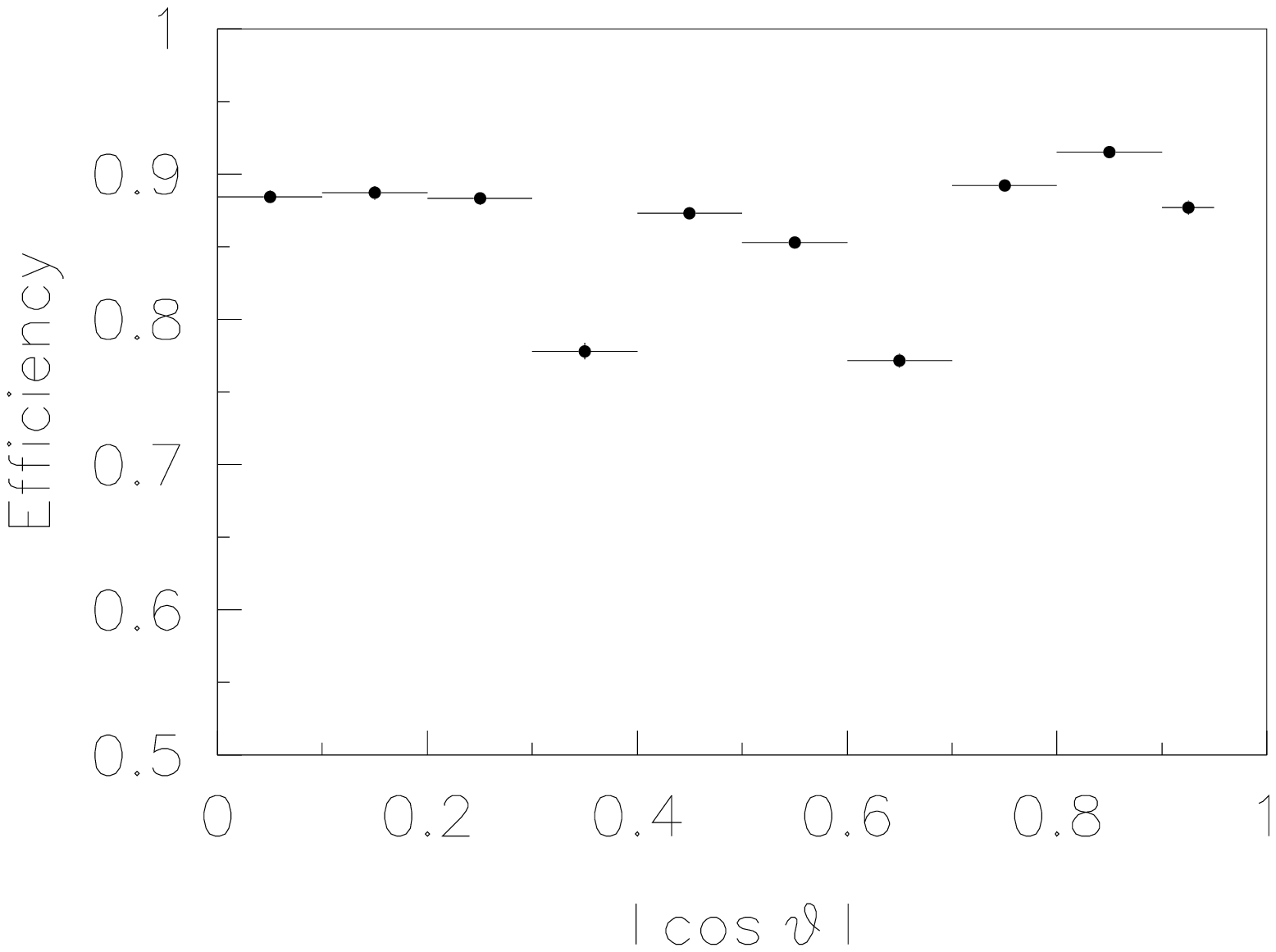}}
\begin{picture}(1,1)(195,-160)
 \put(0,0){(c)}
\end{picture}
\caption[Lepton identification efficiencies]{\label{fig:lepid}{\small 
The efficiencies of the electron identification procedure as a
function of (a) the transverse momentum and (b) the polar angle. The electron
sample used comes from real data at the Z peak selecting photon conversions
with high purity. (c) The efficiency of muon identification using both the HCAL
and muon chambers information as function of the polar angle. The sample
used is again real data $\Z\to\mu^+\mu^-$ where high purity is obtained
by selecting muons with TPC and ECAL information only. From
Ref.~\cite{Buskulic:1995wz}.}} 
\end{center}
\end{figure}

The average efficiency of the procedure has been evaluated to be 65\% in
hadronic events~\cite{Buskulic:1995wz}. See Figs.~\ref{fig:lepid}a
and b for the dependence of the efficiency with the
transverse momentum and the polar angle. No evidence for a strong
correlation between the two is found. 

\subsubsection*{Muon Identification}
Muons have a different hit pattern in the HCAL than hadrons; they only
interact with HCAL material by ionisation, not nuclear interactions,
following a very narrow path, rather than extensive transverse areas. 
Muons are also expected to reach the outer regions of the HCAL, which means
that planes further away from the IP are more likely to be fired by a muon
than a hadron. Finally, the track extrapolated from the TPC, the narrow
hits in the HCAL and the possible hits in the muon chambers aid in the
correct identification of tracks as muons. Furthermore, these patterns are
independent of the muon momentum for muons in the range $5-50\gevc$. 
All muons above 3$\gevc$ are able to penetrate out to the muon chambers.

Tracks are extrapolated from the TPC assuming they are muons defining a
`road' through the HCAL with a width of three times the uncertainty due 
to multiple scattering. A plane in the HCAL is expected to fire if the 
road intersects it. A plane is
said to have fired if a digital hit lies within the width of the road. 
A hit in the muon chambers is assigned to the extrapolated track if it lies
within four times the uncertainty due to multiple scattering. The depth of
penetration of the track can be assessed by looking at the last ten planes,
out of the total expected to be fired, that actually register a hit. 
Thus, a track can be identified as a muon if it satisfies the following
conditions:
\begin{itemize}
\item $N_{\mathrm{fire}}/N_{\mathrm{exp}} \geq 0.4$
\item $N_{\mathrm{exp}} \geq 10$
\item $N_{10} \gt 4$
\item $X_{\mathrm{mult}} \leq 1.5$
\item At least one associated hit in the muon chambers
\end{itemize}
where $N_{\mathrm{fire}}$, $N_{\mathrm{exp}}$, $N_{10}$ and
$X_{\mathrm{mult}}$ are, respectively,  the number of actually firing
planes, the number of expected planes, the number of firing planes within
the last ten expected and the average hit multiplicity per fired plane.

Applying these constraints on a dimuon events sample, an average efficiency
of 86\% is found. Figure~\ref{fig:lepid}c shows the muon identification
efficiency as a function of the polar angle. The two dips are due to the
magnet supporting legs, which are not instrumented with muon chambers.  
 
\subsubsection*{Tau Identification}
\label{tauid}
The tau particle, the heaviest known lepton, can decay both leptonically or
hadronically via the exchange of a W boson. This happens with a mean lifetime
of 290\,fs, or 87$\um$, just enough to be measured at ALEPH for boosted taus.
The main decay modes are listed in Tab.~\ref{tab:taubr}. If the final
state consists of one charged particle it is referred to as
\emph{one-prong}. This is the most likely decay mode with 85\% probability
and gives some relatively clean signal in the leptonic mode.
On the other hand, for Lorentz-boosted taus decaying in the \emph{three-prong}
mode, it is usually very difficult to reconstruct the final particles 
($\pi^{\pm}$ and $\pi^0$ in general) due to the small opening angle between
the three. 

\begin{table}
\begin{center}
\begin{tabular}{|c|l|l|c|} \hline 
 Mode & $X$ & Main Decay & BR(\%) \\ \hline \hline
\multirow{6}{2cm}{One-prong} 
   & $\mathrm{e}\anue$ & $\tau^-\to\ele\anue   \anut$ & 18 \\
   & $\mu\anum$        & $\tau^-\to\mu^-\anum  \anut$ & 17 \\
   & $h$               & $\tau^-\to\pi^-       \anut$ & 13 \\
   & $h+\pi^0$         & $\tau^-\to\pi^- \pi^0 \anut$ & 25  \\
   & $h+2\pi^0$        & $\tau^-\to\pi^-2\pi^0 \anut$ & 10  \\
   & $h+3\pi^0$        & $\tau^-\to\pi^-3\pi^0 \anut$ & 1.5 \\ \hline
\multirow{2}{1.4cm}{3-prong} 
   & $3h$                & $\tau^-\to\pi^-\pi^+\pi^- \anut$     & 9.5 \\
   & $3h+(\geq 1 \pi^0)$ & $\tau^-\to\pi^-\pi^+\pi^-\pi^0 \anut$& 5   \\ \hline
5-prong  & $5h+(\geq 0 \pi^0)$ & $\tau^-\to 3h^-2h^+     \anut$ & 0.1 \\ \hline
\end{tabular}
\end{center}
\caption[Tau branching ratios.]{\label{tab:taubr} \small Tau lepton main
decay modes with the corresponding branching ratio (BR), rounded to nearest
integer. $X$ is the final state for $\tau\to X \anut$. $h$ stands for
charged hadron and in general means a $\pi^\pm$, although K$^\pm$ is also 
possible. Neutral pions decay promptly to two photons and charged pions are
detected in the HCAL. 
The one-prong decay modes add up to 85\%, the three-prong to 15\%
and the five-prong to $\sim$$0.1\%$. Adapted from Ref.~\cite{pdg}. }
\end{table}

%

Tau identification is performed by clustering the energy flow
particles into two or four jets, depending on the analysis, with the Durham
algorithm~\cite{Dokshitzer:1991fc}. A jet will be a tau-jet candidate if it
contains one or three good tracks (or two if it contains an identified
electron, to allow for asymmetric photon conversions) and if the jet invariant
mass is less than 2$\gev$. A cut on the cluster-radius as a function of the
momentum is also available. 

It is important to define well the structure of jets, be it in Higgs, W or
$\tau$ decays. Several algorithms exist that operate by joining adjacent
energy flow objects to form a jet. The Durham jet clustering algorithm used
here, described in Ref.~\cite{Dokshitzer:1991fc}, uses the 
\emph{invariant mass} $M_{ij}$ of two objects $i$ and $j$ as a proximity
measure. 
The procedure is as follows:
\begin{enumerate}
\item all possible pairs between particles are formed and the quantity
  \begin{equation}
     M_{ij}^2 = 2[\min(E_i,E_j)]^2(1-\cos\theta_{ij})
  \end{equation}
  is evaluated, where $E_i$ and $E_j$ are the energies of the objects and
  $\theta_{ij}$ is the angle between them. 
\item the pair with the minimum value of $M_{ij}^2$ is replaced by a
pseudo-particle with energy $E_i+E_j$ and momentum $\p_i+\p_j$
\item steps 1 and 2 are iterated over the set of pseudo-particles and
remaining objects until
\begin{equation}
y_{ij}\equiv\frac{M_{ij}^2}{E_{\rm vis}^2} > y_{\rm cut} 
\label{eq:durycut}
\end{equation}
\end{enumerate}

Thus the parameter $y_{\mathrm{cut}}$
sets the proximity condition for two tracks to be grouped together. 
The higher the value of $y_{\mathrm{cut}}$, the easier it
is to form jets with the available particles: so a larger $y_{\rm cut}$
will give fewer jets. 
The definition of $M_{ij}$ is such that while the separation $\theta_{ij}$
is important, the association of two objects is still possible if the value
$\min(E_i,E_j)$ is sufficiently low. 

For instance, if we want to force the event into four jets, the value
$y_{\mathrm{cut}}$ for which the event is grouped from three into four jets
receives the name of $y_{34}$. 
A genuine four jet event will generally have a large value of $y_{34}$,
thus tightening the value of $y_{34}$ will reject processes that do not
generally produce four jets.  

\section{Event Simulation}
\label{evsim}
Any analysis in searches for new particles requires two event samples to be
generated. One is the signal process we are interested in, for example
slepton pair-production, and the other is the set of all known Standard
Model processes that could mimic the signal process in the
detector, the background. 
Finally, of course, there is the actual data collected with the
detector which will decide if the signal has been found in the data and is
distinguishable from background. 
Monte Carlo (MC) event samples for signal and background are
generated with specific physics generators and processed in ALEPH in four
stages.  
\begin{enumerate}
\item The physics generator emulates $\ee$ collisions and produces the
final state particles and four-momenta of the process or processes in
question. The output is compliant with a set of kinematic rules imposed by
KINGAL, an ALEPH program that will generate the kinematics of all particles
in the final state to simplify further processing. This information is then
passed onto the next stage,
\item GALEPH, the ALEPH detector simulation program, based on
GEANT3~\cite{Brun:1987ma}, which simulates the response of the ALEPH detector
to the final state particles. GALEPH~\cite{galeph} contains a detailed 
and updated description of the geometry and composition of the detector and
 simulates the interactions of the generated particles with the
sub-detectors on their passage through the different materials. 
The output format of GALEPH is the same as a real event. 
\item As for real data, the next stage is the reconstruction program,
JULIA~\cite{julia}, which facilitates the later analysis by forming
physical objects. Unlike data, the reconstructed Monte Carlo files
contain information on the true simulated event, thus enabling useful
comparisons to be performed.
\item The final stage is the analysis itself where, based on the process
under study, a set of variables is calculated for each event to try and
make it distinguishable from the background and provide any necessary
measurement. This is done with ALPHA, the ALeph PHysics Analysis
package~\cite{alpha}.  
\end{enumerate}

Usually, background files are generated and made available for the
collaboration. In the analysis described here, background files were not
generated by the author. 

\thispagestyle{empty}
\chapter{Searches for six-lepton final state topologies in GMSB}
\label{sixl}
\begin{center}
\begin{spacing}{1.5}
\setlength{\fboxsep}{5mm}
\begin{boxedminipage}[tb]{0.9\linewidth} \small
The $\stau_1$ NLSP scenario is favoured in GMSB models with the number of
messenger families greater than one. In this case, pairs of selectrons and
smuons may be produced at LEP2 and decay immediately to final states with
two leptons and four taus and missing energy. This type of decay depends
crucially on the mass difference between the sleptons and staus, but no less
important is the neutralino mass, which enters the process as a mediating
particle either on shell (two-body decays: $\slR \to l\neu \to
l\tau\stau_1$) or off-shell (three-body decays: $\slR \to l\tau\stau_1$). 
Both processes are studied in
this chapter. Up to now, only evidence for two-body decays has been looked
for at colliders. The analysis described in
this thesis is the first experimental search for the three-body process. 

This final state topology is interesting because it covers a very sensitive
area of parameter space, specially in the low gravitino mass range, i.e. for
negligible stau lifetimes. Other searches, for cascade decays of neutralinos
or direct slepton decays, are unable to cover this region of
parameter space and thus the appeal to see if Nature has realised SUSY in
this region. 

After an introduction to the decays and the motivation for the search, the
experimental searches are described in detail and since no evidence for
this process has been found in the highest energy ALEPH data, limits on the
cross section  and sparticle masses are set. 
\end{boxedminipage}
\end{spacing}
\end{center}
\newpage
\section{Phenomenology}
\begin{figure}[bt]
\begin{center}
\includegraphics[width=0.6\linewidth]{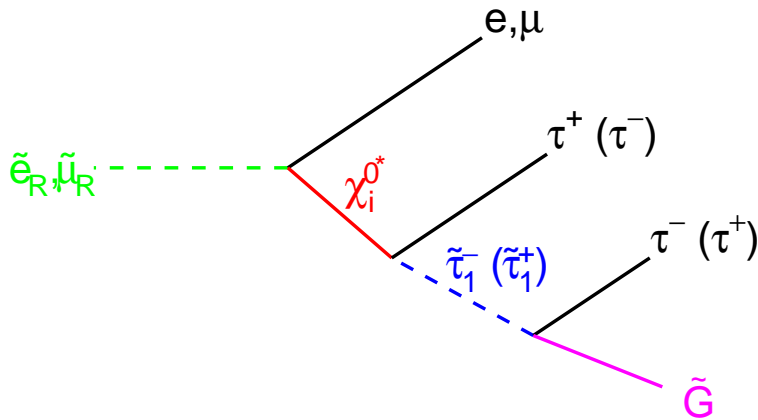}
\caption[Slepton decay to a lepton, tau and stau]
{\label{fig:3bdecay}{\small The decay of selectrons and smuons
when the lightest stau is the NLSP,
can occur via a two-body decay to an on-shell neutralino, or via a
three-body decay directly to the NLSP through a virtual neutralino. Both
cases are addressed in this work.}}
\end{center}
\end{figure}

In the context of Gauge Mediated SUSY Breaking (GMSB) models, final states
with six leptons may appear when the lightest stau $\stau_1$ is the NLSP
and pairs of lightest selectrons $\selR$ or smuons $\smuR$ are
produced. The pair-production of sleptons at LEP is shown in
Fig.~\ref{fig:lep2prod}a. The decay chain under study in this thesis is
shown in Fig.~\ref{fig:3bdecay}. 

The final state is thus two leptons $l$, either electrons or muons with
opposite sign, four taus, and missing energy carried away by the undetected
gravitinos. In the case of a light gravitino, $m_{\grav} < 10\evcc$,
the NLSP will decay promptly after production. The analysis presented here
will be restricted to this case. If the gravitino mass is between 10 and around
$500\evcc$, the $\stau_1$ will present some sizeable decay length in the
detector and the topology would then have kinks or large impact parameters
from the $\stau_1$ decays. This analysis `with lifetime' has been performed
by OPAL in Ref.~\cite{opalgmsb}. 

The decay of the selectron or smuon $\slR$, occurs via a
neutralino. It has usually been assumed in experimental searches that the
neutralino would be on-shell with a mass range between:
\begin{equation}
\label{mchi_2B_constraint}
\renewcommand{\fboxsep}{5pt}
\fbox{$m_{\slR} - m_l > m_{\chi} > m_{\stauO} + m_{\tau}$}
\end{equation}	
where, $l$ stands for e or $\mu$.
And hence the decay would follow two-body kinematics:
\begin{equation}
\label{2bdec}
\mathrm{Two-body:}~\ee \to \slR\:\slR \to l\chi\:l\chi \to
l\tau\stau_1\:l\tau\stau_1 \to l\tt\grav\:l\tt\grav 
\end{equation}

But specially important is the fact that this decay can proceed
through an off-shell neutralino with the special kinematics of
three-body decays:
\begin{equation}
\label{mchi_3B_constraint}
\renewcommand{\fboxsep}{5pt}
\fbox{$m_{\chi} > m_{\slR} - m_{l} >  m_{\stauO} + m_{\tau}$}
\end{equation}
\begin{equation}
\label{3bdec}
\mathrm{Three-body:}~\ee \to \slR\:\slR \to l\tau\stau_1\:l\tau\stau_1 \to l\tt\grav\:l\tt\grav 
\end{equation}

Three-body decays of selectrons or smuons into the lepton, tau and stau
have been theoretically studied in Ref.~\cite{Ambrosanio:1998bq} and
this thesis is the first experimental search for this particular type of
decay. In this case, there is the advantage of eliminating the upper
constraint on the neutralino mass (left hand side
in Eq.~\ref{mchi_2B_constraint}), which can now lie well above the
$\slR$ mass or the beam energy at LEP. This search is therefore
sensitive to an unexplored range of neutralino masses at LEP2. 	

\begin{figure}[tb]
\begin{center}
\includegraphics[width=0.49\linewidth]{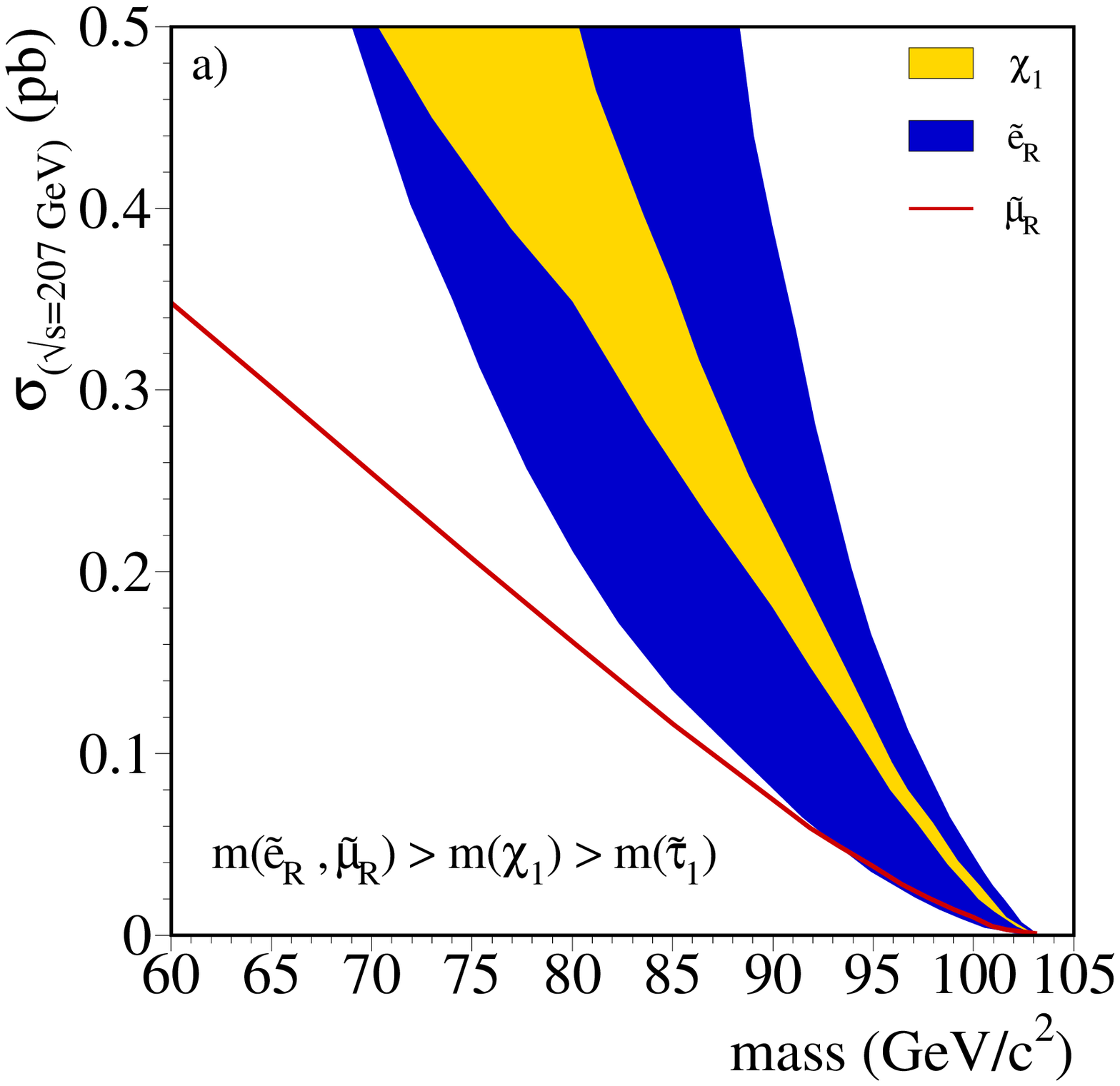}
\includegraphics[width=0.49\linewidth]{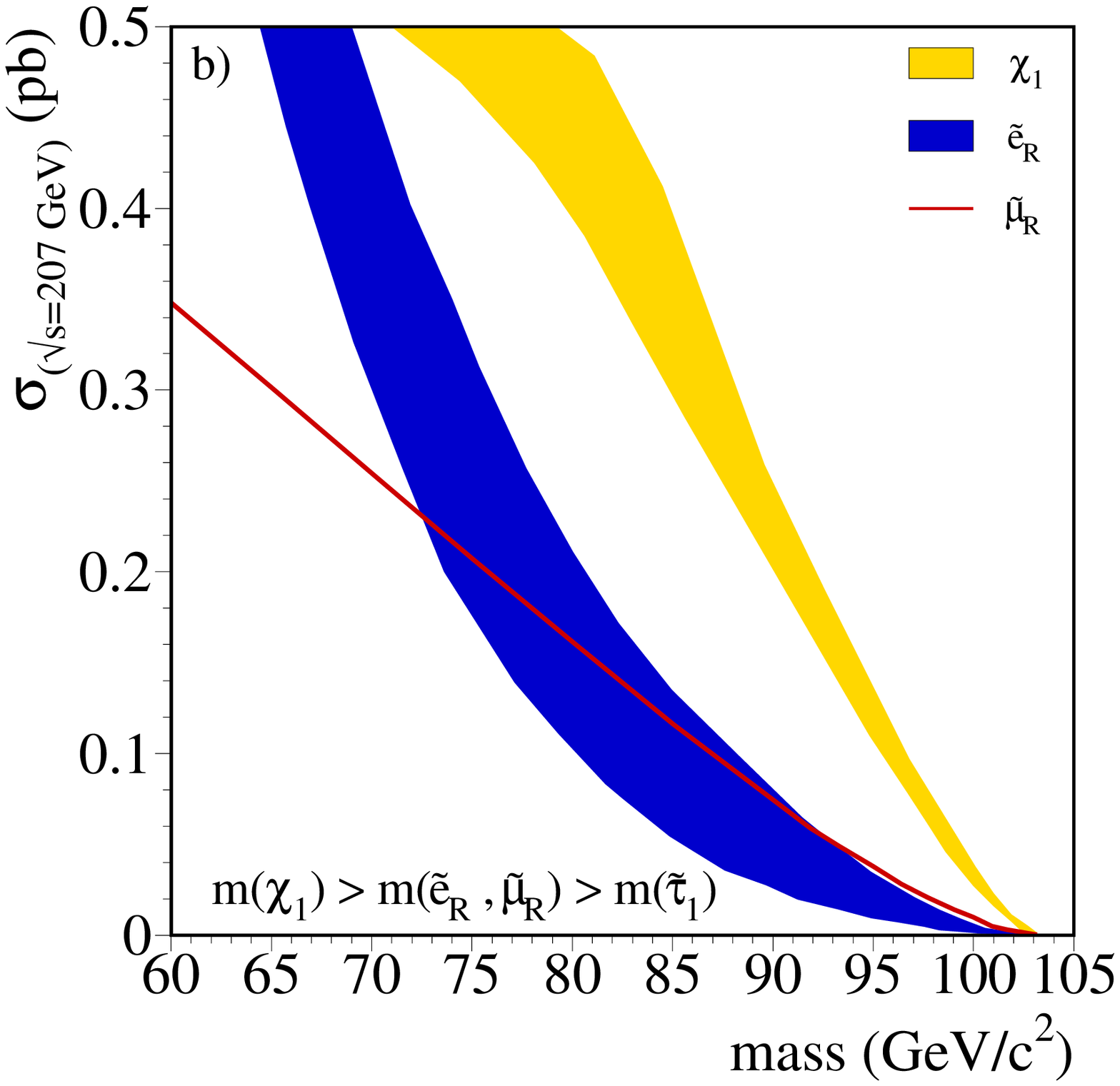}
\caption[Sparticles pair-production cross sections at $\roots = 207\gev$ as
a function of their mass]
{\label{fig:sixl-cs}{\small Pair-production cross sections at
$\roots = 207\gev$ for the lightest neutralino, selectron and smuon in
the (a) two-body  and (b) three-body  scenarios. 
Selectrons (dark shaded
area) may have larger production cross sections than neutralinos (light shaded
area) in the two-body case, but never in the three-body 
case.  The lightest stau production
cross section, not relevant for six-lepton topologies, is very similar to
the smuon cross section since they do not have $t$-channel contributions. 
}} 
\end{center}
\end{figure}

Figure~\ref{fig:sixl-cs} shows the production cross section as a function
of particle mass for neutralinos, selectrons and smuons in the two-body and
three-body scenarios. The $\smuR^+\smuR^-$ production cross section is only
larger than that of $\selR^+\selR^-$ for masses greater than $\sim$$90\gevcc$ in
the three-body scenario. 
So one may expect more $\mu^+\mu^-\tt\stau_1\stau_1$ events
than $\ee\tt\stau_1\stau_1$  with three-body kinematics, although this
is not guaranteed. Smuons are only produced via $s$-channel exchange of a
$\gamma$ or Z, while selectron and neutralino pair-productions are model
dependent via the additional $t$-channel exchange of a
neutralino and a selectron respectively, see Fig.~\ref{fig:lep2prod}. 
The $t$-channel contribution to
the neutralino cross section is constructive, hence the generally higher
neutralino cross section in the three-body case with respect to the 
two-body case, where the exchanged selectron is heavier than the neutralino. 
On the contrary, the interference between $s$ and $t$ diagrams is
destructive in the selectron production, so that light neutralinos cause
the $t$-channel to dominate and reduce the total cross section. 

As a result, three-body $\slR$ decays are disfavoured at the production
level with respect to neutralino production with a slepton NLSP.
The decay of selectrons and smuons with a stau NLSP has to compete in the
available parameter space with direct slepton NLSP production ($\slep \to
\ell\grav$) and neutralino production with a slepton NLSP ($\chi \to \slep\ell \to
\ell\ell\grav$).  
However, the six-lepton final state topology offers a complementary
channel for the discovery of supersymmetry in GMSB models and is sensitive
to exclusive areas of parameter space.  
A discussion on the available parameter space in GMSB models for this type
of topology follows. 

\subsection{Available parameter space}
\begin{figure}[p]
\begin{center}
\includegraphics[width=0.99\linewidth,height=0.6\textheight]{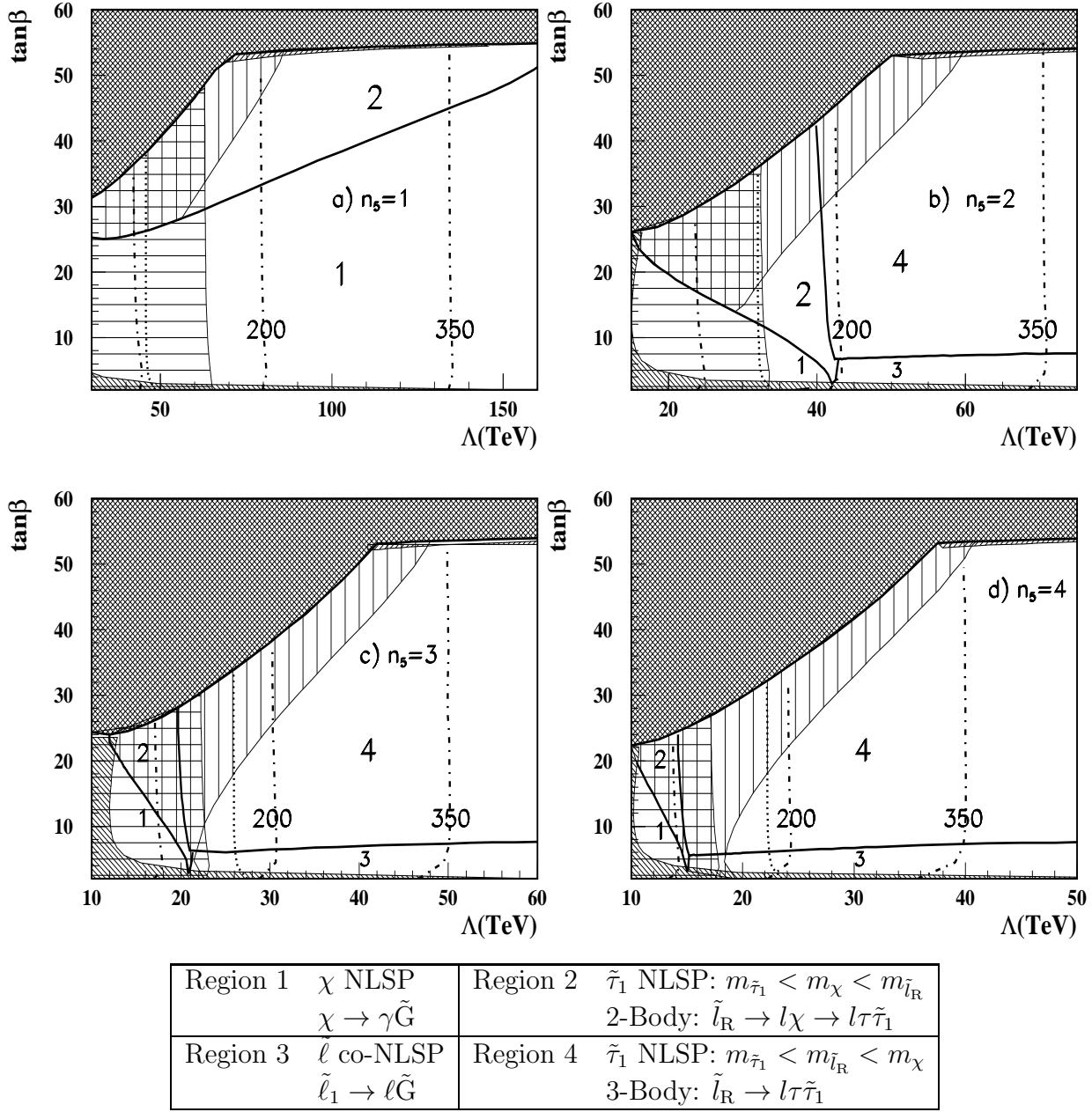}
\begin{tabular}{|ll|ll|} \hline
Region 1 & $\chi$ NLSP & 
Region 2 & $\stau_1$ NLSP: $m_{\stauO} < m_{\neu} < m_{\slR}$ \\
  & $\chi \to \gamma \grav$ & & 2-Body: $\slR \to l\chi \to l\tau\stau_1$ \\ \hline
Region 3 & $\slep$ co-NLSP  & 
Region 4 & $\stau_1$ NLSP: $m_{\stauO} < m_{\slR} < m_{\neu} $\\
  & $\slep_1 \to \ell \grav$ & & 3-Body: $\slR \to l\tau\stau_1$ \\ \hline
\end{tabular}
\caption[Different NLSP regions in the ($\Lambda$,$\tanb$) plane]
{\label{fig:l-tb_avail}{\small In the plane ($\Lambda$,$\tanb$)
and for $\N = 1,2,3,4$ (a, b, c and d respectively), 
regions of $\chi$ NLSP (1), $\slep$ co-NLSP (3) and $\stau_1$ NLSP with
$m_{\chi} < m_{\slR}$ (2) and  $m_{\neu} > m_{\slR}$ (4).
$\Mmess$ has been fixed at 3$\Lambda$ and $\mu$ is positive. 
The heavy solid lines denote the boundaries between these regions. 
Regions 2 and 4 are those relevant for the six-lepton final state
topology, where it has to compete with $\chi \to \tau\stau_1$ and
direct production processes.
The cross-hatched region at high $\tanb$ is excluded because
electroweak symmetry is not correctly broken. The shaded regions are
excluded because $m_{\stau} \leq 76\gevcc$ (vertical shading), 
$m_{\neu} \leq 84\gevcc$
(horizontal shading) or $m_h \leq 95\gevcc$ (or $m_A \leq 83\gevcc$)
(diagonal shading). The dot-dashed contours are where the chargino mass
is 95, 200 or 350$\gevcc$, while the dotted line is the contour of
$m_{\selR} = 90\gevcc$. From Ref.~\cite{Baer:1999tx}.}}
\end{center}
\end{figure}

In GMSB models with $\N > 1$ the $\stau_1$ NLSP scenario dominates. This
can be clearly seen in Fig.~\ref{fig:l-tb_avail}, where Region 1,
corresponding to a $\chi$ NLSP is only dominant for $\N < 2$. Regions 2 and
4 correspond to a $\stau_1$ NLSP scenario, where the
six-lepton final state topology is possible. Three-body decays of
$\slR$ have more parameter space available (Region 4) than two-body
decays (Region 2) for $\N > 1$. It is also noticeable how Region 4 extends
down to relatively low values of $\tanb \sim 5$ independently of
$\Lambda$. For even lower values of $\tanb$, the $\slR-\stau_1$ mass difference 
becomes less than the tau mass and the three sleptons coexist as co-NLSP
(Region 3). 

\begin{figure}[tb]
\begin{center}
\includegraphics[width=0.49\linewidth]{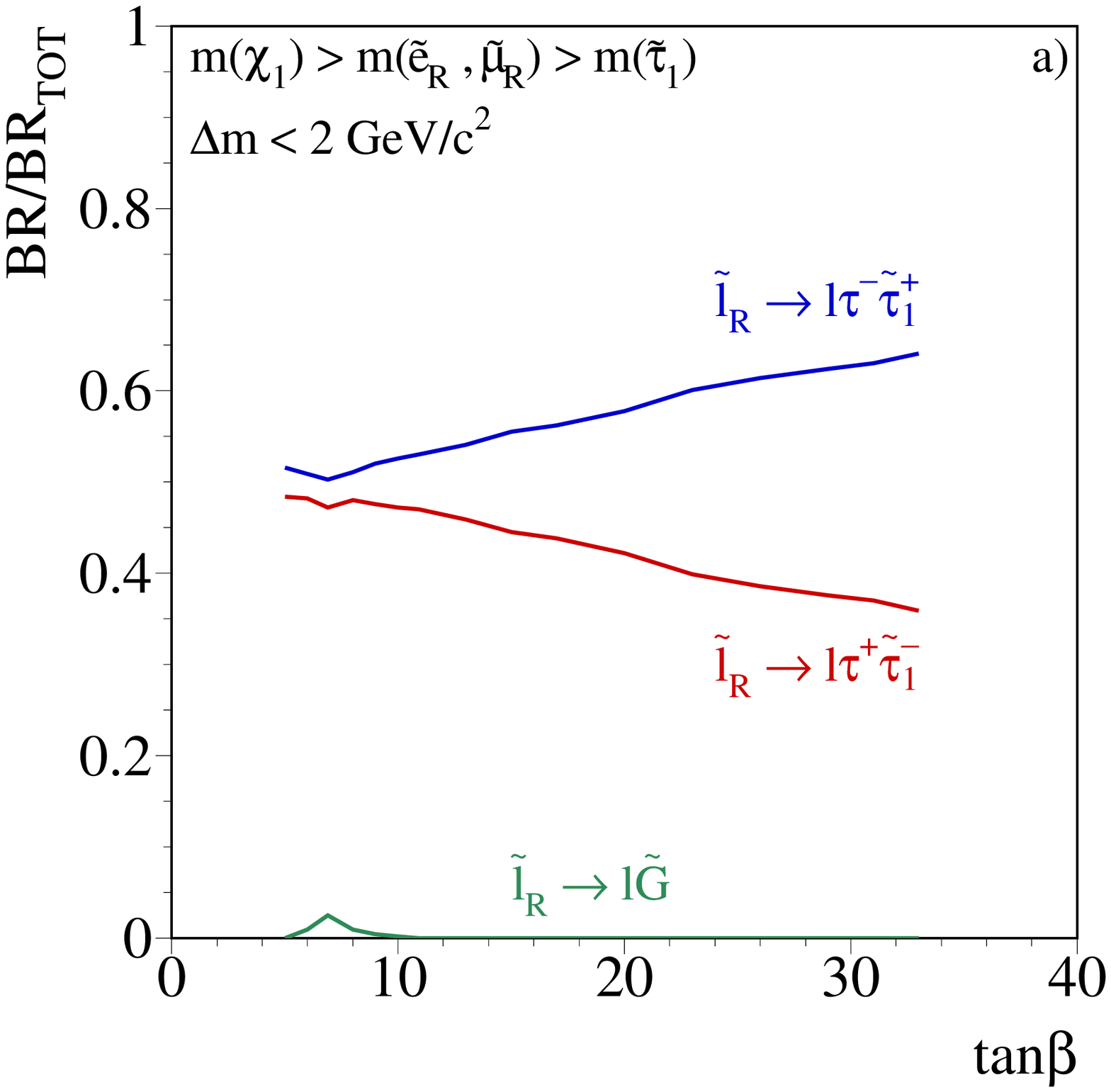}
\includegraphics[width=0.49\linewidth]{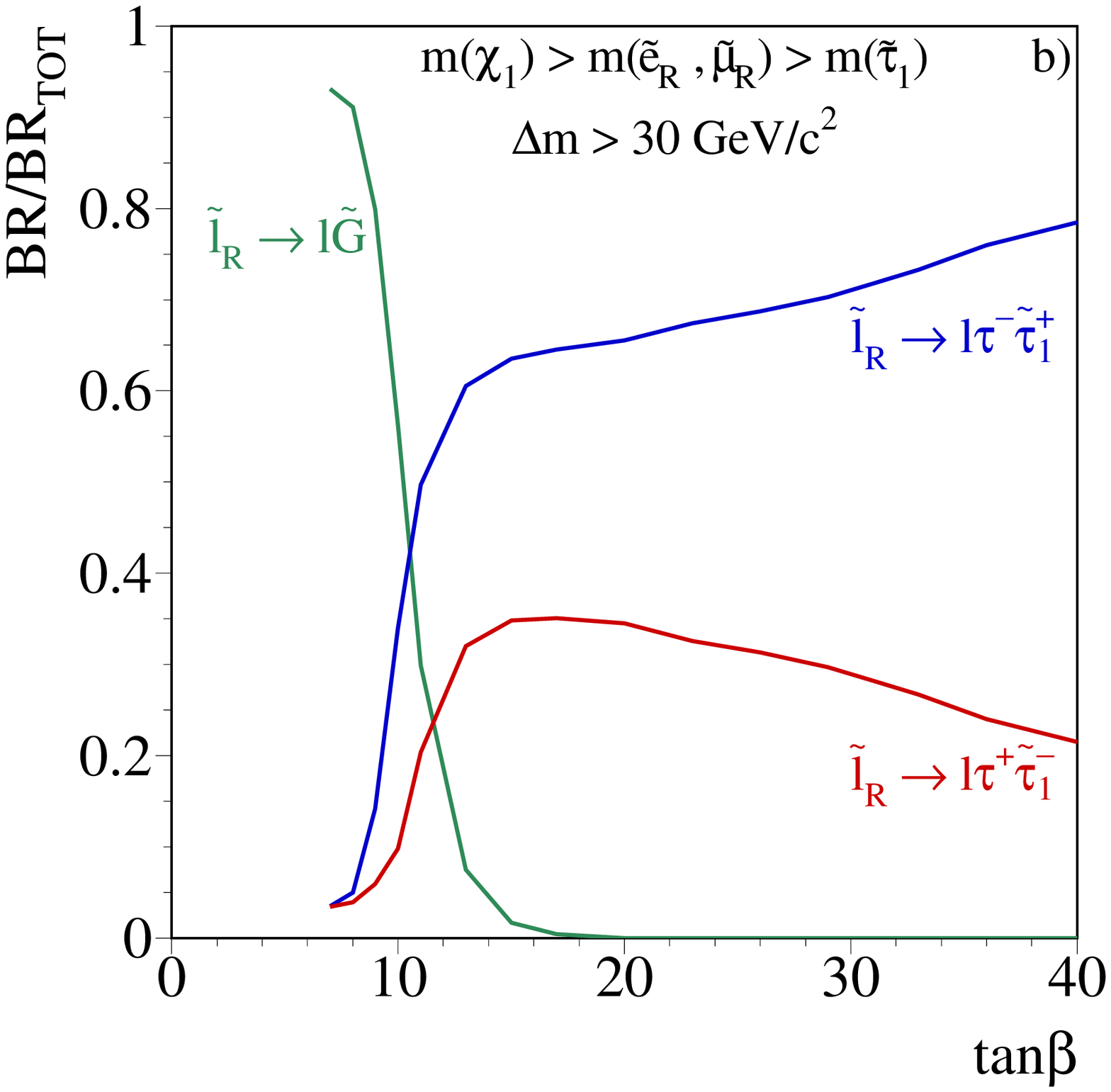}
\caption[Slepton decay modes contribution to the total branching fraction
as a function of $\tanb$]
{\label{fig:3b-br}{\small 
Slepton branching fractions as a function of $\tanb$ in the three-body
scenario for two ranges in $\dm_{\chi\sll}=m_{\neu}-m_{\slR}$. 
Only for highly virtual neutralinos 
($\dm_{\chi\sll} > 30\gevcc$) and $\tanb \leq 10$ (or $m_{\slR}-m_{\stauO}<5\gevcc$) 
may the channel $\slR \to l\grav$ 
be open and dominate. Otherwise the three-body decay modes dominate. 
The asymmetry between $\Gamma(\slR^- \to l^-\tau^-\stau_1^+)$ and
$\Gamma(\slR^- \to l^-\tau^+\stau_1^-)$ is explained in Sec.~\ref{sixlkin}.
The condition $m_{\neu} > m_{\slR}-m_{l} >
m_{\stauO}+m_{\tau}$ is only possible for $\tanb \geq 5$ and $\N \geq 2$.}}
\end{center}
\end{figure}

In the stau NLSP scenario, if the mass ordering is $m_{\slR}-m_{l} >
m_{\chi}$, then the two-body decay $\slR \to l\neu$ dominates over
$\slR \to l\grav$. The latter decay is suppressed by the gravitational
nature of the gravitino. If the mass order is $m_{\neu} > m_{\slR}-m_{l} >
m_{\stauO}+m_{\tau}$ the three-body modes $\slR \to l\tau\stau_1$ will
dominate\footnote{There could also be three-body decays 
$\slR \to \nu_{l}\anut\stau_1$ through off-shell charginos $\chaO$, although they
are strongly suppressed due to the large chargino mass and the very
small coupling of $\slR$ to $\nu_{l}\cha$.} as shown in Fig.~\ref{fig:3b-br}.
The decay to $\slR \to l\neu$ is impossible in this region
of masses, and only $\slR \to l\grav$ may become available in pathological
cases where the slepton mass is very close to the stau mass and the decay
to the stau is thus not possible. Only then the direct decays from sleptons
to gravitinos may be realised overcoming the very low gravitational
coupling. Once $\tanb$ is greater than $\sim$10 the stau mass becomes much
lighter than the
slepton masses due to mixing effects and the three-body decay modes are
restored. It is found that all slepton decays proceed through $\slR \to
l\tau\stau_1$ and $\slR \to l\grav$ is avoided if the difference
between the slepton and the stau masses is greater than $5\gevcc$.

\subsection{Kinematics of two- and three-body {\boldmath$\slR$} decays}
\label{sixlkin}
In the two-body scenario (Eq.~\ref{mchi_2B_constraint}) the mediating neutralino
decays independently into  $\tau^-\stau_1^+$ or $\tau^+\stau_1^-$ with equal
probability. One expects therefore no asymmetry in the charges of the two
staus: there will be equal numbers of like-signed staus and opposite-signed
staus. 
But in the three-body scenario, it is important to distinguish
between the different charge channels, since the decay widths 
$\Gamma(\slR^- \to l^-\tau^+\stau_1^-)$ and $\Gamma(\slR^- \to
l^-\tau^-\stau_1^+)$ can actually be quite different.
This asymmetry is caused by the different coupling of the neutralino to the
stau depending on the stau mixing. If the lightest stau is purely
right-handed, there is no asymmetry and the largest
hypercharge states $\tau^-_{\rm R}$ and $\tau^+_{\rm L}$ are predominantly 
produced in equal amounts as primary taus. 
But if the left-handed stau content of the lightest stau
increases, i.e. the mixing angle $\theta_{\stau}$ increases, then the wino 
($\tilde{\W}_3$) nature of the neutralino will prefer the state with more 
$\tau^-_{\rm L}$ content and the asymmetry will be created.

Assuming a fully bino neutralino, which is generally almost true in
GMSB models, and a near threshold production (small $\slR$ momenta), one
can deduce the following relations\footnote{After some algebra from the 
calculations in Ref.~\cite{Ambrosanio:1998bq}}: 
\begin{equation}
\label{wid_tmstp_approx}
\Gamma(\slR^- \to l^-\tau^-\stau_1^+) \sim
\frac{m_{\slR}^5}{(m_{\neu}^2-m_{\slR}^2)^2}
\left [ \sin\theta_{\stau} +
\frac{m_{\neu}}{m_{\slR}}\cos\theta_{\stau} \right ]^2
\end{equation}
\begin{equation}
\label{wid_tpstm_approx}
\Gamma(\slR^- \to l^-\tau^+\stau_1^-) \sim
\frac{m_{\slR}^5}{(m_{\neu}^2-m_{\slR}^2)^2}
\left [ \cos\theta_{\stau} +
\frac{m_{\neu}}{m_{\slR}}\sin\theta_{\stau} \right ]^2
\end{equation}

Both the `slepton-charge flipping' channel (Eq.~\ref{wid_tmstp_approx})
and the `slepton-charge preserving' decay (Eq.~\ref{wid_tpstm_approx}) are
suppressed for increasing neutralino mass, with other parameters held
fixed. 
Figure~\ref{fig:3b-br} shows the asymmetry between the two decay modes 
as a function of $\tanb$, which is roughly inversely proportional to the
stau mixing angle $\theta_{\stau}$, and as a function of
$\dm_{\chi\sll}=m_{\neu}-m_{\slR}$. It is clear from the plots that:
\begin{equation}
\renewcommand{\fboxsep}{5pt}
\fbox{$\Gamma(\slR^- \to l^-\tau^-\stau_1^+) ~\geq~ \Gamma(\slR^- \to
l^-\tau^+\stau_1^-)$}
\end{equation}
so that same-sign
$\tau^{\pm}\tau^{\pm}\stau_1^{\mp}\stau_1^{\mp}$ states are suppressed
compared to opposite-sign states $\tau^+\tau^-\stau_1^+\stau_1^-$.
This effect will be most significant for large $m_{\neu}/m_{\slR}$ and
for small negative values of $\cos\theta_{\stau}$. For
$m_{\slR}-m_{\stauO} < 10\gevcc$, $|\cos\theta_{\stau}|$ usually ranges
between 0.1 and 0.3 (small mixings). 

\begin{figure}[tb]
\begin{center}
\includegraphics[width=0.46\linewidth]{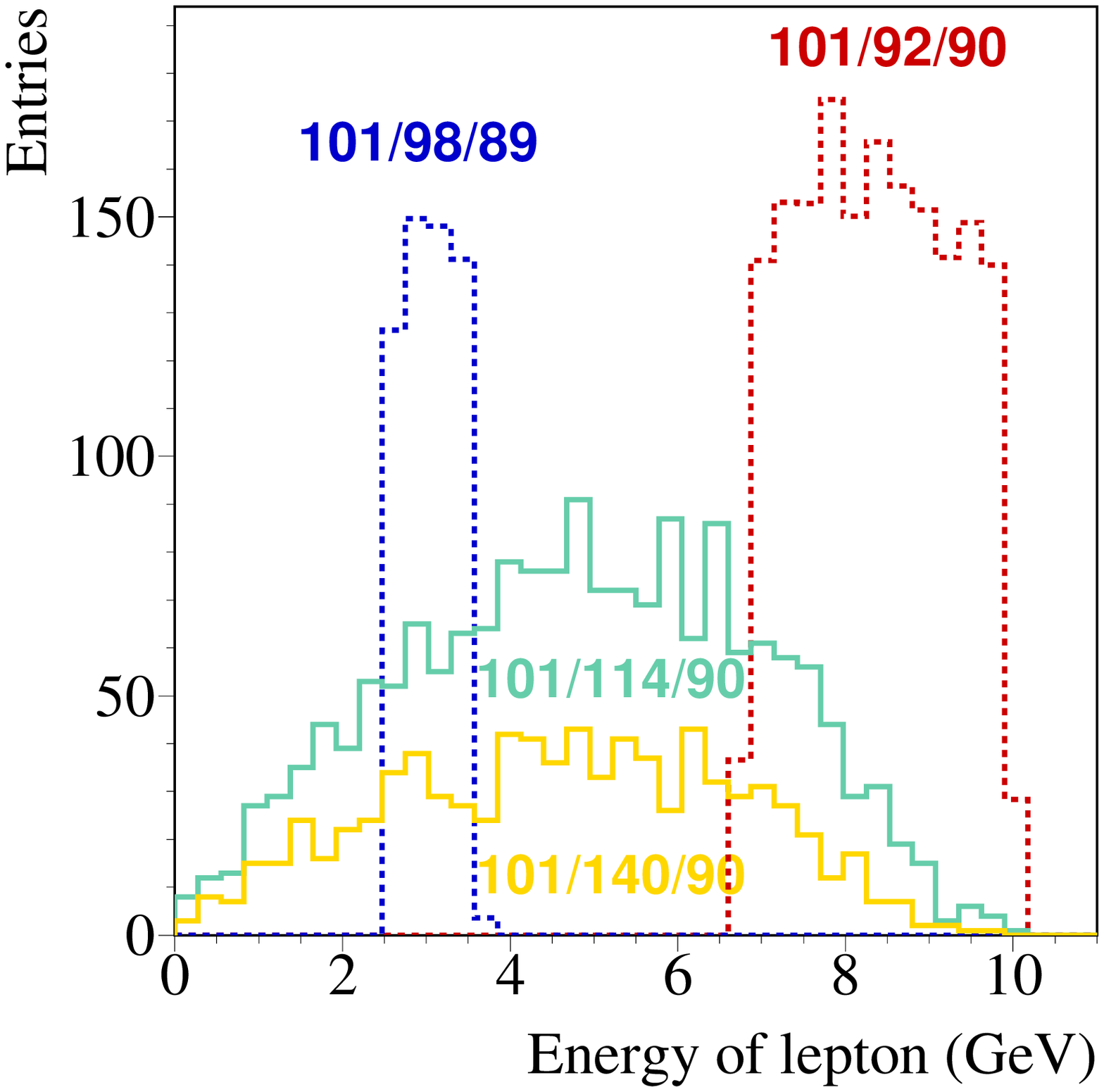} 
\hspace{0.5cm}
\includegraphics[width=0.46\linewidth]{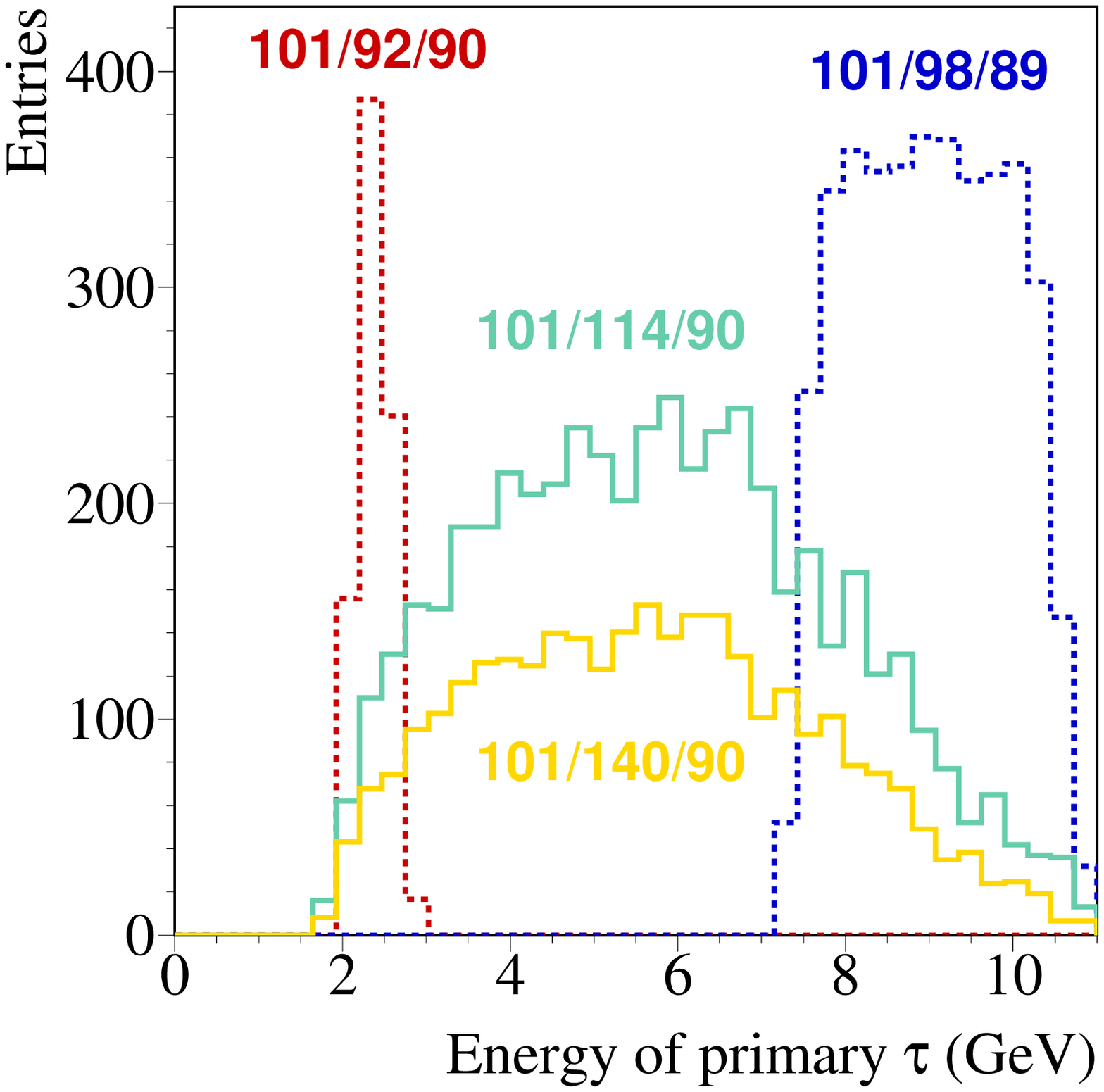} \\
\caption[Energy distributions for the lepton and the primary tau in two-
and three-body scenarios]
{\label{fig:kinem}{\small
Energy distributions, with arbitrary vertical axis, 
for lepton (electron or muon) and primary tau at the generation level.  
The dotted histograms represent the two-body decay modes $\slR \to l\neu$
and the full histograms, the three-body decays $\slR \to l\tau\stau_1$.
Signal points are denoted by: $m_{\slR}/m_{\neu}/m_{\stauO}$. In
both cases $\dm = m_{\slR} - m_{\stauO}$ is fixed at $\sim$$11\gevcc$
and the neutralino mass is changed to see the effect.}} 
\end{center}
\end{figure}

The kinematics of the event will ultimately depend upon the mass
differences: $m_{\slR}-m_{\stauO}$, $|m_{\slR}-m_{\neu}|$ and 
$m_{\neu}-m_{\stauO}$, neglecting lepton masses.
Figure~\ref{fig:kinem} shows the energy
distributions for the lepton and the primary tau in both two- and
three-body decays with $m_{\slR}-m_{\stauO}$ fixed at $\sim$$11\gevcc$. 
In the two-body scenario, as one would expect, the energy of the primary
lepton is fully dependent on the $\slR-\neu$ mass difference whereas the
energy of the tau lepton depends on $m_{\neu}-m_{\stauO}$. 
But in the three-body scenario, these energy distributions prove to be
completely independent of the neutralino mass: i.e., no matter how virtual
($m_{\neu}-m_{\slR}$) the neutralino is, the lepton and primary tau
energies are determined only by the total available energy $\sim$$\Ebeam-m_{\stau}$.

The analysis described here will assume negligible stau lifetime ($m_{\grav}
< 10 \evcc$). Could the sleptons nevertheless have a long lifetime?
The probability that a slepton $\slR$ with energy $E$ will travel a distance
$x$ in the lab frame before decaying is given by:  
\begin{equation}
P(x) = e^{-x/\lambda} ~~~ \mathrm{with} ~~~ \lambda=c\tau\gamma\beta=
0.2 \left ( \frac{1\,\mathrm{meV}}{\Gamma_{\slR}} \right )
\sqrt{\frac{E^2}{m_{\slR}^2}-1} ~ \mm
\end{equation}
It can be shown~\cite{Ambrosanio:1998bq} that the decay length
for sleptons produced at LEP2 (where $E$ is the beam energy, and thus
the Lorentz factor is not much greater than 1) is only
measurable if $\dm=m_{\slR} - m_{\stauO}$ is less than one GeV. 
It is therefore assumed in this analysis, without loss of generality, that
the sleptons will not have a sizeable decay length. 

\begin{figure}[p]
\begin{center}
\vspace{-0.6cm}
\renewcommand{\subfigtopskip}{-10pt}
\renewcommand{\subfigcapskip}{-5pt}
\subfigure[\label{fig:dali2B} {\scriptsize
Two selectrons were produced with $99\gevcc$ which then decayed to two
electrons (tracks 2 and 5) and two neutralinos with mass $90\gevcc$. Of the
four produced taus, three underwent a one-prong decay into pions (tracks 1, 3
and 9) and one decayed into three pions (tracks 4, 7 and 8).}
]
{\includegraphics[width=0.65\linewidth,height=0.65\textheight,angle=-90]
{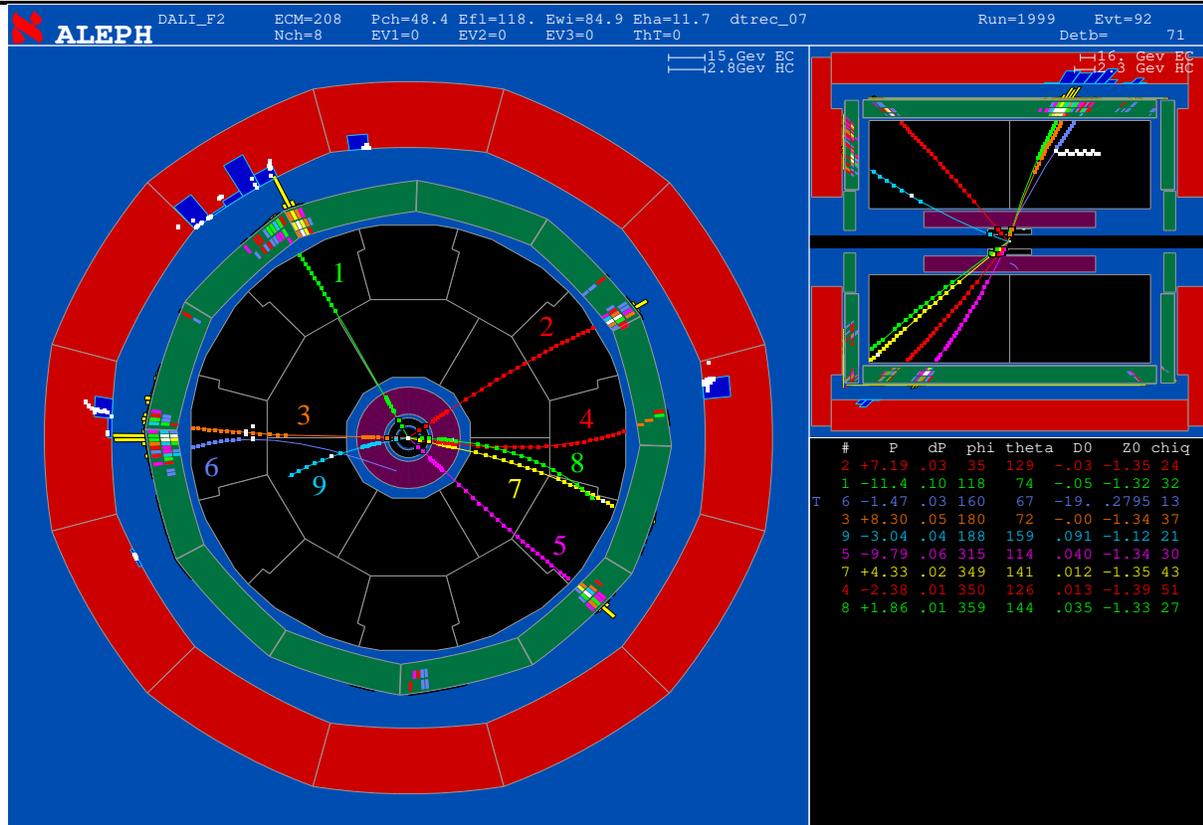}}
\subfigure[\label{fig:dali3B} {\scriptsize
Two smuons decaying via a three-body process. 
The two corresponding muons are tracks 1 and 2. 
One of the primary taus decayed into a $\pi^+$ (track 3) and the other into
three pions (tracks 5, 6 and 7). The secondary taus coming from the staus
decayed into a muon (track 4) and into three pions
(tracks 8, 9 and 10). There is a photon conversion in the inner wall of
the TPC, between tracks 3 and 4.}
]
{\includegraphics[width=0.65\linewidth,height=0.65\textheight,angle=-90]
{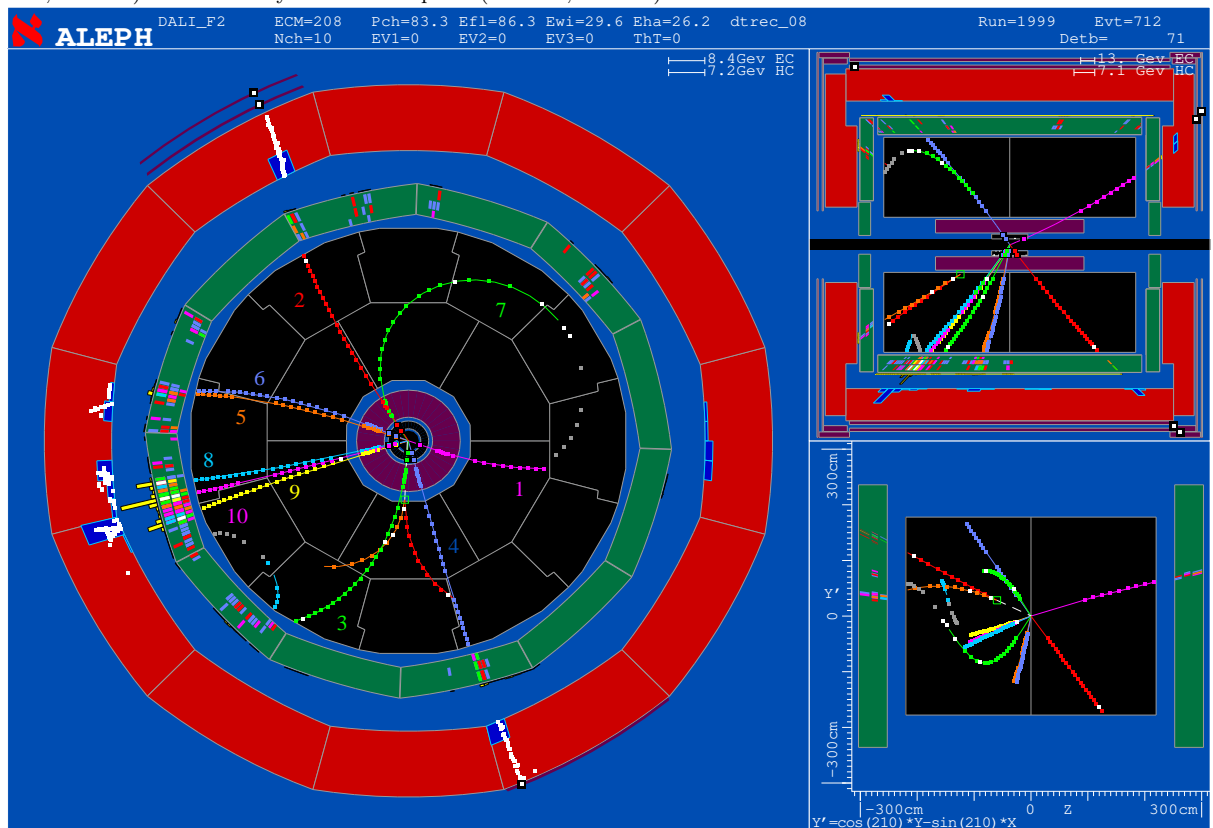}}
\caption[Signal Monte Carlo events at $\roots=208\gev$]
{\label{fig:dalisig}{\small Signal Monte Carlo events at
$\roots=208\gev$. The  detector is shown in the `fish-eye' view to magnify
the tracking relative to the calorimeters.}}
\end{center}
\end{figure}

Examples of how selectron and smuon production with six-lepton final states
might look are shown in Fig.~\ref{fig:dalisig}.

\section{Data sample and Monte Carlo simulation}
Six-lepton events with missing energy in GMSB scenarios constitute 
the events of interest in this analysis, and they are called `signal'
events. Once they have been
described and the new physics behind them is understood, it is necessary to
know how would they look were they to be produced in a collision in
the ALEPH detector. This is performed by means of Monte Carlo simulation. 
The same technique is utilised in generating Standard Model processes that
could mimic or fake the signal events we are interested in. These processes,
arising from known physics, are called `background' events.
A `selection' procedure to distinguish between the two will
be devised such that it retains the maximum number of signal events and
rejects as many unwanted background events as possible.
Finally, the selection is applied to the real data collected by ALEPH. 
A measurement of the degree of evidence for the
signal process under investigation can then be obtained by comparing the
selected number of events in the data and the expected number of events
from the background simulation. 

\subsection{Data sample}
The data analysed in this thesis were collected with the ALEPH detector between
1998 and 2000 with a total integrated luminosity of $628\invpb$ at
centre-of-mass energies between 189 and $209\gev$. Table.~\ref{tab:lumin}
gives the detailed luminosity at each energy. 
The LEP accelerator was pushed to its limits during year 2000 and ALEPH
collected 11.6$\invpb$ of data between 200 and 204$\gev$, 
197.3$\invpb$ between 205 and 207$\gev$ and 7.3$\invpb$ at more than 208 GeV. 
The excellent performance of both the 
accelerator and the detector must be acknowledged.
\begin{table}[htb]
\begin{center}
\begin{tabular}{|c||c|c|c|c|c|c|c|} \hline
Year  & 1998 & \multicolumn{4}{c|}{1999} & \multicolumn{2}{c|}{2000} \\ \hline
$\langle \sqrt{s} \rangle$ (GeV) & 188.6 & 191.6 & 195.5 & 199.5 & 201.6 & 205.0 & 206.7 \\ \hline
$\int \mathcal{L}$dt ($\invpb$)         & 173.6 &  28.9 &  79.9 &  87.0 &  44.4 &  79.5 & 134.3 \\ \hline
\end{tabular}
\caption[ALEPH collected luminosity by centre-of-mass energy bin]
{\label{tab:lumin} {\small Average centre-of-mass energy and
corresponding luminosities of the analysed data sample
for data collected by the ALEPH detector from 1998 to 2000.}}
\end{center}
\end{table}

\subsection{Signal}
The ALEPH SUSY Task Force generally uses SUSYGEN~\cite{susygen2} as the
signal Monte Carlo program for the generation of supersymmetric events in
the simulation. All
searches described in this thesis were performed using SUSYGEN, except the
six-lepton search described in this chapter. SUSYGEN
is unable to produce the type of events where selectrons and smuons
are heavier than the stau and the decay is a three-body mode. 
Instead, ISAJET 7.51~\cite{isajet7.48} which has the full
capability to generate $\slR \to l\tau\stau_1$ events was used. 
ISAJET had to be interfaced with KINGAL to allow the generated events to be
processed by the standard ALEPH reconstruction routines. 

In the GMSB framework, the minimum number of parameters to describe a
complete sparticle spectrum is six: 
$\Lambda$, $\Mmess$, $m_{\grav}$ (or $\sqrt{\rm{F_0}}$), $\N$, $\tanb$ and
sign($\mu$). But phenomenologically there are only three relevant
parameters in the search for slepton decays: 
the masses of the SUSY particles involved, 
$m_{\slR}$, $m_{\neu}$ and $m_{\stauO}$, 
which will completely determine the kinematics of the event.
Instead of an artificial grid with values for the three masses, the scan
over the six theoretical parameters described in Sec.~\ref{sec:scan} was
used to obtain the interesting values for the masses.
Thus having scanned millions of possible GMSB models, the chosen signal
points were obtained by imposing the
appropriate hierarchy in masses (Eqs.~\ref{mchi_2B_constraint}
and~\ref{mchi_3B_constraint}) and the zero 
lifetime condition $m_{\grav} < 10 \evcc$ for the stau NLSP. 

\begin{figure}[tb]
\begin{center}
\vspace{-0.8cm}
\includegraphics[width=0.49\linewidth]{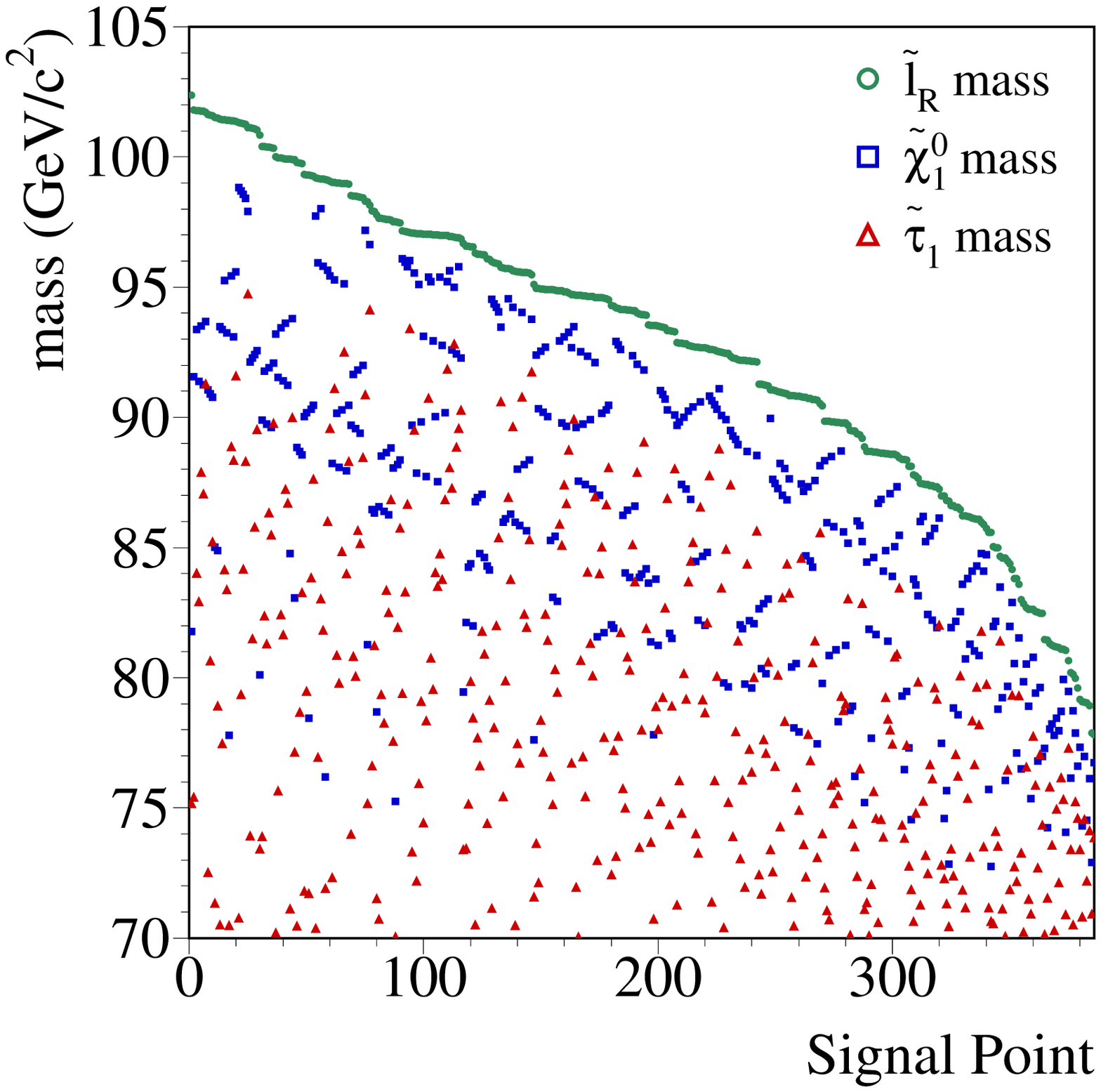}
\includegraphics[width=0.49\linewidth]{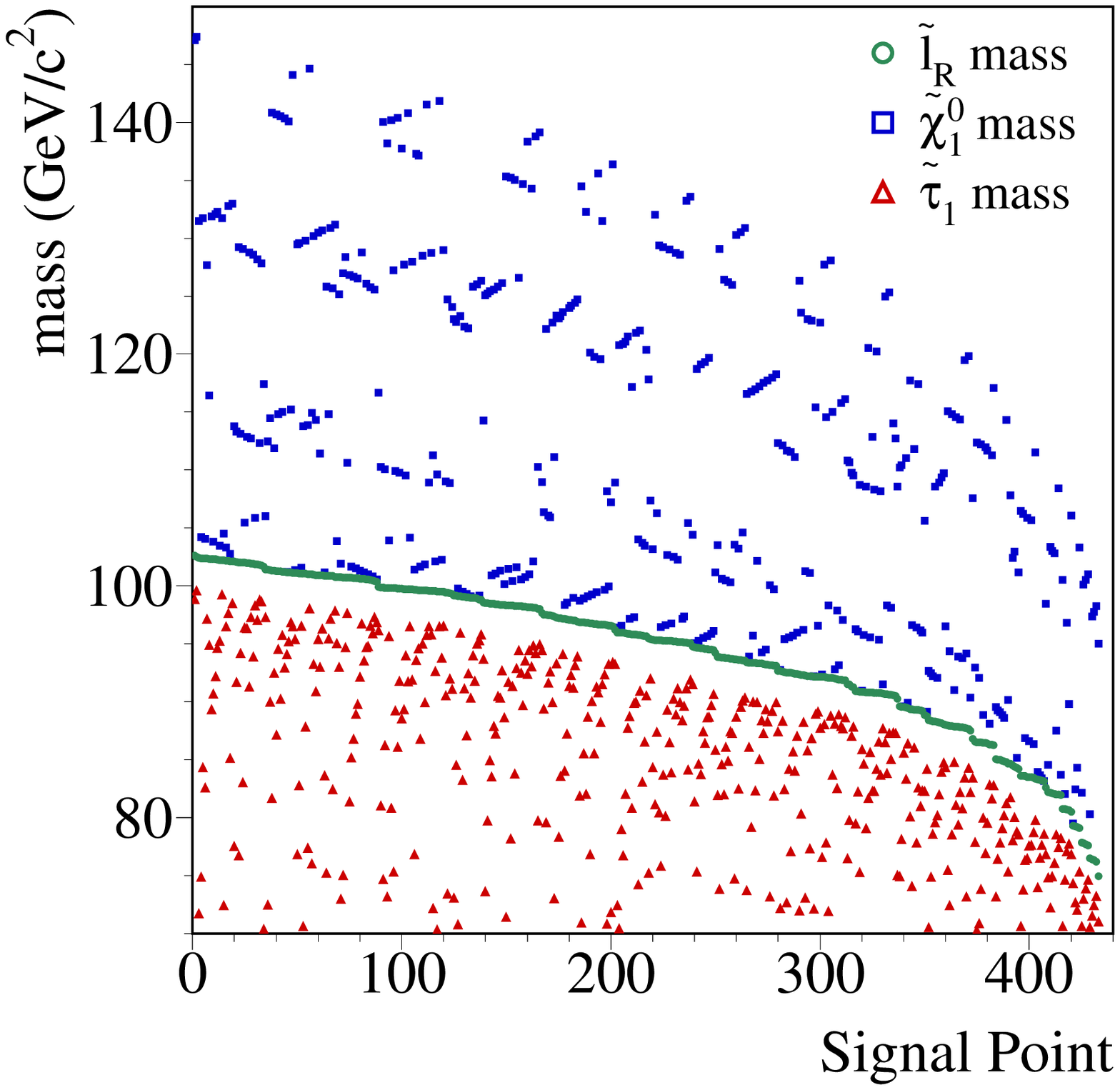} \\
\caption[Values of $m_{\slR}$, $m_{\neu}$ and $m_{\stauO}$ for each signal
point in the two- and three-body scenarios]
{\label{fig:sigpoints}{\small The signal points in GMSB parameter space
were obtained from the scan described in Sec.~\ref{sec:scan}. 
The relevant kinematical
parameters, the mass of sleptons, neutralino and stau, are shown here for
each signal point in the two-body scenario (left) and three-body scenario
(right). In the three body scenario $\dm=m_{\slR}-m_{\stauO} > 5\gevcc$ to
ensure that no slepton decays into $l\grav$. }}
\end{center}
\end{figure}

Some reduction was needed in the number of selected signal points, and this
was performed in the three mass-planes involved $(m_{\neu},m_{\stauO})$,
$(m_{\neu},m_{\slR})$ and $(m_{\slR},m_{\stauO})$ by requiring a separation
between the points of more than $2\gevcc$ in any one plane. A high density
of points was achieved without unnecessary overlapping. Only stau NLSP
masses greater than $70\gevcc$ were used. In Fig.~\ref{fig:sigpoints} the
signal points are ordered with decreasing $\m_{\slR}$ in
the horizontal axis and the values of their three defining masses are shown
in the vertical axis to illustrate the fine coverage of all possible
kinematical regions. 

A total of 385 and 457 points in the minimal GMSB parameter space were
used as signal for the two-body and the three-body scenarios
respectively. For each point, a total of 1000 events were produced at
$\roots = 189$, 206 and $208\gev$. Taking into account the beam energy 
limit in the production of sleptons, this represents a total of 2 million
reconstructed signal events. 

\begin{figure}[ht]
\begin{center}
\vspace{-0.6cm}
\includegraphics[width=0.49\linewidth]{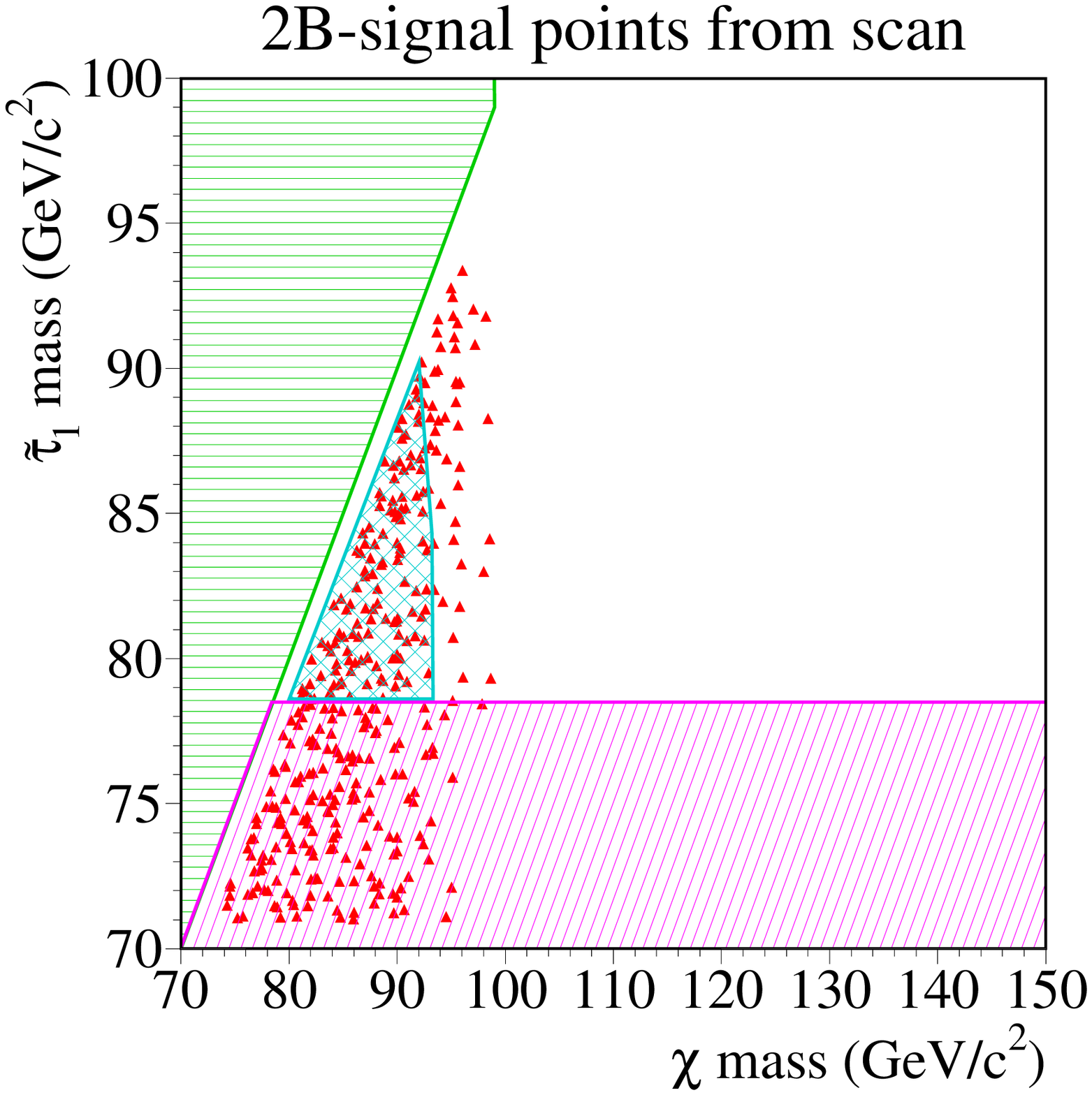}
\includegraphics[width=0.49\linewidth]{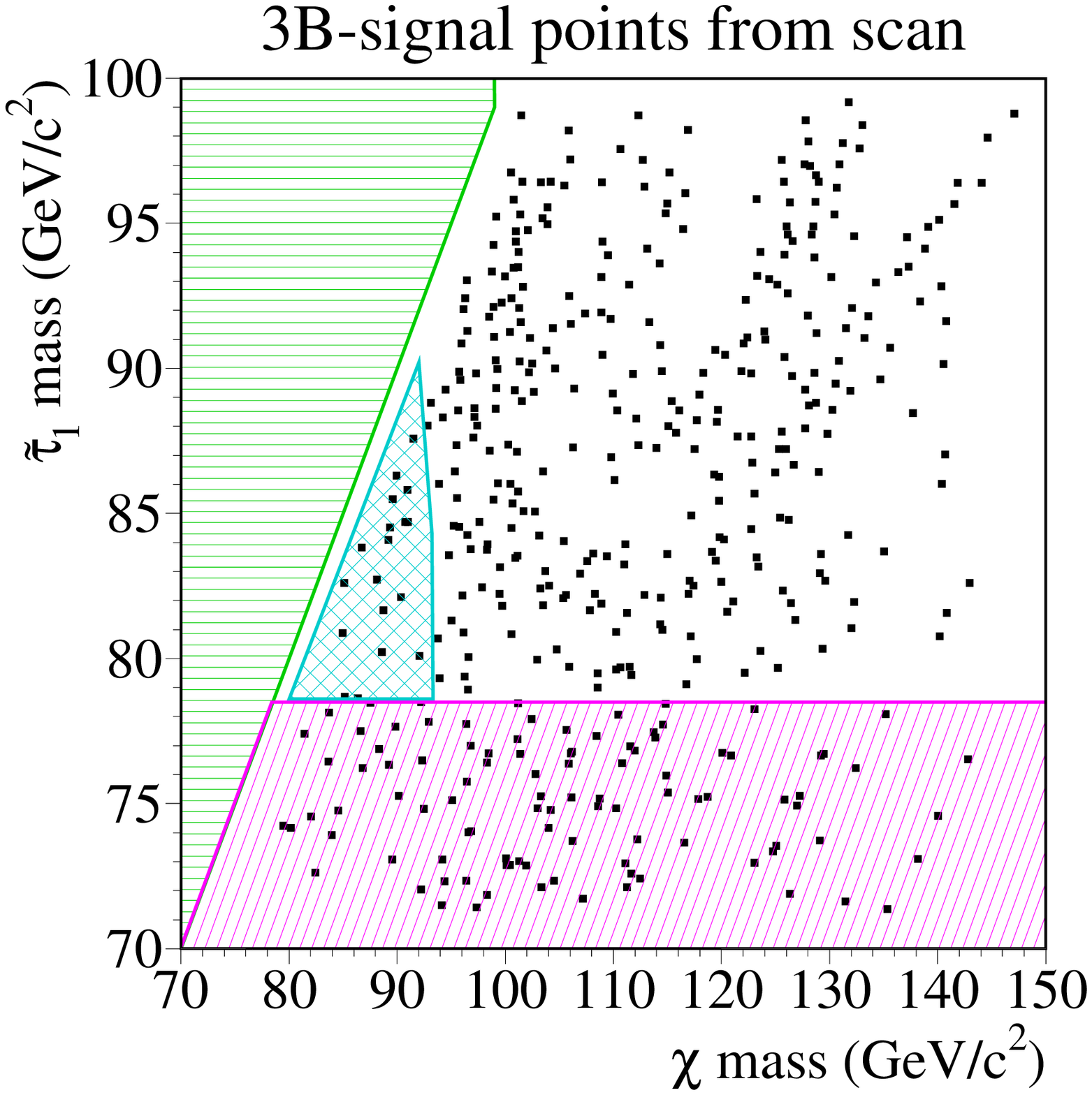} \\
\caption[Distribution of signal points in the $(m_{\stauO},m_{\neu})$ plane]
{\label{fig:neustasig}{\small Signal events were generated at each one of
the two-body (left) and three-body (right) points displayed here in
the neutralino-stau mass plane. The hatched areas mark the present
exclusion from GMSB searches in that plane for $m_{\grav} < 10\evcc$, as
explained in Fig.~\ref{fig:neustau}.}}
\end{center}
\end{figure}
Finally, the simulated signal points in the two- and three- body scenarios
are shown in Fig.~\ref{fig:neustasig} in the $(m_{\neu},m_{\stauO})$
plane. Also shown are the present limits derived from other GMSB
searches (described in Chapter~\ref{gmsbphen}). 
This plot demonstrates the reach
of three-body slepton decays over an area in parameter space where the 
neutralino is heavier than the
beam energy and is not excluded by existing limits. 
This is the main motivation to use this topology and cover that region of
parameter space. 
In the case of two-body
decays, one is limited by the kinematical conditions and is further
constrained by the suppressed slepton production cross section with respect
to neutralino production. 

\subsection{Background}
Standard Model processes have to be simulated and reconstructed in ALEPH in
order to compare data and `known physics' predictions; only when the first
shows disagreement with the latter does the possibility for new physics
arise. It is therefore very important to understand and be able to
simulate correctly all possible processes that could resemble the one
sought. Table~\ref{tab:bkg} lists for each centre-of-mass energy the cross
section, number of reconstructed events and MC generator for each of the
analysed backgrounds. The background generation was performed by other
members of the ALEPH collaboration, corresponding to at least 20 times the
collected luminosity in the data. 

\begin{figure}[tb]
\begin{center}
\includegraphics[width=0.6\linewidth]{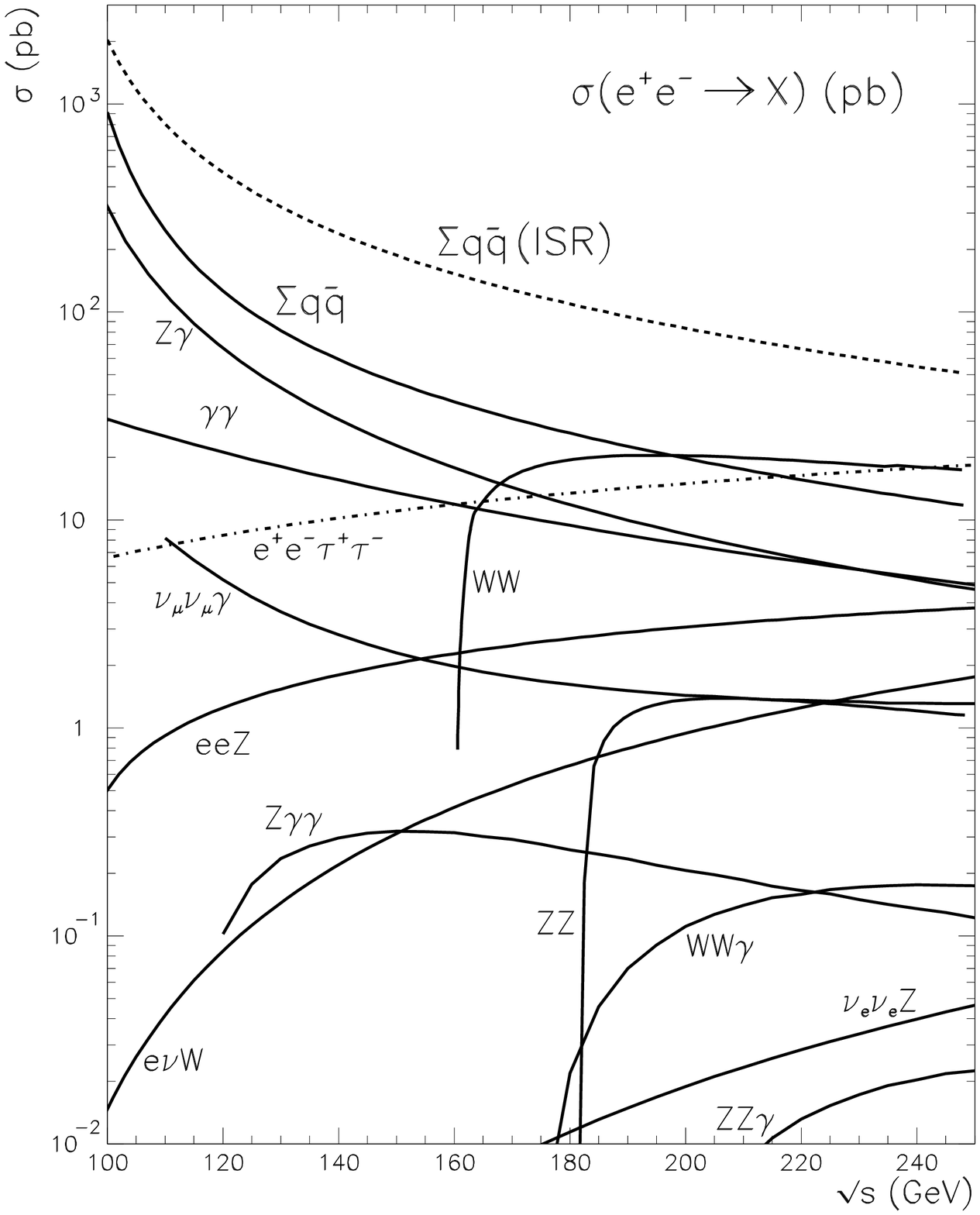}
\caption[Cross sections for Standard Model processes
as a function of the centre-of-mass energy]
{\label{fig:sm_cs}{\small Cross sections for Standard Model processes
as a function of the centre-of-mass energy.  
Taken from Boudjema and Mele in Ref.~\cite{lep2phys}.}}
\end{center}
\end{figure}

Cross sections for some of these
processes at LEP2 energies are shown in Fig.~\ref{fig:sm_cs}. For
comparison, typical SUSY particle cross
sections are $\sim$0.1\,pb or less if the particles are produced at
threshold ($m\sim\roots/2$), as can be seen in Fig.~\ref{fig:sixl-cs}. 
\begin{figure}[p]
\begin{center}
\subfigure[Two-fermion with ISR]
	{\includegraphics[width=0.45\linewidth]{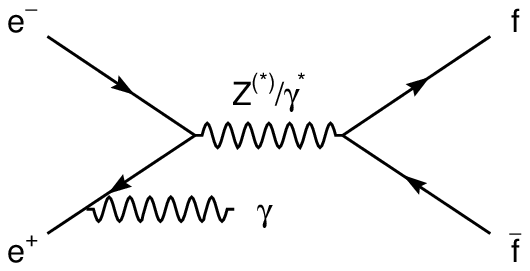}}
\subfigure[$\zz^*$ dominant contribution]
	{\includegraphics[width=0.45\linewidth]{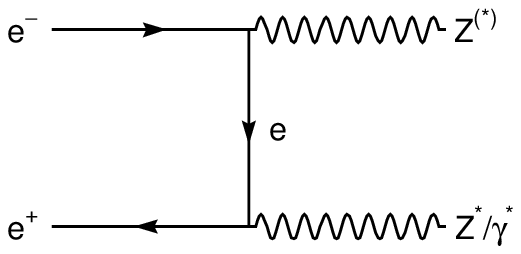}} \\
\subfigure[WW $s$-channel]
	{\includegraphics[width=0.45\linewidth]{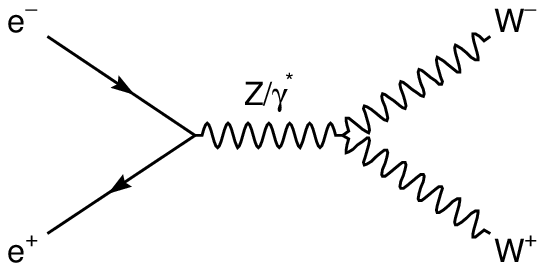}}
\subfigure[WW $t$-channel]
	{\includegraphics[width=0.45\linewidth]{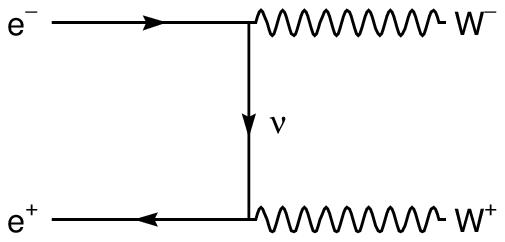}} \\
\subfigure[We$\nu$]
	{\includegraphics[width=0.45\linewidth]{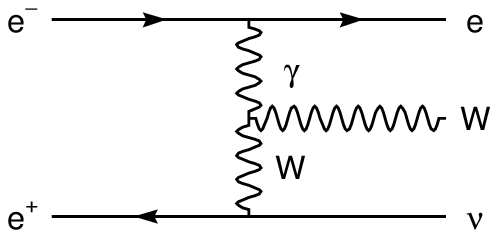}}
\subfigure[Zee]
	{\includegraphics[width=0.45\linewidth]{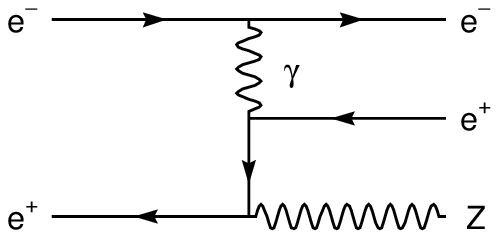}} \\
\subfigure[Two-photon]
	{\includegraphics[width=0.5\linewidth]{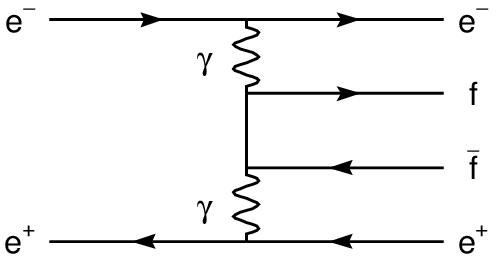}} \\
\caption[Feynman diagrams of SM background processes]
{\label{fig:feynbkg}{\small Feynman diagrams of SM background processes:
(a) two-fermion, (b-f) four-fermion and (g) two-photon .}}
\end{center}
\end{figure}

\begin{table}[tb]
\small
\begin{center}
\vspace{-0.5cm}
\begin{tabular}{|c|cc|cc|cc|cc|c|} \hline
Process &\multicolumn{2}{c|}{189 GeV} & \multicolumn{2}{c|}{196 GeV} 
        &\multicolumn{2}{c|}{202 GeV} & \multicolumn{2}{c|}{206 GeV} & MC \\ 
        \cline{2-9}
& N &$\sigma$(pb)& N &$\sigma$(pb)& N &$\sigma$(pb)& N &$\sigma$(pb)&Generator\\
\hline \hline
ee         & 2000 & 966  & 1700 & 894  & 400 &842  & 480 & 811  & BHWIDE \\
$\mu\mu$   & 75	  & 8.27 & 50   & 7.57 &  50 &7.08 & 50  & 6.80 & KORALZ \\
$\tt$      & 75	  & 8.21 & 50   & 7.54 &  50 &7.04 & 50  & 6.74 & KORALZ \\
$\qq$      & 400  & 99.40& 600  & 89.74& 125 &84.54& 125 & 80.60& KORALZ \\
\hline
$\ww$	   & 108  & 16.59& 100  & 7.61 & 100 &17.46& 100 & 17.54& KORALW \\
We$\nu$	   & 15	  & 0.66 & 20   & 0.75 &  20 & 0.83& 20  & 0.88 & PYTHIA \\
$\zz$	   & 40	  & 2.76 & 50   & 2.85 &  50 & 2.84& 50  & 2.81 & PYTHIA \\
Zee        & 450  & 99.11& 250  & 99.04& 210 &98.94& 400 & 98.72& PYTHIA \\ 
\hline
$\gaga\to$ ee      
           &1200& 3800& 2000& 3840& 2000& 3970& 600 & 190$^{\dagger}$ & PHOT02 \\
$\gaga\to\mu\mu$   
           &1293& 3550& 2000& 3620& 2000& 3660& 600 & 185$^{\dagger}$ & PHOT02 \\
$\gaga\to\tt$ 
           &210 & 431 &  300&  441&  500& 454&  300 & 90.00$^{\dagger}$& PHOT02\\
$\gaga\to\qq$      
           &-   &  -  &  520&  585&  606& 592&  602 & 630 & PHOT02 \\ \hline
\end{tabular}
\caption[Number of events generated and cross sections for the SM
background processes analysed in the six-lepton search] {\label{tab:bkg}{\small Standard Model background
processes analysed in the
six-lepton topologies search. At each centre-of-mass energy, the number of
produced events N in units of $10^3$ and the corresponding cross section is
given for each process. 
In the `tagged' two-photon event generation it is required that the
visible mass is greater than $2.5\gevcc$ and the scattering polar angle
greater than 5\,mrad. 
The $\gaga\to\ell\ell$ subsample at $\roots = 206\gev$
was required to have visible masses greater than $10\gevcc$, hence the lower
cross sections (marked with~$^{\dagger}$).
The MC generators are BHWIDE~\cite{Jadach:1997nk},
KORALZ~\cite{Jadach:1999tr}, KORALW~\cite{Skrzypek:1996wd},
PYTHIA~\cite{Sjostrand:2000wi} and PHOT02~\cite{Buskulic:1993mm}. }}
\end{center}
\end{table}

\subsubsection*{Two-fermion processes}
These are $\ee\to$~Z/$\gamma\to$~f$\fbar$ processes, where f is a quark or
lepton (Fig.~\ref{fig:feynbkg}a). 
They were the dominant background at LEP1 when the centre-of-mass
energy was the Z mass, and thus the cross section was maximal. As LEP
energies have increased away from the Z peak and above the WW threshold, 
the dominant background has
become four fermion processes. At LEP2, $\ee\to$~Z will often be
produced with an ISR (initial state radiation) photon at low angle. This is
the so called \emph{radiative return to the Z} and leads to missing
momentum signatures. Only $\tau^+\tau^-$ and
$\qq$ events could resemble the six-lepton final state topology and
constitute a relevant background to the signal. 
\subsubsection*{Four-fermion processes}
W pair-production $\ee\to$~W$^+$W$^-$ such as shown in 
Figs.~\ref{fig:feynbkg}c and d,
constitutes a source for jets and leptons since W can decay into $\qq$ (70\%)
and $\ell\nu$ (30\%). 
Similarly, $\ee\to\zz^*$ (with Z$^*$ = Z or $\gamma$) is a
major background for SUSY processes. The possible decay modes of the Z boson
$\qq$ (70\%), $\ell^+\ell^-$ (10\%) and $\nu\bar{\nu}$ (20\%) form final
states which overlap with different SUSY signals. 
Also possible are $\ee\to$~We$\nu$ and $\ee\to$~Zee, see
Figs.~\ref{fig:feynbkg}e and f, which both involve
photon radiation from one of the initial beam particles. 
The photon propagator contributes to the cross section with weight 1/$q^2$:
\begin{equation}
\frac{1}{\Ebeam E_{\mathrm{e}} (1-\cos\theta_{\mathrm{tag}})}
\end{equation}
where $E_{\mathrm{e}}$ and $\theta_{\mathrm{tag}}$ are the scattered
electron energy and polar angle. Thus events with low values of 
$\theta_{\mathrm{tag}}$ will have larger cross sections. The 
scattered electron is then more likely to go undetected down the beam-pipe 
leaving its characteristic missing transverse momentum.
Conversely, the neutrino produced in We$\nu$ processes arises from a
virtual W whose propagator is given by
\begin{equation}
\frac{1}{\Ebeam E_{\nu} (1-\cos\theta_{\mathrm{\nu}})-M_{\mathrm{W}}^2}
\end{equation}
This expression resonates for large neutrino scattering angles. So that
characteristic We$\nu$ events will have large missing momentum and a visible
mass close to the W mass. Although the cross section for this type of
process is around one pb, they constitute a difficult background to reject
because six-lepton final states can be quite similar. 
For the same reasons, Zee events are typically identified by a very
energetic electron, visible mass close to the Z mass and missing momentum.
\subsubsection*{Two-photon processes} 
LEP is not only an $\ee$ collider but also a high energy photon collider. 
Beam particles are constantly radiating photons which may
interact with one another with high probability. The cross section for
these $\ee\to\ee$~f$\fbar$ processes is of the order of a nb. 
Figure~\ref{fig:feynbkg}g shows the relevant Feynman diagram.
As discussed above, the majority of these radiative events 
are produced when the two final state $\ee$ continue through the beam-pipe and
are not detected (referred to as `untagged' events). They can be distinguished
by their low visible mass and particle multiplicity. 
Nevertheless if one of the final state electrons does appear in the
detector, generally at very low angles, these events can fake SUSY signals
with large missing energy. 

\section{Signal Selection}
Two analyses have been designed to search for $\slR \to l\neu$ and 
$\slR \to l\tau\stau_1$ in the $\stau_1$ NLSP scenario. 
No independent search has been made for either two- or three-body decay
modes since both decays ultimately depend on the mass difference between
the produced sleptons and the stau: $\dm=m_{\slR}-m_{\stauO}$.

The case of large $\dm$, when $\dm \gtrsim 10 \gevcc$, is the most
favourable since the primary leptons are energetic, 
multiplicity is somewhat larger and these events contain large visible mass.
When $\dm$ is small, $\dm \lesssim 10 \gevcc$, there are
fewer identified leptons, greater missing momentum and much less visible
mass.    


\subsection{Description of the selections}
The final state topology consists of four taus, two of them very energetic,
and two leptons (electrons or muons) which might be soft if the mass
difference between the sleptons and the stau is small. Only in the
large $\dm$ case is it possible to reject events containing fewer than two
identified leptons, which is a very successful cut against two-fermion
backgrounds. In the case of small $\dm$ a more detailed study of the
event is required and a procedure was devised in which the event is
clustered into four tau jets using the Durham algorithm. 

The selection criteria for the large $\dm$ case are summarised in
Tab.~\ref{tab:sel_largeDM} and for small $\dm$ in
Tab.~\ref{tab:sel_smallDM}. The principal rejected background is cited
beside the cut, although there are usually various backgrounds affected
simultaneously. 
\begin{table}[tb]
\begin{center}
\begin{tabular}{|llc|} \hline 
\multicolumn{3}{|c|}{Six-lepton Selection for large $\dm$} \\ \hline  \hline
\multirow{3}{2cm}{Preselection} & Charged tracks     & $4 \lt \nch \lt 12$ \\
                                & Identified leptons & $N_{l} \geq 2$ \\
                                & Visible mass & $\mvis\lt 0.85\sqrt{s}$ \\
\hline 
\multirow{2}{2cm}{Anti-$\gaga$} & Visible mass & $\mvis \gt 25$ \\
                            &Missing $\pt$ & $\ptmiss \gt 0.02\sqrt{s}$ \\
\hline
\multirow{3}{2cm}{Anti-4f} & Hadronic mass & $M_{\rm{vis-2\ell}} \lt 120$ \\
            & \multirow{2}{4.4cm}{Neutral hadronic energy} 
                                   & $E_{\rm{had}^0} \lt 0.17E_{\rm{ch}}$ \\
            &                      & $E_{\rm{had}^0} \lt 0.06\sqrt{s}$ \\
\hline
\multirow{2}{2cm}{Anti-2f,ZZ} & \multirow{2}{4cm}{Jet clustering} & 
                                   $y_{23} \gt 2\times10^{-3}$ \\
            &                    & $y_{34} \gt 1\times10^{-4}$ \\
\hline
Anti-WW & $\qq\ell\nu$ kinematics & $\chi_{\ww} \geq 4.5$ \\ \hline
\end{tabular}
\caption[Selection criteria for six-lepton
final states with large $\dm$]
{\label{tab:sel_largeDM}{\small Selection criteria for six-lepton
final states with large $\dm=m_{\slR}-m_{\stauO}$. Masses are in $\gevcc$,
momenta in $\gevc$. }}
\end{center}
\end{table}

\begin{figure}[p]
\begin{center}
\vspace{-0.6cm}
\includegraphics[width=0.42\linewidth]{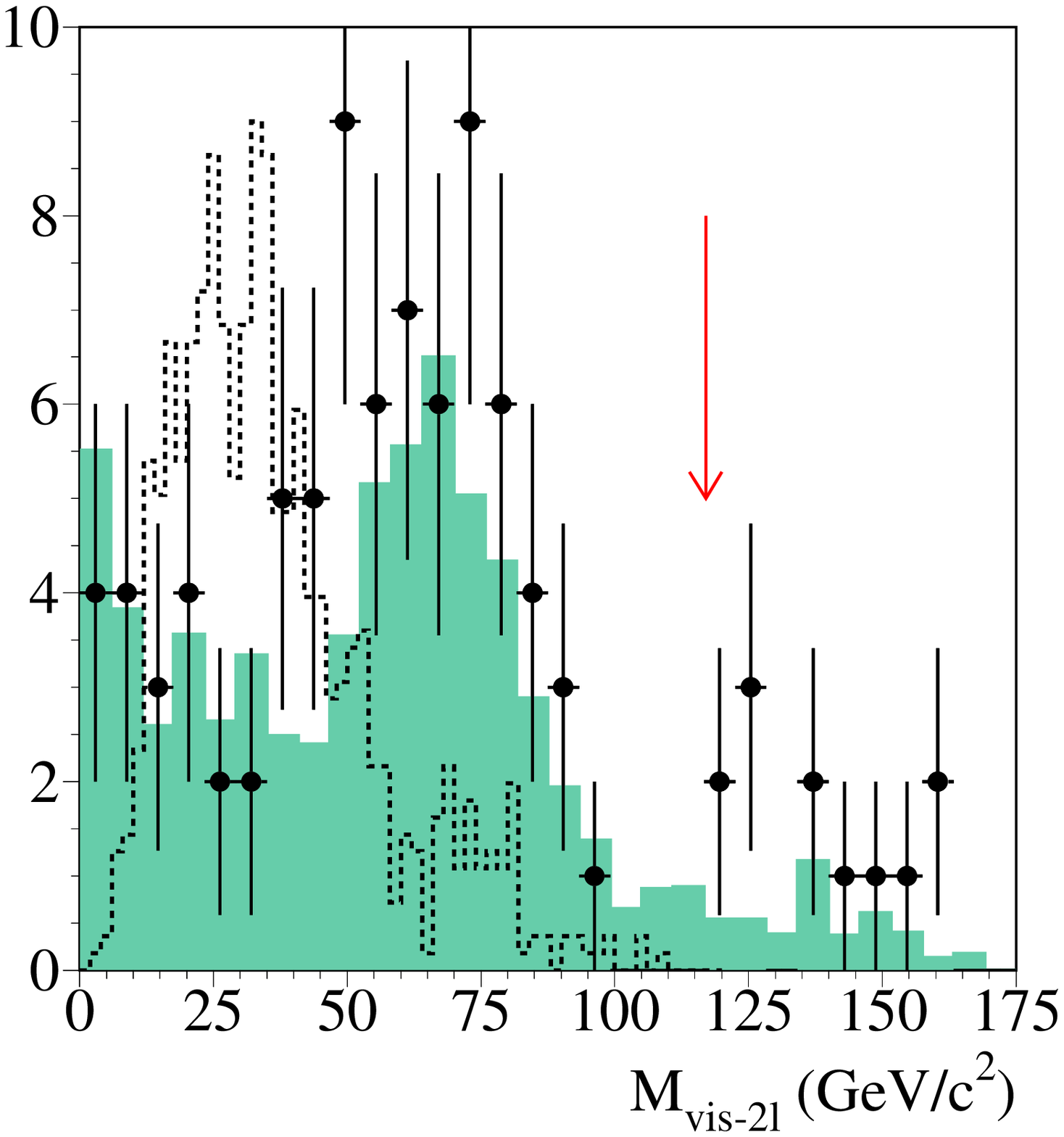} ~~~~
\includegraphics[width=0.42\linewidth]{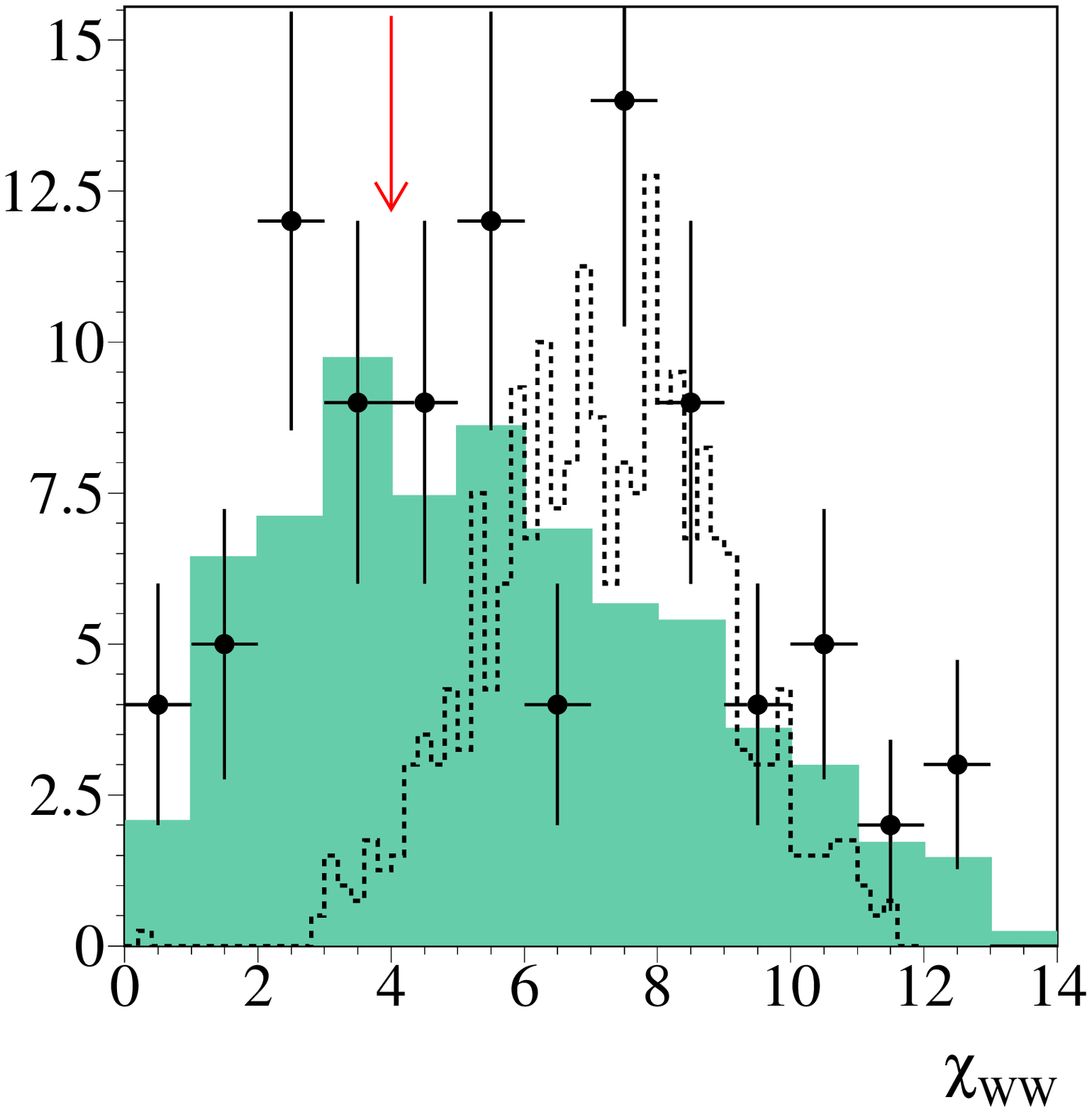} \\
\includegraphics[width=0.42\linewidth]{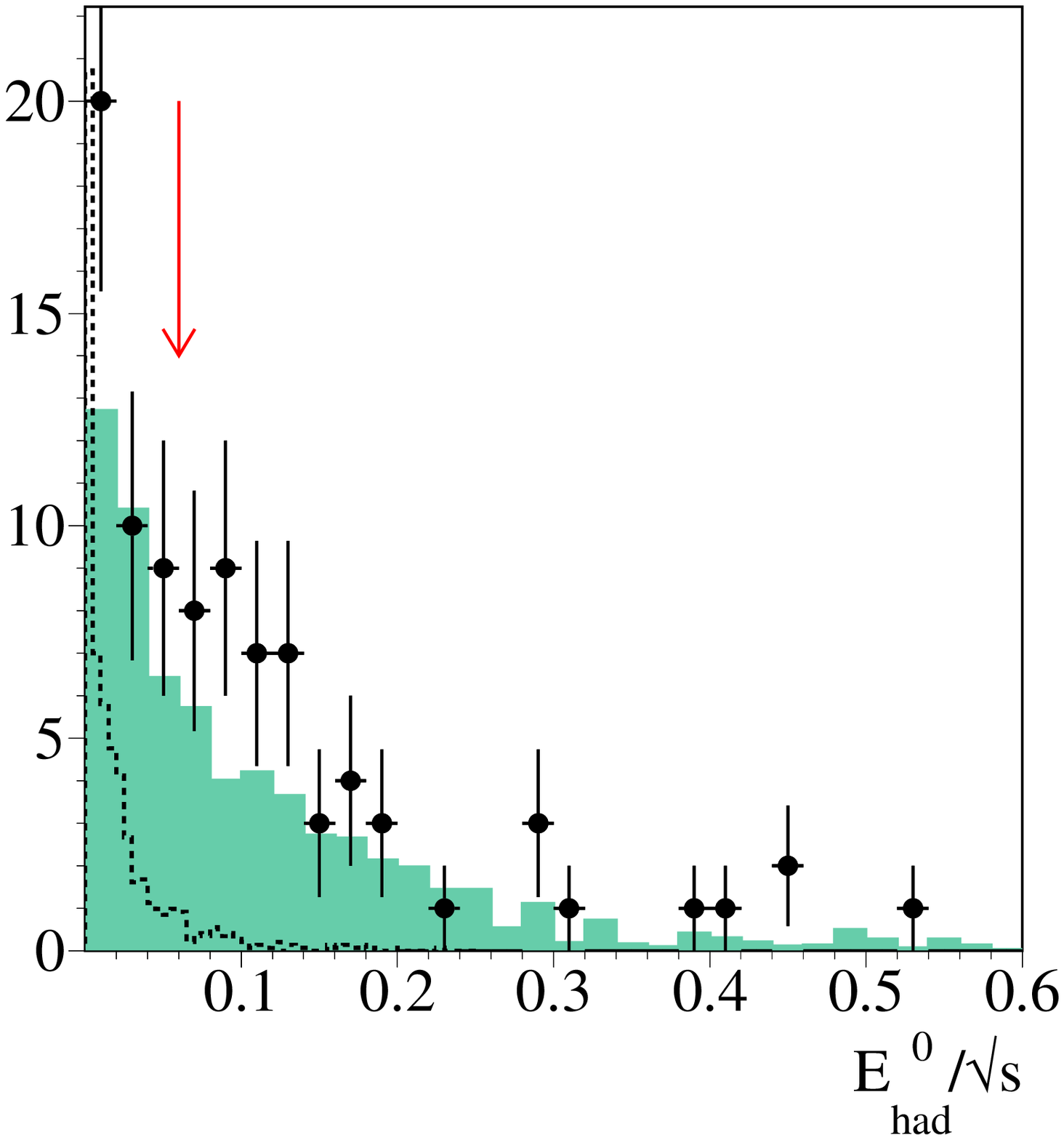} ~~~~
\includegraphics[width=0.42\linewidth]{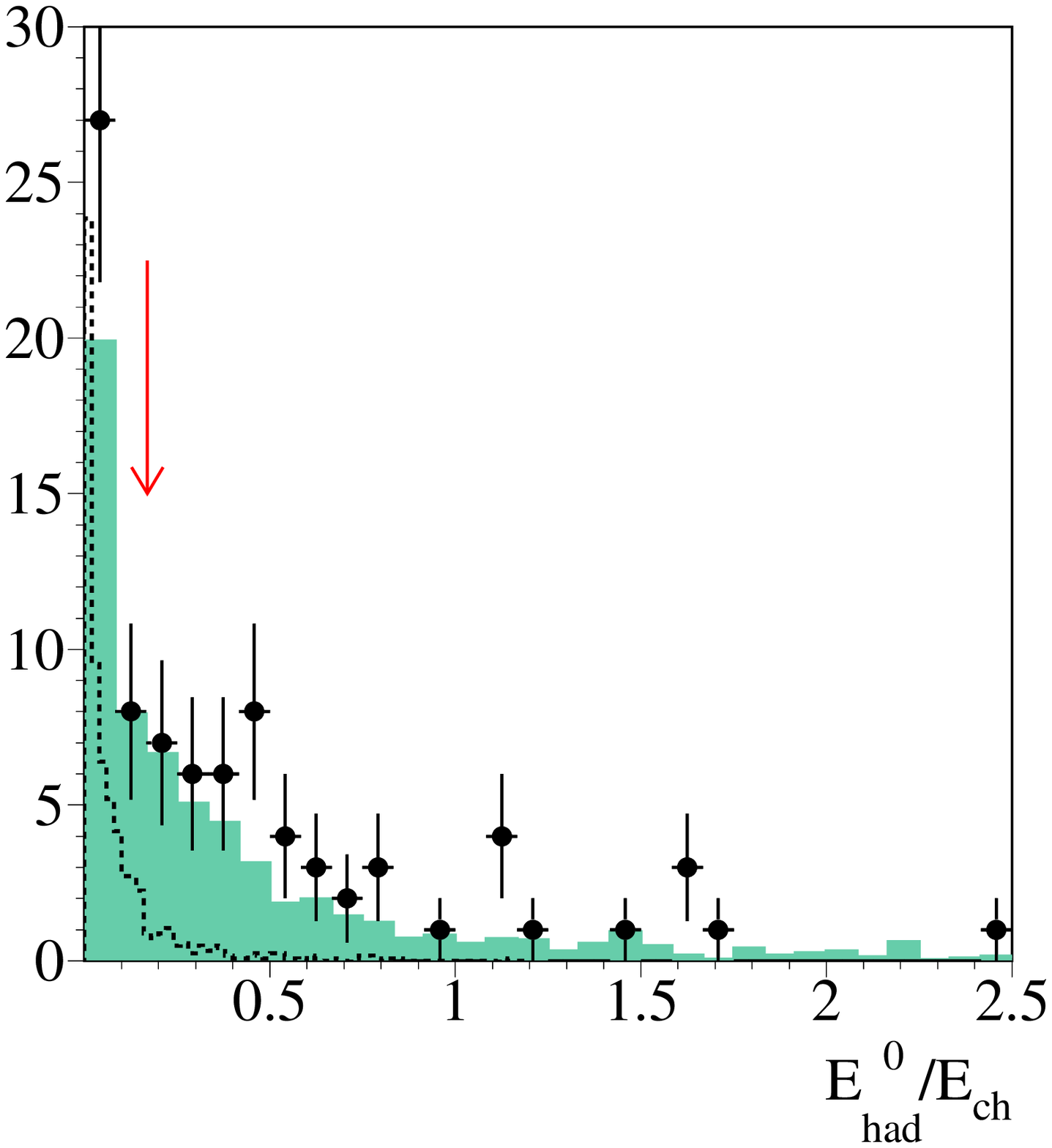} \\
\includegraphics[width=0.42\linewidth]{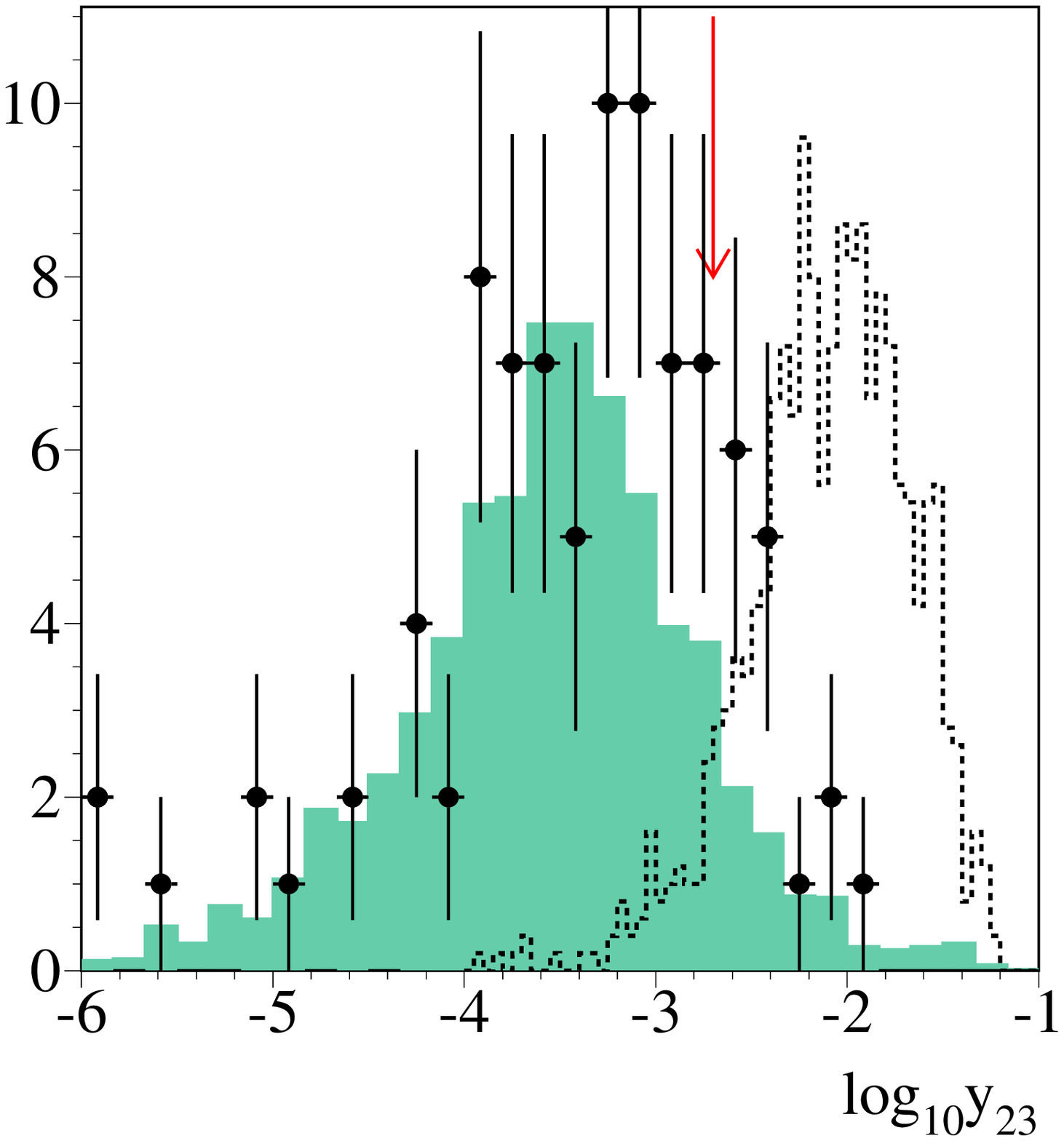}~~~~
\includegraphics[width=0.42\linewidth]{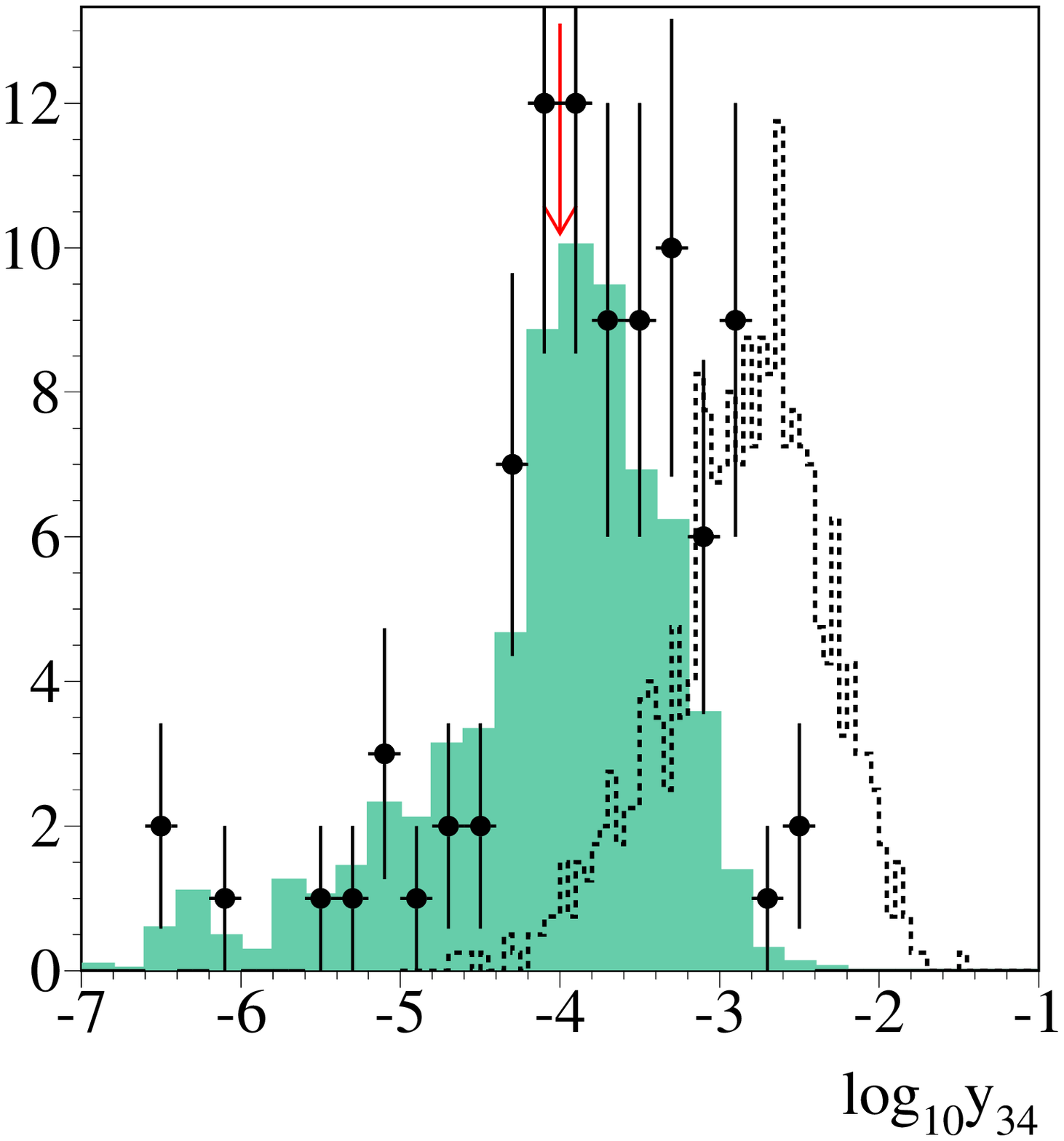}
\caption[Distributions of data, expected background and signal in the
large $\dm$ selection after preselection and anti-$\gaga$ cuts for the
remaining cut variables]
{\label{fig:dbs_large}{\small After applying the preselection and anti-$\gaga$
cuts, the distributions of the remaining variables used in the large $\dm$
selection. The data (dots) at $\roots=206\gev$ are compared to the
background Monte Carlo (filled histograms). The dashed histograms show the
signal distributions (for the values
$m_{\slR}:m_{\neu}:m_{\stauO}=95:87:73\gevcc$) in an arbitrary normalisation.
The location of the cut is indicated with an arrow. The cut values were
optimised and fixed before analysing the data, as described in
Sec.~\ref{optim}.}}
\end{center}
\end{figure}

The variables used in both the small  and large $\dm$ analyses are
described below.
\begin{itemize}
\item The number of charged tracks $\nch$ and good tracks $\ngood$

The number of charged tracks, also known as \emph{event multiplicity}, peaks
at around 20 for four-fermion events. Signal events have much lower
multiplicity. The number of good tracks (as defined in
Sec.~\ref{trkreco}) restricts $\nch$ to tracks with a minimum of four TPC
hits, small impact parameter and $p_t \gt 2$\,MeV/$c$. 
\item Number of identified leptons $N_{l}$

Total number of electrons and muons in the event. Lepton identification was
described in Sec.~\ref{lepid}. This is a very sensitive parameter to
$\dm$. In the large $\dm$ case it helps in rejecting We$\nu$, ZZ, $\gaga$, and
two-fermion processes which all peak at zero $N_{l}$ if taus are involved. 
\item Variables $\mvis$, $\ptmiss$, $E_{\rm{ch}}$, 
$E_{\rm{had}^0}$, $M_{\rm{had}}$, and $M_{\rm{had-2\ell}}$ 

The visible mass $\mvis$ is calculated by adding the energies and momenta
of all energy flow particles in the event. The missing momentum
is identical in magnitude and opposite in direction to the total
momentum. Its projection onto the transverse plane defines $\ptmiss$. 
Two-photon events have very small values of $\mvis$ and $\ptmiss$ and these
variables are very effective against such processes.
$E_{\rm{ch}}$ is the contribution to the visible energy coming from charged
particles, $E_{\rm{had}}$ from all hadrons and $E_{\rm{had}^0}$ from neutral
hadrons. To cut against four-fermion events, the $M_{\rm{had-2\ell}}$
variable is defined as the total hadronic mass with the energy of the two most
energetic leptons subtracted. 
The normalised $\mvis$ against $M_{\rm{had}}$ distribution for the
WW background is displayed in Fig.~\ref{fig:bkgplane}a. The small $\dm$
signal is shown to have much less hadronic energy than four-fermion events
and significantly less visible mass. 
\begin{table}[tb]
\begin{center}
\vspace{-0.6cm}
\begin{tabular}{|llc|} \hline 
\multicolumn{3}{|c|}{Six-lepton Selection for small $\dm$} \\ \hline \hline
\multirow{2}{2cm}{Preselection} & Good tracks & $2 \lt \ngood \lt 10$ \\
                                & Visible mass & $\mvis\lt 0.10\sqrt{s}$ \\
\hline 
\multirow{2}{2cm}{Anti-$\gaga$} 
   & Visible mass & $\mvis \gt 0.05\sqrt{s}$ \\
   &Missing $\pt$ & $\ptmiss \gt 0.10\sqrt{s}$ \\
\hline                      
\multirow{3}{2cm}{Anti-4f} & Hadronic mass & $M_{\rm{had}} \lt 50$ \\
   & Hadronic mass & $M_{\rm{had-2\ell}} \lt 0.25\sqrt{s}$ \\
   & Charged tracks in two tau jets & $N_{\rm{ch}}^{\tau_1+\tau_2} \lt 6$ \\
\hline
\multirow{3}{2cm}{Anti-$\tt$} 
   & $\tau$ jets with good tracks & $N_{\tau}^{\rm{good}} \gt 2$ \\ 
   & Charged tracks not in tau jets & $N_{\rm{ch}}^{\rm{no 2\tau}} \geq 2$\\
   & Thrust & Thrust $\lt 0.95$ \\
\hline
\end{tabular}
\caption[Selection criteria for six-lepton
final states with small $\dm$]
{\label{tab:sel_smallDM}{\small Selection criteria for six-lepton
final states with small $\dm=m_{\slR}-m_{\stauO}$. Energies are in $\gevcc$,
momenta in $\gevc$.}}
\end{center}
\end{table}
\item Cut values of the jet clustering algorithm $y_{23}$ and $y_{34}$

The Durham jet clustering algorithm~\cite{Dokshitzer:1991fc} 
was described in Sec.~\ref{tauid}.
Here, the sensitive variables are $y_{23}$ and $y_{34}$, which are  
the $y_{\rm cut}$ values (from Eq.~\ref{eq:durycut}) necessary to force the
event from two to three jets and from three to four jets respectively.
An example of the distribution of the $\tt$ and ZZ background along with
the signal for the large $\dm$ case is shown in Fig.~\ref{fig:bkgplane}b in
the plane ($y_{23}$,$y_{34}$).
\begin{figure}[tb]
\begin{center}
\includegraphics[width=0.49\linewidth]{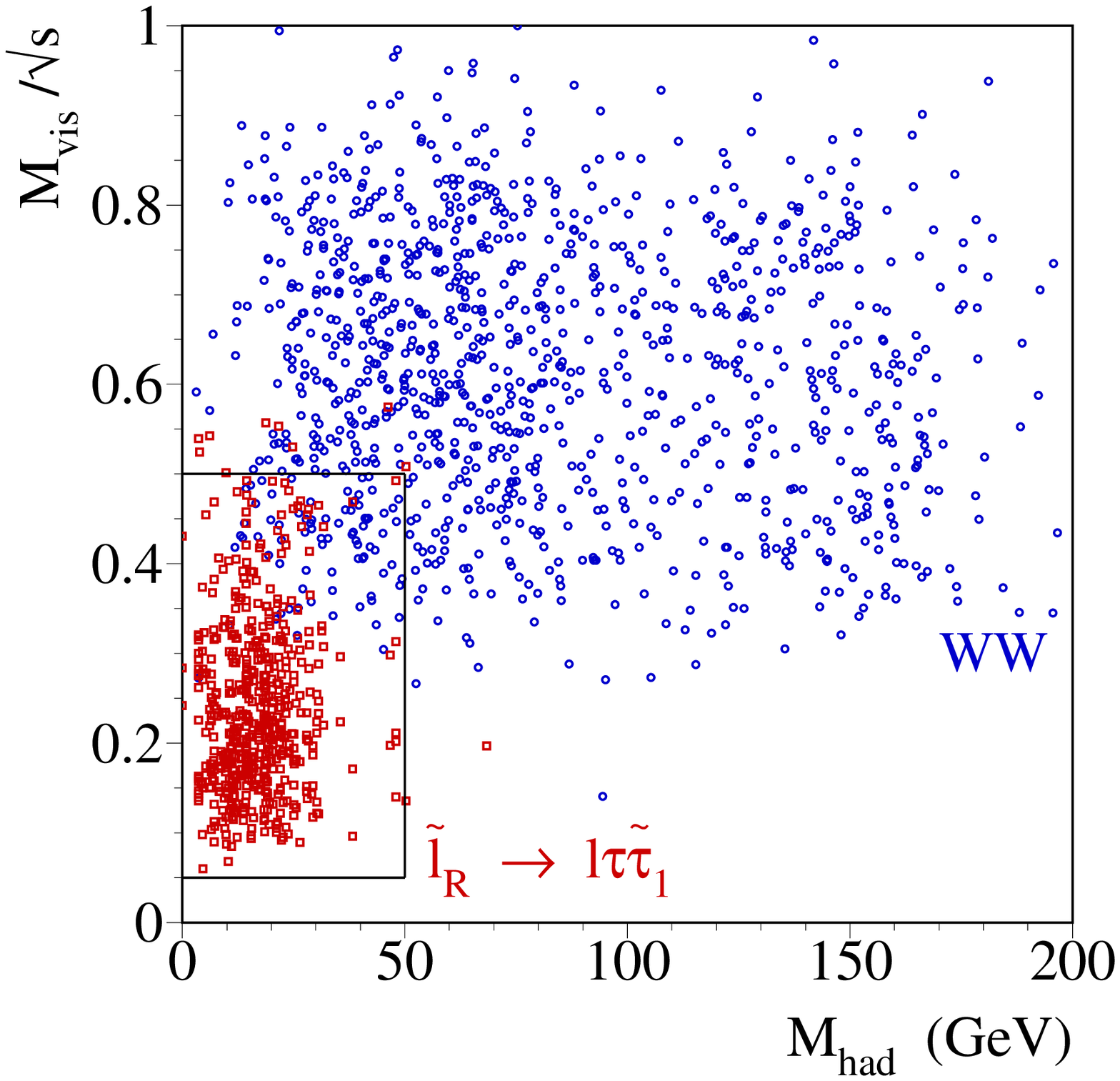} 
\includegraphics[width=0.49\linewidth]{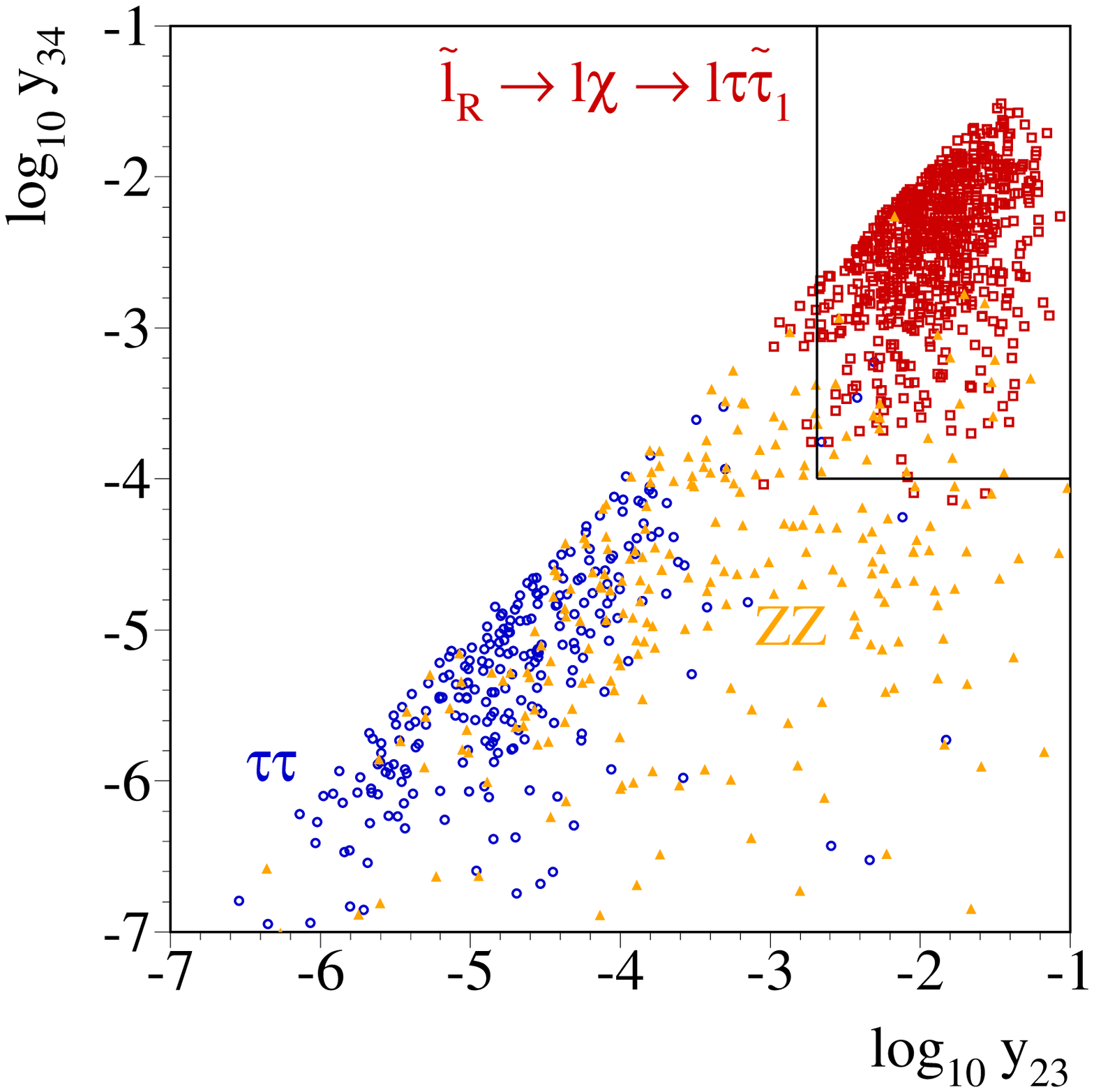}\\
\caption[Background distributions in $(\mvis,M_{\rm had})$ and $(y_{23},y_{34})$]
{\label{fig:bkgplane}{\small (a) 
Visible and hadronic mass distributions for WW background (open circles)
and six-lepton signal points (three-body decay mode
$m_{\slR}$:$m_{\chi}$:$m_{\stauO}=$~97:123:94$\gevcc$) in the small $\dm$
selection after a subset of cuts has been applied.
(b) Background distributions of $\tt$ (open
circles) and ZZ (triangles) events in the $(y_{23},y_{34})$ plane before the
cuts on those variables in the large $\dm$ selection. 
The six-lepton signal is shown as empty squares, in
a two-body decay mode:
$m_{\slR}$:$m_{\chi}$:$m_{\stauO}=$~95:87:73$\gevcc$. }}
\end{center}
\end{figure}

\item $\ww \to \qq\ell\nu$ chi-squared $\chi^2_{\ww}$

The kinematics of these events can be exploited to suppress the background
by defining:
\begin{equation}
\chi^2_{\ww} = \left ( \frac{m_{\qq}-\MW}{10\gevcc} \right )^2 + 
               \left ( \frac{m_{l\nu}-\MW}{10\gevcc} \right )^2 + 
               \left ( \frac{p_l+E_{\rm{miss}}-94\gevc}{10\gevc} \right )^2 
\end{equation}
where $m_{\qq}$ is the hadronic mass, i.e. the mass of the event after
removing the leading lepton, $m_{l\nu}$ is the mass of the leading lepton
and the missing momentum, $p_l$ is the momentum of the leading
lepton and $E_{\rm{miss}}=\roots-E_{\rm{vis}}$ is the missing mass. WW
events are likely to occur at small $\chi^2_{\ww}$, and can therefore be
rejected by requiring a minimum $\chi^2_{\ww}$ for events to be
selected. See  Fig.~\ref{fig:dbs_large}.
\item Four-tau jets variables $N_{\rm{ch}}^{\tau_1+\tau_2}$,
$N_{\tau}^{\rm{good}}$ and $N_{\rm{ch}}^{\rm{no 2\tau}}$ 

These variables are defined by forcing the event into four tau jets. 
$N_{\rm{ch}}^{\tau_1+\tau_2}$ is the number of charged tracks inside the
two most energetic tau jets. It is used against We$\nu$ and ZZ
backgrounds. Two very effective anti-$\tt$ variables are:
$N_{\tau}^{\rm{good}}$, which is the number of tau jets that
contain \emph{good} tracks; and $N_{\rm{ch}}^{\rm{no 2\tau}}$, the number
of charged tracks not associated with any of the two most energetic tau
jets.  
\item Thrust

The thrust is defined as:
\begin{equation}
\mathrm{Thrust} = \max_{\hat{n}}\frac{\sum_i |\vec{p}_i \cdot \hat{n}|}
{\sum_i |\vec{p}_i|} 
\end{equation}
where the sum runs over all reconstructed particles. 
The thrust ranges from 0.5 for `open' (spherical) events to one for events
with two well collimated jets. It is helpful in rejecting $\tt$ events with
thrust values very close to one. 
\end{itemize}

\begin{figure}[p]
\begin{center}
\vspace{-0.6cm}
\includegraphics[width=0.42\linewidth]{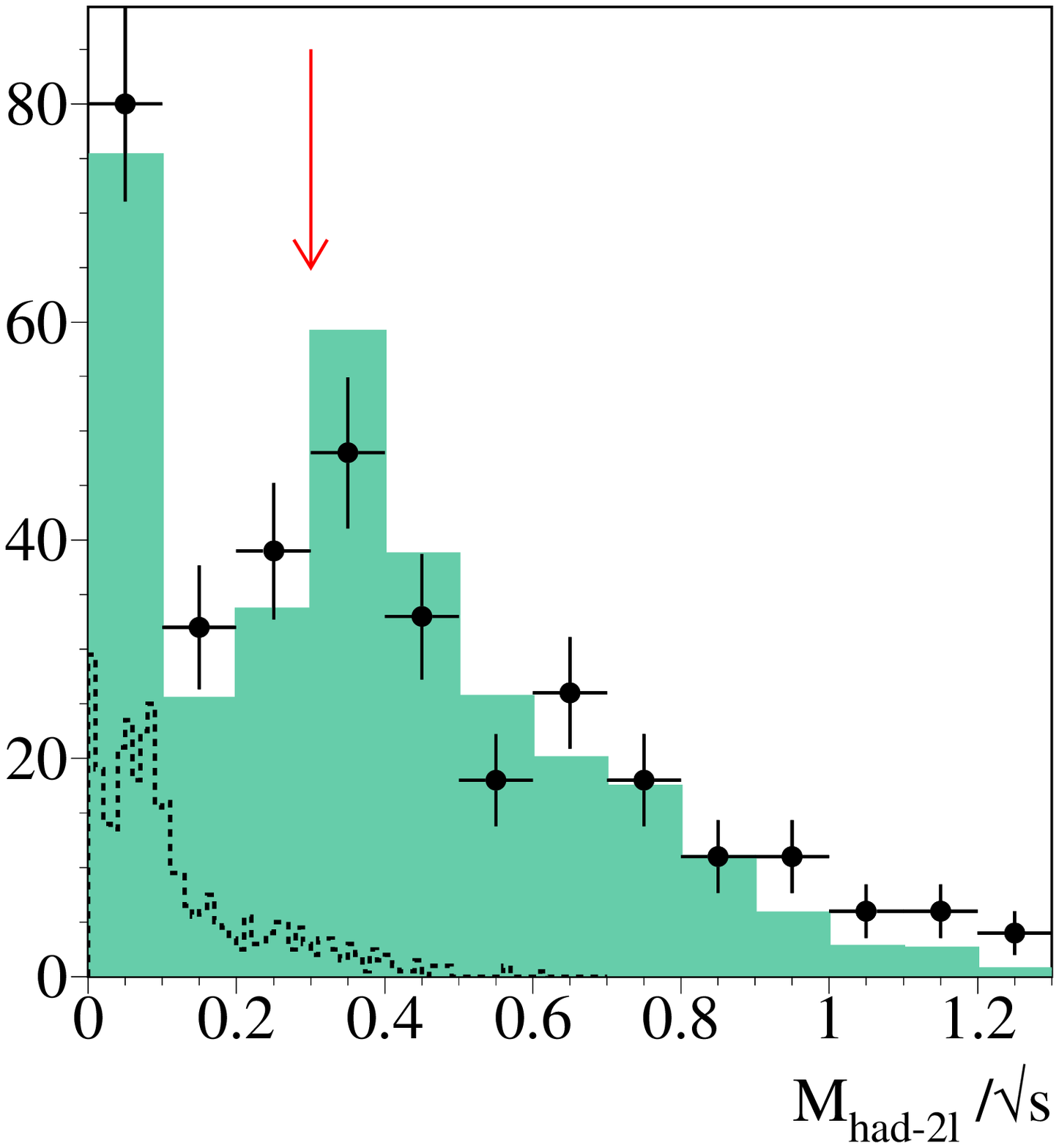} ~~~~
\includegraphics[width=0.42\linewidth]{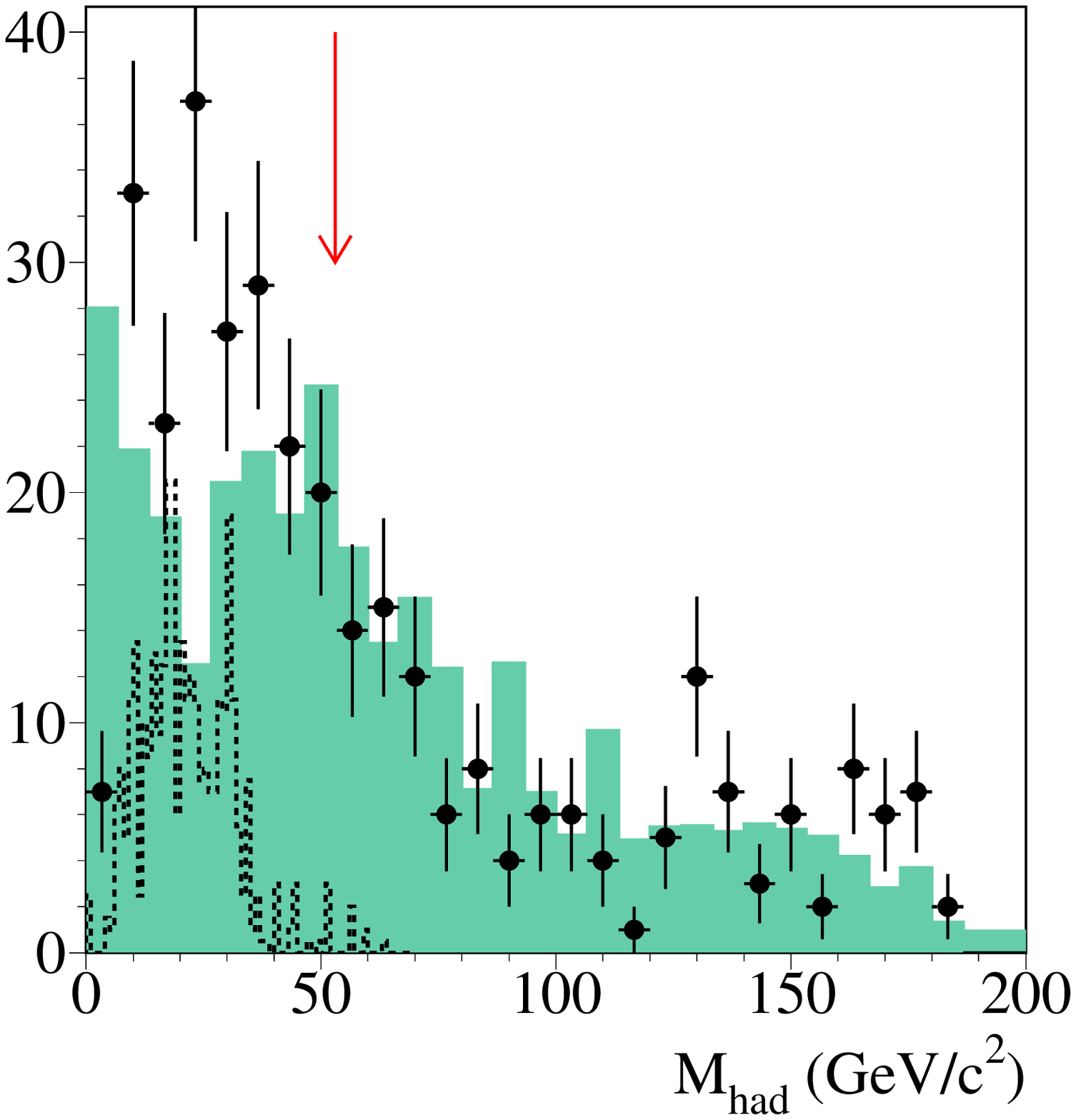} \\
\includegraphics[width=0.42\linewidth]{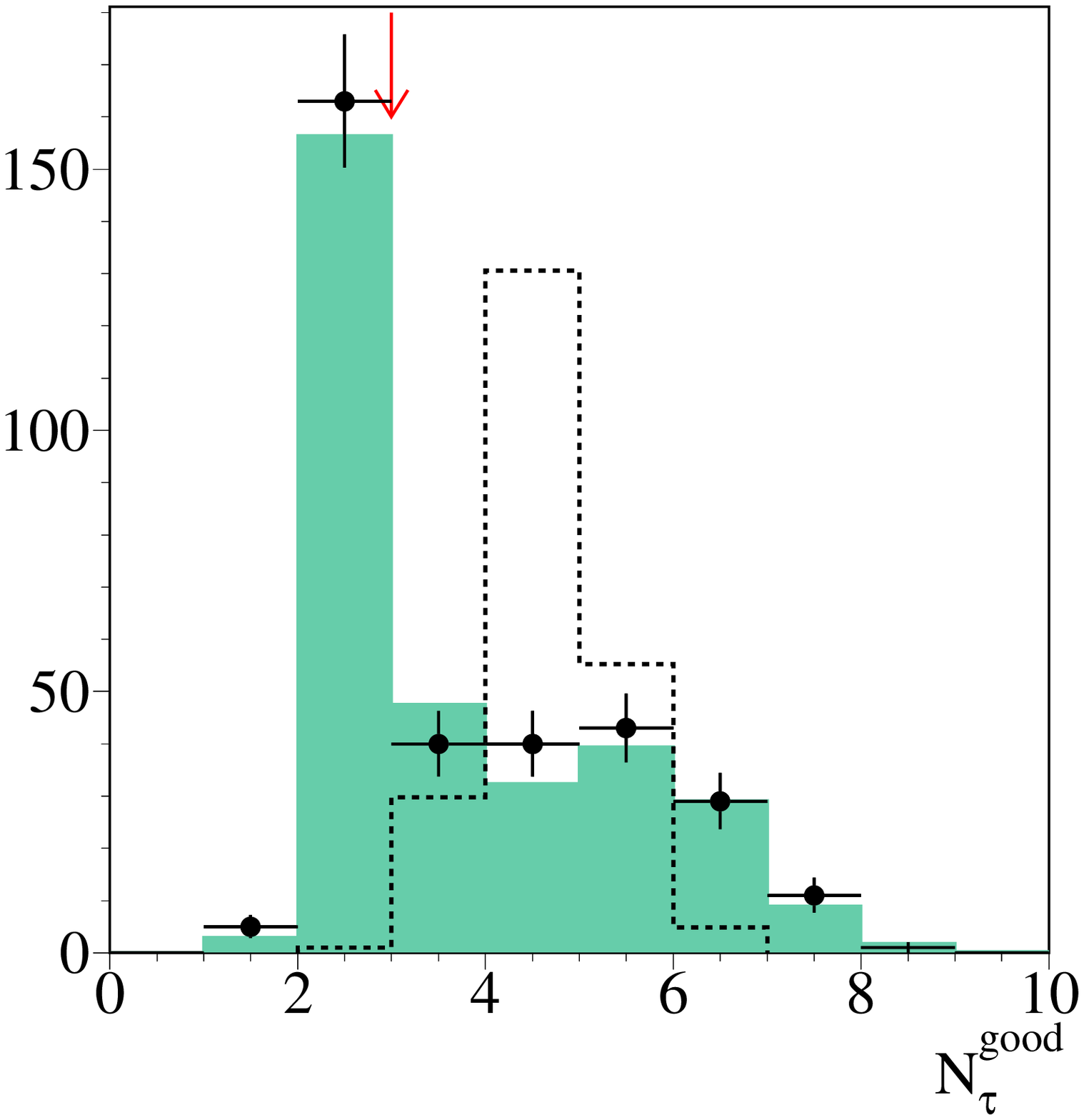} ~~~~
\includegraphics[width=0.42\linewidth]{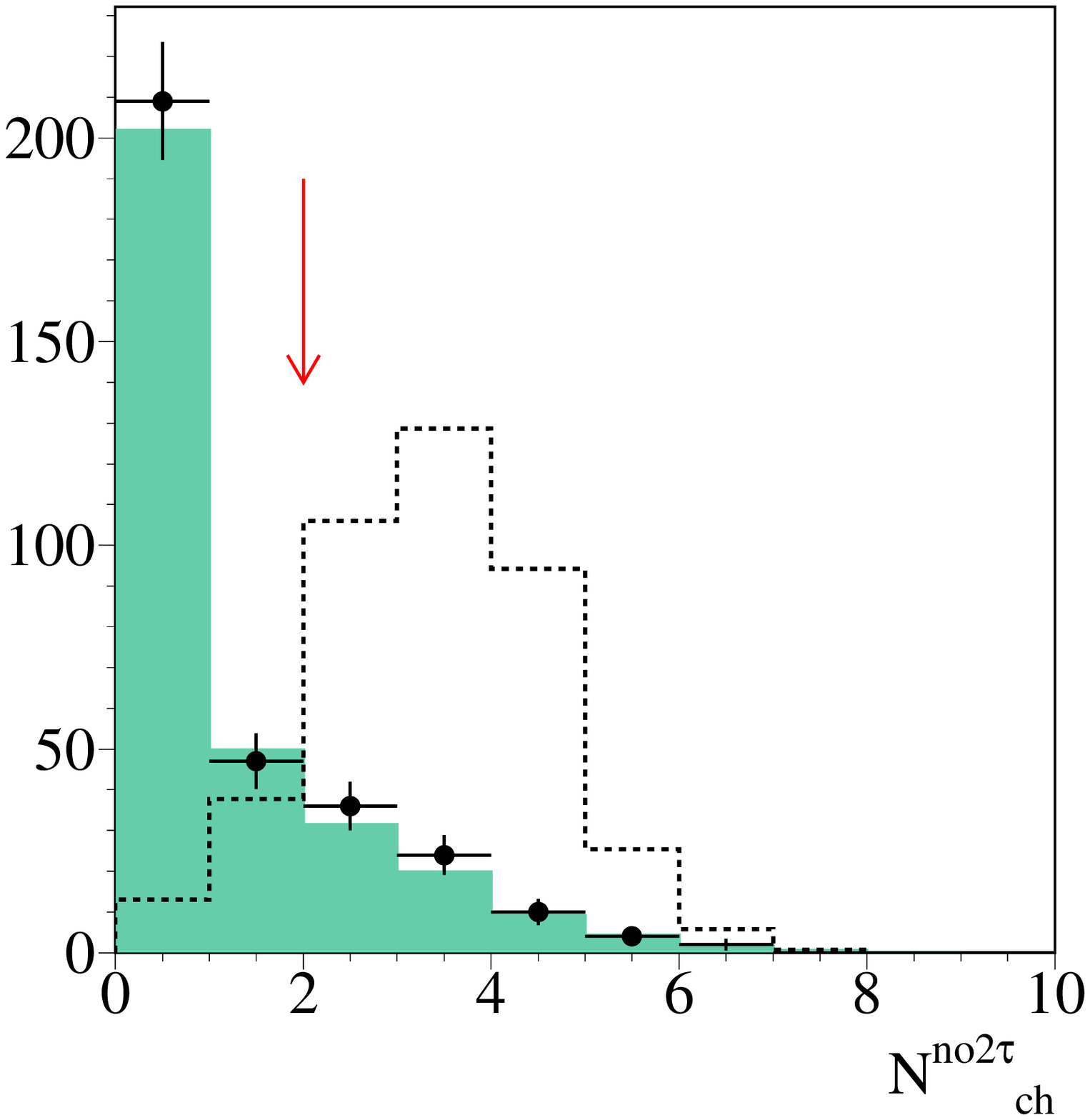} \\
\includegraphics[width=0.42\linewidth]{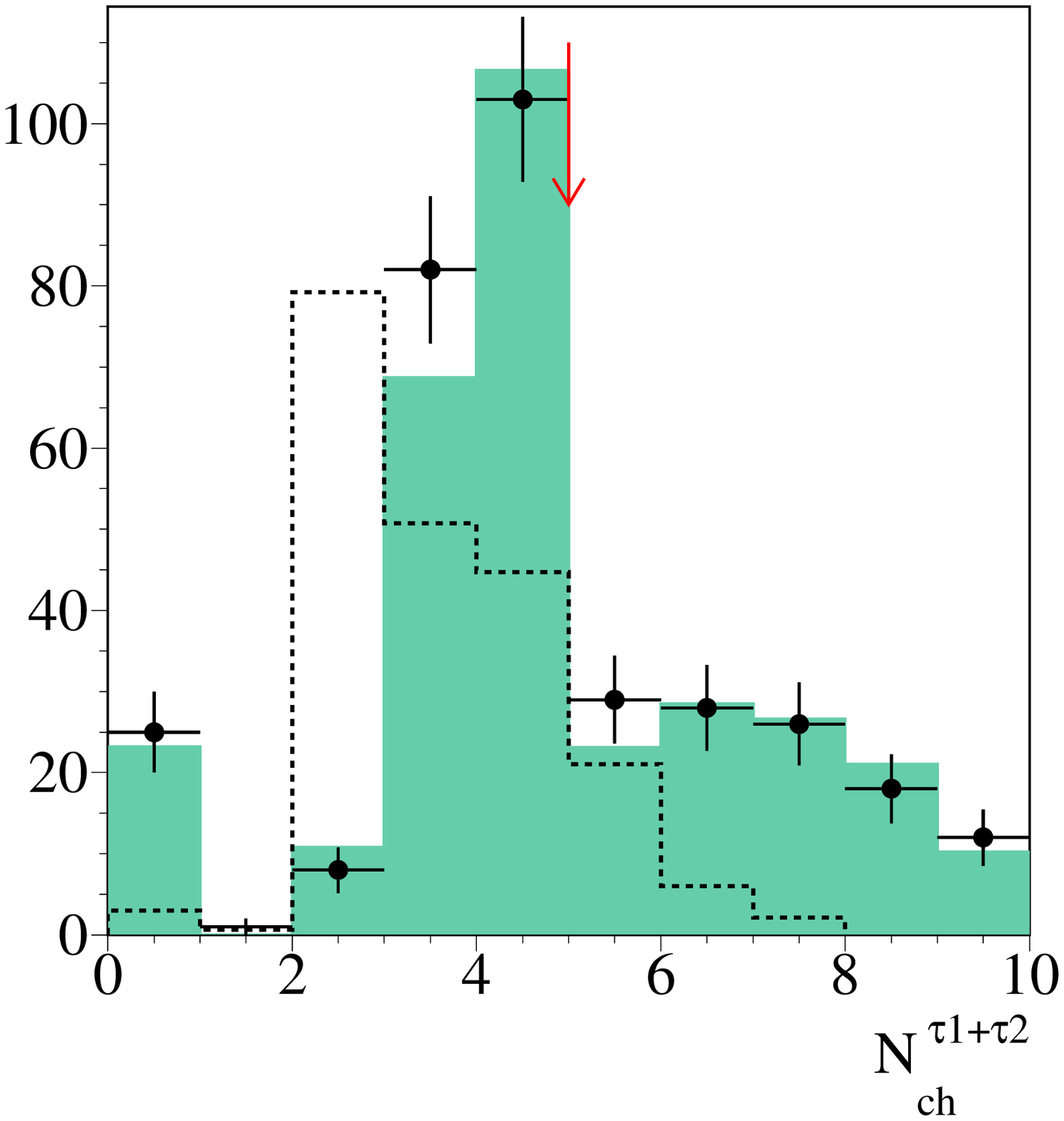} ~~~~
\includegraphics[width=0.42\linewidth]{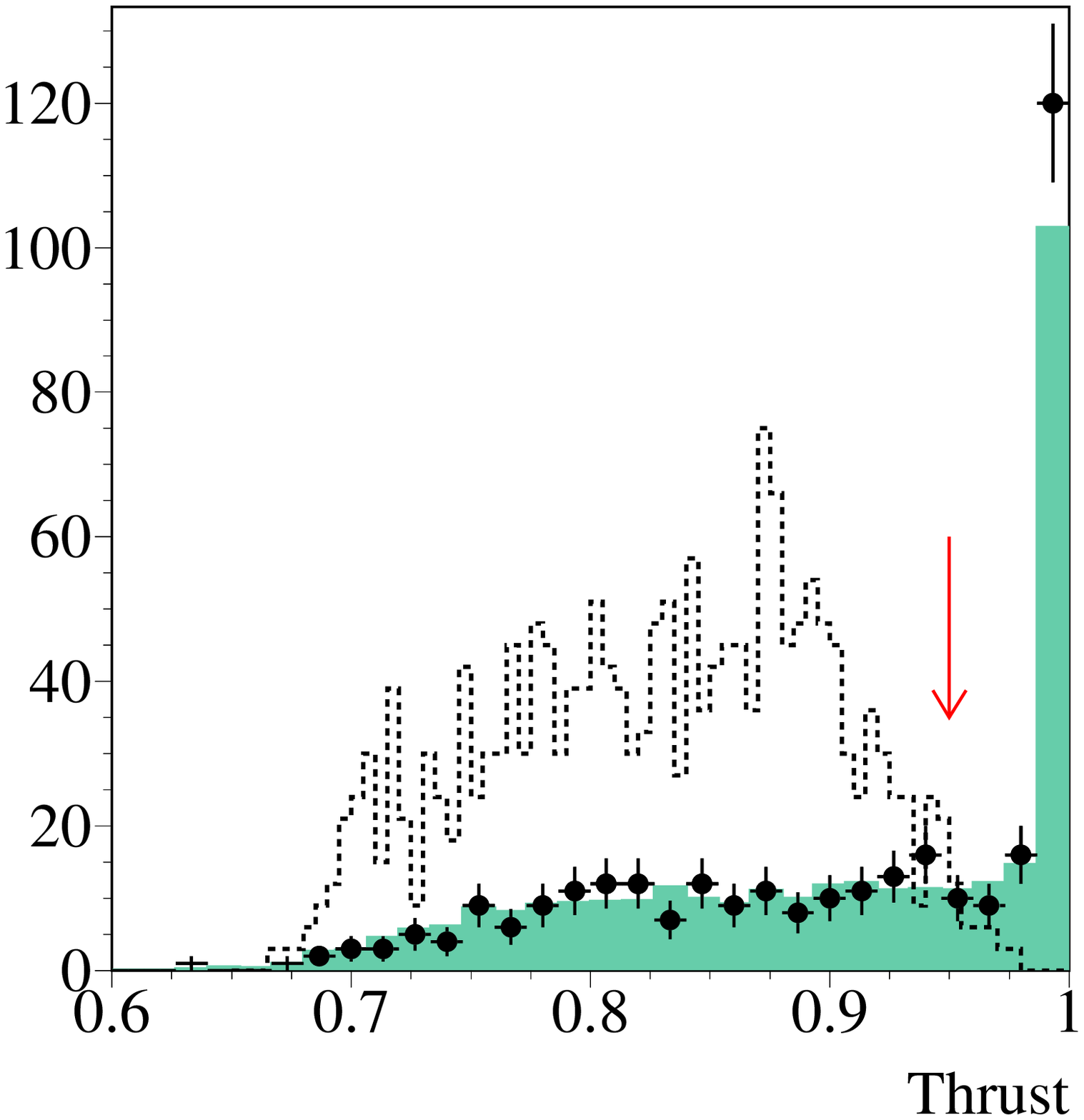} 
\caption[Distributions of data, expected background and signal in the
small $\dm$ selection after preselection and anti-$\gaga$ cuts for the
remaining cut variables]
{\label{fig:dbs_small}{\small After applying the preselection and anti-$\gaga$
cuts, the distributions of the remaining variables used in the small $\dm$
selection. The data (dots) at $\roots=206\gev$ are compared to the
background Monte Carlo (filled histograms). The dashed histograms show the
signal distributions (for the values
$m_{\slR}:m_{\neu}:m_{\stauO}=97:123:94\gevcc$) in an arbitrary normalisation.
The location of the cut is indicated with an arrow. The cut values were
optimised and fixed before analysing the data, as described in
Sec.~\ref{optim}.}}
\end{center}
\end{figure}

The distributions of total background, data and signal after the
preselection and anti-$\gaga$ cuts for the large  and small $\dm$
selections are shown in Figs.~\ref{fig:dbs_large} and~\ref{fig:dbs_small} for
$\roots=206\gev$ with an integrated luminosity of $\Lum=126.5\invpb$.

\subsection{Optimisation}
\label{optim}
As can be seen in Fig.~\ref{fig:dbs_large}, to decide where to place the
cut, one has to reach a compromise
between the number of accepted background events, which should be small,
and the efficiency in selecting signal events, which should be kept high. 
Such a compromise can be obtained analytically by means of the `$\sbNF$
prescription', described in Ref.~\cite{Grivaz:1992nt}.

The number $N$ of expected events from any given process with production
cross section $\sigma$ if the experiment records $\Lum$ integrated
luminosity and the selection efficiency is $\eff$, is given by
\begin{equation}
N=\eff\sigma\Lum
\end{equation}
Nevertheless, in a search experiment there are two independent
contributions to the number of observed events: the expected background $b$
and the expected signal $s$. The number of `candidate' events $n$ selected in
the data after the selection has been applied can be described by Poisson
statistics~\cite{glen}: 
\begin{equation}
\label{pnsb}
P(n;s+b)=\frac{(s+b)^n}{n!}e^{-(s+b)}
\end{equation}
which gives the probability to observe $n$ events in the data if we expect
$s+b$. If $b$ is small and $n$ is compatible with this expectation, then it is
reasonable to assume that the observed candidate events come from
background and not from signal. Thus an upper limit on the number of
possible signal events can then be calculated. 
Usually the upper limit $a_n$ for $n$ candidates observed, is given with a
95\% confidence level (C.L.), by solving: 
\begin{equation}
\label{an95}
P_n(a_n) = e^{-a_n}\sum_{k=0}^{n}\frac{a_n^k}{k!}=0.05
\end{equation}
This is the probability to see $n$ or less candidates if we expect $a_n$. 
The coefficients $a_n=3.00,4.74,6.30\ldots$ are the solutions
to Eq.~\ref{an95} with $n=0,1,2,\ldots$ For example, if no candidate events
are observed in the data, every signal leading to 3.00 or more expected
events is excluded at 95\% C.L. This follows from the fact that the number 
of observed events from a signal process described by a Poisson
distribution with a mean of three, has a 5\% chance of fluctuating down to
zero. 

The selection procedure should be optimised to yield the signal efficiency
and the background expectation corresponding to the best obtainable limit. 
As regards the background expectation, the position $x$ of the cut on the
discriminating variable being optimised is determined by minimising
$\nbNF$. This is the expectation value of $a_n$ in the absence of any
signal, i.e. if all candidates are background: 
\begin{equation}
\label{n95}
\bar{N}_{95}(x) =
e^{-b(x)}
\sum_{n=0}^{\infty}a_n\frac{b(x)^n}{n!}=
e^{-b(x)}
\left [ 3.00 + 4.74b(x)+6.30\frac{b^2(x)}{2!}+7.75\frac{b^3(x)}{3!}+\ldots
\right ]
\end{equation}
where $x$ is the position of the cut to be optimised, 
$b(x)$ is the expected number of background events surviving the cut 
and $n$ is the number of observed events. 
The number $b(x)$ is determined
after all other cuts on the selection have been applied to the total
background Monte Carlo sample. 
Equation.~\ref{n95} is just the sum of the coefficients $a_n$,
weighted by the Poisson probability to observe $n$ events if $b$
background events are expected.

However, if the number of expected background is large it is useful to 
recalculate the coefficients $a_n$ taking into account the background.
This procedure is referred to as `subtracting the background', and
generally only applies to well modelled and understood backgrounds.
It is possible to only subtract one type of background
events and not the total, i.e. only WW expected events or even only 80\% of
them if one is cautious about the systematic uncertainties in the
estimation. 
In this case, if $b_{sub}$ is the number of `subtracted' background
events, Eq~\ref{an95} is modified to become~\cite{pdg}: 
\begin{equation}
\label{an95bsub}
\frac{P_n(a_n+b_{sub})}{P_n(b_{sub})}=0.05
\end{equation}

\begin{figure}[tb]
\begin{center}
\vspace{-0.8cm}
\includegraphics[width=0.36\linewidth]{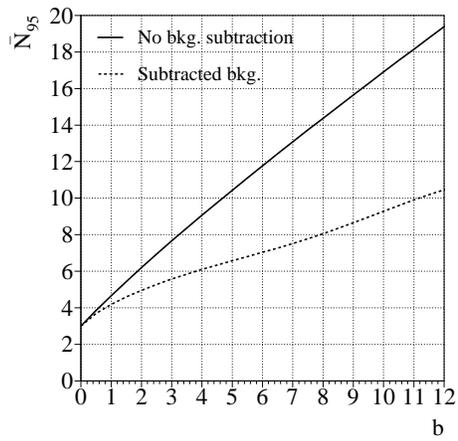}
\caption[$\nbNF$ as a function of the expected background $b$]
{\label{fig:nbar95}{\small The expected limit on the signal events as a
function of the expected background $b$, if the background is subtracted,
and if no background subtraction is performed.}}
\end{center}
\end{figure}

The expected limits when background is subtracted are stronger than the
limits produced when background is not subtracted. 
Fig.~\ref{fig:nbar95}
shows the different $\nbNF$ as a function of the expected background
events, if background is fully subtracted or not.
The background level in the selections developed in this chapter is low
and there is no need to perform background subtraction. 

\begin{figure}[!ht]
\begin{center}
\vspace{-0.8cm}
\includegraphics[width=0.49\linewidth]{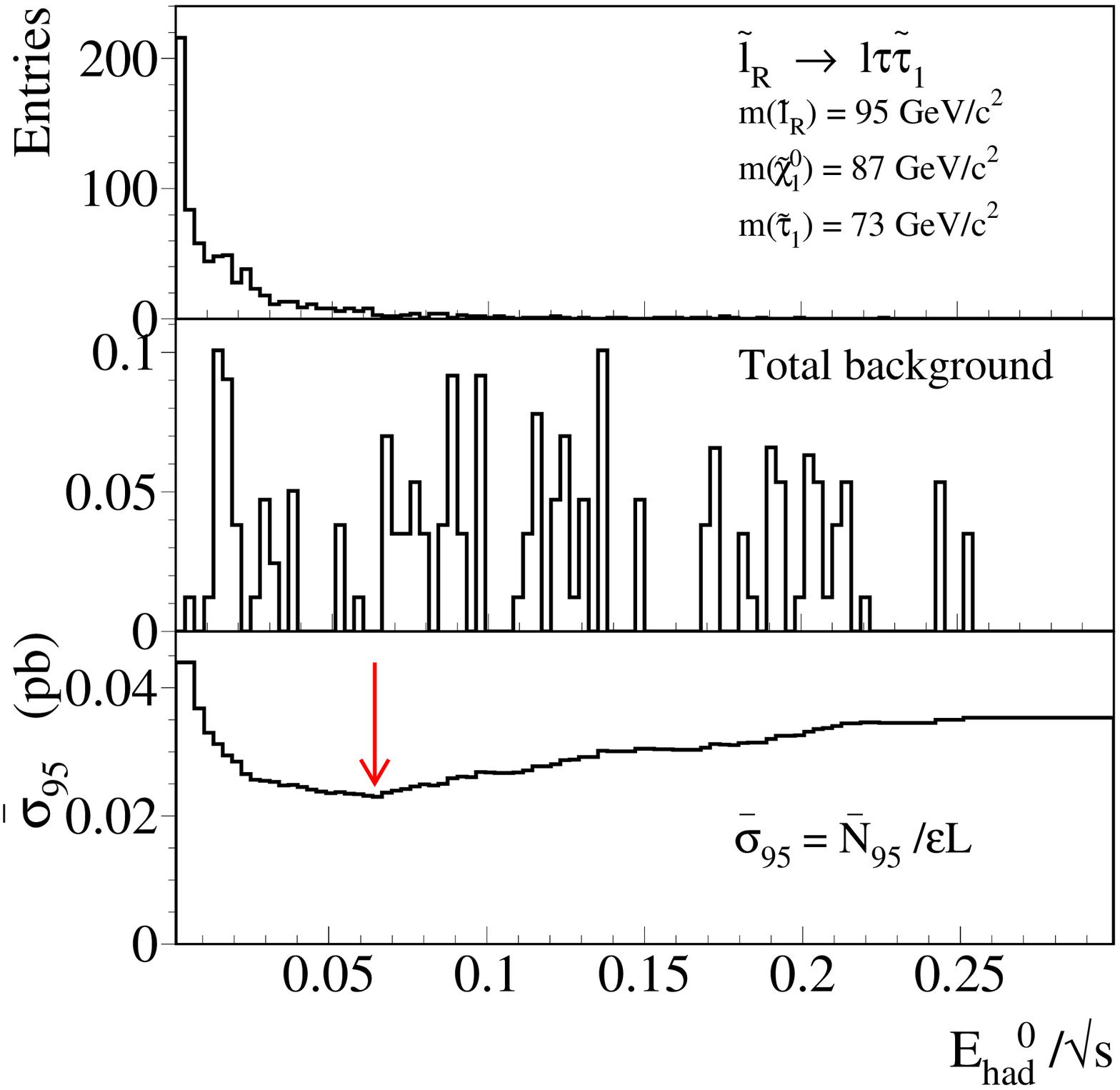}
\includegraphics[width=0.49\linewidth]{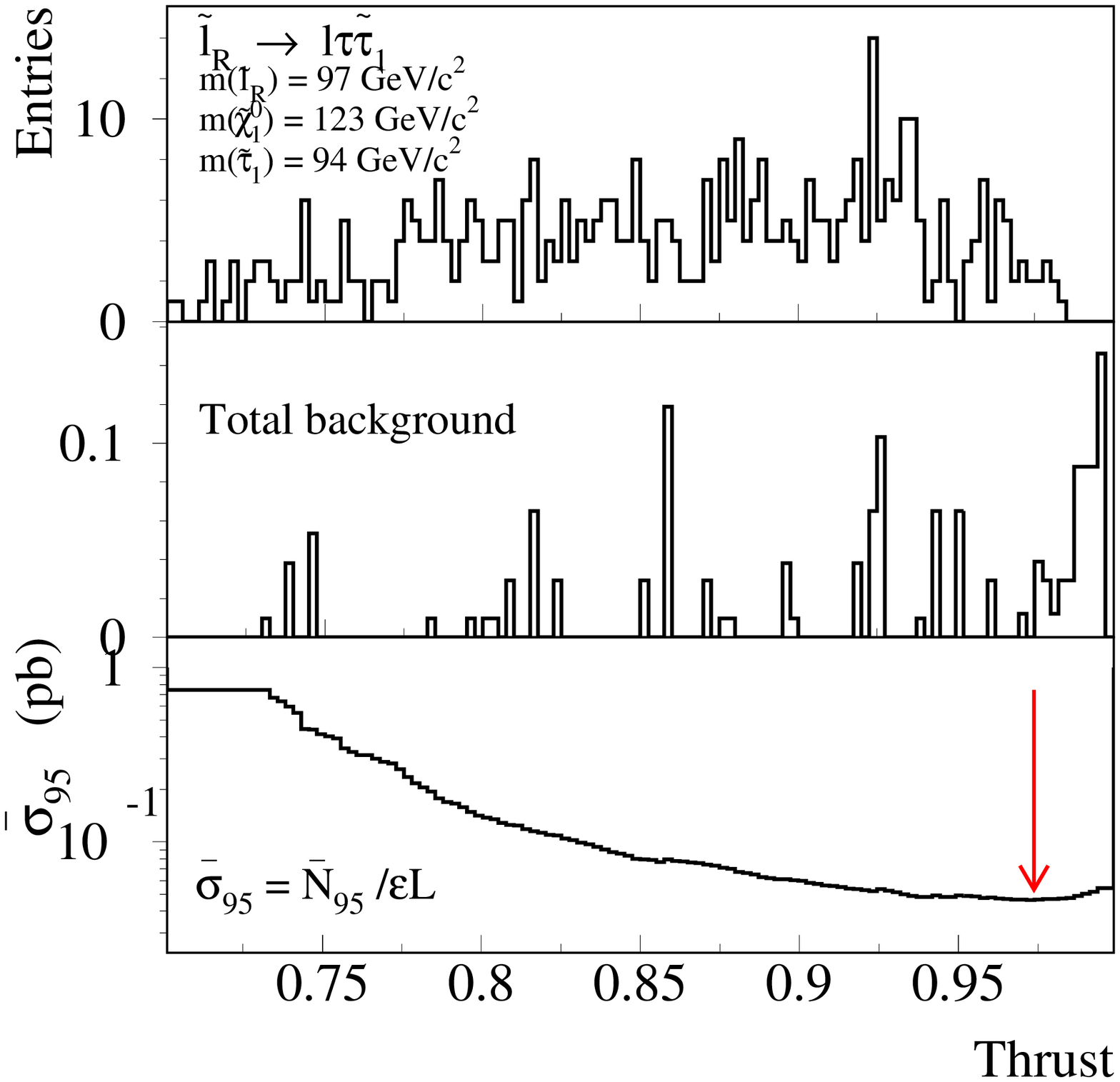}
\caption[Optimisation procedure]
{\label{fig:optim}{\small The optimisation procedure for the neutral
hadronic energy in the small $\dm$ selection (left) and the thrust in the
large $\dm$ selection (right). 
The signal distribution is shown on the upper third,
the background in the middle and the minimised $\sbNF$ at the bottom. All
other cuts have been applied and the arrow indicates the optimal position
of the cut on these variables.}}
\end{center}
\end{figure}

Finally, the best position $x$ for the cut on the variable under
optimisation is such that the expected excluded cross section: 
\begin{equation}
\sbNF = \frac{\nbNF(x)}{\eff(x)\Lum}
\end{equation}
is minimal. 
While the numerator depends only on the number of accepted
background events, the denominator only depends (cumulatively) 
on the number of selected signal events. 
Thus the best possible limit is calculated at the Monte
Carlo level without using the data. 
Figure~\ref{fig:optim} shows this process on two selected variables.
First, the cut values are set approximately by looking at background and
signal distributions as shown in Figs.~\ref{fig:dbs_large}
and~\ref{fig:dbs_small}. Once the chosen cuts give a good
background rejection, the list of cuts is frozen and each one is optimised
sequentially applying the others.
\vspace{1cm}

\subsection{Selection efficiencies}
The main systematic uncertainties on the efficiencies
come from the number of generated events in the simulated samples 
($\sim$3\%). The total systematic uncertainty introduced by lepton
identification variables has been estimated to be $<2\%$~\cite{gmsbpaper}.
Selected data events and background MC were
compared to see any difference in the performance of the lepton
identification algorithms. The estimators were shifted by the difference
between data and MC, and new efficiencies were calculated to see the
impact.  
These effects have been taken into account by conservatively reducing the 
selection efficiency by one standard deviation. 

The generated signal data contained selectron and smuon events with the
appropriate cross sections and branching ratios calculated by ISAJET.
Thus no separation between selectron and smuon events in the selection
efficiency is possible. The efficiency values are then for the topology as
a function of $\dm$ rather than independent measurements of the flavour of
the event.  

Signal events were only generated at 189, 206 and 208$\gev$, 
thus an interpolation for intermediate energies
was needed. A check on some signal efficiencies was performed at these
intermediate energies to parametrise the efficiency as a function of
$\roots$.
It was found that a linear interpolation between centre-of-mass
values was appropriate since the calculated values were well within the
systematic errors on the efficiencies of the generated points. Furthermore,
only a few points had variations larger than $10\%$ between the efficiency
at 189 and 208$\gev$. 

\begin{figure}[ht]
\begin{center}
\includegraphics[width=0.48\linewidth]{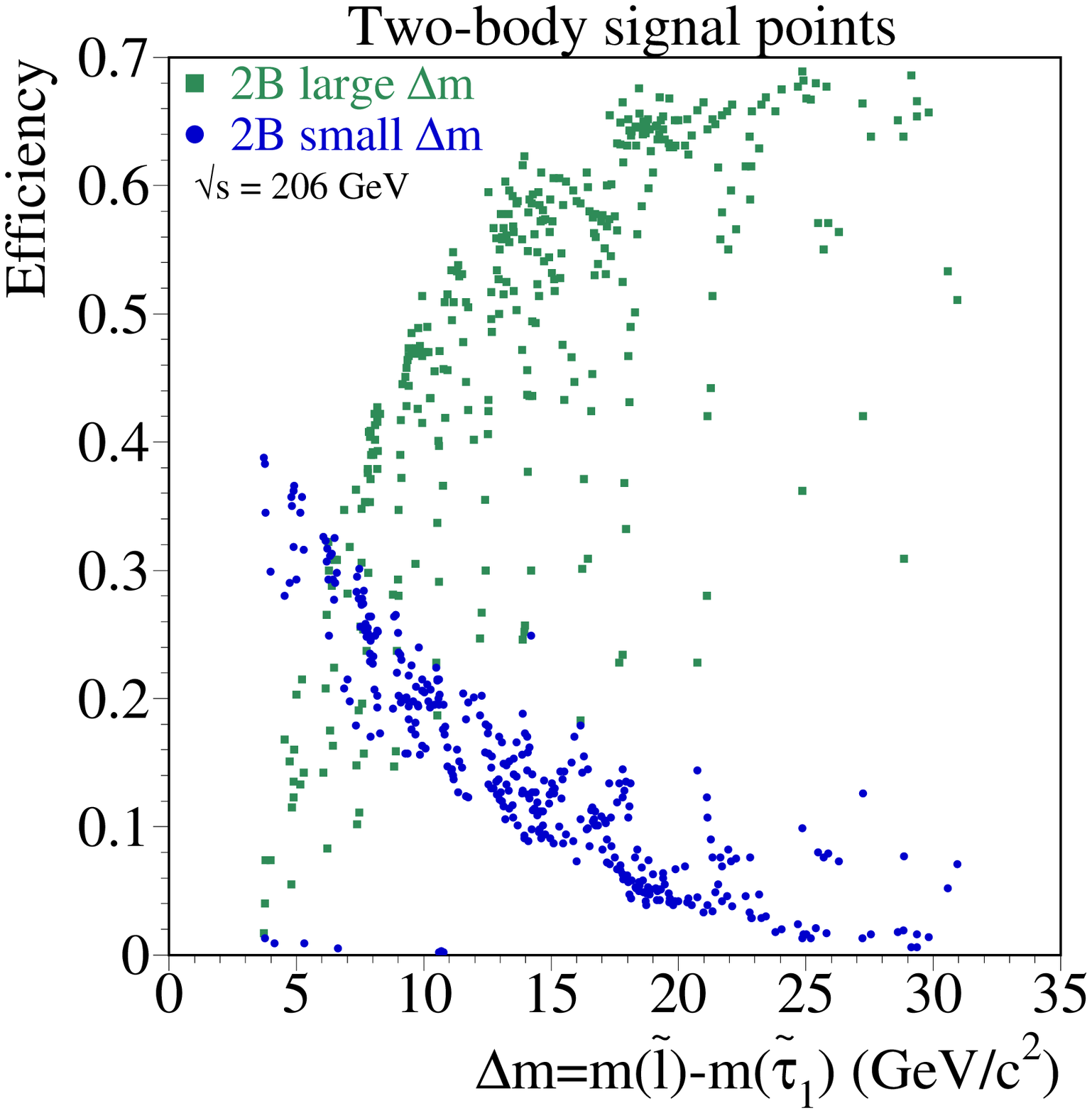}
~
\includegraphics[width=0.48\linewidth]{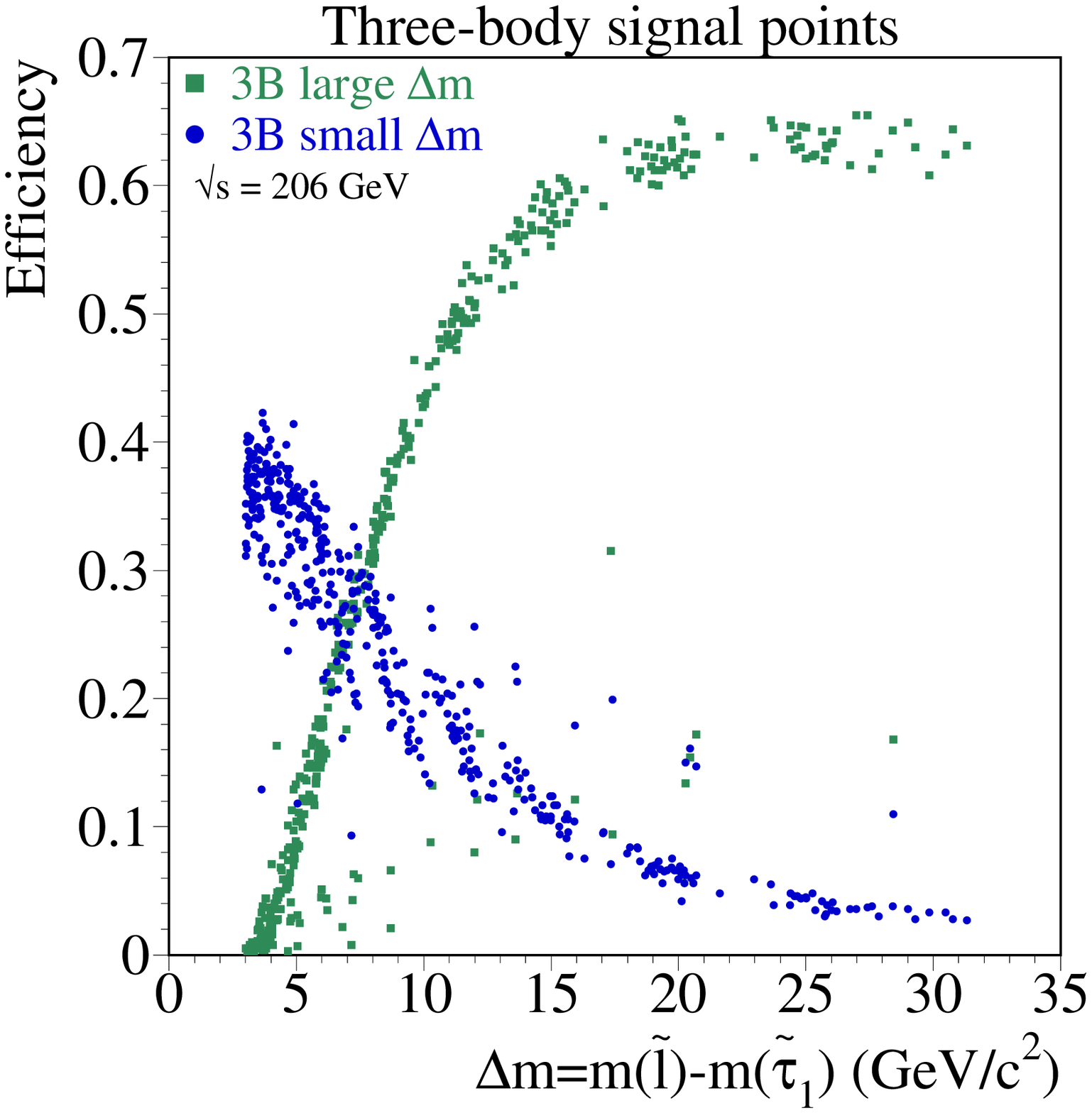}
\caption[The efficiency of the small and large $\dm$
selections as a function of $\dm$ for all analysed
signal points]
{\label{fig:eff-dm}{\small The efficiency of the small and large $\dm$
selections as a function of $\dm$ for all analysed
signal points.}}
\end{center}
\end{figure}

Finally, the selection efficiencies are presented in
Fig.~\ref{fig:eff-dm} as a function of
$\dm=m_{\slR}-m_{\stauO}$ for both two- and three-body signal points.
Efficiencies greater than 30\% are assured over the full range in $\dm$. 
The spread in the bands is due to the range in slepton masses generated
from the kinematical limit down to $70\gevcc$. 
The spread in the large $\dm$ selection when applied over
two-body signal points arises from small or large slepton-neutralino mass
differences. It is noticeable here that the same selection applied over
three-body signal points shows no dependence on the virtuality of the
neutralino. 

\section{Results}
In this section the above selections are applied to the data and
the background MC samples. The selected candidate events are briefly
discussed and the significance of the observed results is described. Limits
on the production cross section for selectrons and smuons are derived and
conservative limits on the masses set. 
\subsection{Events selected in the data}
The number of observed candidates in the data
along with the SM background expectation at each centre-of-mass energy is
 presented in Tab.~\ref{tab:data}. 
Generated Monte Carlo samples for background processes were only available
at the energies detailed in Tab.~\ref{tab:bkg}. To obtain an expected
background value for intermediate energies, the two closest available
values were scaled to the luminosity of the desired energy and linearly
interpolated between them. 

In the $628\invpb$ of ALEPH data between 189 and 209$\gev$, 
the search for six-lepton topologies with small $\dm$ selects three events 
when 2.6 are expected and the search for large $\dm$ finds one candidate 
event with 1.1 expected. 
\begin{table}[tb]
\begin{center}
\begin{tabular}{|cr|cc|cc|} \cline{3-6}
\multicolumn{2}{c}{} & \multicolumn{2}{|c|}{6$\ell$ small $\dm$} &
 \multicolumn{2}{c|}{6$\ell$ large $\dm$} \\ \hline
$\langle \sqrt{s} \rangle$ & $\int\Lum dt$ 
& data & bkg & data & bkg  \\
\hline \hline
189 & 173.6 & 0 & 0.73 & 0 & 0.25 \\
192 &  28.9 & 1	& 0.12 & 1 & 0.04 \\
196 &  79.9 & 0	& 0.34 & 0 & 0.12 \\
200 &  87.0 & 1	& 0.37 & 0 & 0.13 \\
202 &  44.4 & 1	& 0.14 & 0 & 0.23 \\
205 &  79.5 & 0	& 0.34 & 0 & 0.12 \\ 
206 & 126.5 & 0	& 0.54 & 0 & 0.18 \\
208 &   7.8 & 0	& 0.03 & 0 & 0.01 \\
\hline \hline        
TOT:& 627.6 & 3 & 2.60 & 1 & 1.07 \\ \hline
\end{tabular}
\caption[Number of candidate events in the six-lepton selections]
{\label{tab:data} {\small After all cuts have been applied to the
data and the background Monte Carlo the numbers of events for each
selection at the different centre-of-mass
energies. The P-values are: $P(k\geq3;b=2.60,s=0)=0.48$ and 
$P(k\geq1;b=1.07,s=0)=0.65$.}}
\end{center}
\end{table}
What can one then say about the signal process under study? 
To quantify how likely it is to have signal production in the data,
i.e. $s \neq 0$ in Eq.~\ref{pnsb}, one can calculate the probability to find
$n$ events or more if background only was expected:
\begin{equation}
\label{pvalue}
P(k \geq n;b,s=0) = \sum_{k=n}^{\infty} \frac{b^k}{k!}e^{-b} 
= 1 - \sum_{k=0}^{n-1} \frac{b^k}{k!}e^{-b} 
\end{equation}
This is sometimes referred to as the confidence level
($\clb$) or P-value. 1-$\clb$ measures the compatibility of the observation
with the background hypothesis. 
As a convention, it is agreed that one has 
\emph{discovered} a signal if 1-$\clb$ is less than $\sim$$6\times
10^{-7}$. 
That is to say, the probability to observe
such an unexpected result if the only source is the `known'
background, would have to be $\sim$$10^{-7}$ to be convinced that the signal
is present in the data. The median expectation for pure background is
1-$\clb$=0.5. Therefore if $\clb$ is smaller than 0.5 there is a deficit in
the background expectation.  
The above equation yields a P-value of 0.48 and 0.65 for the small  and
large $\dm$ selections, respectively. Thus the data is in perfect agreement
with the expected value from the Standard Model\footnote{ 
Of course this argument relies heavily on the background estimation, 
so it should be used quoting estimation errors. In the case of a 10\%
fluctuation in the background expectation, the above results for the
P-value would still be compatible with the background only hypothesis. 
}.

The composition of the expected background for the
small $\dm$ selection is as follows: 
WW and $\gaga\tt$ dominate with 40\% each, the rest has equal contribution
from $\tt$, We$\nu$ and Zee. In the search for
large $\dm$ topologies, the dominant background is ZZ with 35\%, followed
by  $\qq$, $\tt$ and Zee. All other background sources are completely
eliminated. 

\begin{figure}[p]
\begin{center}
\vspace{-0.3cm}
\renewcommand{\subfigtopskip}{-5pt}
\renewcommand{\subfigcapskip}{3pt}
\subfigure[Large $\dm$]
          {\includegraphics[angle=0,width=0.49\linewidth,height=0.47\textheight]
          {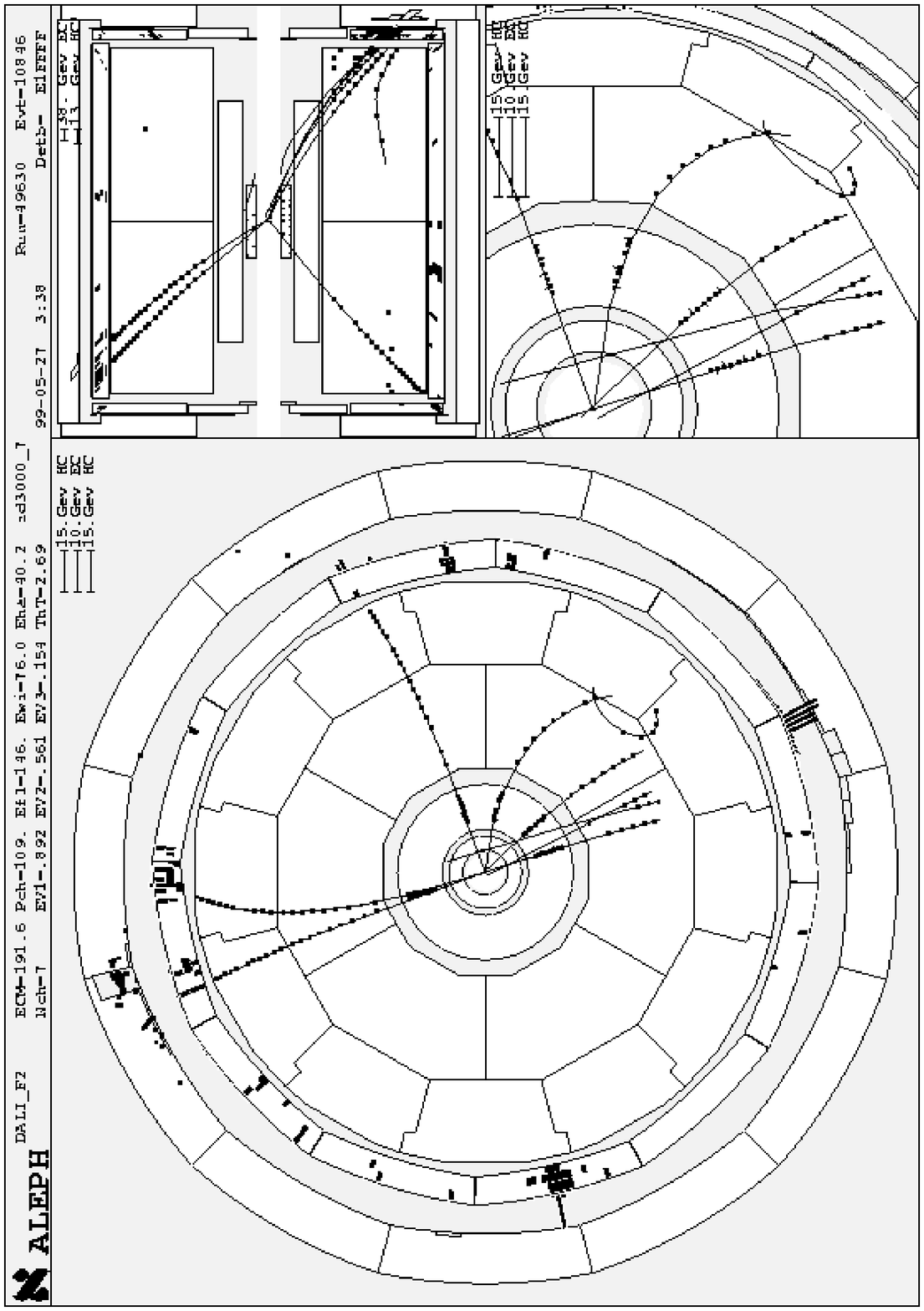}}
\subfigure[Small $\dm$]
          {\includegraphics[angle=0,width=0.49\linewidth,height=0.47\textheight]
	  {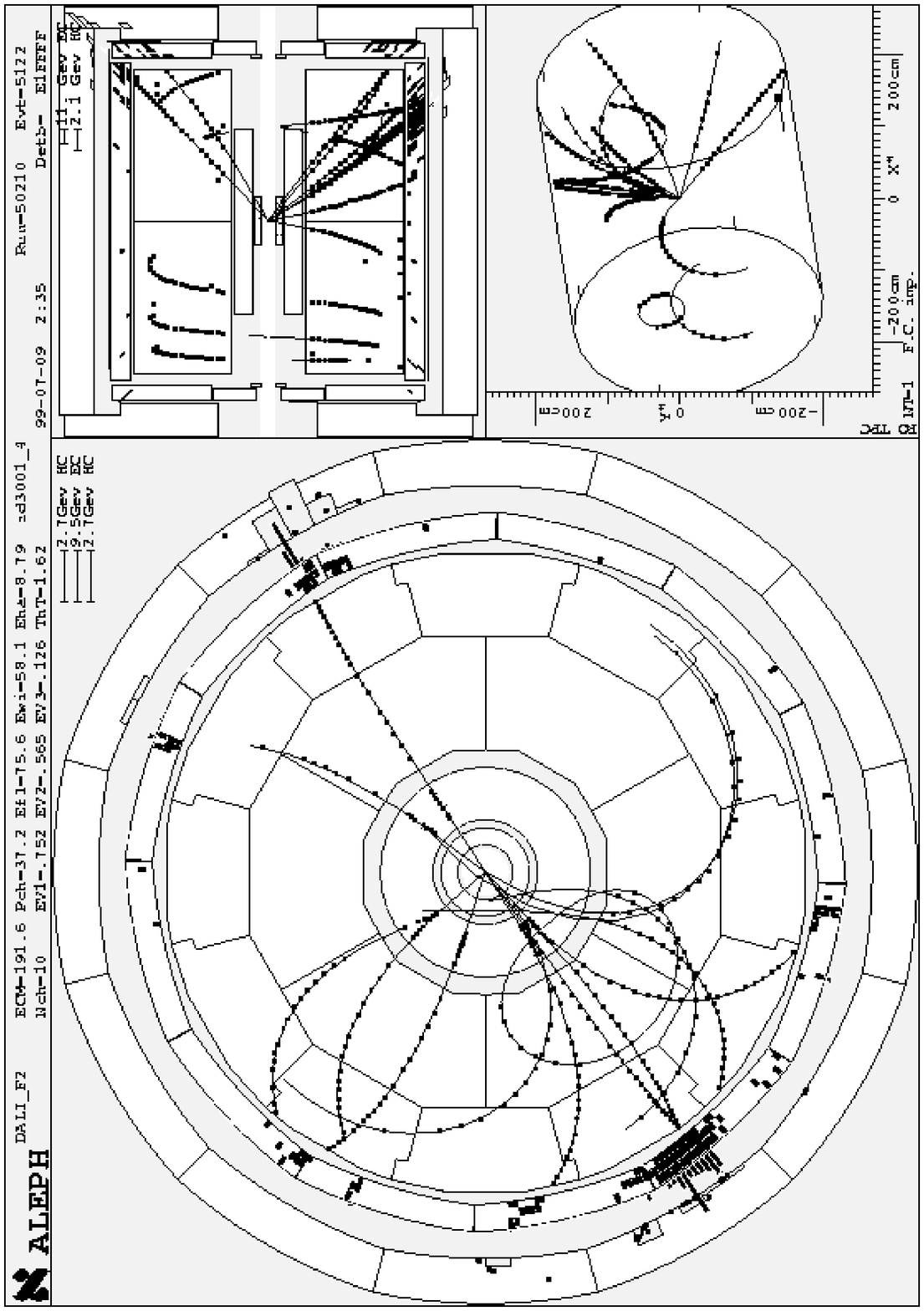}}
\subfigure[Small $\dm$]
          {\includegraphics[angle=0,width=0.49\linewidth,height=0.47\textheight]
	  {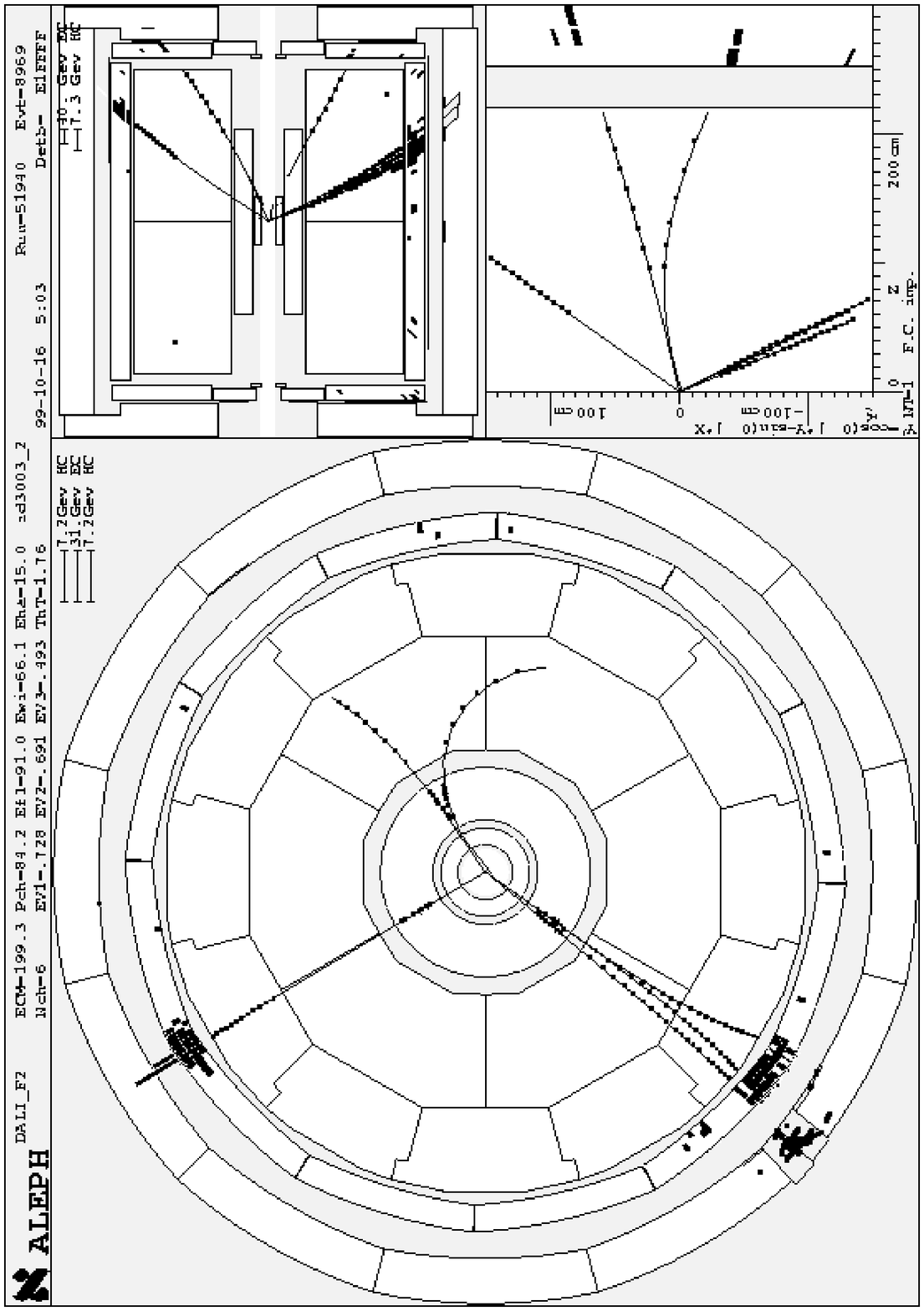}}
\subfigure[Small $\dm$]
          {\includegraphics[angle=0,width=0.49\linewidth,height=0.47\textheight]
	  {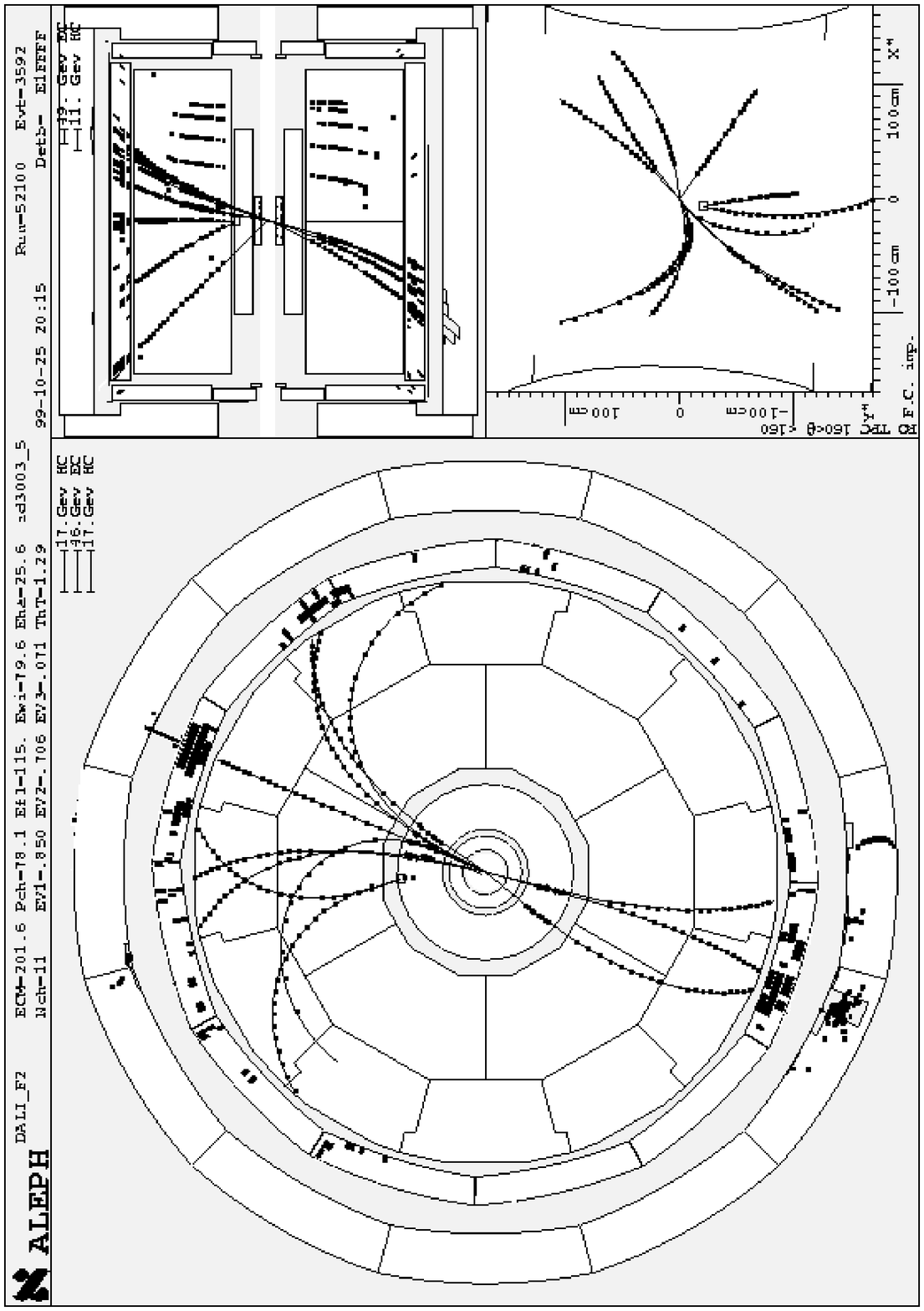}}
\caption[Event displays of the selected data]
{\label{fig:cands}{\small Candidate events.}}
\end{center}
\end{figure}

The four selected data events are all compatible with SM sources as
listed in Tab.~\ref{tab:dataevents}. They are displayed in
Fig.~\ref{fig:cands}. 
The event selected by the large $\dm$ selection at
$192\gev$, shown in Fig.~\ref{fig:cands}a, has two identified
electrons. One is isolated with momentum $6.5\gevc$ and the other has
$66\gevc$ momentum, with $27\gev$ of hadronic energy in a cone of $30\degs$
around its direction. There is large missing momentum in the $z$ direction,
suggesting that one particle went through the beam-pipe. It could therefore
be a $\gaga$ event with two low multiplicity jets, one of them with a
semileptonic decay. 
Of the three candidate events selected by the small $\dm$ selection, 
the two recorded at 192 and $202\gev$, in Figs.~\ref{fig:cands}b and d,
present a very energetic electron of 15 and $66\gev$ respectively, and large
missing longitudinal momentum. Both invariant masses are compatible
with $m_{\Z}$, at 59 and 98$\gevcc$ respectively. 
Therefore they are possibly
Zee events where one of the electrons goes undetected. Finally, the event
at $200\gev$ shown in Fig.~\ref{fig:cands}c is probably a WW event with a
nuclear interaction very close to the IP. There is an identified electron
with 50$\gev$ energy and a three-prong tau candidate with an invariant mass
of 1.9$\gevcc$, suggesting $\ww\to\mathrm{e}\nu\tau\nu$. The remaining two
tracks have together an invariant mass of 0.1$\gevcc$. 

\begin{table}[tb]
\begin{center}
\begin{tabular}{|cccc|} \hline
$\roots$ & Run & Event & SM interpretation \\ \hline \hline
191.6 & 50210 & 5122 & Zee \\
199.3 & 51940 & 8969 & WW \\ 
201.6 & 52100 & 3592 & Zee \\ \hline \hline
191.6 & 49630 & 10846 &  $\gaga\qq$ \\ \hline
\end{tabular}
\caption[List of candidate events]
{\label{tab:dataevents} {\small Candidate events selected by the
small $\dm$ selection (first three) and the large $\dm$ selection (bottom).}}
\end{center}
\end{table}

\subsection{Limits on slepton production}
In the absence of any evidence for a signal, it is possible to constrain
the parameters of the theory that would produce such a signal. However, the
statistical nature of the counting experiment makes it impossible to rule
out with 100\% certainty any hypothetical process. Thus all limits quoted
here will be expressed at 95\% confidence level, which should be understood
as: `if the experiment were to be performed again, 
the probability to obtain a result which is in as bad or in worse agreement
between the expectation and the observation is less than 5\%'. 

\subsubsection{Mass limits}
The ratio 
\begin{equation}
\cls=\frac{\clsb}{\clb} = \frac{\prod\limits_{i}P(k \leq n;s+b)}
                               {\prod\limits_{i}P(k \leq n;b)} 
= \frac{\prod\limits_{i}\sum\limits_{k=0}^{n}e^{-(s+b)}(s+b)^k/k!}
       {\prod\limits_{i}\sum\limits_{k=0}^{n}e^{-b}b^k/k!}
\end{equation}
is used to derive a lower limit on the
hypothetical slepton mass. $\clb$ was defined in Eq.~\ref{pvalue} and
$\clsb$ follows from the same equation when the number of expected signal
events $s$ is also included. $\clsb$ is thus a measure of the
compatibility with the signal+background hypothesis. 
The product over $i$ channels makes reference to the different energy bins
where data was taken. Thus $n$, $s$ and $b$ change with the centre-of-mass
energy $i$ (see Tab.~\ref{tab:data}). To perform the calculation of the
observed $\cls$ and the average expected without looking at the data, 
the FORTRAN package described in Ref.~\cite{Junk:1999kv} was used. 

\begin{figure}[hb]
\begin{center}
\vspace{-1.0cm}
\includegraphics[width=0.72\linewidth]{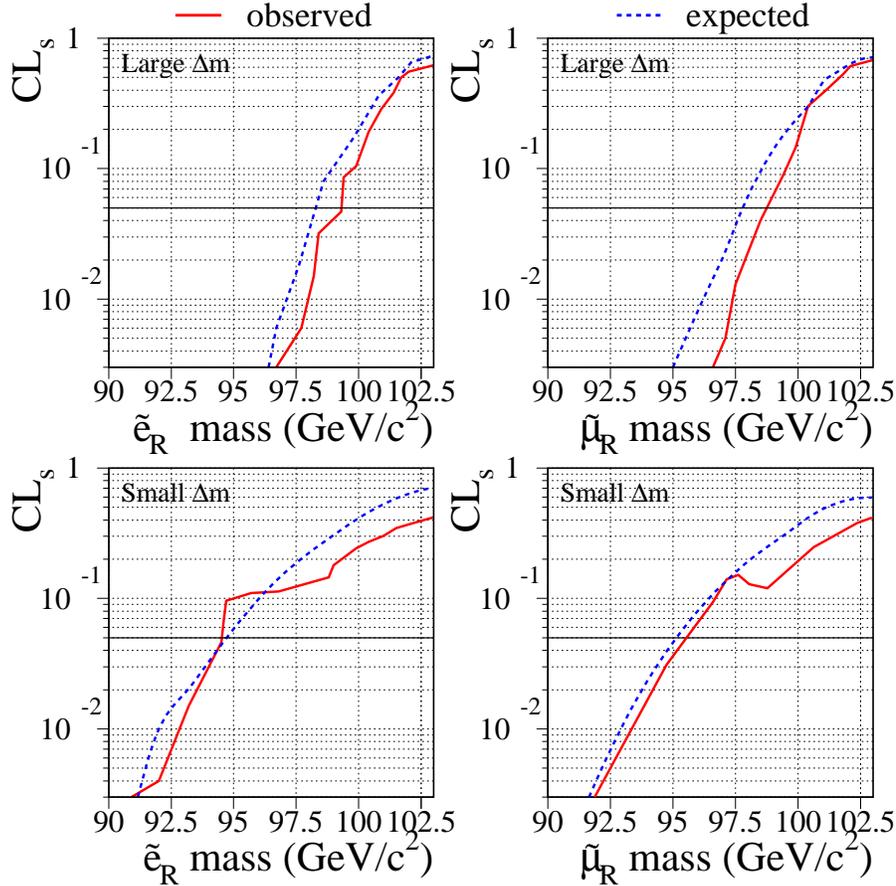}
\caption[$\cls$ for the two selections as a function of the selectron and
smuon mass]
{\label{fig:cls}{\small Lower mass limits for selectrons and smuons from
the two six-lepton selections. The large $\dm$ selection confidence level
$\cls$ is plotted in the upper row for selectron and smuon masses. 
The small $\dm$ selection in the lower plots. 
Solid lines give the observed limit and dashed lines the average expected
limit for background.}}
\end{center}
\end{figure}

The limits on the masses are set when the confidence limit ($\cls$) reaches
0.05, i.e. the probability that a signal-like experiment is less compatible
with the signal than what is observed is below 5\%. 
The confidence level for the different selections as a function of
the selectron and smuon masses is shown in Fig.~\ref{fig:cls}.

The limits on the slepton masses derived from the above confidence level
are the following:
\begin{center}
\fbox{
\begin{tabular}{cccc} 
Large $\dm$: & $m_{\sel} > 99\gevcc$ & ~ & $m_{\smu} > 98\gevcc$ \\
Small $\dm$: & $m_{\sel} > 94\gevcc$ & ~ & $m_{\smu} > 95\gevcc$ \\
\end{tabular}
}
\end{center}

\subsubsection{Cross section limits}
The calculation of a model independent cross section limit is somewhat more
problematic given the span of centre-of-mass energies of the recorded data.
To provide a combined limit, the respective
luminosities must take into account the appropriate phase space for slepton
production at each energy. 
The best approach for the combination is to scale the
luminosity of the lower energy data by the ratio of the cross section at
that energy to the cross section at $\roots=208\gev$. Thus the upper limit
on the production cross section at the 95\% C.L. can be written as:
\begin{equation}
\sNF = \frac{\nNF}{\sum_{\roots}^{}\eff(\roots)\Lum(\roots)
\frac{\sigma(\roots)}{\sigma(208)}}
\end{equation}
where $\nNF=3.00,4.74,6.30\ldots$ from Eq.~\ref{an95} 
depends only on the number of observed candidates when no background
subtraction is necessary as is the case here. 
This expression takes into account the variation of the efficiency with
$\roots$ as well as the change in luminosity if the considered slepton mass
cannot be produced at that centre-of-mass energy. 
Using the ratio of cross sections to perform this adjustment in the
phase space, is more accurate than using the simple $\beta^3/s$ factor
expected for sfermion production. Cross sections are calculated taking
into account radiative corrections and other model dependent factors
whereas the $\beta^3/s$ scale is only valid at the Born level and for
$s$-channel production.
Clearly, $\sNF$ has a dependence on the sparticle mass through the
efficiency function and the cross sections. Thus for every possible slepton
mass a value of $\sNF$ is calculated. If the production cross section for a
slepton at that mass is greater than the upper limit $\sNF$, that test mass and
more generally the model that produced that slepton are excluded. Or
inversely, one can say that only those models of the theory which
produce sleptons with $\sigma < \sNF$ are possible and still realisable in
Nature. 

\begin{figure}[tb]
\begin{center}
\includegraphics[width=0.7\linewidth]{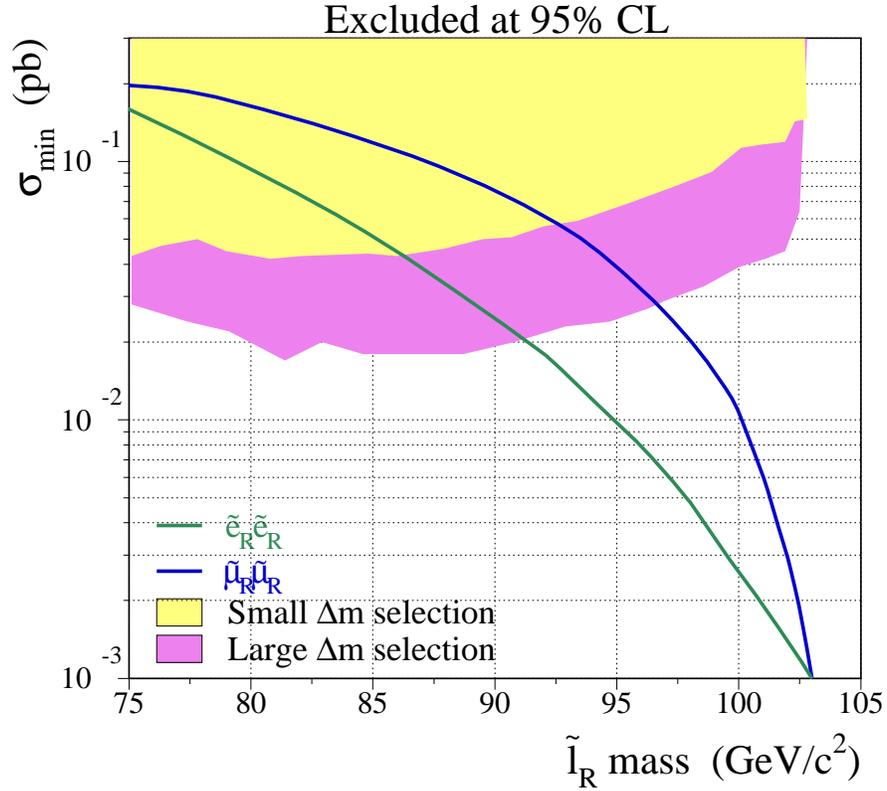}
\caption[95\% C.L. excluded cross section areas as a
function of the slepton mass]
{\label{fig:cslim}{\small 95\% C.L. excluded cross section areas as a
function of the slepton mass for the small $\dm$ (light shaded
area) and large $\dm$ (dark shaded area) selections.
The minimum production cross section for pairs of
selectrons and smuons covering all possible
six-lepton topologies is also shown for comparison.}}
\end{center}
\end{figure}

Figure~\ref{fig:cslim} shows the excluded cross sections as shaded areas
determined by the upper limit $\sNF$ for the six-lepton selections as a
function of the hypothetical slepton mass. The minimal possible production
cross section is also shown. 

The interpretation of these results in the wider context of GMSB models is
left for Chapter~\ref{scan}, where these limits and the limits imposed by
other searches in the GMSB framework are analysed in the most general scenario. 

\thispagestyle{empty}
\chapter{GMSB searches and present limits}
\thispagestyle{empty} 
\label{gmsbphen}
\vspace{-1.0cm}
\begin{center}
\setlength{\fboxsep}{5mm}
\begin{boxedminipage}[tb]{0.9\linewidth} {\small
\begin{spacing}{1.5}
Gauge mediated Supersymmetry breaking models could be realised at LEP
in the form of many different topologies. 
These arise from just two
factors: whether the NLSP is the lightest neutralino or the lightest stau,
and whether the NLSP lifetime is negligible (prompt decays), short
(decays inside the detector) or long (decays outside the detector). 
Searches to cover all possible NLSP types and lifetimes, considering also
indirect production, have been developed by other members of the ALEPH
collaboration and are described here. Their results will be used in the
next chapter to constrain possible GMSB models. 
The possible final states under study are: two energetic photons, 
non-pointing single photons, two acoplanar leptons, large impact parameter
leptons, detached slepton decay vertices, heavy stable charged sleptons and
multi-leptons plus missing energy final states.
\end{spacing}
{\small 
\begin{center}
\vspace{-0.8cm}
\begin{tabular}{cccc}
  & Negligible lifetime & Short lifetime & Long lifetime \\
\rotatebox{90}{\hspace{1.0cm}Neutralino NLSP} &
\includegraphics[width=0.25\linewidth]{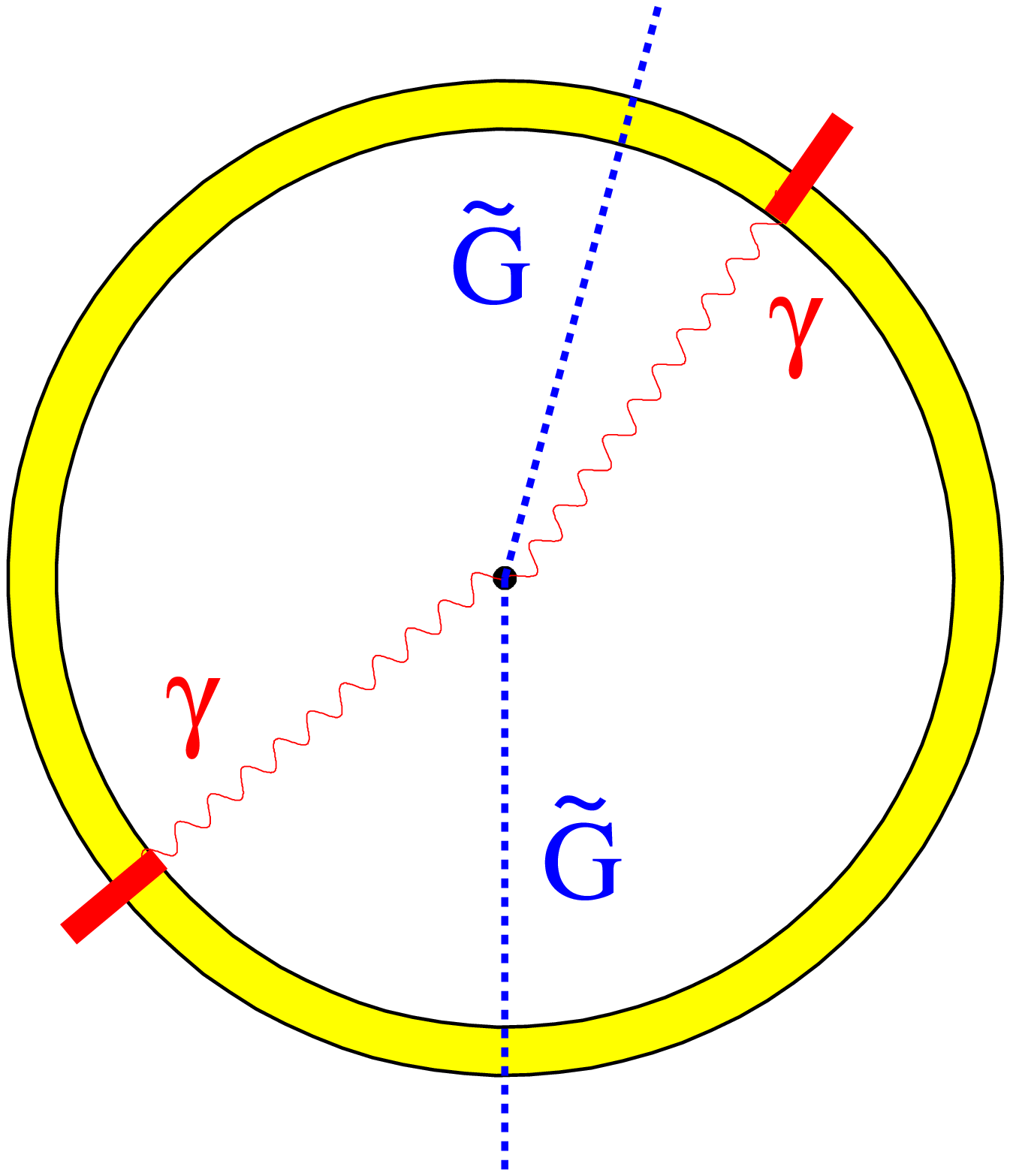} &
\includegraphics[width=0.25\linewidth]{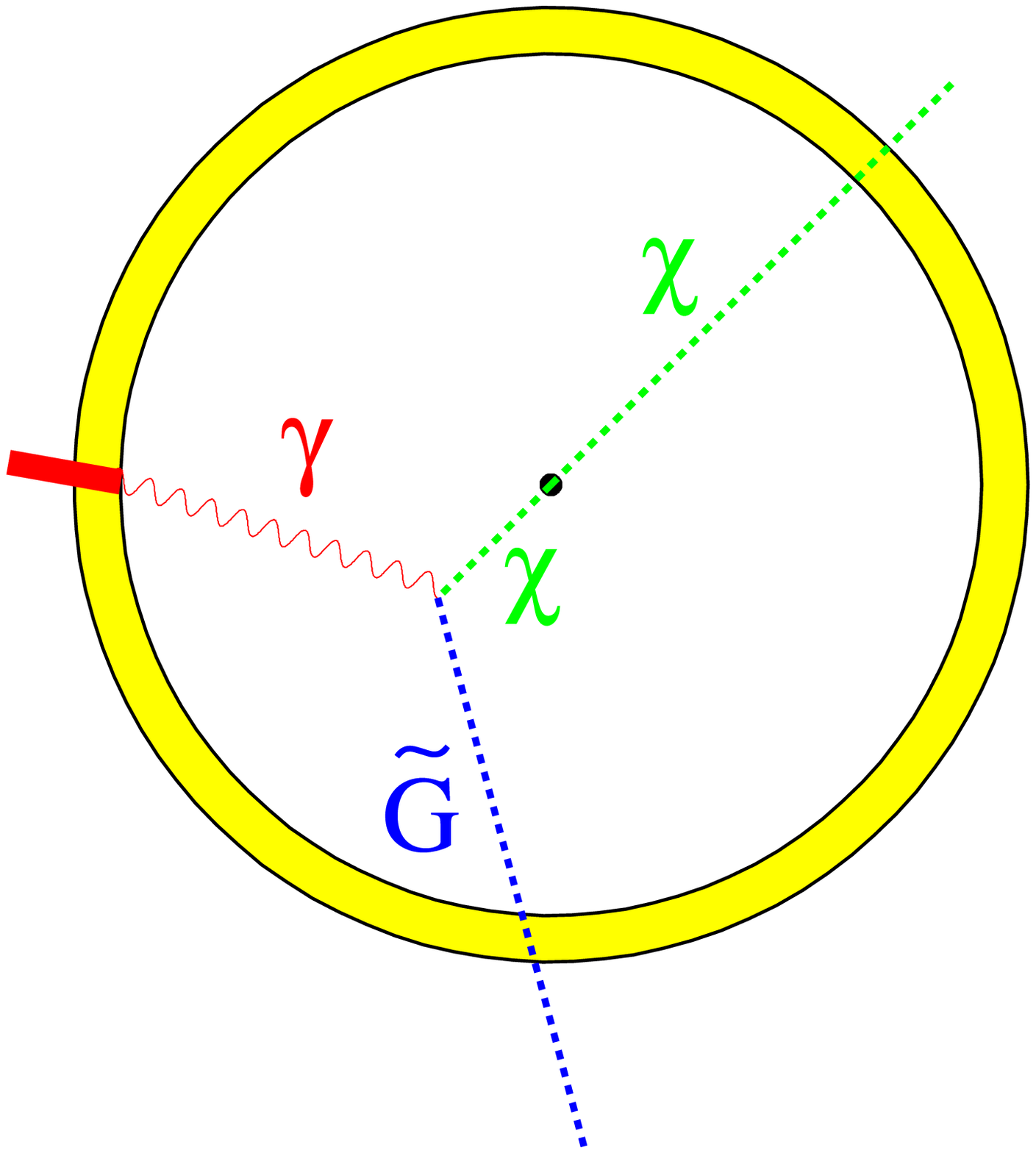} & 
\raisebox{14.0ex}{(Invisible)} \\
\rotatebox{90}{\hspace{1.0cm}Slepton NLSP} & 
\includegraphics[width=0.25\linewidth]{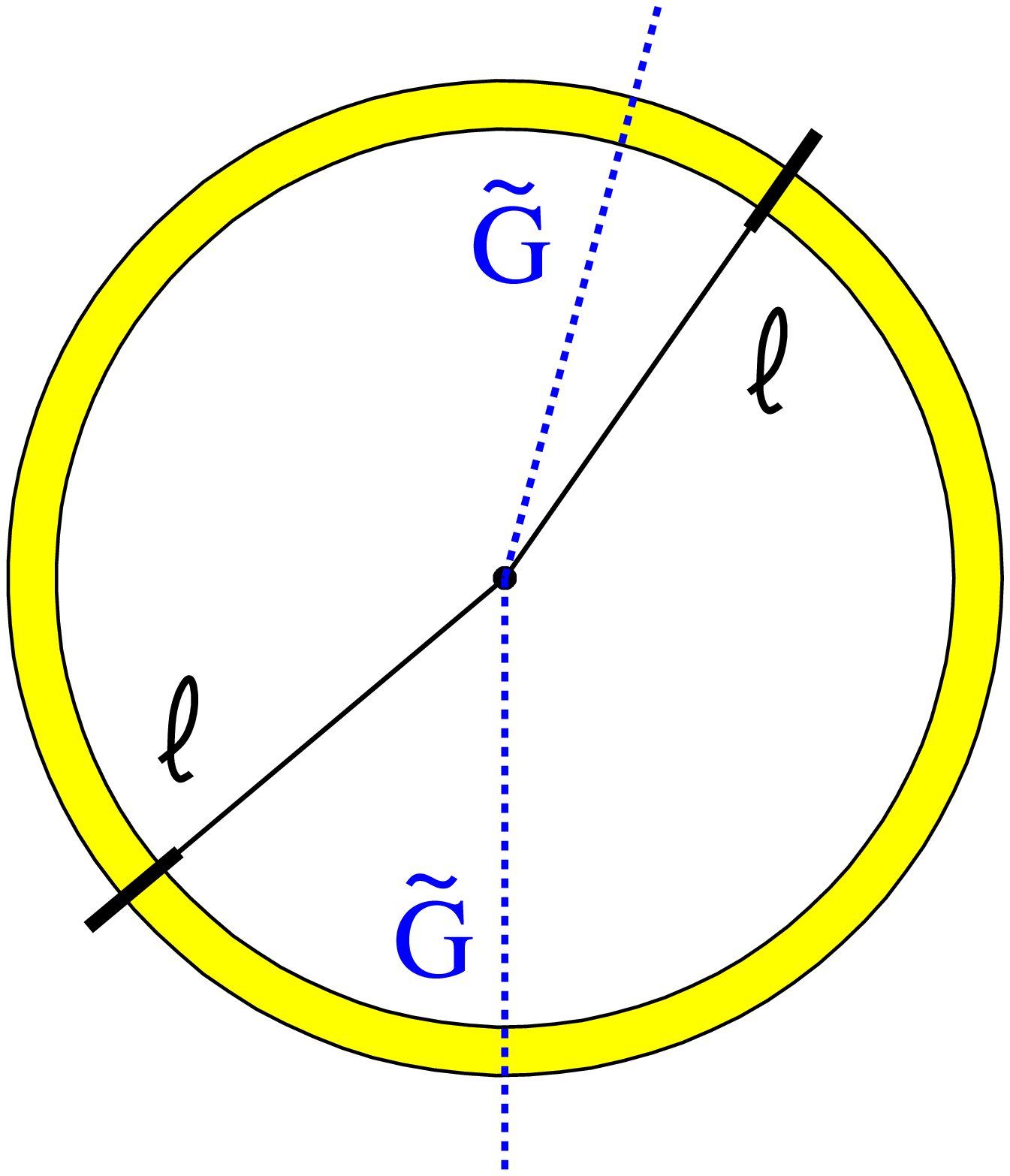} &
\includegraphics[width=0.27\linewidth]{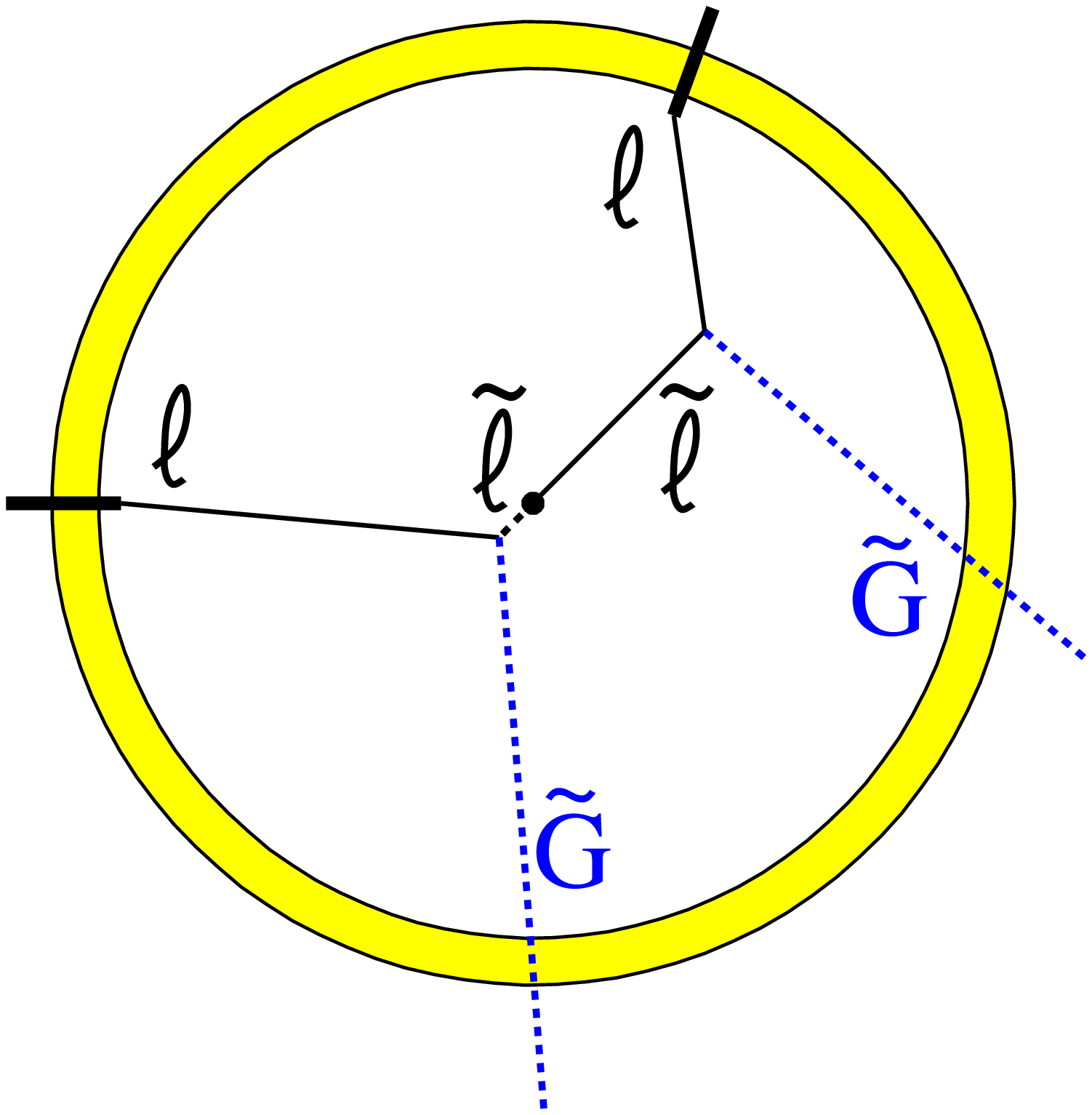} &
\includegraphics[width=0.25\linewidth]{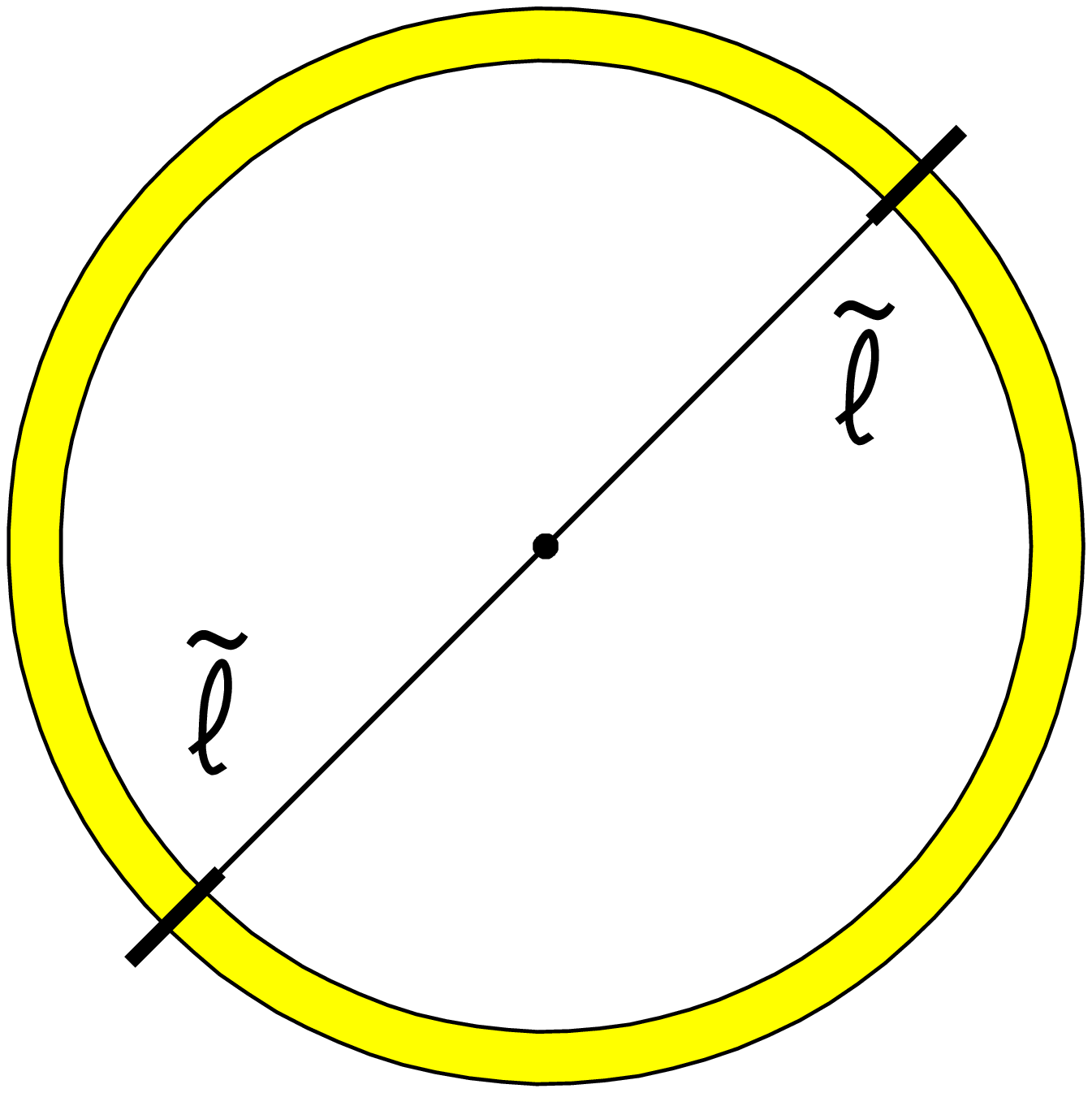}
\end{tabular}
\end{center}
}}
\end{boxedminipage}
\end{center}
\newpage
\section{Description of GMSB topologies}
There is a breadth of topologies to study at LEP2
to cover all possible GMSB scenarios if one considers not only the NLSP type but
also its lifetime~\cite{Ambrosanio:1997rv}. 
Table~\ref{tab:topo} lists all possible searches in the
different NLSP scenarios.\footnote{OPAL and DELPHI also include searches for
charginos in $\ee\to\chi^+_1\chi^-_1\to\W^*\chi\W^*\chi$ and
$\chi^+_1\chi^-_1\to\slep\nu\slep\nu$ with the NLSP decaying inside the
detector~\cite{opalgmsb}. But charginos may only be
produced at LEP2 if $\Mmess$ is very large. DELPHI has also searched for
sgoldstino production and decay to $\gaga\gamma$ and $gg\gamma$, and is able
to set limits on the scale of SUSY breaking $\sqrt{\rm{F_0}}$ from this
search~\cite{delphigmsb}.} 
\begin{table}[tb]
\begin{center}
\rotatebox{0}{
\begin{tabular}{|c|l|l|c|l|l|}
\hline
 NLSP & $\ee\to$ & Decay mode & Signal & $\lambda=c\tau\gamma\beta$ 
                                       & Expected topology \\
\hline \hline
\multirow{5}{0.5cm}{$\neu$} & & & & $\lambda \ll \ldet$ & Acoplanar photons \\
 &  $\neu \neu$ & $\neu \to \gamma \grav$  & $\gamma(\gamma)\Emiss$ & 
    $\lambda \sim \ldet$ & Non-pointing photon  \\
 &  &  &  &  $\lambda \gg \ldet$ & [Indirect search] \\
\cline{2-6}
 & \multirow{2}{0.5cm}{$\slep \slep$} &
            \multirow{2}{2.75cm}{$\slep\to\ell\neu\to\ell\gamma\grav$} 
          & \multirow{2}{1.3cm}{$\ell\ell\gaga\Emiss$} &
            $\lambda \ll \ldet$ & Leptons and photons  \\
 & & & & $\lambda \sim \ldet$ & Leptons and displaced $\gamma$ \\
\hline \hline
\multirow{5}{0.9cm}{co-$\slep$ or $\stauO$} &  &  &  &
            $\lambda \ll \ldet$ & Acoplanar leptons \\
 & $\slep \slep$ & $\slep \to \ell \grav$   & $\ell\ell\Emiss$ & 
                   $\lambda \sim \ldet$ & Kinks and large $d_0$  \\
 &  &  &  & $\lambda \gg \ldet$ & Heavy charged particles   \\
\cline{2-6}
 & \multirow{2}{0.5cm}{$\neu \neu$} & 
            \multirow{2}{2.75cm}{$\neu\to\ell\slep\to\ell\ell\grav$} &  
            \multirow{2}{1.8cm}{$\ell\ell^{\prime}(\ell\ell^{\prime})\Emiss$} & 
            $\lambda \ll \ldet$  & Four leptons   \\
&  &  &  & $\lambda \sim \ldet$  & Kinks and large $d_0$ \\
\hline \hline
$\stauO$ & $\slR\slR$ & $\slR\to l\tau\stauO$ & 
          $\tau\tau(l l \tau\tau)\Emiss$ & $\lambda \ll \ldet$ & Six leptons \\ 
\hline
\end{tabular}
}
\caption[Possible final state topologies in GMSB]
{\label{tab:topo}{\small All possible final state topologies
according to the NLSP scenario and decay length $\lambda$.
Particles in brackets may be soft or even undetected. Only 
$\ell\ell\gaga\Emiss$ signals were not studied in the $189-209\gev$ ALEPH
data set.}}
\end{center}
\end{table}

Depending on the gravitino mass, the following topologies are 
expected:
\begin{itemize}
\item If the gravitino mass is below a few $\evcc$, the NLSP decays
immediately after its production before the tracking subdetectors.
In the neutralino NLSP scenario, two very energetic
photons and missing energy are expected from direct neutralino
production. If the sleptons are accessible, the cascade decay
$\slep\to\ell\neu$ with two
photons and two leptons in the final state is a background free search. 
Nevertheless, two photon searches alone are able to exclude most of the 
neutralino NLSP parameter space in the `no lifetime' region. In the slepton
NLSP scenario, if sleptons are directly produced, SUGRA searches for two
acoplanar\footnote{The acoplanarity
$\alpha$ is defined as the angle between the projections of two tracks into
the transverse plane.}
 leptons can be used. Also cascade decays $\neu\to\ell\slep$ are
possible here, if the neutralino pair-production is kinematically
allowed. Indeed, its production cross section will be greater than the slepton
production cross section because of the $\beta^3/s$ suppression factor on
scalars. Finally, in the purely stau NLSP case, where selectrons and smuons
are heavier than the stau, final states with two leptons and four taus are
possible, as described in the preceding chapter.
\item For $m_{\grav}$ between a few $\evcc$ and a few hundred $\evcc$, the
NLSP will decay somewhere inside the detector (see Fig.~\ref{fig:NLSPnature})
The decay products of the NLSP, either photons or leptons, will present
large impact parameters or in the case of slepton NLSP when the slepton
is reconstructed in the tracking volume, the slepton and the lepton will
form a `kinked' track. 
\item For $m_{\grav}$ above a few hundred $\evcc$ the NLSP is stable for
detector searches as it decays outside the detector volume. Only in the
slepton NLSP case are direct searches possible: highly ionising
back-to-back tracks are formed by slow moving sleptons. The situation in
the neutralino NLSP reverts to the usual SUGRA phenomenology with
a neutralino LSP escaping the detector, and only searches for indirect
production of neutralinos are possible. 
\end{itemize}

\section{Searches for neutralino NLSP}
Searches for $\neu\to\gamma\grav$ rely on good photon identification. The
highly segmented ALEPH ECAL makes it possible to measure the energy of
isolated photons with an excellent resolution (Eq.~\ref{eres}). 
The main SM background source for this search is $\ee\to\nu\bar{\nu}$, 
from diagrams with $s$-channel Z exchange or $t$-channel W exchange and
one or two initial state radiation (ISR) photons. 
The invariant missing mass $M_{\rm{miss}}^2\equiv
(p_{\rm{e^+}}+p_{\rm{e^-}}-p_{\gamma_1}-p_{\gamma_2})^2$, will thus have a
pronounced peak at the Z mass and a long tail from the contribution of the W
exchange. 

The simulation of the background was performed using the KK
generator~\cite{Jadach:1999vf}, which was checked with an independent generator
NUNGPV~\cite{Montagna:1998ce}. Signal processes were simulated using
SUSYGEN~\cite{susygen2}.   
\subsection{Acoplanar photons}
\begin{figure}[tb]
\begin{center}
\vspace{-0.5cm}
\includegraphics[width=0.49\linewidth]{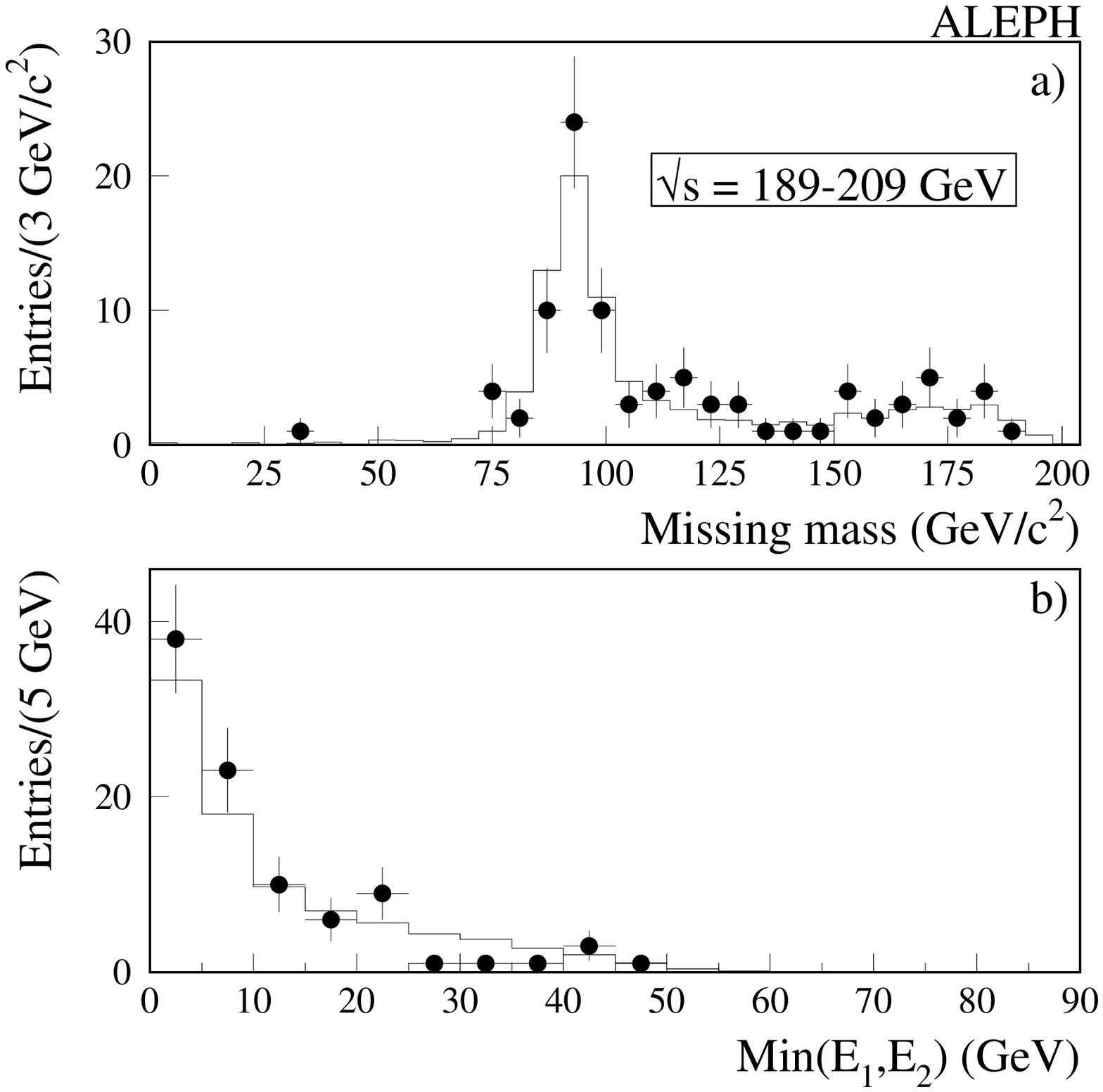}
\includegraphics[width=0.49\linewidth]{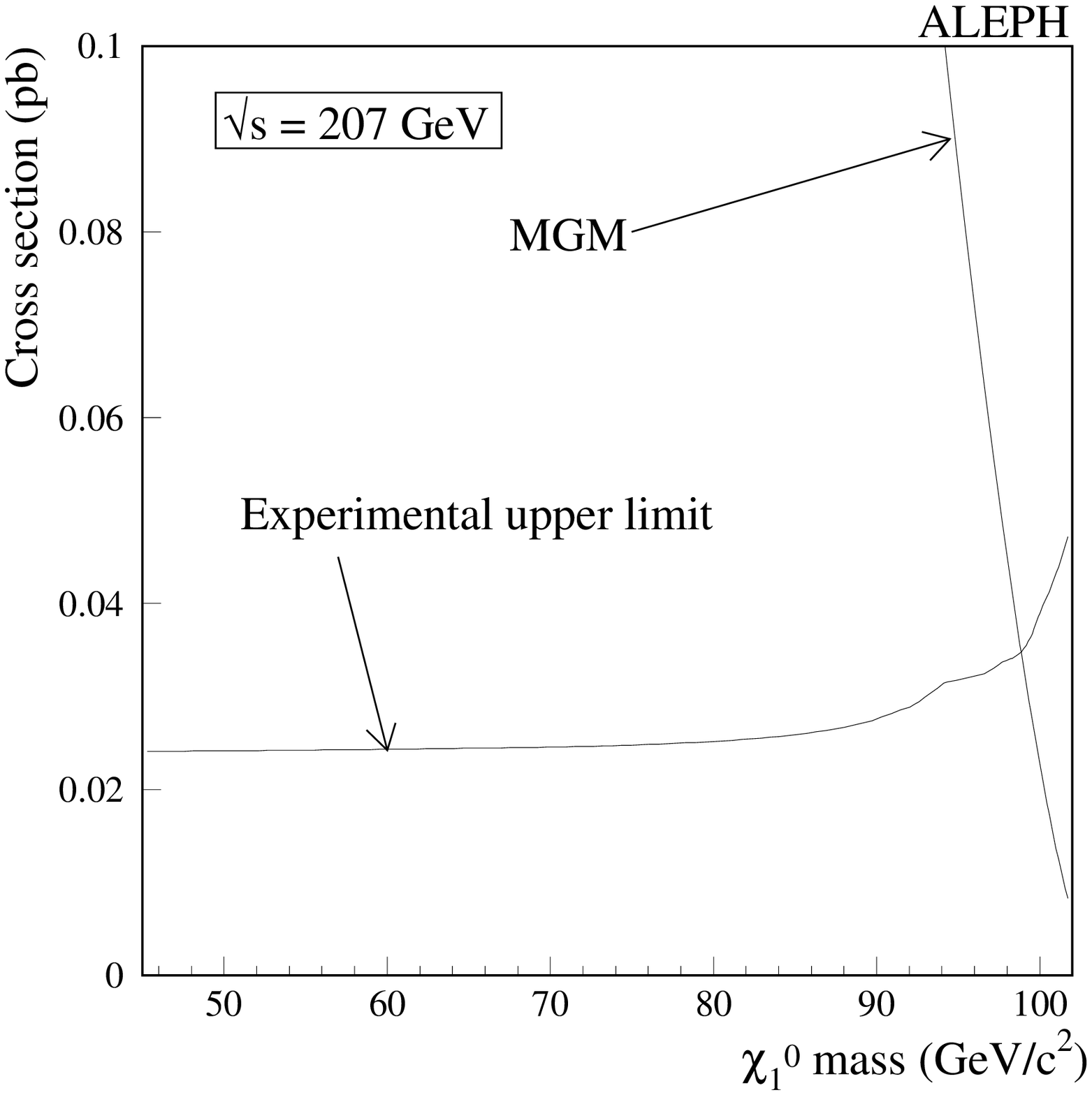}
\caption[Two-photon analysis results]
{\label{fig:2phot} {\small The distributions of (a) the invariant
mass of the system recoiling against the photon candidates and (b) the
energy of the second most energetic photon, for the two photon search
before the cut on the energy of the second most energetic photon. The
SM expectation is the histogram and dots with errors are data. On the
right, the 95\% confidence level upper limit on the $\ee\to\neu\neu$ cross
section in the neutralino NLSP scenario when the neutralino lifetime is
negligible: $\tau_{\neu}<3\ns$. The expected cross section for the MGM
model~\cite{Dimopoulos:1996va} (where the neutralino is purely bino and
$m_{\selR}=1.1m_{\neu}$) is also shown. From Ref.~\cite{Heister:2002ut}.}}
\end{center}
\end{figure}

Each neutralino is produced with the beam energy $\roots/2$, and since the
gravitino is nearly massless and the decay $\neu\to\gamma\grav$ is isotropic
in the neutralino rest frame, the photon energy spectrum is expected to be
flat $E_{\rm{min}}<E_{\gamma_1},E_{\gamma_2}<E_{\rm{max}}$,
with~\cite{Ambrosanio:1997rv}: 
\begin{equation}
\label{photendpoints}
E_{\rm{max,min}}=\frac{1}{4}\left ( \sqrt{s}\pm\sqrt{s-4m_{\neu}^2} \right )
\end{equation}
Thus imposing a threshold cut of $E_{\rm{min}}=37\gev$ on the energy of the
least energetic photon is very effective against the SM background, where
the photons are produced via bremsstrahlung. The cut
value was optimised for models with a pure bino neutralino and
$m_{\selR}=1.1m_{\neu}$~\cite{Dimopoulos:1996va}. 
The preselection procedure is based on the requirement of two
acoplanar photons and cuts to reduce the $\ee\to\gaga(\gamma)$ events, like
$\alpha < 177\degs$ and no additional energy greater than $1\gev$. To
reduce the doubly radiative Bhabha events $\ee\to\ee\gaga$, a cut on the
total transverse momentum of the photon system $\pt$ is very
effective. The complete set of cuts for this search is summarised in
Tab.~\ref{tab:neucuts}. The distribution of data and background for the missing
energy and $E_{\gamma_2}$ is shown in Fig.~\ref{fig:2phot} before the cut
on $E_{\gamma_2}$. 

A total of four candidate events are selected in the data while 4.9 are
expected from background. After full subtraction of the expected
background, a 95\% C.L. upper limit on the neutralino production cross
section can be set for an NLSP laboratory lifetime of less than $3\ns$, as
shown in Fig.~\ref{fig:2phot}. Systematic uncertainties, mainly due to
the photon reconstruction efficiency and the level of background, only
change the cross section limit by less than 1\%~\cite{Heister:2002ut}. 
\begin{table}[tb] 
\begin{center} 
\begin{tabular}{|c|c|} 
\hline 
Two photons & Non-pointing photon \\ \hline \hline 
$N_{\gamma}$=2 or ($N_{\gamma} \geq 3$ and $E_{\rm{miss}} > 0.4 \sqrt{s}$) & 
$N_{\gamma}$=1 and $d_0>40\cm$\\ 
Accept conversions & No conversions \\  
Acoplanarity $< 177\degs$ & $E_{14}=0$ \\ 
$E_{\gamma_2} \geq 37\gev$ & $E_{\rm{ECAL}} \geq 0.4 E_{\gamma}$ and  
                             $E_{\rm{ECAL}}^{15\cm}=0$ \\ \hline  
\multicolumn{2}{|c|}{Additional energy $<$ 1 GeV} \\ 
\multicolumn{2}{|c|}{$|\cos\theta_\gamma|<0.95$} \\ 
\multicolumn{2}{|c|}{Total $\pt > 0.0375 E_{\rm{miss}}$}  \\ \hline 
\end{tabular} 
\caption[List of cuts used in searches for a $\neu$ NLSP] 
{\label{tab:neucuts} {\small List of cuts for the neutralino NLSP 
scenario in the no lifetime case (two acoplanar photons and $\Emiss$) and 
the intermediate lifetime case (non-pointing photon).}} 
\end{center} 
\end{table} 

\subsection{Non-pointing photons}
If the neutralino NLSP has medium decay length such that it can decay
inside the detector, the most probable topology is a single photon
which does not point to the interaction region. 
Therefore one of the produced neutralinos is more likely to escape the detector 
while the other decays before the ECAL. Here,
the calorimeter granularity is crucial to be able to reconstruct the shower
axis and thus accurately determine the displaced vertex distance to the
IP.  The basic selection criteria are to select only one identified photon 
with an impact parameter greater than $40\cm$ in the acceptance region
$|\cos\theta_\gamma|<0.95$ and require no charged
tracks (which excludes photon conversions). To eliminate radiative
Bhabha events $\ee\to\ee\gamma$, there must be no energy deposited within
$14\degs$ of the beam axis and less than $1\gev$ in additional energy. To
further reduce ECAL noise and cosmic muons, at least 40\% of the photon
energy must be recorded in the ECAL, there must be no activity in the muon
chambers or within a transverse distance of $15\cm$ from the photon candidate
in the ECAL. 

The efficiency for this selection reaches a maximum of 10\% at neutralino
decay lengths of around 8\,m. After all cuts, 0.8 events are expected
from $\ee\to\nu\bar{\nu}\gamma$ and $0.2\pm0.2$ events from cosmic rays and
detector noise. Two events are selected in the data. 

\begin{figure}[h]
\begin{center}
\includegraphics[width=0.48\linewidth]{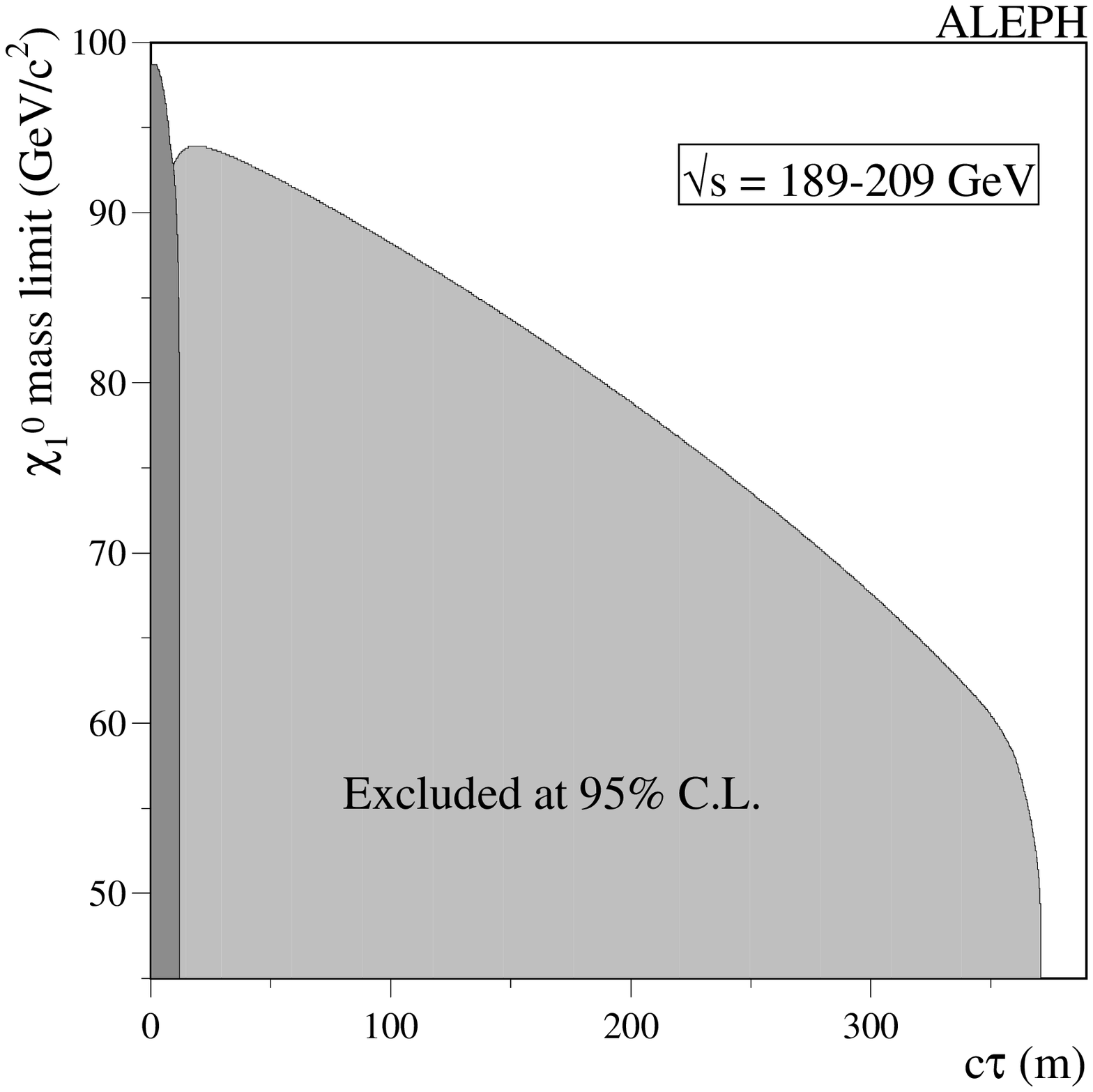}
~
\includegraphics[width=0.46\linewidth]{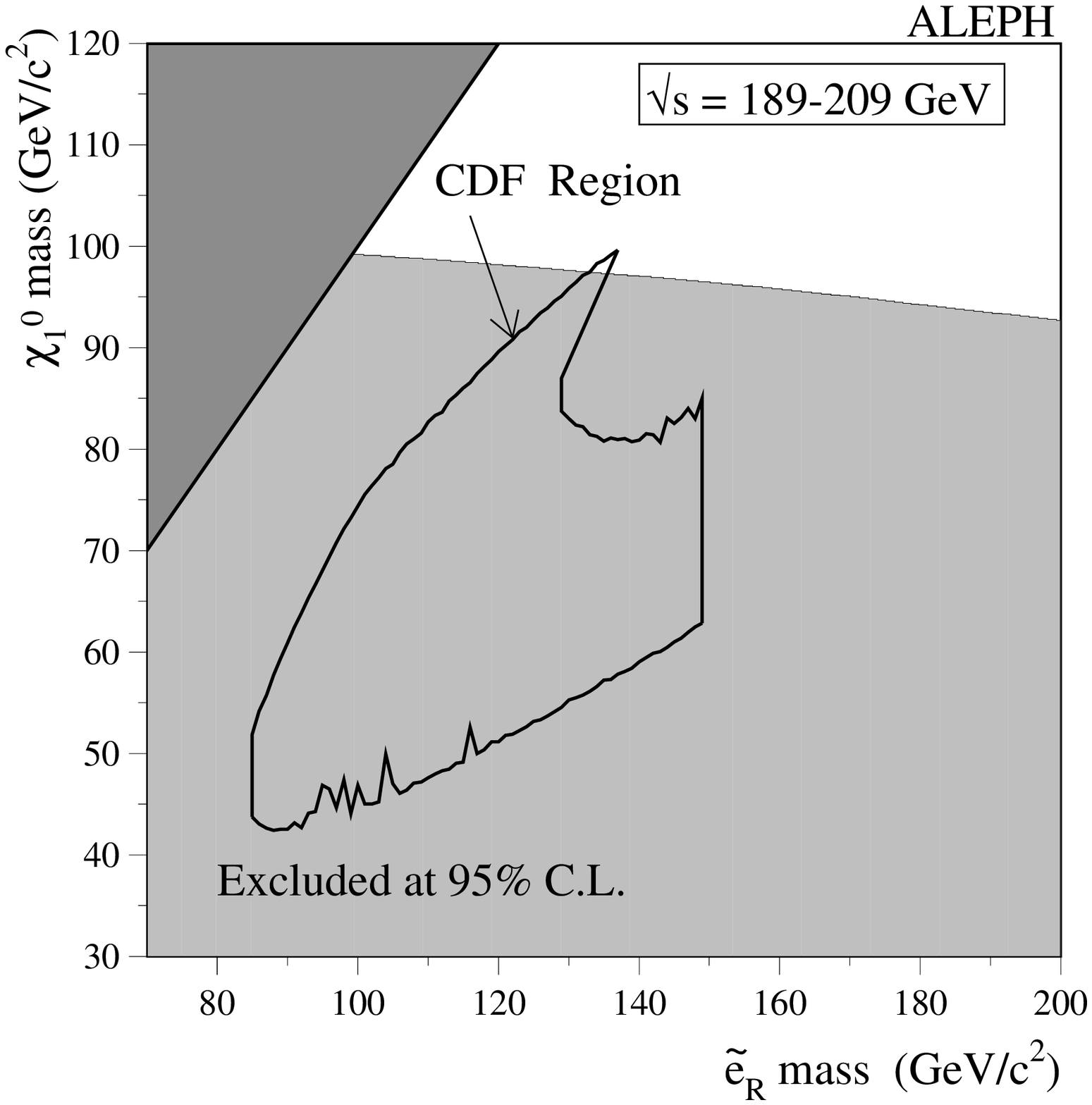}
\caption[Excluded neutralino mass as a function of its lifetime and as a
function of the selectron mass]
{\label{fig:neulim} {\small Left, the excluded neutralino mass as
a function of the proper lifetime for the two-photon search (dark shade) and
non-pointing photon search (light shade). On the right, in the
neutralino-selectron mass plane the 95\% C.L. excluded area by photon
searches assuming a pure bino neutralino NLSP. Overlaid is the region
favoured by the CDF event~\cite{Abe:1998ui}, interpreted as
$\qq\to\selR\selR\to\ee\neu\neu\to\ee\gaga\grav\grav$~\cite{Ambrosanio:1996zr}. 
The dark shaded region corresponds to $m_{\selR}<m_{\neu}$. From
Ref.~\cite{Heister:2002ut}.}} 
\end{center}
\end{figure}

A limit on the neutralino mass as a function of its decay length is
obtained under the assumptions of the MGM (minimal gauge mediated)
model~\cite{Dimopoulos:1996va},  
with a maximum of $98.8\gevcc$ at 95\% C.L. when both two-photon and 
non-pointing photon searches are included. 
As will be seen in the next chapter, this limit
is reduced to $\sim$$94\gevcc$ if the conditions of the MGM model are removed. 
The CDF detector at the Tevatron recorded an event with two opposite charge
electrons with transverse energy of 63 and $36\gev$, two photons of 36 and
$30\gev$ and a total missing transverse energy of $55\gev$~\cite{Abe:1998ui}. 
This type of signature has a very low background expectation ($\sim$$10^{-3}$)
and could be interpreted as a GMSB process in the neutralino NLSP
scenario:
$\qq\to\selR\selR\to\ee\neu\neu\to\ee\gaga\grav\grav$~\cite{Ambrosanio:1996zr}.
At LEP, the plane ($m_{\selR},m_{\neu}$) is directly probed by the direct
production of neutralinos through the selectron $t$-channel exchange, thus
the searches described above are sensitive to the same parameter
space. Figure~\ref{fig:neulim} shows the excluded area in that plane. The
region favoured by the CDF event is now completely excluded with 95\% C.L.

\subsection{Indirect searches}
For neutralinos with decay lengths greater than the detector size, 
indirect searches have to be used to obtain a limit on the neutralino
mass. This scenario is identical to gravity mediated models with a 
stable neutralino LSP escaping the detector.
Thus searches for sleptons $\ee\to\slep\slep\to\ell\neu\:\ell\neu$ and
charginos $\ee\to\chaOp\chaOm\to\W^{*+}\neu\:\W^{*-}\neu$, as developed
for the SUGRA case, are utilised here\footnote{
If the sleptons are lighter than the charginos, two-body
decays of charginos will dominate: $\chaOp\to\slep^+\nu$ or even
$\chaOp\to\ell^+\snu$.
}.
The relationship between the $\neu$ mass and the 
slepton and chargino masses in these processes can be exploited to set
limits on the neutralino mass. 
Searches for two acoplanar leptons and missing energy are described
in Ref.~\cite{Heister:2001nk}. The exclusion limits in the
slepton-neutralino mass plane can be seen in Fig.~\ref{fig:acoplim}.  

Charginos in the MSSM can either decay to a W boson and a neutralino
(generally referred as three-body chargino decay), to a sneutrino and the
corresponding lepton (two-body decay), or to a slepton and the corresponding
neutrino (slepton decay mode). 
Chargino searches in the three-body decay mode 
are split into three, according to the decays of the virtual W: 
hadronic $\chaOp\to\qq\neu$ (four jets
and missing energy), leptonic $\chaOp\to\ell^+\nu\neu$ (two acoplanar
leptons and large missing energy) and mixed (two jets and one lepton). 
The final results are given in Ref.~\cite{charginos} and the selections are
described in Ref~\cite{Barate:1999fs}.

\section{Searches for slepton NLSP}
\subsection{Acoplanar leptons}
If the slepton lifetime is negligible, of the order of a few mm or less,
the GMSB signature arising from slepton pair-production $\slep\to\ell\grav$
does not differ from that of models with gravity mediated SUSY breaking and
neutralino LSP with a very small mass $\slep\to\ell\neu$ . Two energetic
acoplanar leptons and missing energy are expected. The search used to
cover this topology is developed in the SUGRA framework but its results can
be interpreted in GMSB taking $m_{\neu}\simeq 0$ (the gravitino is at least
six orders of magnitude lighter than the neutralino). Details of the
selection criteria are described in Ref.~\cite{Heister:2001nk}. 
The largest background is due to W pair-production followed by leptonic W
decay: $\W\to\ell\nu$ which is kinematically very similar to
$\slep\to\ell\grav$ for slepton masses around the W mass. Cuts on
$M_{\rm{vis}}$ and the momentum of the most energetic lepton are needed to
reduce this type of background\footnote{
The lepton's energy distribution is flat with the same endpoints
$E_{\rm{min}},E_{\rm{max}}$ as Eq.~\ref{photendpoints} with $m_{\slep}$. 
Thus leptons are expected to be energetic and, specially smuons and staus
with only $s$-channel production, 
to have symmetric distributions in the polar angle $\theta$.}.

\begin{figure}[p]
\begin{center}
\renewcommand{\subfigtopskip}{-10pt}
\subfigure{\includegraphics[width=0.45\linewidth]{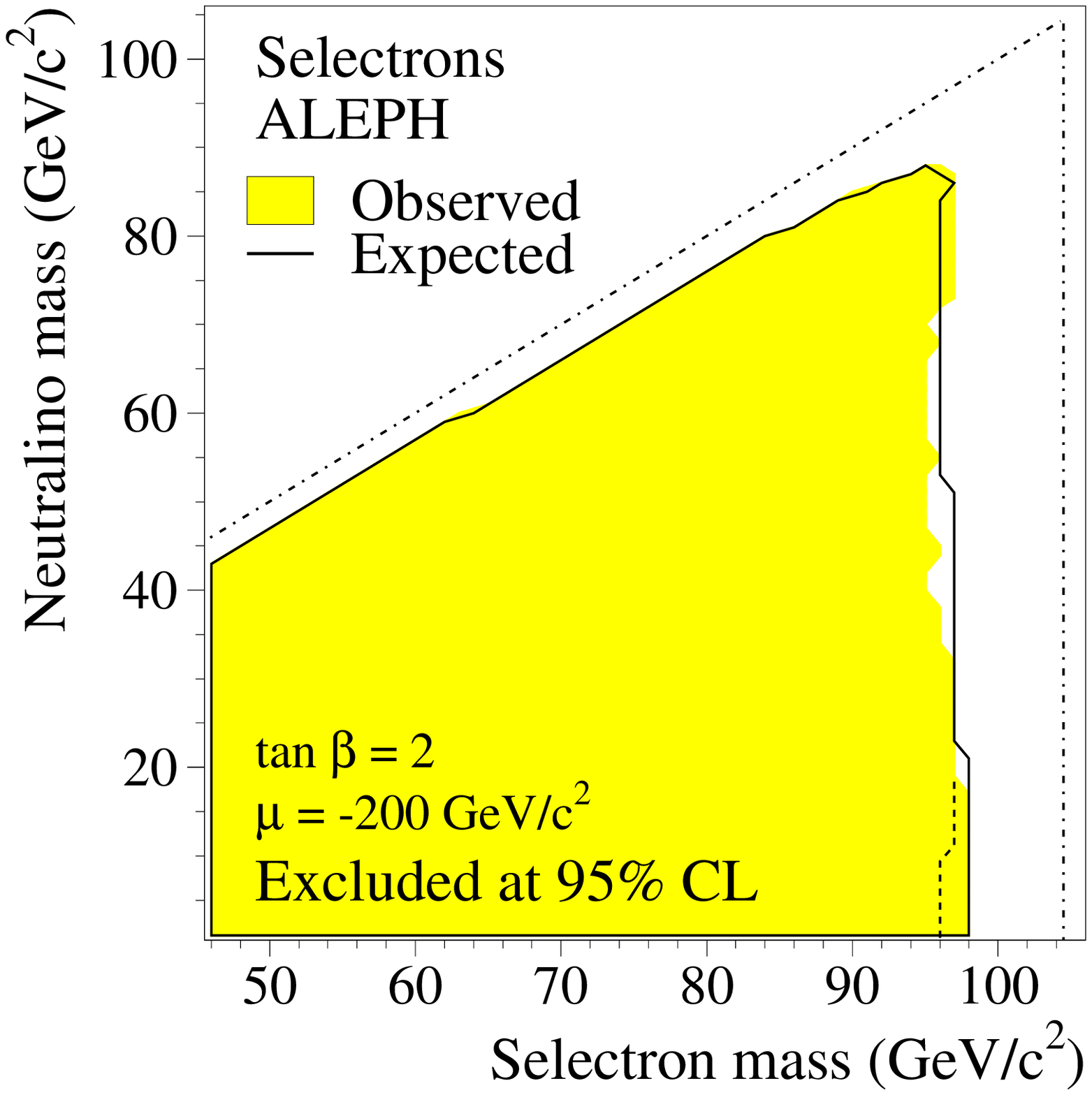}} 
\subfigure{~~~~~\includegraphics[width=0.45\linewidth]{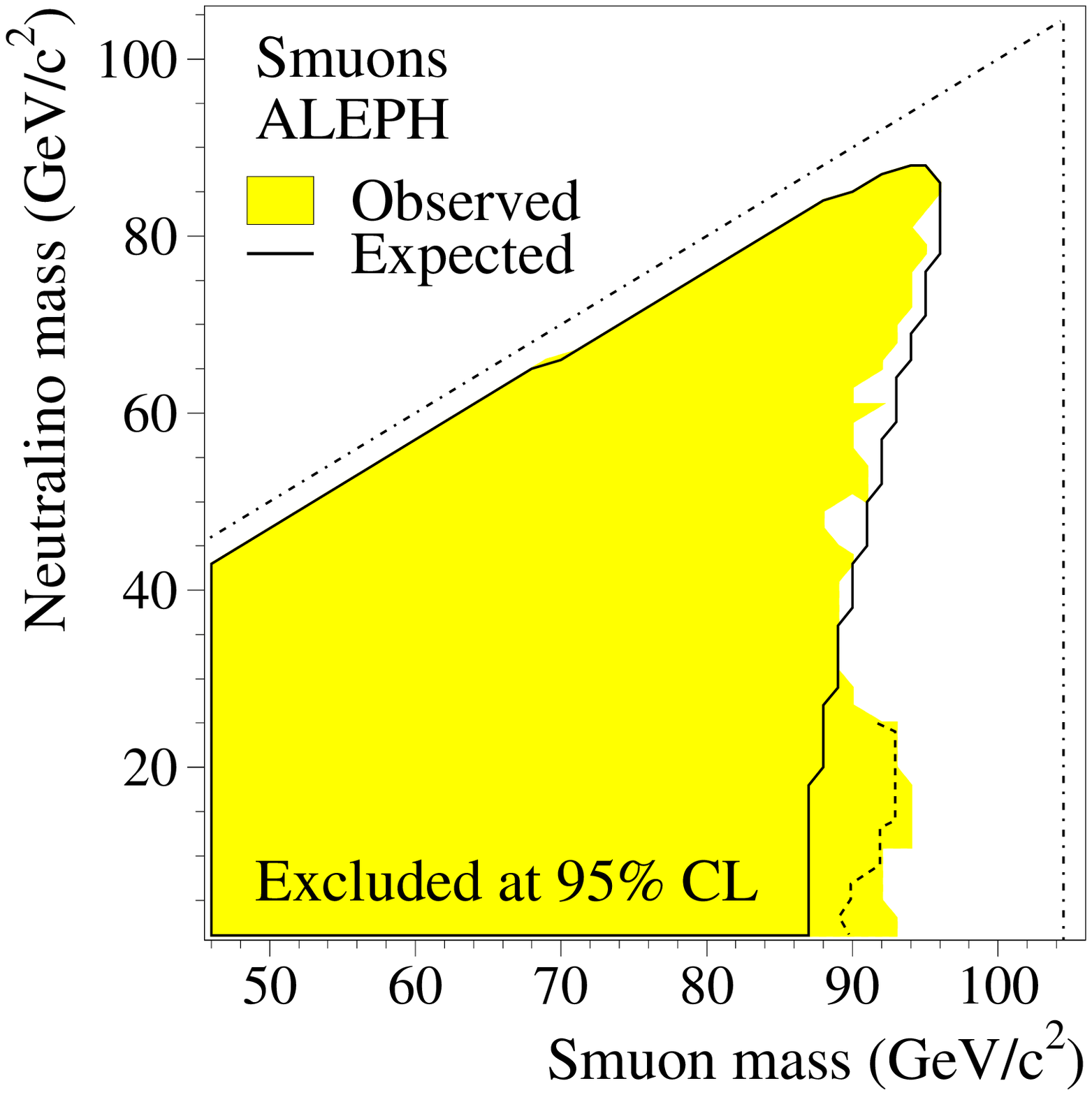}} \\ 
\subfigure{\includegraphics[width=0.45\linewidth]{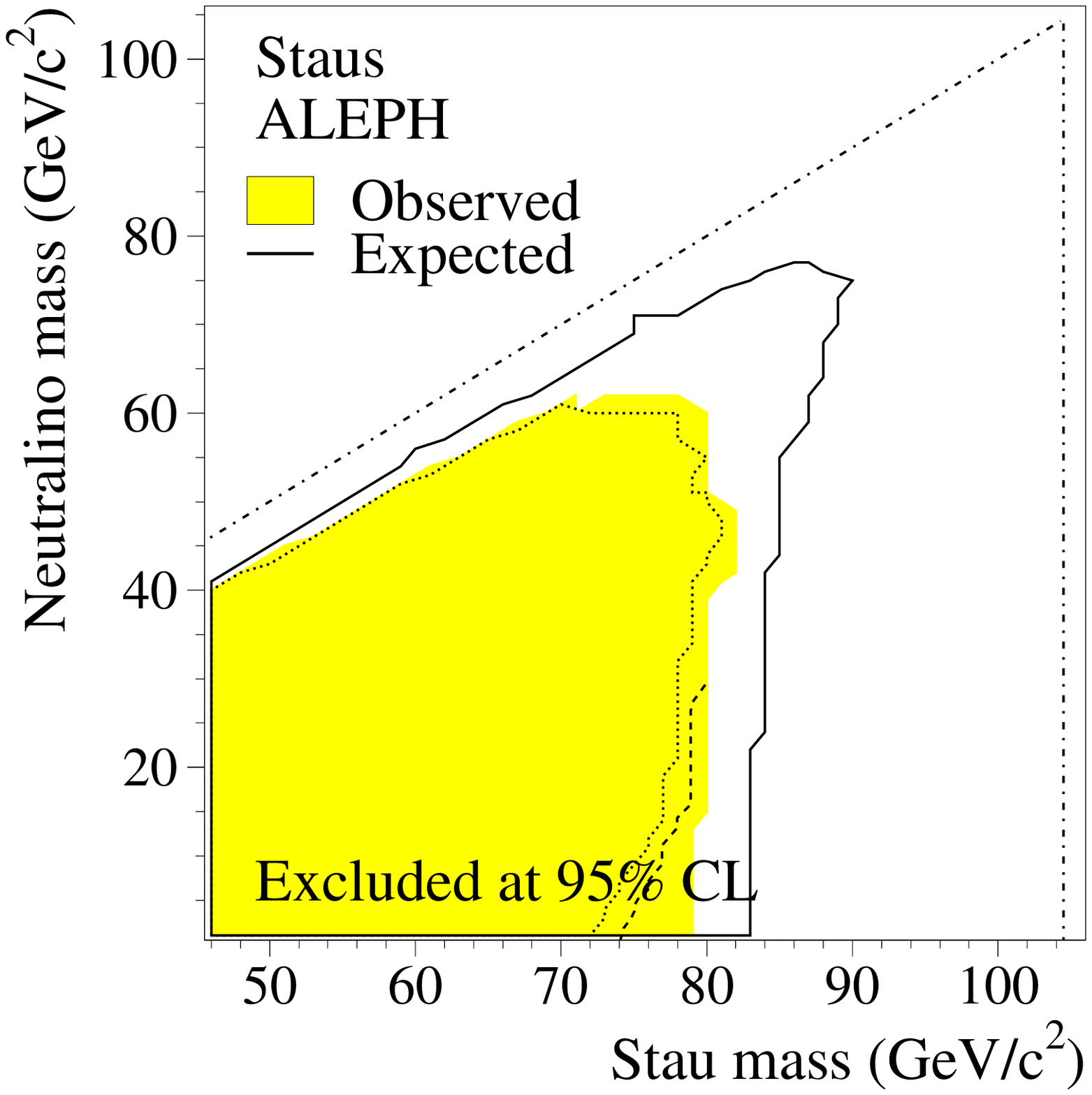}}
\subfigure{\hspace{0.48\linewidth}}
\begin{picture}(1,1)(200,-100)
\small \begin{tabular}{|c|cc|}\hline
Channel & Observed limit & Expected limit \\
        &  $m_{\slep}~(\gevcc)$ & $m_{\slep}~(\gevcc)$ \\ \hline \hline
$\selR$ & 98 & 98 \\
$\smuR$ & 93 & 87 \\
$\stauR$& 79 & 83 \\
$\stau$(min) & 72 & 77 \\ \hline
\end{tabular}
\end{picture}
\caption[Excluded regions at 95\% C.L. in the
$(m_{\slepR},m_{\neu})$ plane in the acoplanar leptons SUGRA
search]
{\label{fig:acoplim}{\small Excluded regions at 95\% C.L. in the
$m_{\slepR}$ versus $m_{\neu}$ plane in the acoplanar leptons SUGRA
search~\cite{Heister:2001nk}.
The observed (shaded area) and expected (solid curve) limits are given for
BR($\slepR\to\ell\neu$)=1. The dashed curves give the observed limits when
slepton cascade decays (into $\ell\neu_2$) are taken into account,
generally for very small neutralino masses.
The dashed-dotted curves mark the kinematically accessible region
in the MSSM at $209\gev$ for a neutralino LSP. The table lists the lower
limits on the slepton masses for $m_{\neu}\simeq 0$. For the stau,
the minimum limit (shown as the dotted curve in the stau plot) 
is calculated with a mixing angle of $\theta_{\stau}\simeq 52\degs$ 
which minimises the production cross section.}}
\end{center}
\end{figure}

For each lepton flavour, the number of data events selected in year 2000
data is 39, 39 and 11 for electrons, muons and taus, to be compared with
respectively 38.7, 34.7 and 12.2 expected events from background. The
contribution from the WW background is around $80\%$. 
Thus good agreement between the SM prediction and the experimental results
is obtained in this topology. 
The expected and observed mass limits (with WW background
subtraction) for each slepton flavour
are shown in Fig.~\ref{fig:acoplim}. 

\subsection{Large impact parameters and kinks}
Two distinctive topologies with very small background level may arise for
intermediate slepton lifetimes. If the slepton decays before having
produced a reconstructible track, i.e. for $\lambda$ between a few mm and a
few cm, the lepton tracks will present large distances to the beam axis
(impact parameters). Otherwise, if the slepton reaches the tracking devices
and is reconstructed as a charged track decaying with $\lambda\gtrsim
40\cm$, the typical signature consists then of kinks formed between the
slepton track and its lepton decay product.  

The preselection procedure is the same for both large impact parameter and
kinks searches. Anti-$\gaga$, dilepton and cosmic rays cuts are applied as
listed in Tab.~\ref{tab:lipk}.  
\begin{table}[tb] 
\begin{center}
\small
\begin{tabular}{|c|c|c|}\cline{2-3}  
\multicolumn{1}{c|}{} & Large impact parameter & Kinks \\ \hline \hline
\multirow{2}{1.5cm}{Anti-$\gaga$} 
& \multicolumn{2}{|c|}{1 or more tracks with 6 TPC hits and $p_{\rm{t}}>0.03\roots$} \\
&  \multicolumn{2}{|c|}{ $\Evis>0.03\roots$} \\ \hline 
Anti-$\ell\ell$, & \multicolumn{2}{|c|}{Remove conversions}  \\
cosmics,         & \multicolumn{2}{|c|}{ECAL timing with single helix fit} \\
Bhabha           & $\Evis<0.65\roots$   & $\Evis<0.90\roots$ \\ \hline
\multirow{7}{1.5cm}{Tracks quality cuts}
& 2 or 4 tracks with $\geq 4$ TPC hits
                      & $\leq 6$ tracks with TPC hits. \\
&                     & Same-charge tracks \\
&                     & with $d_{r\phi}<5\cm$ and $d_{z}<20\cm$.\\  
& \underline{Track \#1}: $d_0>1\cm$ and $p_1>0.01\roots$ 
                       & \underline{Inner track}: $d_0^{(1)}<0.5\cm$. \\ 
& \underline{Track \#2}: $d_0^{\prime}>0.025\cm$  
                       & \underline{Outer track}: $d_0^{(2)}>1\cm$ \\
&                      & with at least 4 TPC hits, \\
&                      & $p>0.015\roots$ and $\alpha_{2,z}>18.2\degs$  \\ \hline 
\multirow{3}{1.5cm}{Anti-$\gaga$, $\ell\ell$ and had.\,ints.} 
& Acoplanarity $< 175\degs$  & $\alpha_{1,2}>5\degs$ or $\alpha_{1,2}>10\degs$\\
&                      &  if outer track has no TPC hits. \\
& Acollinearity $< 11.5\degs$ & $E_5^1<5\gev$ \\ \hline
 \end{tabular}
\caption[List of cuts for the large impact
parameter and kinks selections]
{\label{tab:lipk}{\small List of cuts for the large impact
parameter and kinks selections.}} 
\end{center}
\end{table}
\subsubsection*{Search for large impact parameters}
Events containing exactly two or four charged tracks are selected to
attempt tau identification. At least one track (track 1) is required to
have an impact parameter greater than $1\cm$ and total momentum of at least
$1\%\roots$.  
Only one or three additional tracks with at least four TPC hits are
allowed. If there are three tracks, they must be consistent with a
three-prong tau decay. These tau triplets are treated as one track (track
2) calculating the mean value of the three impact parameters as the tau
$d_0$ and the sum of the momenta as the tau direction. Track 2 is required
to have an impact parameter of at least $0.025\cm$. The acoplanarity and
acollinearity\footnote{The acollinearity is defined as the opening angle
between two tracks.} of tracks 1 and 2 are used to suppress dilepton and $\gaga$
background.  
\subsubsection*{Search for kinks}
Kinks are reconstructed by searching for same-sign tracks crossing in the
$r\phi$ plane (transverse to the beam axis) or  approaching each other in
this projection closer than $5\cm$ inside the tracking volume. At this
point, the separation between the two in the $z$ direction should be less
than 20$\cm$. 

In the search for kinks the most important sources of background are hard
bremsstrahlung and hadronic interactions in the material of the detector
from K$_{\rm{s}}^0$, K$^{\pm}$ or $\pi^{\pm}$ decays. To reduce the former,
the angle between the primary and secondary track $\alpha_{1,2}$ must be
greater than $5\degs$ at the kink vertex: kinks from bremsstrahlung are
peaked at low values of $\alpha_{1,2}$. Finally, the residual background
comes from ditau events with hadronic interactions in the material,
i.e. when the kink is found between 26 and $34\cm$ radius. These events are
characterised by accompanying collinear particles to the interacting
hadron, thus the energy deposited in the calorimeters within $5\degs$
around the axis from the IP to the kink must be less than $5\gev$ to reject
these events.  

\subsubsection*{Combination}
Combined efficiencies above 20\% and with a maximum of $\sim$$65\%$ for stau
samples are obtained in the NLSP lifetime range covered by the large impact
parameter and kinks searches from $\sim$$10^{-10}$ to $10^{-7}$\,s.  
The efficiency is higher for selectron and smuon samples due to the absence
of three-prong decays and the higher momenta of the leptons.  
In the $189-209\gev$ data sample only one event is selected by the combined
search with 1.1 expected from SM backgrounds. The event is recorded at
$189\gev$ and is compatible with a three-prong tau decay in the large
impact parameter selection~\cite{gmsb189}.  

\subsection{Heavy stable charged particles}
Long-lived sleptons can be detected as two back-to-back charged particles
with unusually high values of $dE/dx$ for their momentum. They are not
expected to interact hadronically (they do not have colour charge) nor
electromagnetically (since they are very heavy). Thus they should be very
similar to dimuon events only distinctive by their high specific
ionisation.  In the kinematical limit, where the sleptons are expected to
be very slow, the energy loss becomes high enough to saturate the TPC
electronics. To cover this interesting mass region an additional selection
based on calorimeter information and the pattern of saturated hits was
developed~\cite{Barate:1997dr}.

\begin{table}[h]
\begin{center}
\begin{tabular}{|c|c|}\hline
\multicolumn{2}{|c|}{ Stable sleptons} \\ \hline \hline
\multicolumn{2}{|c|}{ $N_{\rm{ch}}=2$ each with $p_{\rm{t}}>0.1\Ebeam$} \\
\multicolumn{2}{|c|}{ $|cos\theta|<0.9$ with more than 1 ITC hit }\\
\multicolumn{2}{|c|}{ $|d_{01}|+|d_{02}|<0.3\cm$} \\
\multicolumn{2}{|c|}{ $|z_{01}|+|z_{02}|<5\cm$} \\
\multicolumn{2}{|c|}{ No identified electron} \\
\multicolumn{2}{|c|}{ Acollinearity $> 160\degs$} \\
\multicolumn{2}{|c|}{ $|p_1-p_2|<3\sqrt{\sigma_{p1}^2+\sigma_{p2}^2}$} \\
\multicolumn{2}{|c|}{ $E_{\gamma}<250$\,MeV} \\
\multicolumn{2}{|c|}{ $E_{\rm{ECAL1}}+E_{\rm{ECAL2}}<20\gev$ and} \\
\multicolumn{2}{|c|}{ $E_{\rm{HCAL1}}+E_{\rm{HCAL2}}<50\gev$} \\ \hline
$0.52<m_{1,2}/\Ebeam < 0.80$  & $0.80<m_{1,2}/\Ebeam < 0.98$ \\ 
Acollinearity $> 174\degs$ & $R_{I 1}+R_{I 2} > 10$ \\ \hline
 \end{tabular}
\caption[List of cuts for the heavy stable charged particle selection for
intermediate-high masses]
{\label{tab:hscp}{\small 
List of cuts for the heavy stable charged particle selection for
intermediate-high masses. The region close to the kinematic limit
($m_{1,2} \geq 0.95\Ebeam$) is explored by the search for saturated TPC hits
described in Ref.~\cite{Barate:1997dr}.}} 
\end{center}
\end{table}
Table~\ref{tab:hscp} describes the cuts used in the selection of
medium-high masses.
The track quality criteria, the first four cuts in Tab.~\ref{tab:hscp},
are imposed to reject events with poorly
reconstructed tracks and cosmic rays. The electron ID veto and the angle
acceptance reduce the large Bhabha background. The acollinearity cut
effectively reduces the radiative Z returns, $\gaga$ and $\tau^+\tau^-$
events. The momentum cuts also reject $\gaga$ and $\tau\tau$
events. Requiring no photons with energy greater than 250\,MeV
considerably reduces on-shell Z events and Bhabhas. Finally the ECAL and
HCAL energy deposits help in Bhabha rejection. After these cuts are
applied, the high mass of the signal has not been used and the background
is dominated by $\mu^+\mu^-$ events. If the reconstructed mass of the
particles $m_{i}=\sqrt{\Ebeam^2-p^2_i}$ is low the momenta of the tracks
are high enough to produce an energy loss similar to that of ordinary
particles. Only in the high mass range will the $dE/dx$ information be
useful in terms of the estimator $R_I=(I-\langle I_{\mu}
\rangle)/\sigma_{I}$. Here one compares the measured $dE/dx$, $I$, to that
expected for a muon $\langle I_{\mu}\rangle$, where $\sigma_{I}$ is the
expected resolution of the measurement. 

\begin{figure}[tb]
\begin{center}
\vspace{-0.6cm}
\includegraphics[width=0.46\linewidth]{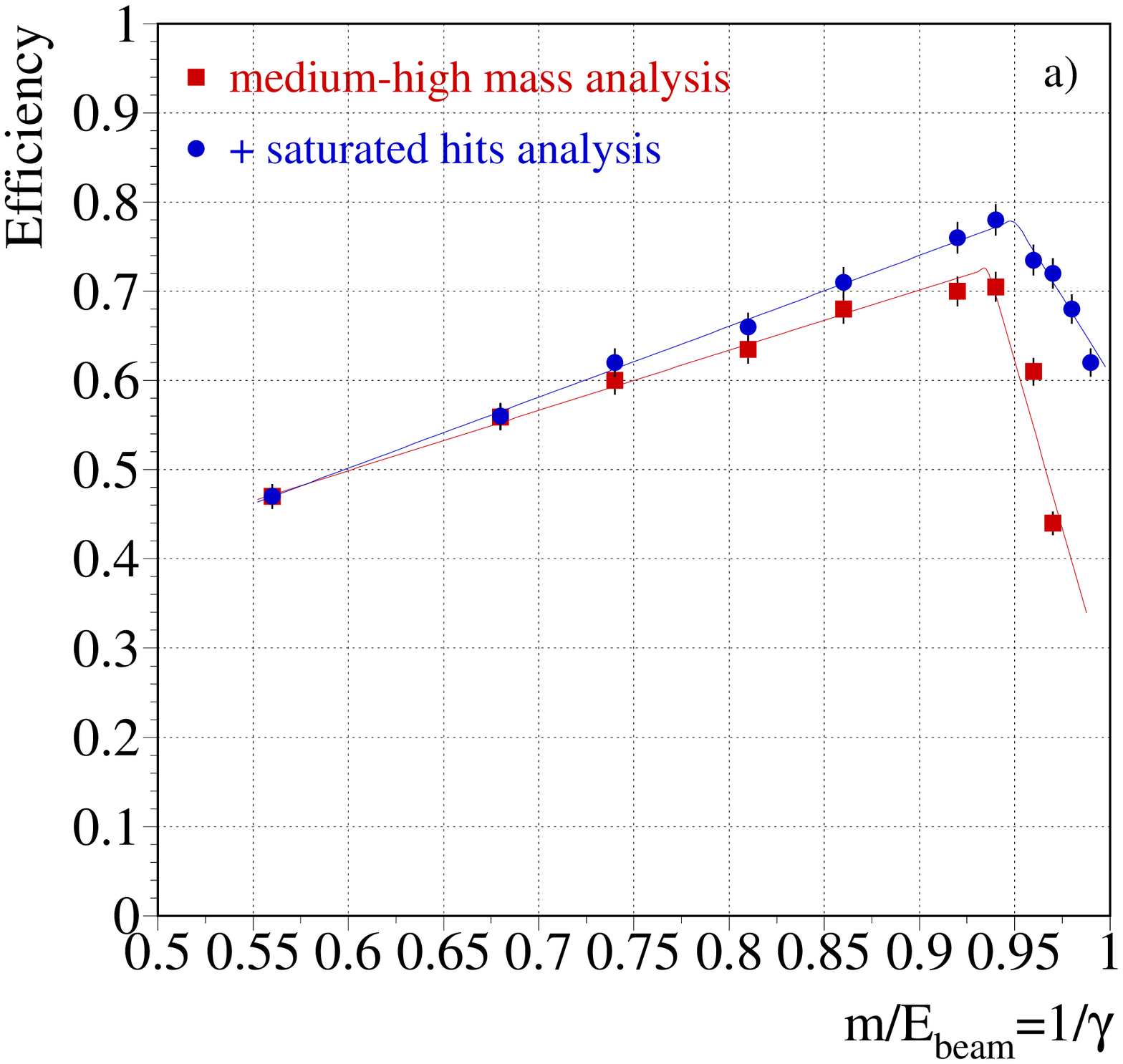}
\includegraphics[width=0.46\linewidth]{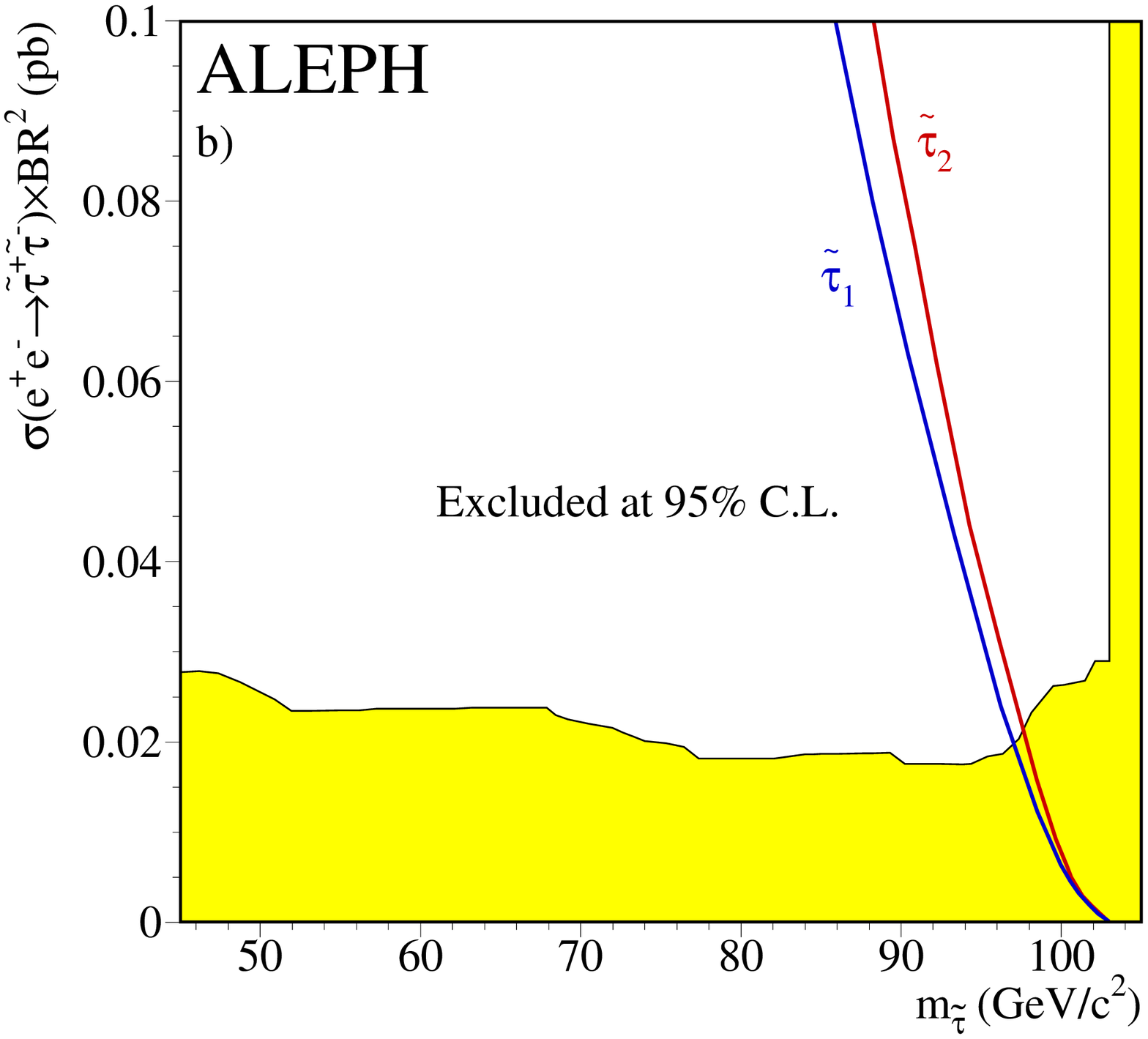}
\caption[Efficiency of the search for heavy stable charged particles and
its excluded cross section]
{\label{fig:stable}{\small (a) Efficiency of the search for pair-production 
of massive, spin-0 stable particles as a function of $m/\Ebeam$ 
with (circles) and without (squares) analysis of events with no good tracks
due to saturation. (b) The maximum allowed cross section calculated at
95$\%$ confidence level with data taken in year 2000 at centre-of-mass
energies of 202--209 GeV (filled area) with the heavy stable charged
particles search (including the analysis of saturated hits). 
The solid lines represent the minimum
expected cross section times branching ratio for the lightest and heaviest
$\stau$ pair-production at $\roots$ = 207 GeV. Thus stau masses of up to
$97\gevcc$ are excluded by this search alone in the stau NLSP scenario.}}
\end{center}
\end{figure}

The case of masses $m_{i}$ above $0.95\Ebeam$ is described in
Ref.~\cite{Barate:1997dr} where the selection makes use of the position of
the two 
most energetic calorimeter objects and the IP to perform a single-helix fit
of the transverse momentum and polar angle of the two track
candidates. Thus the trajectories of the particles can be estimated without
ITC or TPC information and are then compared  with the saturated and
unsaturated hits.  
The efficiency ranges roughly from 50 to 70\% for the intermediate-high
mass, between 0.5 and 0.93 $m/\Ebeam$. If the analysis of saturated hits is
included as described in Ref.~\cite{Barate:1997dr}, the efficiency is
improved and maintained with an approximately flat distribution (within
$\pm10\%$) reaching the highest values of $m/\Ebeam$. The efficiencies for
both medium-high and ultra-high masses selections are shown in
Fig.~\ref{fig:stable}a. 

The combination of intermediate-high mass searches and the saturated hits analysis
selects one data event while 2.3 are expected from background, mainly
dimuons. Systematic effects have been studied with independently selected
dimuon samples from the data to check the performance of the tracking
system. Specific ionisation effects have been checked with electrons, muons
and pions over a large range in momentum. A total systematic error of less
than 5\% is estimated and conservatively applied to the selection
efficiency. Figure~\ref{fig:stable}b shows that cross sections as low as
0.02\,pb are excluded using only the year 2000 data set for stau
pair-production.

\subsubsection{Combination of direct slepton NLSP production searches}
\begin{figure}[tb]
\begin{center}
\vspace{-0.5cm}
\includegraphics[width=0.6\linewidth]{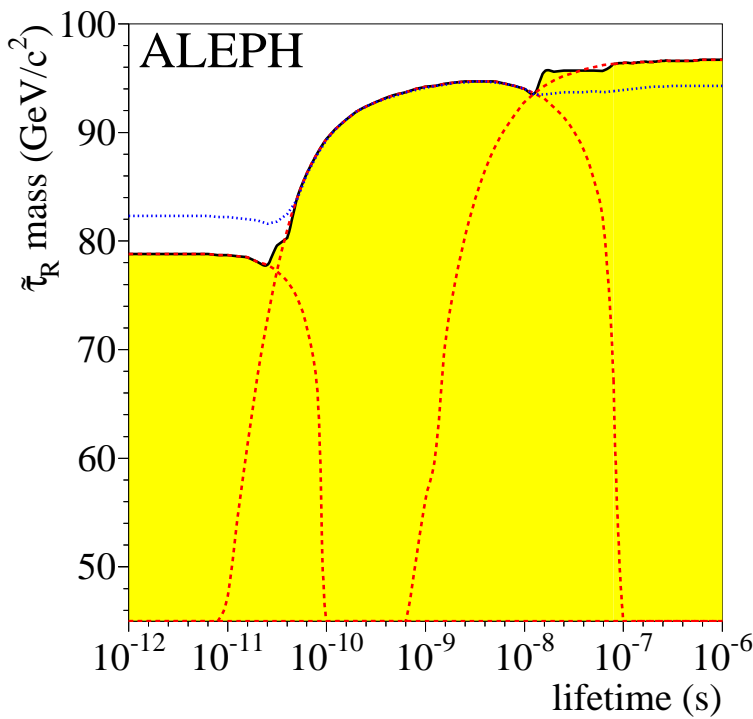}
\caption[Excluded stau mass as a function of its lifetime in the stau NLSP
scenario]{\label{fig:sleplife}{\small Excluded stau mass
at 95$\%$ confidence level as a function of its lifetime (shaded area) from
direct searches. Dashed curves give the limits from 
the different topologies. The search for acoplanar leptons covers the 
case of small lifetimes, 
searches for tracks with large impact parameter and for kinks are
used in the intermediate range, whereas for large lifetimes a 
search for heavy stable charged particles is performed. 
The dotted curve gives the expected limit. From Ref.~\cite{gmsbpaper}.}}
\end{center}
\end{figure}

When all the above selections are combined including all lifetimes, 
the 95$\%$ confidence level lower limits on the
right-slepton masses, independent of lifetime, are set at 
83, 88 and 77$\gevcc$ for selectron, smuon and stau, respectively.
The selectron mass limit is obtained neglecting the
$t$-channel exchange contribution to the production cross section.
Figure~\ref{fig:sleplife} gives the excluded stau mass as a function of its
lifetime when all the above described searches are combined. The dotted lines
represent the independent limits from searches for negligible lifetime
(acoplanar leptons from SUGRA with $m_{\neu}\sim 0$), intermediate lifetime
(large impact parameters and kinks) and long lifetime (heavy stable charged
particles). The .OR. of the searches is the solid line~\cite{gmsbpaper}.

\subsection{Cascade decays of neutralinos}
Searches for slepton NLSP pairs when they are produced directly at LEP and
decay to a lepton and a gravitino have been described above. However, if
the $\neu\neu$ production is allowed, it generally provides the dominant
signal in the slepton NLSP scenario. This is because the production cross
section for neutralinos is larger than for sleptons since it is not
$\beta^3$ suppressed, even for $m_{\neu}>m_{\slep}$ (see
Fig.~\ref{fig:sixl-cs}b). 
The expected signal is therefore four leptons and missing
energy from the cascade decay:
$\ee\to\neu\neu\to\ell\slep\:\ell^{\prime}\slep^{\prime}
\to\ell\ell\grav\:\ell^{\prime}\ell^{\prime}\grav$.
Two of the leptons may be very soft depending on the mass difference
between the neutralino and the slepton. However, in half of the cases the
two hard leptons may have the same charge because of the Majorana nature of
the neutralino. This makes this type of decay a very clean discovery
signal in models with small $\N$ and $\tanb$ not too large.  
In ALEPH, searches for $\ell\ell^{\prime}(\ell\ell^{\prime})\Emiss$ have
been carried out for both negligible slepton lifetimes and intermediate
slepton lifetimes, when two of the tracks present kinks or large impact
parameters~\cite{gmsbpaper}.  
Since the neutralinos decay independently of one another, there are six
possible topologies for the slepton co-NLSP, namely: $\sel\sel$, $\smu\smu$,
$\stau\stau$, $\sel\smu$, $\sel\stau$ and $\smu\stau$. In the stau NLSP
case, only the $\stau\stau$ topology is relevant.  
\begin{table}[h]
\begin{center}
\begin{tabular}{|c|c|c|cccccc|}\hline
\multicolumn{3}{|c|}{Selection} & $\sel\sel$ & $\smu\smu$ & $\stau\stau$ & 
$\sel\smu$ & $\sel\stau$ & $\smu\stau$ \\ \hline  \hline

\multicolumn{2}{|l|}{\multirow{2}{4cm}{
Negligible Lifetime}} &  
        obs. & 6      & 1       & 22      & 2       & 8      & 5      \\
\multicolumn{2}{|l|}{} & exp. & 5.15 & 0.44 & 16.49 & 3.65 & 5.52 & 5.72 \\ \hline
\multirow{4}{4cm}{Intermediate Lifetime} & 
       \multirow{2}{1.5cm}{Short $\lambda$} & obs. & \multicolumn{6}{c|}{5} \\
       & & exp. & \multicolumn{6}{|c|}{5.25}  \\ \cline{2-9}
    &  \multirow{2}{1.5cm}{Long $\lambda$} & obs. & \multicolumn{6}{c|}{4} \\
       & & exp. & \multicolumn{6}{|c|}{1.51}  \\ \hline
\end{tabular}
\caption[Number of observed and expected events
in the searches for four leptons and missing energy]
{\label{tab:fourl}{\small  Number of observed and expected events
in the searches for four leptons and missing energy in the $189-209\gev$
data. In the intermediate lifetime case there is large overlap between the
different selections and the results are given for the total.}} 
\end{center}
\end{table}

\subsubsection{Four leptons and \boldmath{$\Emiss$}}
In the `no lifetime' case for topologies not involving taus, the presence of
at least three identified electrons or muons is required but not more than
four charged tracks. This helps to reject $\qq$ and four-fermion events
with high multiplicity. For each topology, at least two energetic leptons
are required which must be acoplanar, acollinear and their energies must be
in the range allowed by the signal kinematical properties. These cuts
reject effectively the dilepton background. The kinematic cuts reduce also
leptonic WW decays and We$\nu$ background.  

For topologies involving taus $\sel\stau$ and $\smu\stau$, the number of
charged tracks per event and the mean number of neutrinos carrying away
missing energy and momentum are considerably bigger than in  $\sel\sel$,
$\smu\smu$ and $\sel\smu$ topologies. Thus four, five or six charged tracks
are allowed with at least one identified lepton. Finally, for $\stau\stau$
topologies at least three identified tau jets are required with not more
than two other charged tracks. Cuts on the Durham cluster thresholds, the
energy of the event and the momentum of the most energetic lepton are used
to reduce the main background contributions, WW and $\tau\tau$ events.  The
selection cuts are described in detail in Ref.~\cite{gmsbpaper}.  

Efficiencies of up to 80\% are achieved in the $\sel\sel$ and $\smu\smu$
selections. For topologies with taus, the efficiency can be as low as 30\%
for small differences between the neutralino and the stau, reaching a
maximum of 60\% for energetic taus. Figure~\ref{fig:casceffs}a displays the
efficiencies as a function of $\dm_{\neu\slep}$. 
The number of selected events in the $189-209\gev$ data is shown in
Tab.~\ref{tab:fourl} for each topology. 

\subsubsection{Four leptons and \boldmath{$\Emiss$} with lifetime}
A detailed description of this search can be found in Ref.~\cite{Jones:2001te}. 
The topology consists of two tracks with large impact parameter or kinks
and two tracks originating from the interaction point. Thus the basic
selection requirements rely on tracks with more than four hits in the TPC
and high-$d_0$. The main background sources are nuclear interactions,
photon conversions, ECAL `splashbacks' and cosmic muons.  
Table~\ref{tab:fourl} shows the comparison between expected background and
selected data events in the large impact parameter and kinks searches. All
bar one of the selected events can be assigned to a known source of
background.  

\begin{figure}[!h]
\begin{center}
\includegraphics[width=0.45\linewidth]{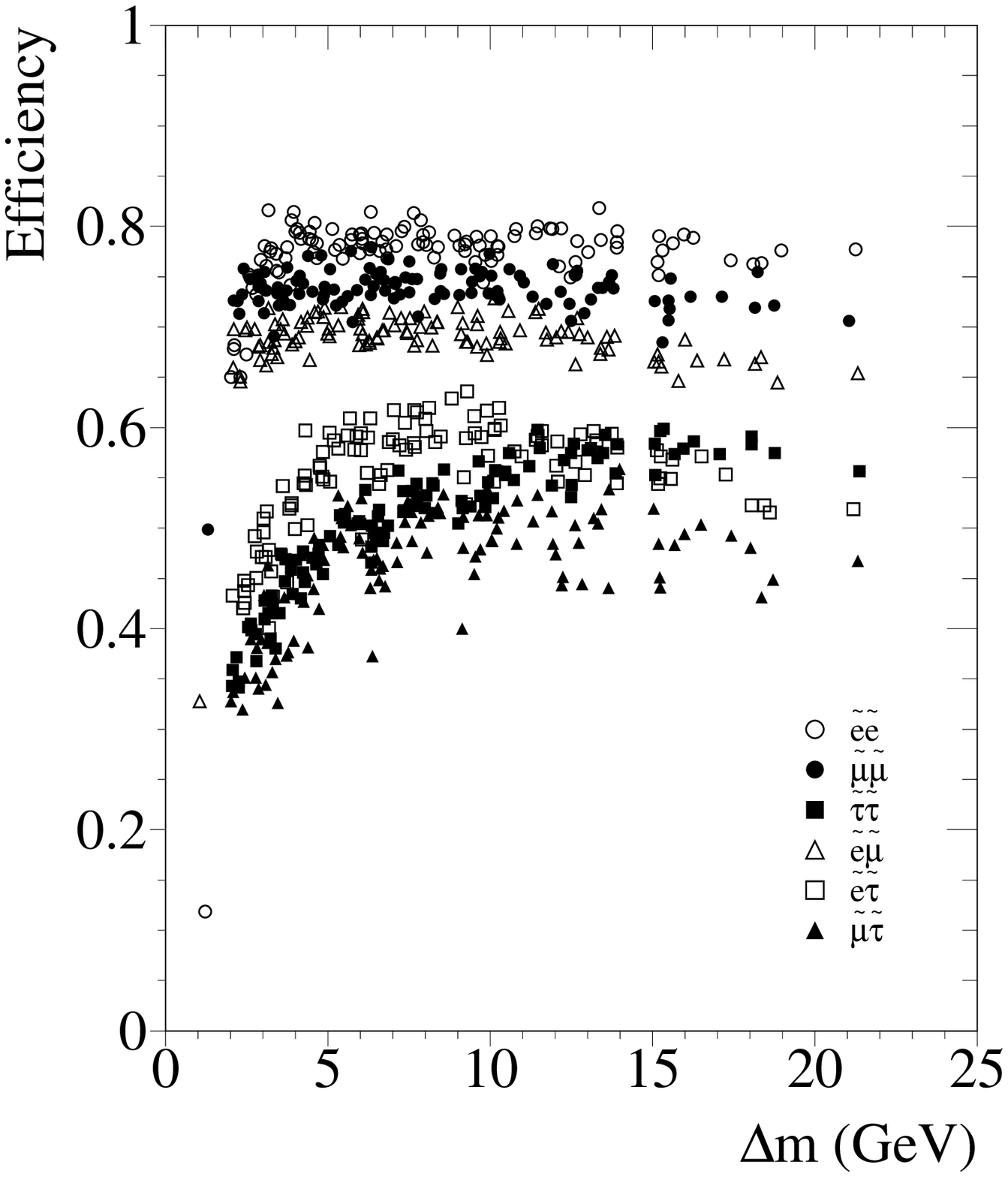}
\includegraphics[width=0.52\linewidth]{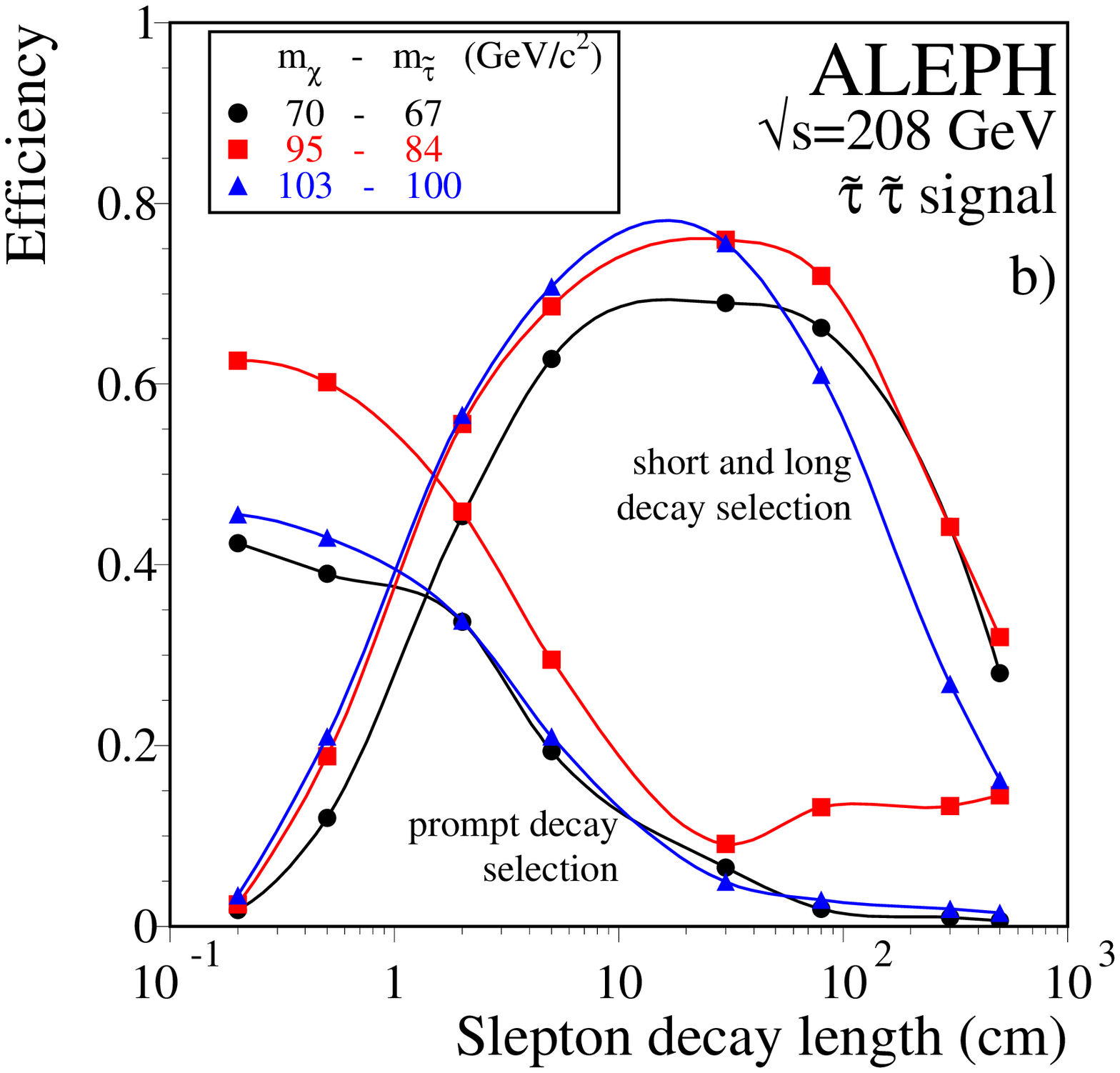}
\begin{picture}(1,1)(290,-220)
 \put(0,0){a)}
\end{picture}
\caption[Efficiencies in the cascade decays with zero lifetime and the
combined efficiency as a function of lifetime]
{\label{fig:casceffs}{\small (a) Selection efficiencies for the
six different event topologies versus $\dm_{\neu\slep}=m_{\neu}-m_{\slep}$
in the negligible slepton lifetime case. The spread
of points observed for a given topology is due to different 
values of neutralino and slepton masses.
(b) Probability for a signal stau-pair event
to be selected by at least one of the six topological searches versus
slepton decay length. 
The set of curves with higher efficiency in the 0.1\,cm area
corresponds to the prompt decay selection. Those peaking at $\sim$10\,cm
correspond to the short and long decay length selection.  
Different lines correspond to different points in the
($m_{\neu}$,$m_{\stau}$) space as indicated.}}
\end{center}
\end{figure}

The efficiency of the long decay selection (kinks) as a function of
lifetime is presented in Fig.~\ref{fig:casceffs}b for the $\stau\stau$
signal. A peak of 80\% efficiency is achieved at decay lengths around
$10\cm$ deteriorating for small $\dm_{\neu\stau}=m_{\neu}-m_{\stau}$. Also
shown in that figure is the efficiency for the negligible lifetime
selection when applied to signal data with slepton lifetime. Overall,
an efficiency in excess of 10\% is maintained for slepton decay lengths
from $\sim$$1\mm$ to $\sim$$10\m$ for all channels~\cite{Jones:2001te}.  

\section{Neutral Higgs boson searches in the MSSM}
Searches for neutral MSSM Higgs bosons can also be used to constrain the GMSB
parameter space. Although these searches are performed in the
framework of gravity mediated scenarios, the Higgs sector is intrinsically
the same in GMSB models. Thus limits from one can be applied to the other
taking into account the different parameters involved in the determination
of the Higgs masses and couplings. 

At LEP2 the lightest CP-even $h^0$ and CP-odd $A^0$ neutral Higgs bosons
can be produced through \emph{Higgsstrahlung} $\ee\to \Z^{*}\to h\Z$ and
\emph{associated production} $\ee\to \Z^{*}\to hA$ (see
Fig.~\ref{fig:lep2prod}d).
The cross sections of these two reactions are complementary, Higgsstrahlung
is dominant at low $\tanb$; while $hA$ has larger cross
sections at large $\tanb$. 
This is due to the MSSM suppression factors $\sin^2(\beta-\alpha)$
and $\cos^2(\beta-\alpha)$~\cite{pdg}:  
\begin{eqnarray}
\sigma(\ee\to h\Z) \sim \sin^2(\beta-\alpha) \sigma_{\rm{SM}} \\
\sigma(\ee\to hA)  \sim \cos^2(\beta-\alpha) \sigma_{\rm{SM}} 
\end{eqnarray}
where $\sigma_{\rm{SM}}$ is the production cross section for the SM
Higgs boson radiated off from a Z, and $\alpha$ is the mixing angle of the
two Higgs fields that produce $h$ and $H$. 

From now on $h$ denotes the MSSM as well as the SM Higgs boson. 
The two most relevant decays of $h$ and $A$ are to $b\bar{b}$ and
$\tau^+\tau^-$. 
Thus four main signal topologies arise: four jets, missing energy, leptonic
and tau final states, as listed in Tab.~\ref{tab:htopo}.  
\begin{table}[ht]
\begin{center}
\begin{tabular}{|c|ccc|cc|} \hline
Topology  & Z decay & $h$ decay & $A$ decay & BR($h$Z$\to$ ) & BR($hA\to$ )\\
\hline \hline
Four jets &  $\qq$  &  $\bb$  &  $\bb$  & 70\% & 86\% \\
Missing 
energy    & $\nu\bar{\nu}$ & $\bb$ & 
            $\neu\neu,\grav\grav,\ldots$ & 20\% & small \\ 
Leptonic  & $l^+l^-$ & $\bb$ & $-$ & 7\% & $-$ \\
\multirow{2}{1.9cm}{With taus} & $\qq$& $\tau\tau$ & $\bb$ 
          & \multirow{2}{0.5cm}{3\%} & \multirow{2}{0.7cm}{13\%} \\
          & $\tau\tau$ & $\bb$      & $\tau\tau$ & & \\ \hline
\end{tabular}
\caption[$h$Z and $hA$ decays]
{\label{tab:htopo} {\small The four main topologies considered in
neutral Higgs boson searches, as derived from $h$Z or $hA$ decays. The
branching ratios for $h$Z decays are given for the Standard Model Higgs
with mass $115\gevcc$. In $hA$ decays, the listed branching fractions are valid
for maximal $h$ masses and $\tanb<30$.}} 
\end{center}
\end{table}

Searches for all the above topologies have been performed in ALEPH and the
selections are described in Ref.~\cite{Barate:2000zr} with the final
results given in Ref.~\cite{Heister:2001kr}, here only a general
description is given.  

The four-jet final state is the most likely signal given its large
branching fraction, but suffers from irreducible backgrounds such as ZZ,
$\bb$ with FSR hard gluon splitting to a pair of $b$-quarks and WW with 
$b$-quark misidentification. Nevertheless, the four jets in the signal are
expected to be contained in a plane if the Z and $h$ or $h$ and $A$ are
produced almost at rest. In the first case, the invariant mass of two of
the jets should be close to $m_{\Z}$ and the other two contain $b$ flavour,
while for the second case all four jets contain $B$ hadrons.  

The missing energy final state is characterised in $h$Z production by
$M_{\rm{miss}}\sim m_{\Z}$ and substantial missing transverse momentum. In
the case of $hA$ production, if other SUSY particles are kinematically
accessible or 
the bosons undergo a WW fusion process into $\bb\nue\anue$, a similar event
topology is obtained.  

If the Z decays to electrons or muons, the two leptons must reconstruct the
Z mass and the two jets should be tagged as $b$-jets. $m_{h}$ can then be
reconstructed accurately from the system recoiling against the two leptons
with a typical mass resolution of about $1.5\gevcc$.  
 
\begin{figure}[!hb]
\begin{center}
\includegraphics[width=0.60\linewidth]{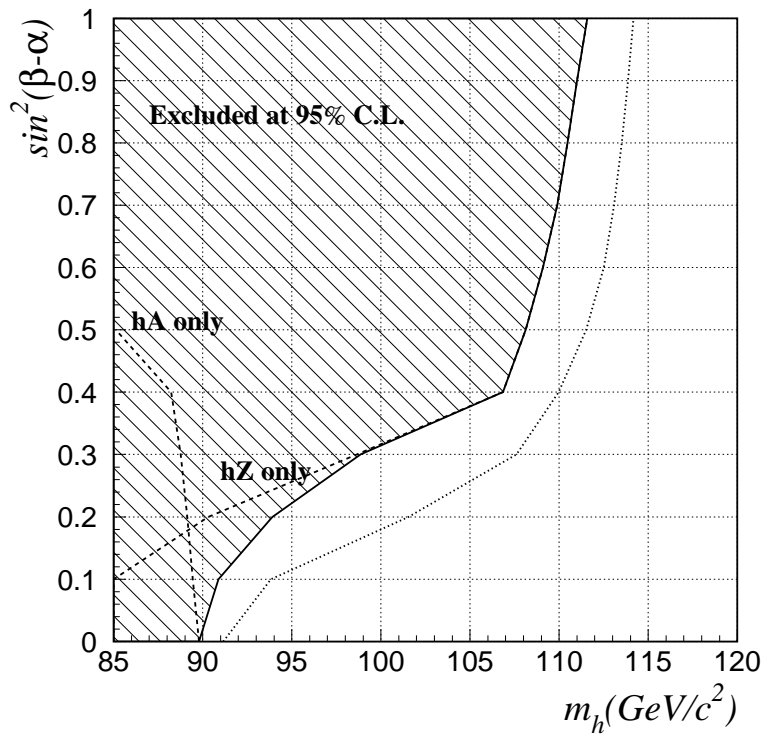}
\caption[Results of the neutral Higgs boson searches in the
$(\sin^2(\beta-\alpha),m_{h})$ plane] 
{\label{fig:hlim} {\small The 95\% confidence level exclusion
contours for the $h$Z and $hA$ searches as a function of
$\sin^2(\beta-\alpha)$ (dashed lines). The combined exclusion is shown by
the hatched area and the dotted line indicates the expected exclusion. 
From Ref.~\cite{Heister:2001kr}.}}
\end{center}
\end{figure}

The relevant limit for GMSB searches is obtained by the combination of the $h$Z
and the $hA$ searches in the plane given by the $h$ mass and the factor
$\sin^2(\beta-\alpha)$. This plane is model independent and the present ALEPH
exclusion is shown in Fig.~\ref{fig:hlim}. 

\thispagestyle{empty}
\chapter{Interpretation of results}
\label{scan}
\begin{center}
\begin{spacing}{1.5}
\setlength{\fboxsep}{5mm}
\begin{boxedminipage}[tb]{0.9\linewidth}\small
The previous two chapters describe all the searches performed in the ALEPH
collaboration in the search for GMSB topologies. No hint for a signal is
observed in any of them. This chapter is therefore devoted to understand
what the lack of evidence for such signals may tell about the theoretical
models that predict them. Understanding the effect of `not seeing'
any signal at LEP is important to design new searches at future colliders
and maybe address new theoretical models. 

The interpretation of the search results in this work has been limited so
far to a specific model or constrained by simplifying assumptions made
to derive the limits. 
To extrapolate these results to the more general case, a scan
is performed over the six parameters that fully determine GMSB
phenomenology. It is then possible to test the robustness of the previous 
results, checking how they may change when no assumptions whatsoever are
made, and to interpret them in a wider context. 
Moreover, each of the above searches  will contribute to exclude different
regions in the minimal GMSB parameter space, thus constraining the
different parameters of the model distinctively. 
It will be important to determine which searches
set the most effective limits on the theoretical parameters.

The final ALEPH combined limits on the parameters set in this chapter are
in most cases very close to the sensitivity of the detector to such
signals. A discussion on the allowed ranges of the parameters is presented,
to asses the validity of these limits if the scan ranges were enlarged.
\end{boxedminipage}
\end{spacing}
\end{center}
\newpage

\section{The scan}
\label{sec:scan}
The entire sparticle mass spectrum and all possible phenomenological
topologies can be determined by only six parameters in GMSB
models~\cite{Ambrosanio:1997rv}. 
A scan over these parameters has been performed to interpret the results of
the experimental searches. The aim is to understand which searches
contribute to exclude which regions in the parameter space and to be able
to set a lower limit on the mass of the NLSP and on the universal mass scale
$\Lambda$, independent of the NLSP lifetime (i.e., for all gravitino
masses).  
The program ISASUSY 7.51~\cite{isajet7.48} was used to calculate masses, cross
sections, and branching ratios for all SUSY particles at each point in the
parameter space. Radiative corrections to chargino and neutralino masses
were applied. The Higgs bosons' masses and couplings were calculated using
the improved Haber et al. treatment at two loops implemented in HZHA
3.0~\cite{hzha} with a top mass of $175\gevcc$.  

\subsection{Scan ranges}
The ranges of the scan are given in Tab.~\ref{tab:ranges} for each
parameter as well as the step size. In total, over 2.3 million points in
the minimal GMSB parameter space have been tested.
If the NLSP mass is greater than the highest beam energy
(104.5 GeV) the point is not considered (and cannot be excluded). The
scan is not exhaustive but it covers a large portion of the allowed
range of the minimal GMSB parameter space, as discussed next.
\begin{table}[h]
\begin{center}
\begin{tabular}{|c|ccc|}
\hline
 Parameter  & Lower limit     & Upper limit & Steps \\ \hline \hline
$\Mmess$    & $10^4\gevcc$    &  $10^{12}\gevcc$         & 7\\
$m_{\grav}$ & $10^{-1}\evcc$  &  $10^{5}\evcc$           & 15\\
$\Lambda$   & $10^{3}\gevcc$  &  $\min(\rootF,\Mmess)$ & 100\\ 
$\N$        & 1               &  5                       & 5\\
$\tanb$     & 1.5             &  40                      & 22\\
sign($\mu$) & $-$             &  +                       & 2\\
\hline
\end{tabular}
\caption[Minimal set of parameters and their
ranges of variation in the scan]
{\label{tab:ranges}{\small Minimal set of parameters and their
ranges of variation used in the scan.}}
\end{center}
\end{table}

\subsubsection*{Validity of the ranges}
\begin{description}
\item[{\boldmath$\Mmess$}] 
An approximate upper bound on the common messenger mass 
is imposed by flavour universality. This can only be maintained
if gravity mediated contributions to the observable masses (of the order
$\rm{F_0}/\MP$) are kept small $\sim$$\mathcal{O}(10^{-3})$. Thus to avoid large
flavour violating transitions~\cite{Giudice:1998bp}: 
\begin{equation} 
\Mmess\lesssim\frac{1}{10^{\frac{3}{2}}}\frac{\alpha}{4\pi}\MP\sim10^{15}\gevcc
\end{equation}
Furthermore, NLSP decays in the early Universe must not spoil the
predictions from Big Bang nucleosynthesis. These bounds can further reduce
the upper limit on $\Mmess$ as described in Ref.~\cite{Gherghetta:1998tq}.
The lower limit is determined by $\Lambda$ with the condition $\Mmess >
\Lambda$.   

\item[\boldmath{$m_{\grav}$}] 
The gravitino mass range is deduced from Cosmology
and from indirect searches. Direct collider searches $\ee\to\grav\neu$ set
a lower limit on $m_{\grav}$ of $\sim$$10^{-5}\evcc$ (see
Sec.~\ref{coll.sign}), but previous indirect limits suggest that
$m_{\grav}>10^{-2}\evcc$~\cite{gmsb189}. In any case, the gravitino mass
controls the lifetime of the NLSP. Phenomenologically, setting $m_{\grav}$
below $10^{-1}\evcc$ would not improve the scan, since that is enough to
cover all cases with negligible NLSP lifetime $\sim$$10^{-12}$\,s. As
regards the upper limit of $100\kevcc$, it is imposed by standard Big Bang
constraints~\cite{Dubovsky:1999xc}. 
A heavier gravitino would overclose the Universe at an early stage, because
$\rho_{3/2}(T)\equiv m_{\grav}n_{\grav}(T)>\rho_{c}(T)$ where $n_{\grav}$ is the
number density of gravitinos, which depends on scattering processes off
thermal radiation and the decay of heavier particles into gravitinos,
and $\rho_{c}$ is the critical density. Even
assuming dilution mechanisms such as inflation, an overproduction of
heavier gravitinos in the reheating epoch can lead to cosmologically
unacceptable values of the relic density~\cite{Moroi:1995fs}.  

\item[{\boldmath$\Lambda$}] 
The universal mass scale parameter is constrained
from above by the mass of scalar messengers.  
\begin{eqnarray}
\label{scalarmessmass}
(m_{0}^{\rm{mess}})^2 = \Mmess^2 \pm \Fm>0 \Longleftrightarrow \Fm < \Mmess^2 \\
\label{lambdaMrootF}
\Lambda \equiv \frac{\Fm}{\Mmess} \Longrightarrow 
\left\{ \begin{array}{l} \Lambda < \Mmess \\
                         \Lambda < \rootF \\
\end{array} \right.
\end{eqnarray}
This upper limit $\Lambda<\min(\Mmess,\rootF)$ 
means that values up to $100\tevcc$ were scanned. 
Higher values would drive the NLSP mass beyond $1\tevcc$.  

\item[{\boldmath$\N$}] 
The number of messenger families is allowed to vary up to
five. The messenger sector affects how the gauge couplings unify at the
GUT scale. 
In order to maintain corrections to the GUT gauge coupling small,
and thus allow for unification, $\N$ and $\Mmess$ should
obey~\cite{Giudice:1998bp}: 
\begin{equation}
\N \leq \frac{150}{\ln \left ( \frac{\rm{M_{GUT}}}{\Mmess}\right )}
\end{equation}
As can be seen in Fig.~\ref{fig:NLSPnature}b, for $\Mmess$ as low as
$100\tevcc$ $\N$ can be at most five. Yet, for
$\Mmess=10^{10}\gevcc$ $\N$ as large as ten is allowed. However, values of
$\N=6$ do not change significantly the results as will be proved later.  

\item[{\boldmath$\tan\beta$}]
The ratio of the two Higgs doublets vacuum
expectation values has been changed from 1.5 up to 40 in the scan. 
Lower values of $\tanb$ are strongly disfavoured by Higgs boson searches,
while calculations of SUSY parameters in the ultra-high regime, beyond
$\sim$$40$, become unreliable. 

\item[{\boldmath$\mathrm{sign(\mu)}$}] 
The value of the mass mixing parameter
between the two Higgs doublets can be computed by enforcing correct
electroweak symmetry  breaking (Eq.~\ref{muEWB}), thus only an 
ambiguity in its sign remains. 
\end{description}

\subsection{Exclusion procedure}
Once the possible GMSB models are available, the exclusion procedure takes place:
\begin{itemize}
\item First, the contribution from unknown processes to the Z width cannot 
exceed 4.7\,MeV.
The extremely precise measurement of $\Gamma_{\Z}$ from LEP1 allows one to
exclude every new particle that could contribute to the Z decay width
beyond the small experimental error. The procedure to determine
$\Delta\Gamma_{\Z}$ is sketched in Fig.~\ref{fig:zwidth}.
\begin{figure}[ht]
\begin{center}
\includegraphics[width=0.3\linewidth]{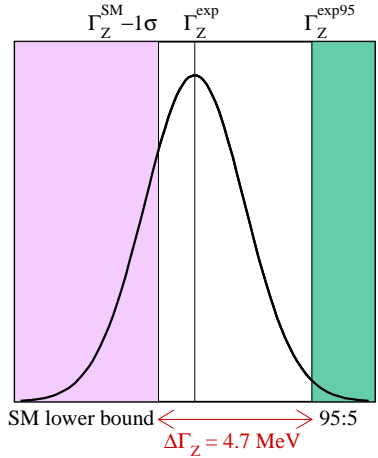}
\caption[Procedure to determine the allowed SUSY
contribution to $\Gamma_{\Z}$]
{\label{fig:zwidth}{\small Procedure to determine the allowed SUSY
contribution to $\Gamma_{\Z}$.}}
\end{center}
\end{figure}
The Z width depends on the Higgs mass $m_{h}$, the strong coupling constant
$\alpha_{\rm{s}}$, the electroweak coupling constant $\alpha_{\rm{em}}$ and
the top- and bottom-quark masses.   
Thus all these parameters, and their experimental errors, enter in the
calculation of the predicted
$\Gamma_{\Z}^{\rm{SM}}\simeq 2496.1 \pm 1.2$\,MeV, which
defines a one-sigma lower bound (2494.9\,MeV). 
This was calculated with the program ZFITTER~\cite{Bardin:1999yd} and the
latest experimental values of the parameters~\cite{pdg}.  
Then a Gaussian distribution is assumed centered on the experimental value
$\Gamma_{\Z}^{\rm{exp}}\simeq 2495.2\pm 2.3$\,MeV. The 95\%
C.L. upper bound is calculated by dividing the remaining area under the
gaussian curve in the ratio 95:5, to obtain
$\Gamma_{\Z}^{\rm{exp95}}=2499.6$\,MeV. The difference between
the SM lower bound and the 95\% C.L. upper limit $\Gamma_{\Z}^{\rm{exp95}}$
is $\Delta\Gamma_{\Z}=4.7$\,MeV.
\item Then the LEP1 limits are applied. The upper limit on the production
cross section derived from heavy stable charged particles
$\sigma_{95}=0.296$\,pb~\cite{Decamp:1992uy} was used. Which generally
means that NLSP masses below $\sim$$m_{\Z}/2$ are excluded by LEP1. 
\item Finally, cross section limits from GMSB searches (and SUGRA searches
for sleptons and charginos) as described in the preceding chapters are
imposed on every point for which the relevant topology is accessible. In
addition, limits from neutral Higgs boson searches are imposed on every
scan point taking the exclusion in the plane $(\sin^2(\beta-\alpha),m_{h})$
from Fig.~\ref{fig:hlim}.
\end{itemize}

Thus a point in the parameter space is excluded if it is kinematically
accessible and the excluded cross section from at least one search is less
than the production cross section for that topology, taking into account
the branching ratio. Cross sections for SUSY particles were calculated at
seven different averaged centre-of-mass energies, as listed in
Tab.~\ref{tab:lumin}. 

\section{Lower limit on the NLSP mass}
In the stau NLSP scenario, the 95\% C.L. limit on $m_{\stau}$ set by
direct searches (Fig.~\ref{fig:sleplife}) is almost unaffected by the
stau mixing and is $77\gevcc$ for any stau lifetime, over the full scan
range. Thus Fig.~\ref{fig:sleplife} from direct slepton production
searches sets the minimum allowed mass for the lightest stau as a function
of the stau lifetime, over the full scan range. Searches in the stau NLSP
scenario for neutralino production (for four leptons and missing energy,
with prompt or short decays) are not able to improve this limit. 

\begin{figure}[tb]
\begin{center}
\includegraphics[width=0.6\linewidth]{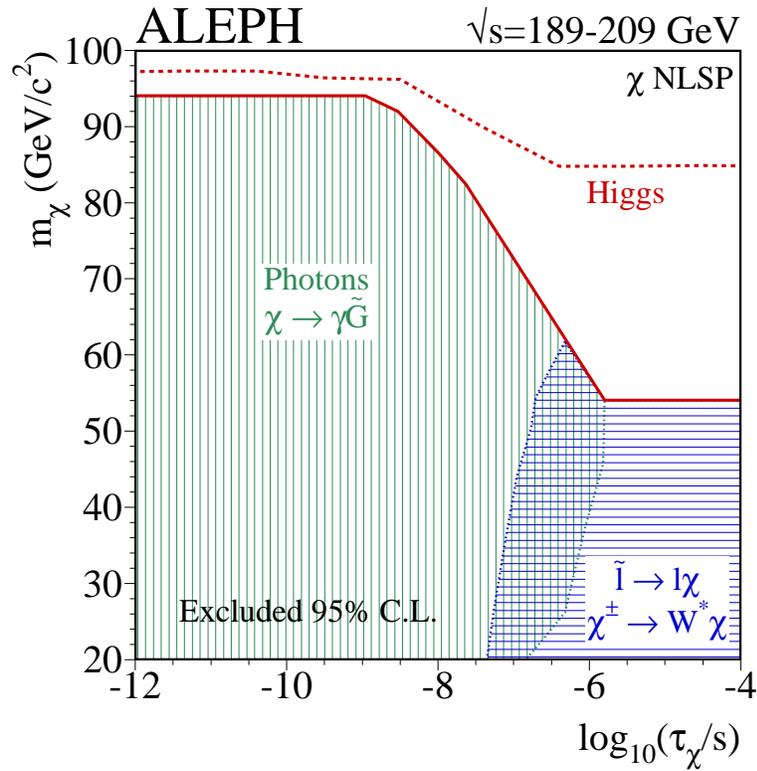}
\caption[Absolute lower limit on $m_{\neu}$ as a function of the $\neu$
NLSP lifetime]
{\label{fig:neuscan}{\small Absolute lower limit on the
neutralino mass in the $\neu$ NLSP scenario from GMSB searches as derived
from the scan. The dashed line gives the lower limit obtained when searches
for $h\Z$ and $hA$ decays are included.
The areas covered by searches for photons, and
sleptons and charginos in SUGRA, are also shown.
The search for single non-pointing photons extends
into very long NLSP lifetimes, since the probability for only one neutralino
decaying inside the detector is still large for lifetimes $\sim$$300\m$. }}
\end{center}
\end{figure}

Limits on $m_{\neu}$ in the neutralino NLSP scenario were shown in
Fig.~\ref{fig:neulim} assuming $m_{\sel}=1.1 m_{\neu}$. 
When this condition is relaxed
and the full scan is tested, $m_{\neu}<94\gevcc$ remains excluded for short
neutralino lifetimes. 
The dependence of this limit on the neutralino lifetime can
be seen in Fig.~\ref{fig:neuscan} where searches for sleptons and
charginos have been included to cover the long lifetime case.
The absolute limit independent of lifetime is set at $54\gevcc$ by indirect
searches for sleptons and charginos in the long lifetime case. 

\begin{figure}[tb]
\begin{center}
\includegraphics[width=0.49\linewidth]{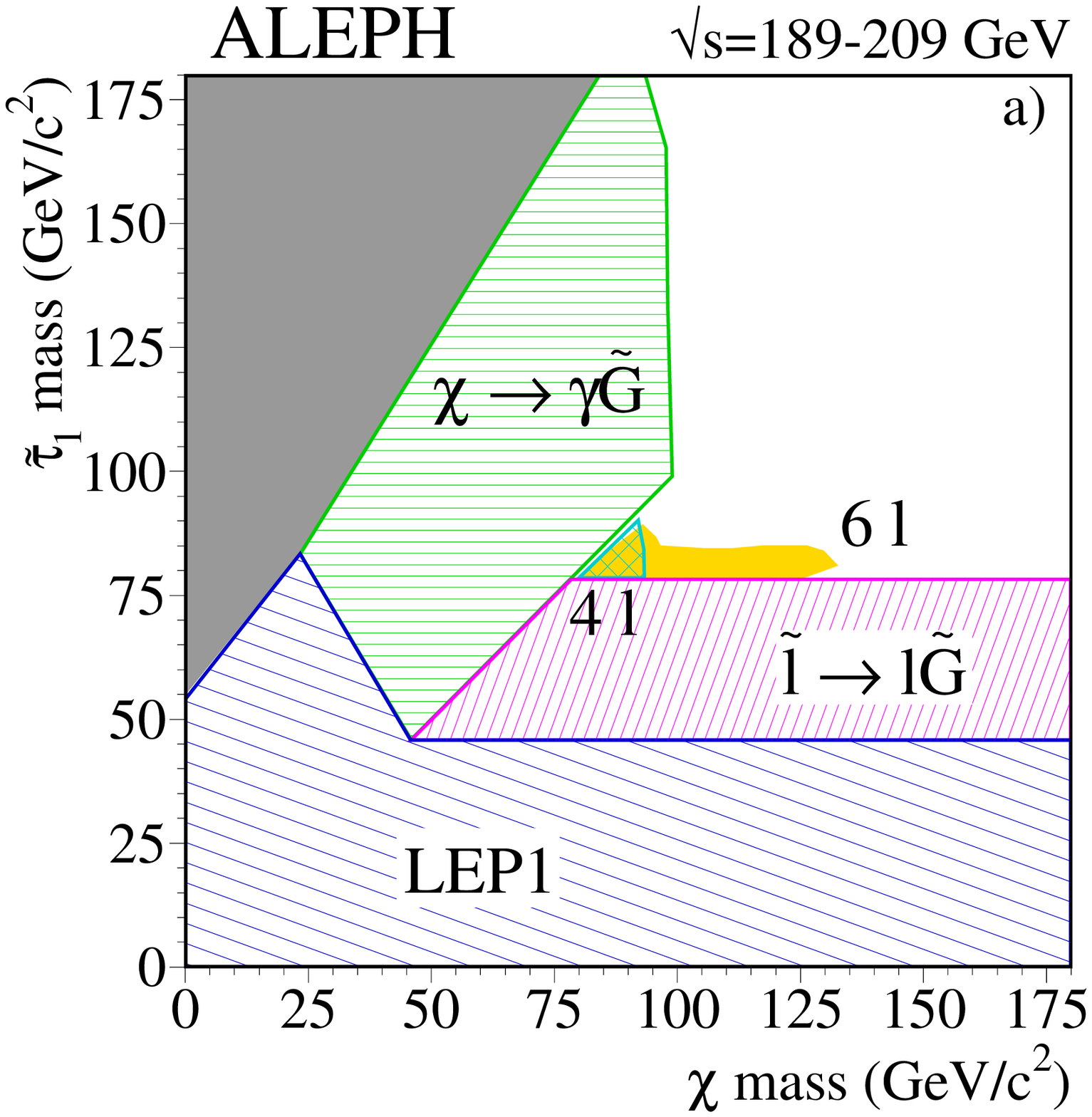}
\includegraphics[width=0.49\linewidth]{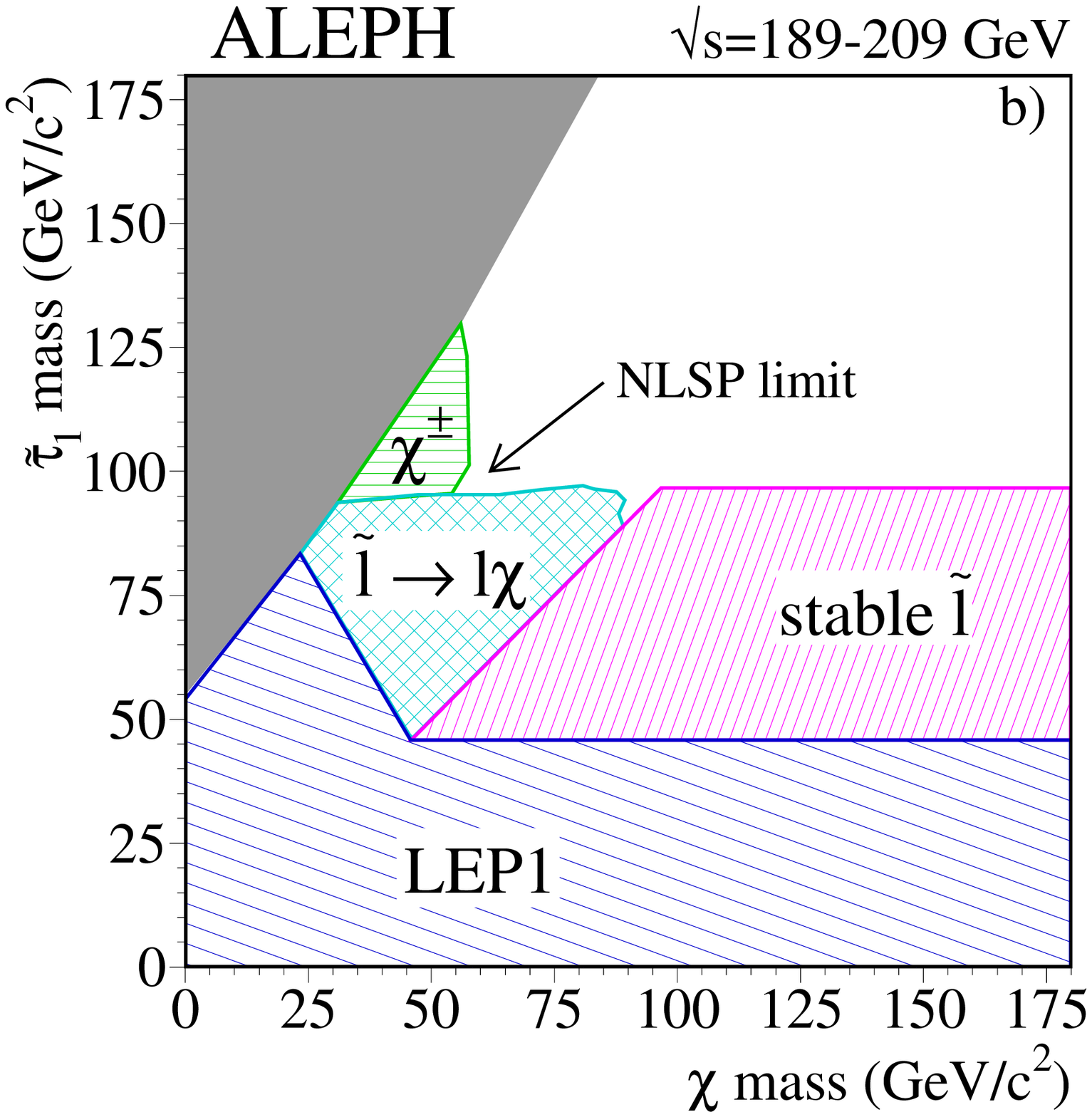}  
\caption[Excluded area in the $(m_{\neu},m_{\stau})$ plane]{\label{fig:neustau}{\small Region excluded by the different analyses
described in the text at 95\% confidence level in the 
$(m_{\neu},m_{\stau})$ plane for (a) short NLSP lifetimes ($m_{\grav} \leq
10\evcc$) and (b) long NLSP lifetimes ($m_{\grav} \geq 1\kevcc$). 
Points in the dark region are not accessible to the scan. 
The absolute NLSP mass limit is set at $54\gevcc$ in b by
the intersection of chargino and slepton searches.}}
\end{center}
\end{figure}

The interplay of the different searches in the
$(m_{\neu},m_{\stau})$ plane is shown in Fig.~\ref{fig:neustau}. For short
NLSP lifetimes (Fig.~\ref{fig:neustau}a) searches for two photons, two leptons,
four leptons and six leptons contribute to exclude points in this plane. 
It can be seen here how searches for four leptons and missing energy are able to
extend the limit from acoplanar leptons searches in the slepton NLSP
scenario up to  $94\gevcc$ in neutralino mass. Nevertheless, the search for
six leptons and missing energy, described in Chapter~\ref{sixl} in this
work, covers most of that area and is also able to exclude $\stauO$ masses below
$84\gev$ for $m_{\neu}\leq 130\gev$.

In Fig.~\ref{fig:neustau}b the case of long NLSP
lifetimes is presented. Limits in the
neutralino NLSP region are less constraining than those
for the short neutralino lifetimes, due to the use of topologies with 
indirect neutralino production. The absolute lower limit on the NLSP mass of
\begin{equation}
\renewcommand{\fboxsep}{5pt}
\fbox{$m_{\rm{NLSP}} \geq 54\gevcc$} 
\end{equation}
is visible in this plot, determined by the chargino and sleptons searches. 
This point is found at $\N=1$, $\tanb=3$, $\Lambda=39\tevcc$,
$\Mmess=10^{10}\gevcc$ and $m_{\grav}=10^{5}\evcc$, where
the neutralino is the NLSP with the $\slep$ masses around $96\gevcc$ and
all other supersymmetric particles above threshold.

\begin{figure}[tb]
\begin{center}
\includegraphics[width=0.49\linewidth]{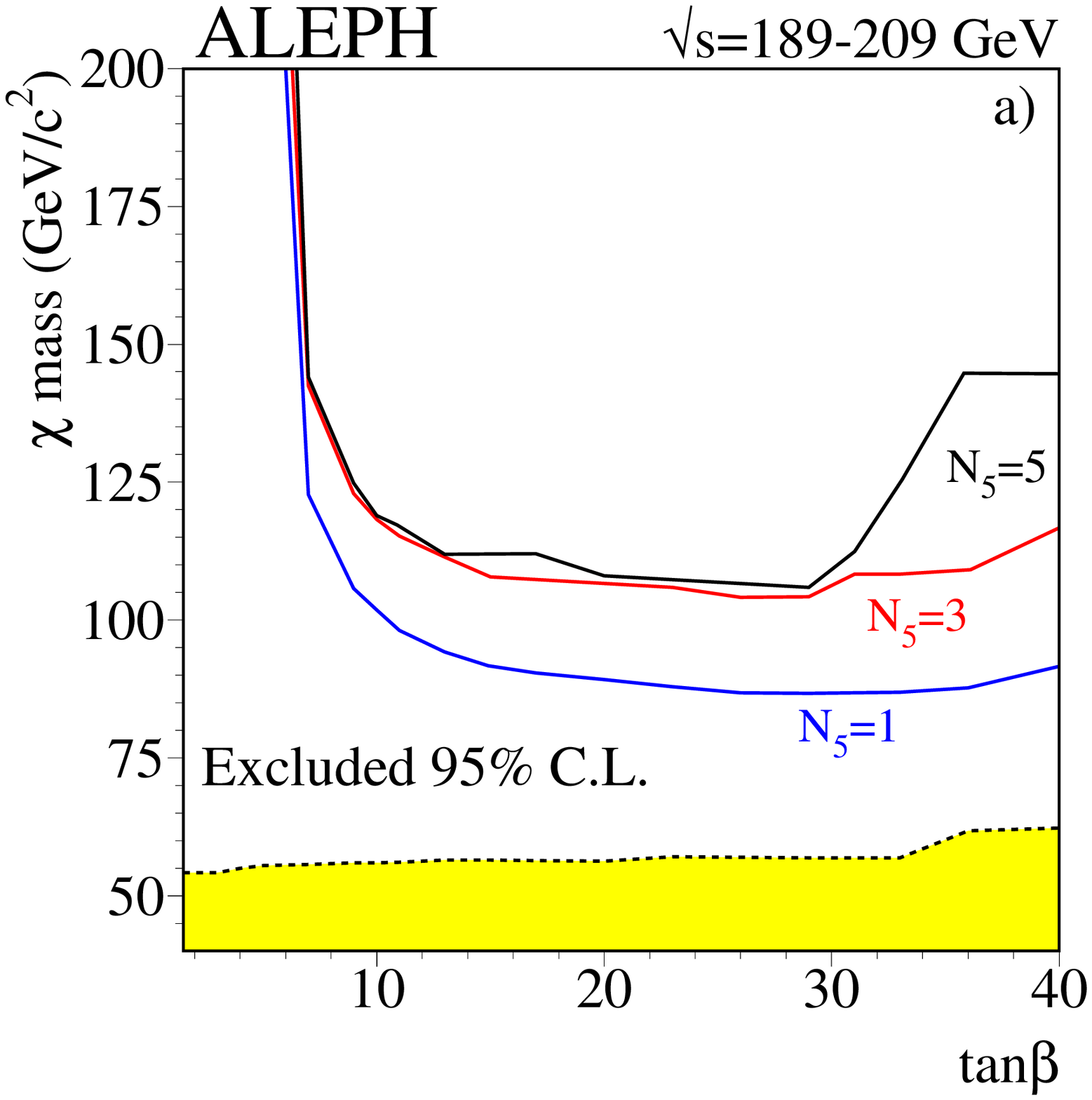}
\includegraphics[width=0.49\linewidth]{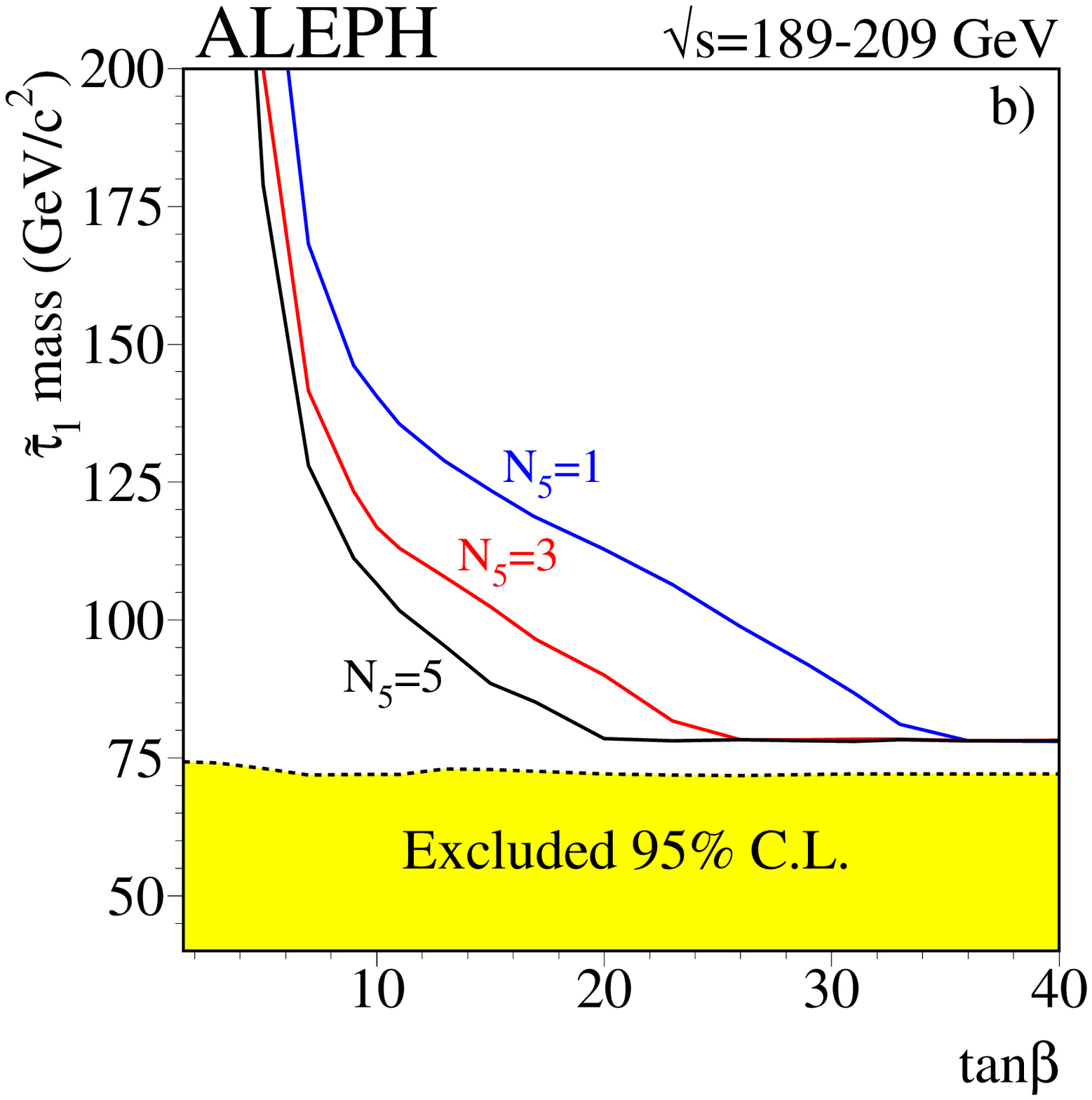}
\caption[Lower limits on $m_{\stauO}$ and $m_{\neu}$ as a function
of $\tanb$]
{\label{fig:mnslphiggs}{\small 
Lower limits on the masses of (a) $\neu$ and (b) $\stau_1$ as a function
of $\tanb$, for different values of $\N$, as set by the Higgs boson
searches. The shaded area represents the minimum excluded area, for any $\N$, 
as derived from GMSB searches alone.}}
\end{center}
\end{figure}

The impact of the neutral Higgs boson searches on the neutralino and stau 
mass limits is shown in Fig.~\ref{fig:mnslphiggs} as a function of $\tanb$. 
In this case, the NLSP absolute mass limit is $77\gevcc$ obtained for large 
$\tanb$ and in the stau NLSP scenario. Small values of $\tanb$ are excluded
almost independently of $\N$. 
 
\section{Lower limit on the mass scale parameter $\Lambda$}
\begin{figure}[ptb]
\begin{center}
\includegraphics[width=0.49\linewidth]{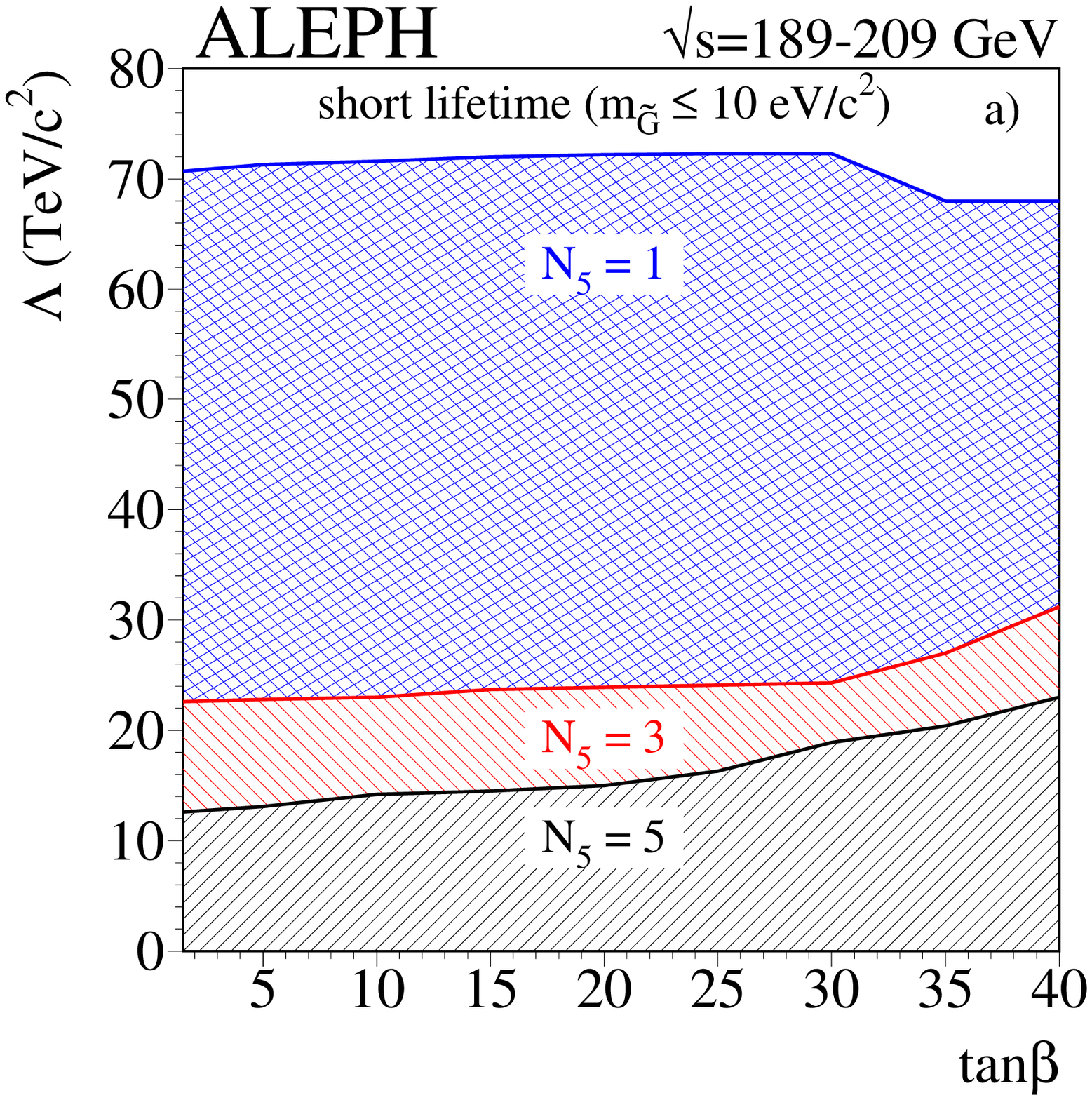}
\includegraphics[width=0.49\linewidth]{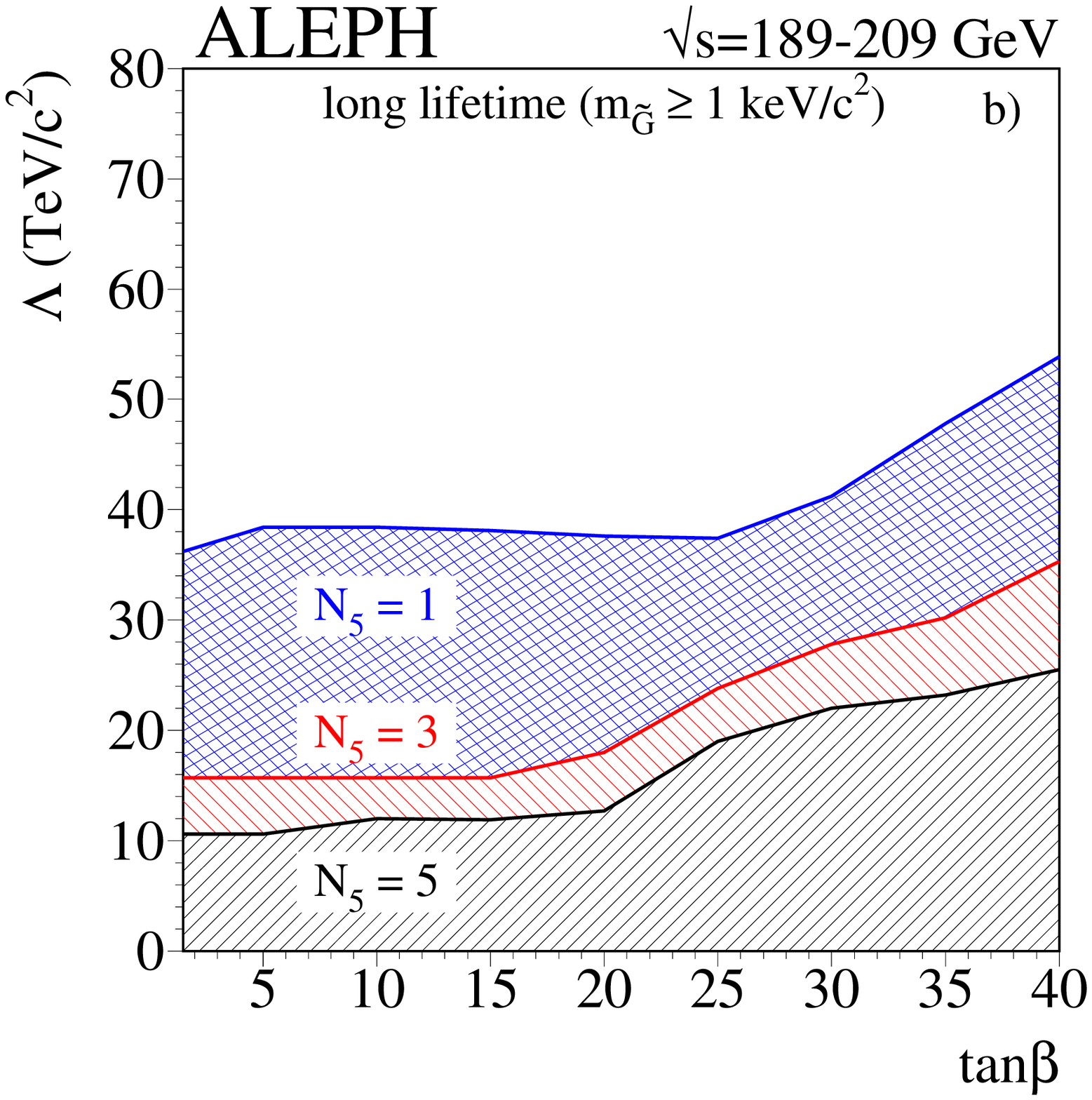} 
\includegraphics[width=0.49\linewidth]{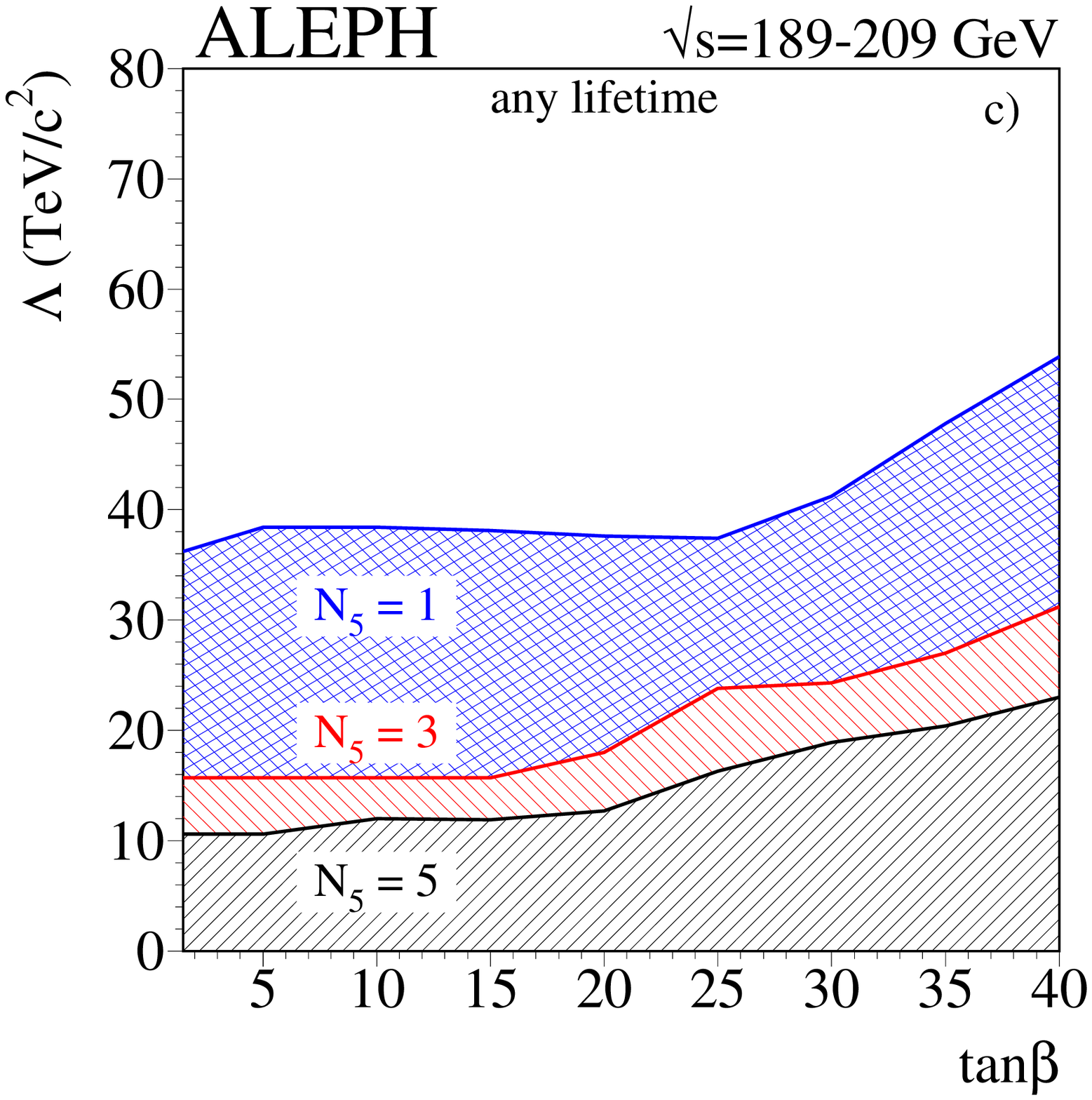}
\includegraphics[width=0.49\linewidth]{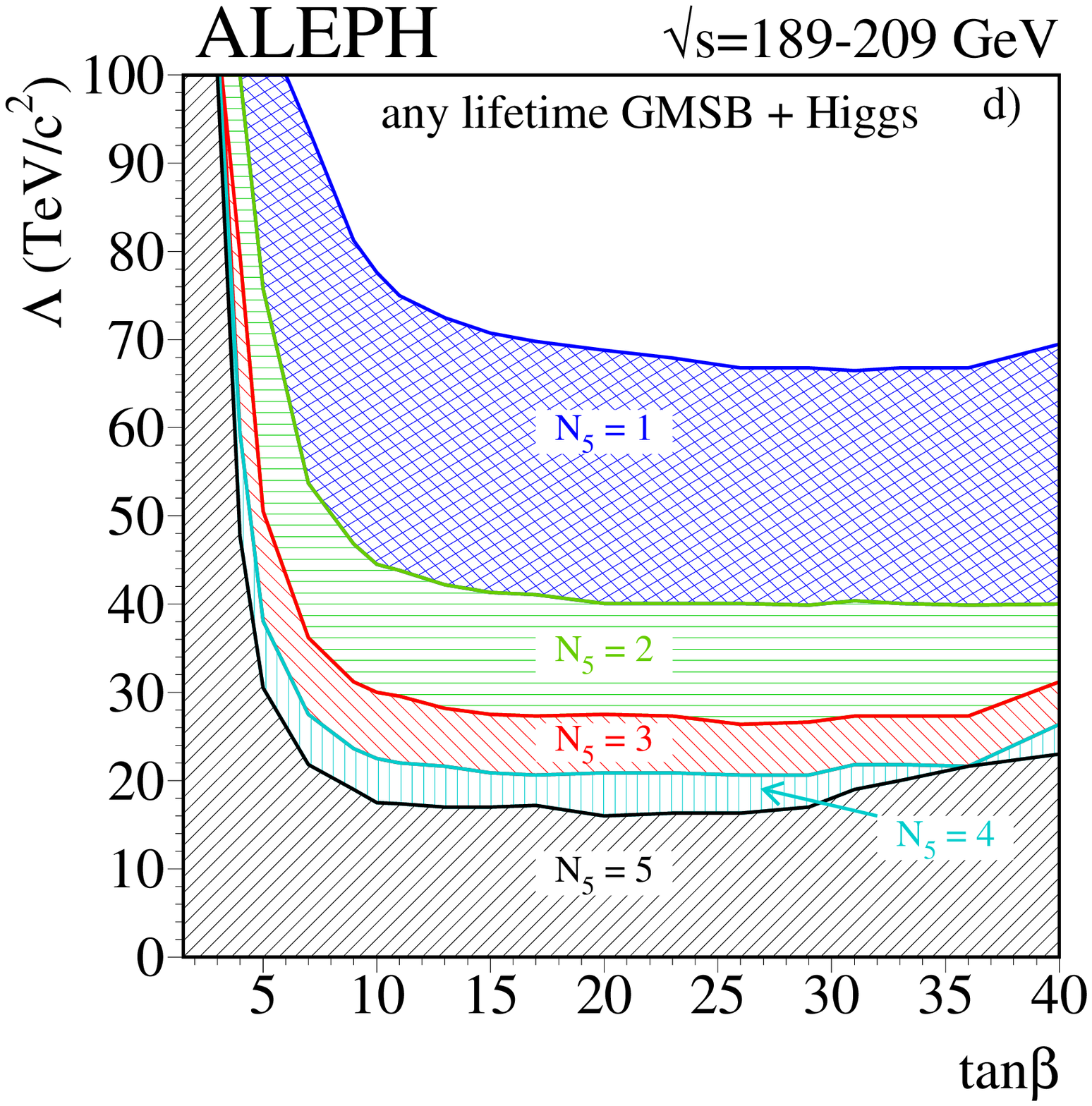}
\caption[Region excluded at 95\% C.L. in the 
($\Lambda,\tanb$) plane]
{\label{fig:latan}{\small Region excluded at 95\% C.L. in the 
($\Lambda,\tanb$) plane for (a) short, (b) long and (c) any NLSP lifetime. 
The impact of the Higgs search is included in d. 
Values of $\tanb$ less than 3 are excluded for large $\N$ at
any mass parameter $\Lambda$, while $\tanb$ up to 6 can be excluded for
$\N=1$.}}
\end{center}
\end{figure}

The parameter $\Lambda$ represents the energy scale at which the
messenger particles couple to the visible sector and hence fixes
the universal mass scale of SUSY particles. Hence a lower limit on the
NLSP mass implies a lower limit on $\Lambda$. Since the relation
between the masses and $\Lambda$ depends on $\N$ (see
Eqs.~\ref{gauginomass} and~\ref{scalarmass}), it is interesting to see
how this limit changes with $\N$. 
The excluded values for the parameter $\Lambda$ as a function of $\tanb$
are shown in Fig.~\ref{fig:latan} for different values of $\N$. 
In the short lifetime case the lower limit on $\Lambda$ appears around
$12\tevcc$ for $\N=5$. 
The absolute limit from the full scan is set at 
\begin{equation}
\renewcommand{\fboxsep}{5pt}
\fbox{$\Lambda \gtrsim 10\tevcc$} 
\end{equation}
in the long NLSP lifetime case. 
This limit is set at $\N=5$, $\tanb=1.5$, 
and $\Mmess=10^{12}\gevcc$. 
The neutralino is the NLSP here with a mass of $73\gevcc$, slepton masses
are around $76\gevcc$ and all other sparticles are above threshold. 
The excluded $\Lambda$ increases with $\tanb$ as slepton NLSP searches
become more relevant and improve on the indirect limits from neutralino
searches alone.  

Figure~\ref{fig:latan}c shows the lower limit on $\Lambda$ for any NLSP
lifetime as a function of $\tanb$. When searches for neutral
Higgs bosons are included, the excluded
area significantly increases, as is shown in Fig.~\ref{fig:latan}d,
specially for low $\tanb$. For example, for $\N=1$, $\Lambda$ up to
67$\tevcc$ and $\tanb$ up to 6 are excluded. 

Finally, the lower limit on $\Lambda$ independent of lifetime and $\tanb$ is
shown in Fig.~\ref{fig:lambdan5}a as a function of $\N$. If the Higgs exclusion
is taken into account the limit on $\Lambda$ increases up to $66,39,26,20$
and $16$ for $\N=1,2,3,4,5$ respectively, assuming $m_{t}=175\gevcc$.

The implication of possible extensions of the scan ranges is discussed
next. 

\subsection{Validity of the limits}
Two additional small scans are performed. One to check the limits if the
number of families is increased to six, and the other to see the effect of
a $180\gevcc$ top mass on the limits derived with $m_{t}=175\gevcc$.

The lower limit on $\Lambda$ for $\N=6$ is 9.3 and $14.4\tevcc$ for GMSB
searches alone and Higgs boson searches, respectively. Thus the limit on the
mass parameter $\Lambda$ is reduced if more families are included in the
messenger sector. But it does so by a small amount, 
from around 10 to around $9\tevcc$, compared to the decrease from four to
five messenger families. 

The top mass is measured directly to be $174.3\pm 5.1\gevcc$~\cite{pdg}.
The general scan was performed assuming $175\gevcc$. 
If this is increased by the experimental error to $180\gevcc$, 
the lower limit on $\Lambda$ for $\N=5$ from neutral
Higgs boson searches is reduced from 16 to $15\tevcc$. 

A change in the ranges of other parameters would not
affect these limits on $\Lambda$ because they are independent
of the NLSP lifetime ($m_{\grav}$ is completely covered) and larger  
values of $\tan\beta$ ($> 40$) may only produce lighter $\stau_1$
which are already excluded. 

\section{Lower limit on the gravitino mass}
\begin{figure}[tb]
\begin{center}
\includegraphics[width=0.49\linewidth]{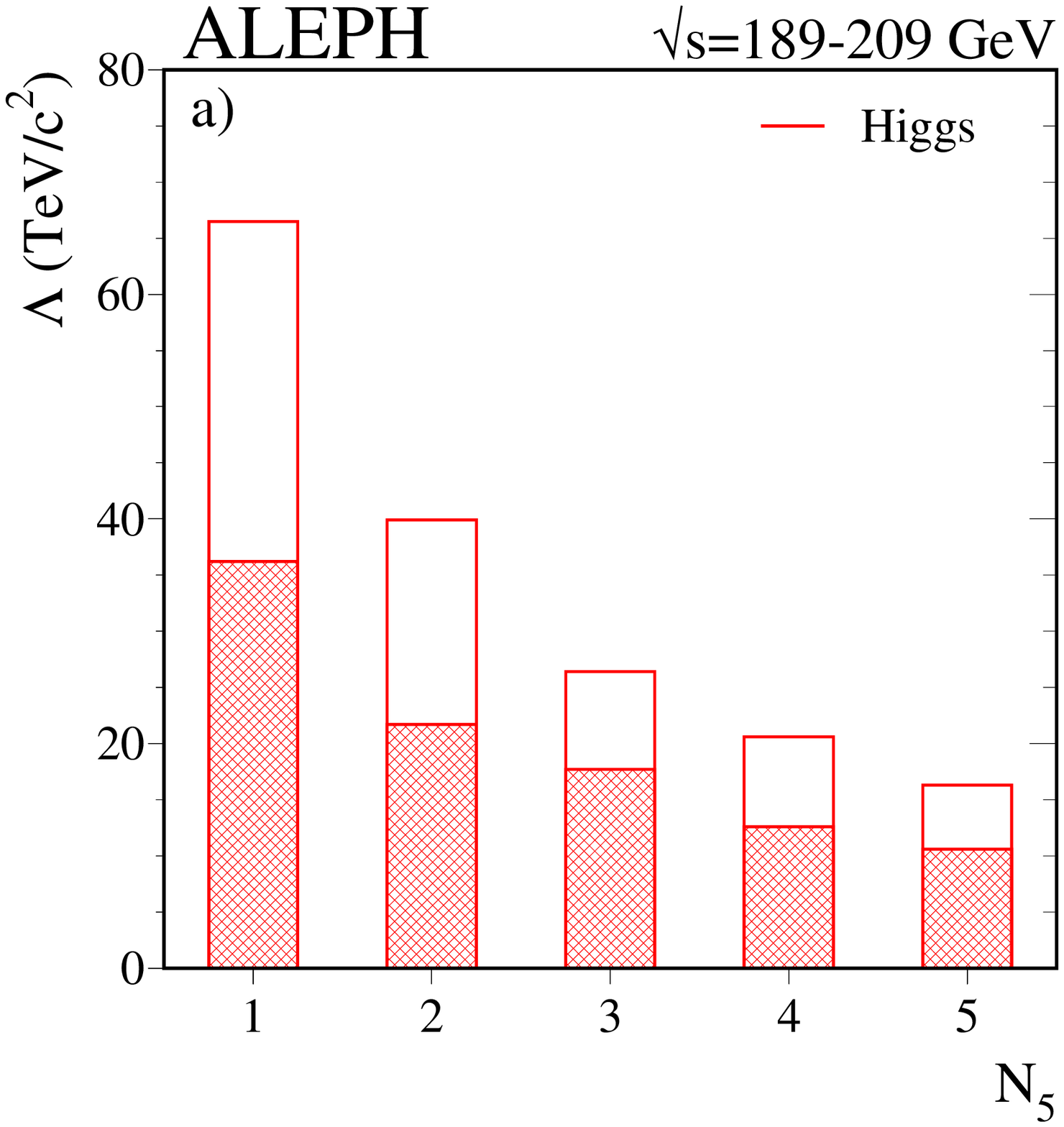}
\includegraphics[width=0.49\linewidth]{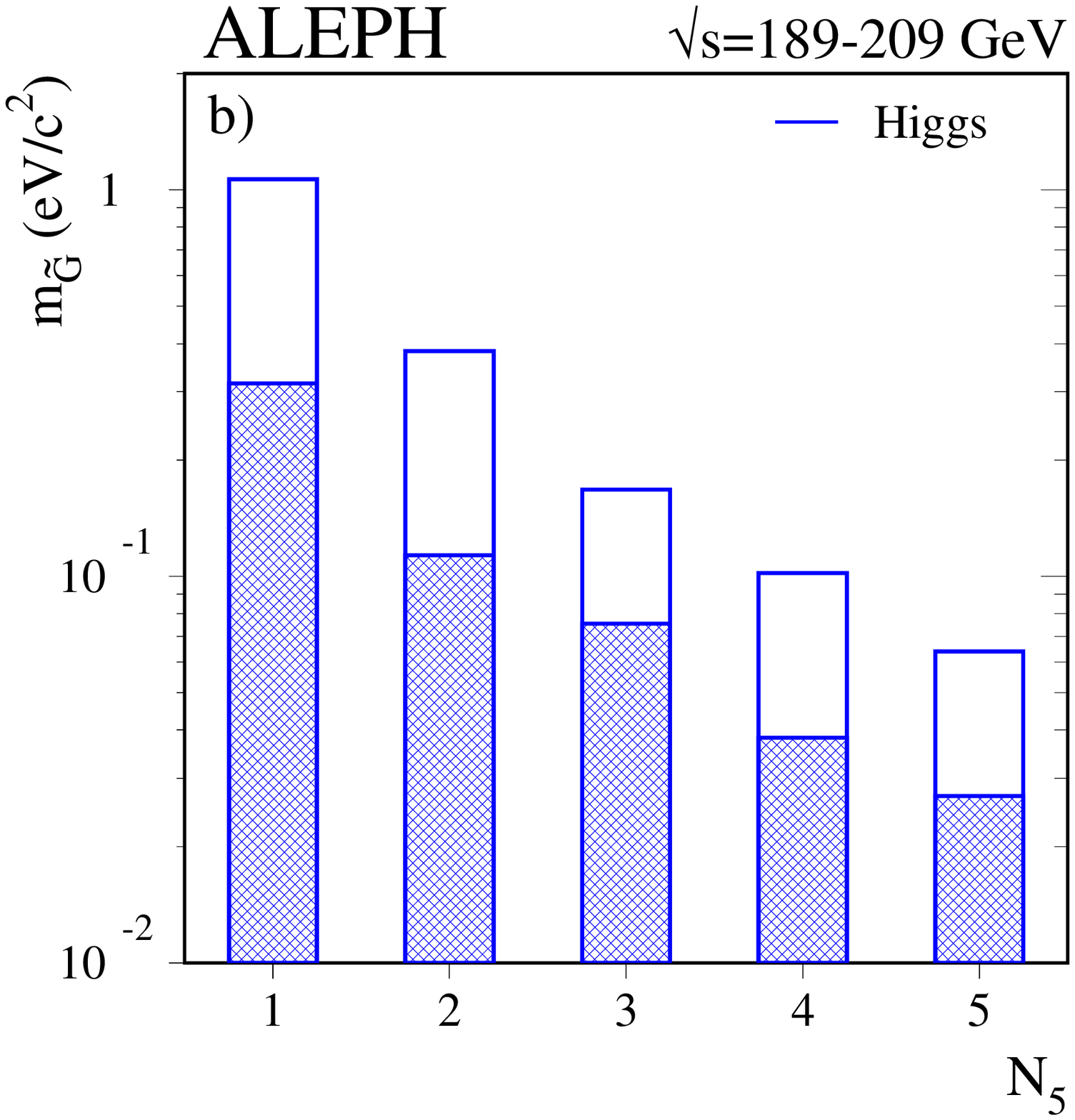} 
\caption[Excluded $\Lambda$ and $m_{\grav}$ as a function of $\N$]
{\label{fig:lambdan5}{\small Exclusions at 95\% confidence
level (a) for $\Lambda$ and (b) for $m_{\grav}$ as a function of $N_5$, 
derived from the minimal GMSB scan (shaded). 
The unshaded bars represent the excluded region when the neutral Higgs
boson exclusion is applied.}}
\end{center}
\end{figure}

The equation that relates the gravitino mass to the scale of SUSY
breaking $\rm{F_0}$ can be exploited to put an indirect limit 
on the gravitino mass. Using Eq.~\ref{lambdaMrootF} and the fact that the
messengers `feel' the SUSY breaking at a lower scale: $\rm{F_0}>\Fm$, 
\begin{equation}
m_{\grav} = \frac{\rm{F_0}}{\sqrt{3}\MP} \Longrightarrow m_{\grav} >
\frac{\Lambda^2}{\sqrt{3}\MP}
\end{equation}
Thus the lower limit on $\Lambda$ can then be converted into an indirect
limit on the gravitino mass. 
The dependence of $m_{\grav}$ on $\N$ is illustrated in
Fig.~\ref{fig:lambdan5}b, and the lower limit of $\Lambda > 10\tevcc$
($16\tevcc$ if the Higgs limits are included) implies a lower limit 
on $m_{\grav}$ of 0.024~(0.061)$\evcc$. 
Therefore,
\begin{equation}
\renewcommand{\fboxsep}{5pt}
\fbox{$m_{\grav} \gtrsim 2 \times 10^{-2}\evcc$} 
\end{equation}
This limit is set for $\N=5$, but for only one messenger family and, if
the $hA$ and $h\Z$ limits are included, it is increased to $1\evcc$.

\thispagestyle{empty}
\chapter{Summary and discussion}
\label{concl}

\section{This work}
Models of supersymmetry where the MSSM masses are generated by radiative
corrections due to messenger particles coupling through the gauge interactions
offer an extremely rich phenomenology. Several differences arise with
respect to the usual approach of gravity mediated SUSY breaking, as listed in
Tab.~\ref{tab:sugra_vs_gmsb}. 
\begin{table}[h]
\begin{center}
\renewcommand{\arraystretch}{1.2}
\begin{tabular}{|c|c|} \hline
mSUGRA         & mGMSB \\ \hline \hline
$m_{1/2},~m_0,~A_0,~\tanb,~{\rm sign(\mu)}$ &
    $\Mmess,~m_{\grav},~\Lambda,~\N,~\tanb,~{\rm sign(\mu)}$ \\ \hline
\multicolumn{2}{|c|}{Gaugino unification mass relations (Eq.~\ref{unirel})} \\
\multicolumn{2}{|c|}{Radiative EWSB (Eq.~\ref{muEWB})} \\ \hline
$\rootF\sim 10^{11}\gev$ & $\rootF\simeq 10^{4}-10^{10}\gev$ \\
$\Mmess \sim \MP$        & $\Mmess\simeq 10^{3}-10^{15}\gev$ \\
possible FCNC & no FCNC \\
$A_0$ = free parameter & $A \simeq 0$ \\ 
$m_0^2\gg m_{1/2}^2\Rightarrow m_{\tilde{\rm q}}\simeq m_{\slep} > m_{\neu}$ 
     & \multirow{2}{6.7cm}{\small 
    $m_{\tilde{\rm q}}\!:m_{\tilde{\ell}_{\rm L}}\!:m_{\slepR}\!:m_{\neu} =
    11\!:2.5\!:1.1\!:\sqrt{\N}$}\\
$m_0^2\ll m_{1/2}^2\Rightarrow m_{\tilde{\rm q}} > m_{\slep} \simeq m_{\neu}$ & \\
$m_{\grav}\sim100\gevcc$ & $m_{\grav}<1\kev$ \\
\multirow{2}{1.5cm}{$\neu_1^0$ LSP} & 
         $\grav$ LSP with $\neu_1^0$ or $\slep_1$ NLSP, \\
       & NLSP has lifetime \\ \hline
\end{tabular}
\caption{\label{tab:sugra_vs_gmsb}{\small Comparison between minimal
supergravity and minimal gauge mediated models.}}
\end{center}
\end{table}

If SUSY breaking is generated at $\Mmess$ as
opposed to $\MP$, the gravitino mass is drastically reduced and it becomes the
LSP. The decay width of the next lightest sparticle to the gravitino 
$\Gamma_{\rm NLSP}\propto m_{\rm NLSP}^5m_{\grav}^{-2}\MP^2$ can then be small 
enough to
produce displaced vertices inside detectors. The nature of the NLSP can be
read off from Tab.~\ref{tab:sugra_vs_gmsb} to be the neutralino if only one
family of messengers exists, or the lightest slepton (usually the stau) in
any other case. 

In this last type of scenario, where the $\stauO$ is the NLSP and the
other sleptons may be even heavier than the lightest neutralino $\neu$,
signatures with two soft electrons or muons, four taus and missing energy
are expected in ALEPH. A search for this topology has been developed and,
for the first time, special attention has been placed on three-body decays
of selectrons and smuons to the corresponding lepton, a tau and the stau NLSP. 
A dedicated selection covered the region of very soft leptons, when the
mass splitting between the sleptons and the NLSP is small. 
This search alone conservatively excludes selectron and smuon masses of up
to 94 and $95\gevcc$ respectively, with 95\% confidence level in the case
that the stau lifetime is negligible. 
The existing limits for selectron and smuon NLSP direct
production were 98 and $93\gevcc$, respectively.

But the importance of this search becomes apparent when a full scan of all
possible GMSB models is performed and the very many different searches are
combined to understand the exclusions in a wide context. It is then proven
to significantly exclude an as yet underestimated area in parameter space
over that ruled out by cascade decays of neutralinos to sleptons and direct
slepton decays (see Fig.~\ref{fig:neustau}a).
Searches to cover almost every possible signature have been developed in 
ALEPH and are detailed here, including the neutral Higgs bosons searches. 
The interplay
between the parameters of the models and the sensitivity of the limits to
its variation has been studied. From the combined exclusion, and allowing
the models to extend in the parameter space as much as physically possible,
three model independent limits are set:
\begin{itemize} 
\item The mass of the NLSP is excluded up to $77\gevcc$ (see
Tab.~\ref{tab:finmlim} for details).
\begin{table}[h]
\begin{center}
\begin{tabular}{|c|c|l|} \hline
NLSP         &  mass limit (95$\%$ C.L.) & validity \\ \hline \hline
             & 92 \gevcc & short $\neu$ lifetime ($m_{\grav} \leq 10\evcc$)\\ 
\raisebox{1.9ex}[-1.5ex]{$\neu$} & 54 \gevcc     & any lifetime \\ \hline
$\stau_1$    & 77 \gevcc & any lifetime \\ \hline
any          & 77 \gevcc & Higgs exclusion \\ \hline
\end{tabular}
\caption{\label{tab:finmlim}{\small NLSP mass limits from GMSB searches,
as derived from the scan.}}
\end{center}
\end{table}
\item The universal mass scale parameter $\Lambda$ is excluded up to
$16\tevcc$ for $\N=5$. This limit is not expected to be affected much for
larger $\N$. Values of $\Lambda$ in excess of
$\sim\frac{100}{\N}\tevcc$ would  produce very heavy sparticles, destroying the
solution to the hierarchy problem.    
\item The mass of the gravitino can be indirectly excluded up to
$6\times10^{-2}\evcc$, or equivalently, the SUSY breaking scale $\rootFo$
has to be greater than $16\tev$. Cosmology bounds $\rootFo$ from above at
$\sim$$1000\tev$. 
\end{itemize}

The above limits are calculated with neutral Higgs bosons results and GMSB
results as described in Chapter~\ref{gmsbphen}.

\section{GMSB after LEP}
Supersymmetry has escaped detection at LEP, but the techniques and limits
from this machine will be valuable for future experiments which will surely
discover it, if it exists. 
The LEP collaborations have all developed very similar searches
for possible GMSB signals, as described in
Refs.~\cite{opalgmsb,delphigmsb,l3susy}. The OPAL collaboration sets lower
limits on the SUSY mass scale of $\Lambda>40,27,21,17,15\tevcc$ for $\N=1,2,3,4,5$
respectively, taking all NLSP lifetimes into account in a similar scan to
the one described in this work. 
DELPHI quotes a limit
of $\Lambda>17.5\tevcc$ for $\N=4$ and negligible NLSP lifetime. Only the
ALEPH collaboration has included neutral Higgs boson searches into the GMSB
analysis, rendering better limits. 
As regards the NLSP mass, OPAL excludes them below 53.5, 87.4 and
$93.7\gevcc$ in the neutralino, stau and slepton NLSP scenarios
respectively. DELPHI has also looked for direct
evidence of a massive 
sgoldstino S (the bosonic superpartner of the goldstino, with even $R_P$
and produced via: $\ee\to S\gamma$),
in topologies with three energetic photons ($S\to\gaga$) or two jets and one
photon ($S\to{\rm gg}$). This search allows direct exclusion of the
SUSY breaking scale $\rootFo$ up to $650\gev$ for a light sgoldstino. 

\begin{figure}[p]
\begin{center}
\vspace{-1cm}
\includegraphics[width=0.49\linewidth]{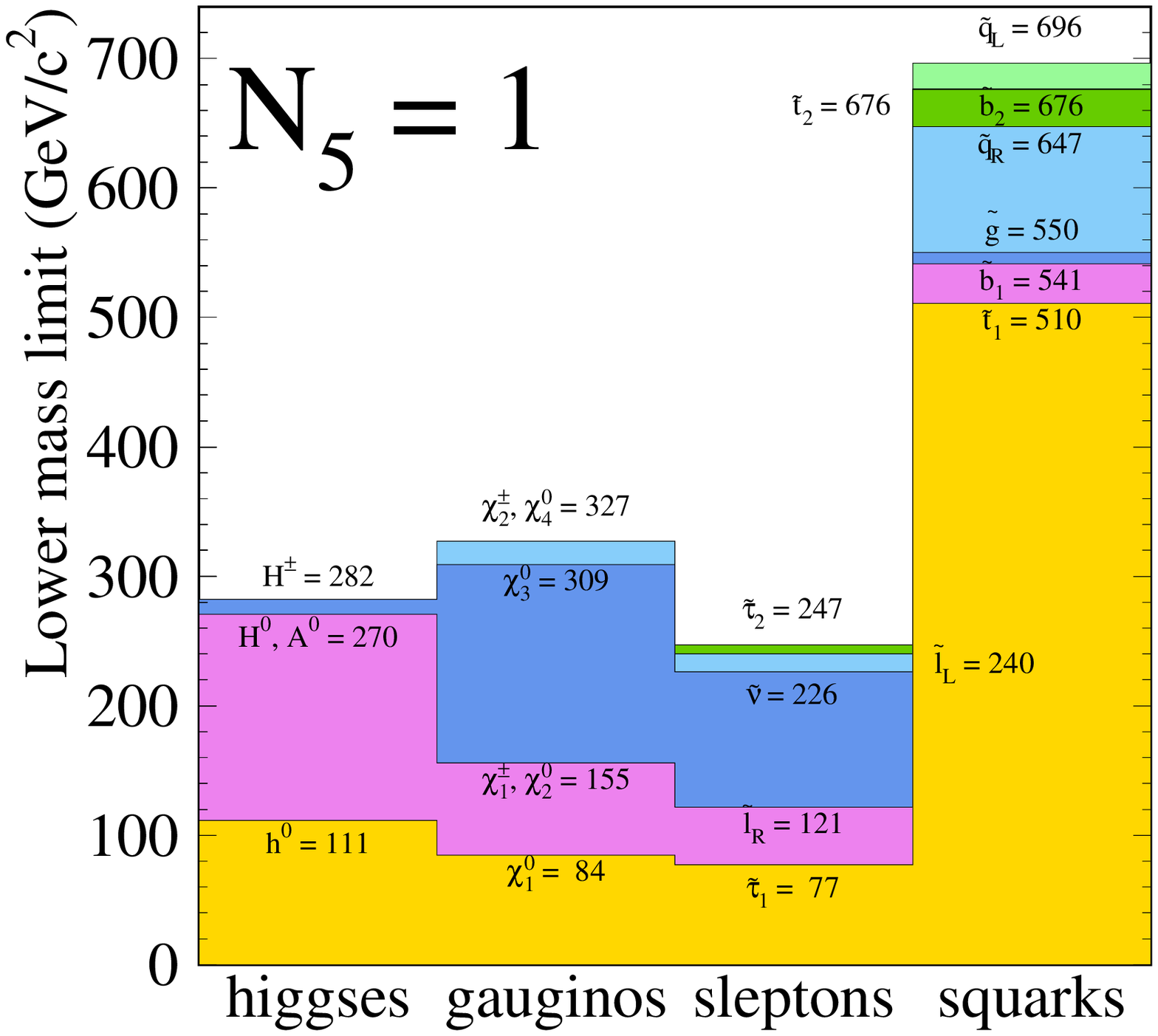}~~
\includegraphics[width=0.49\linewidth]{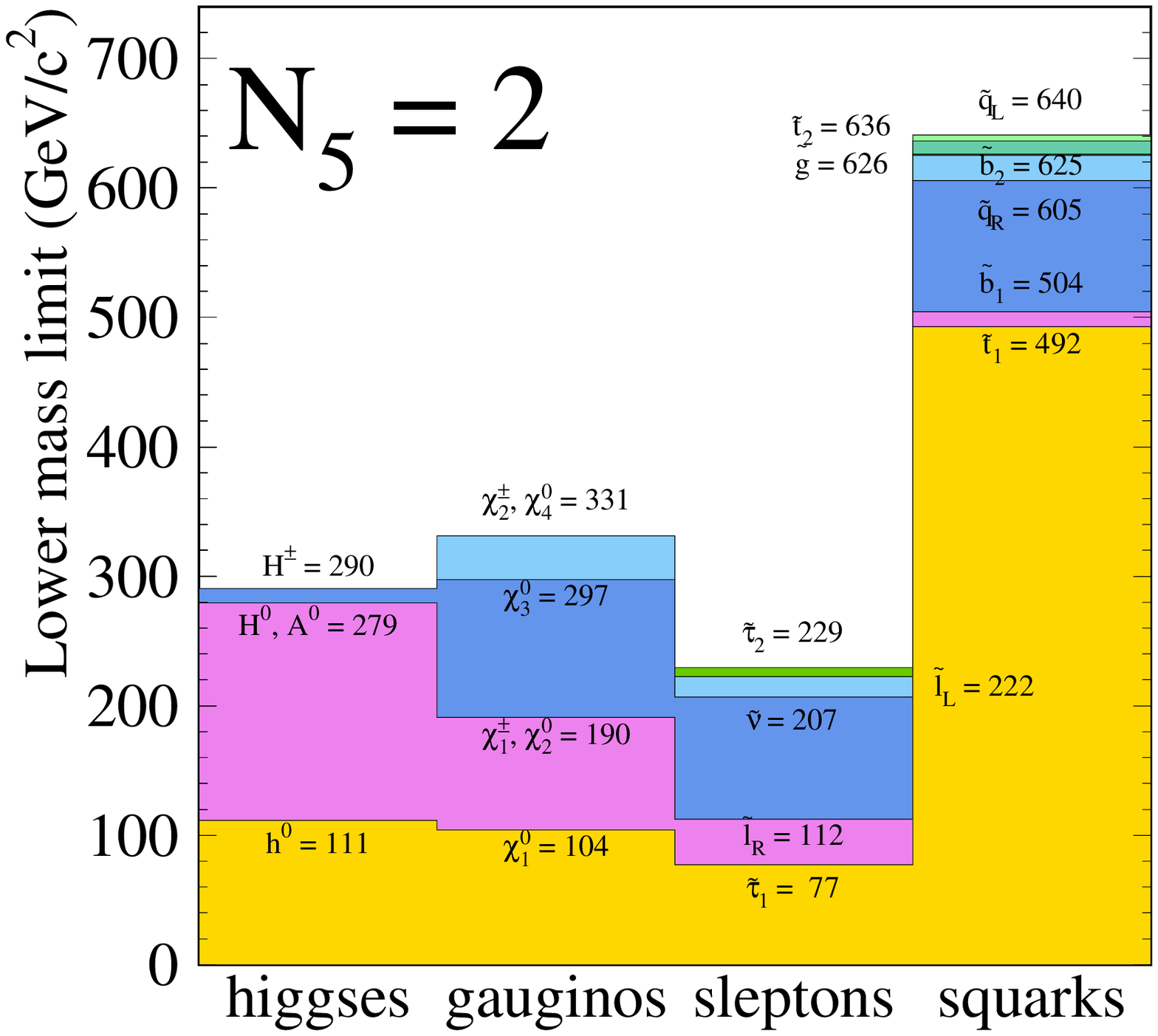}\\
\includegraphics[width=0.49\linewidth]{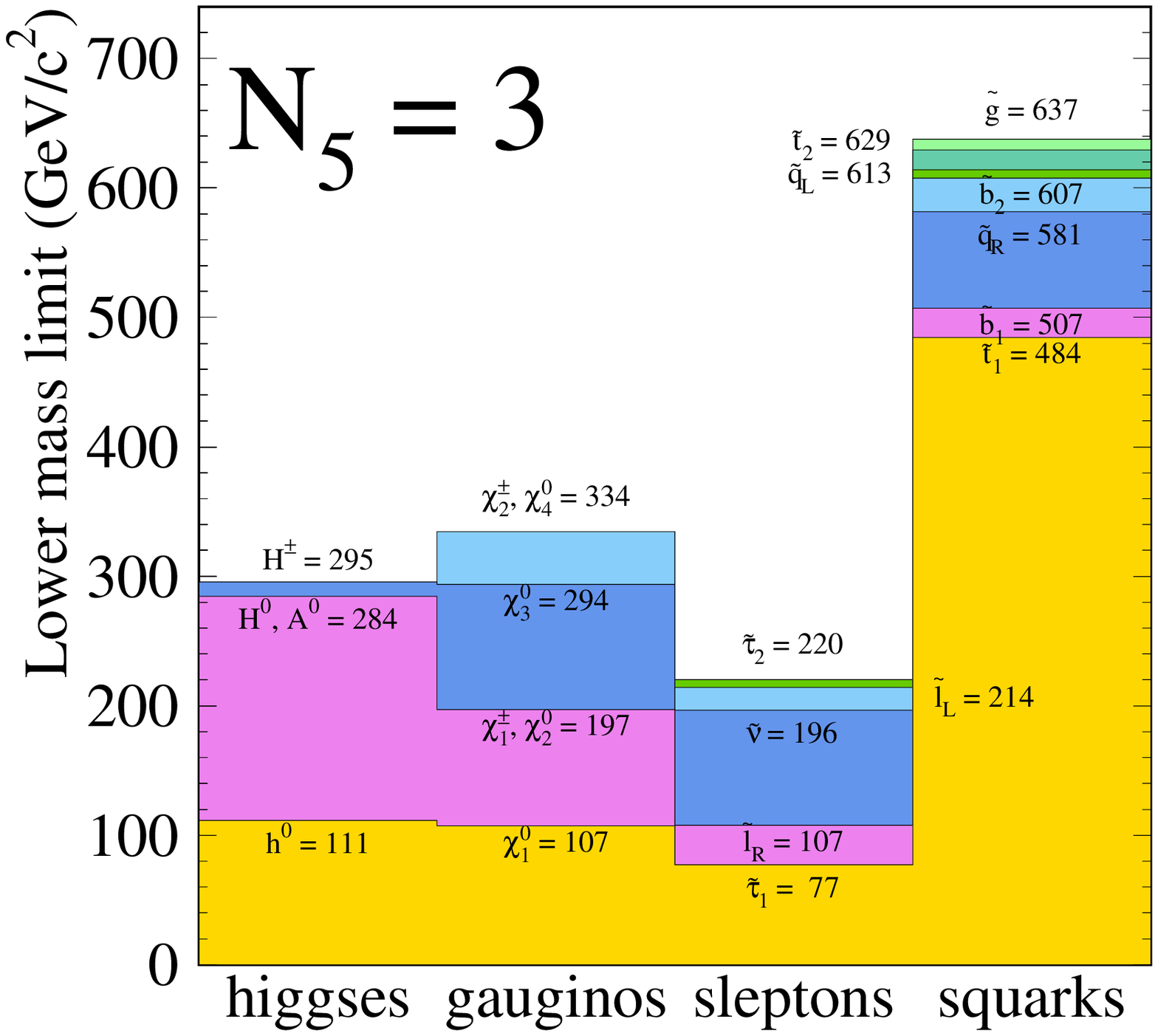}~~
\includegraphics[width=0.49\linewidth]{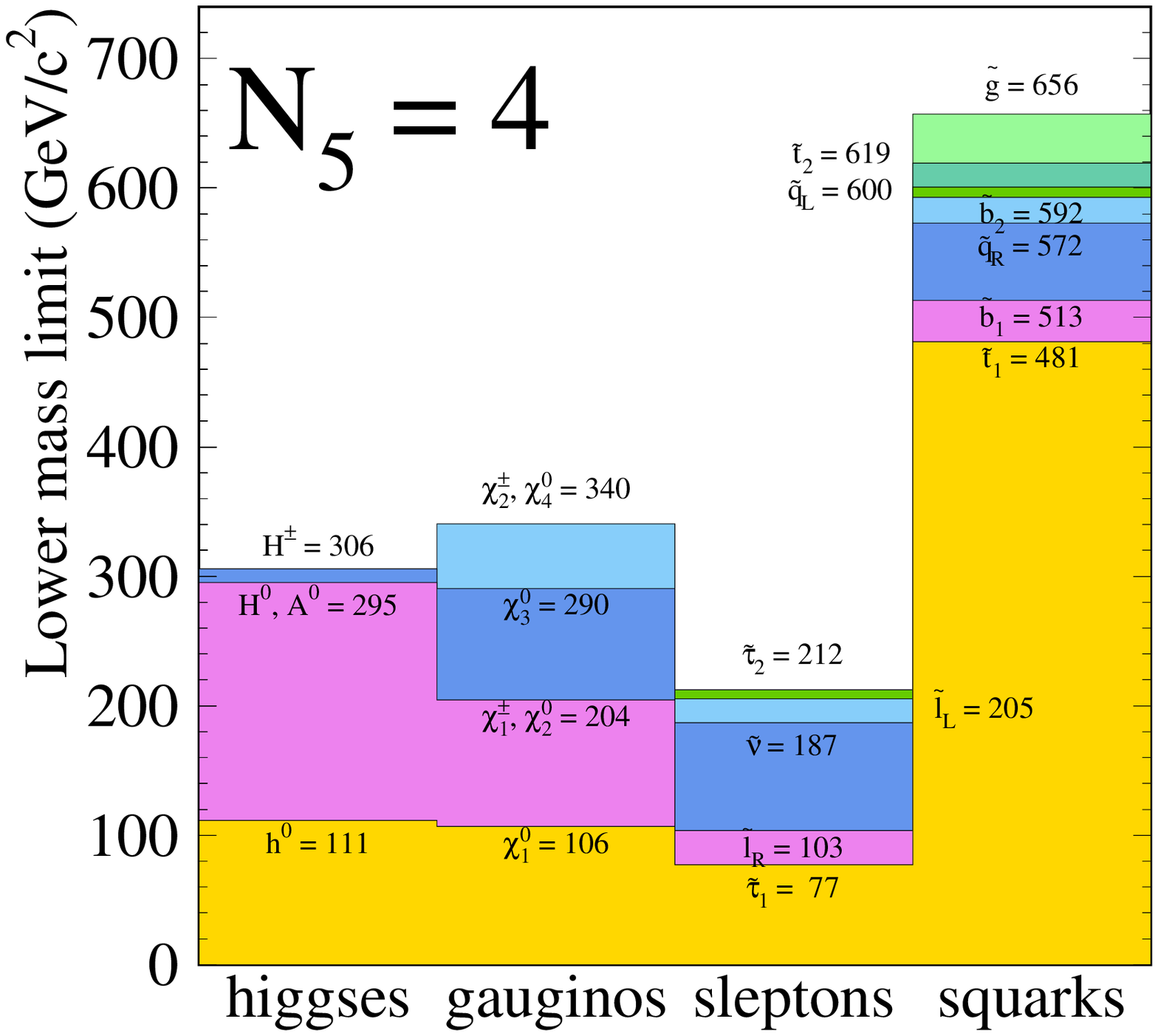}\\
\includegraphics[width=0.49\linewidth]{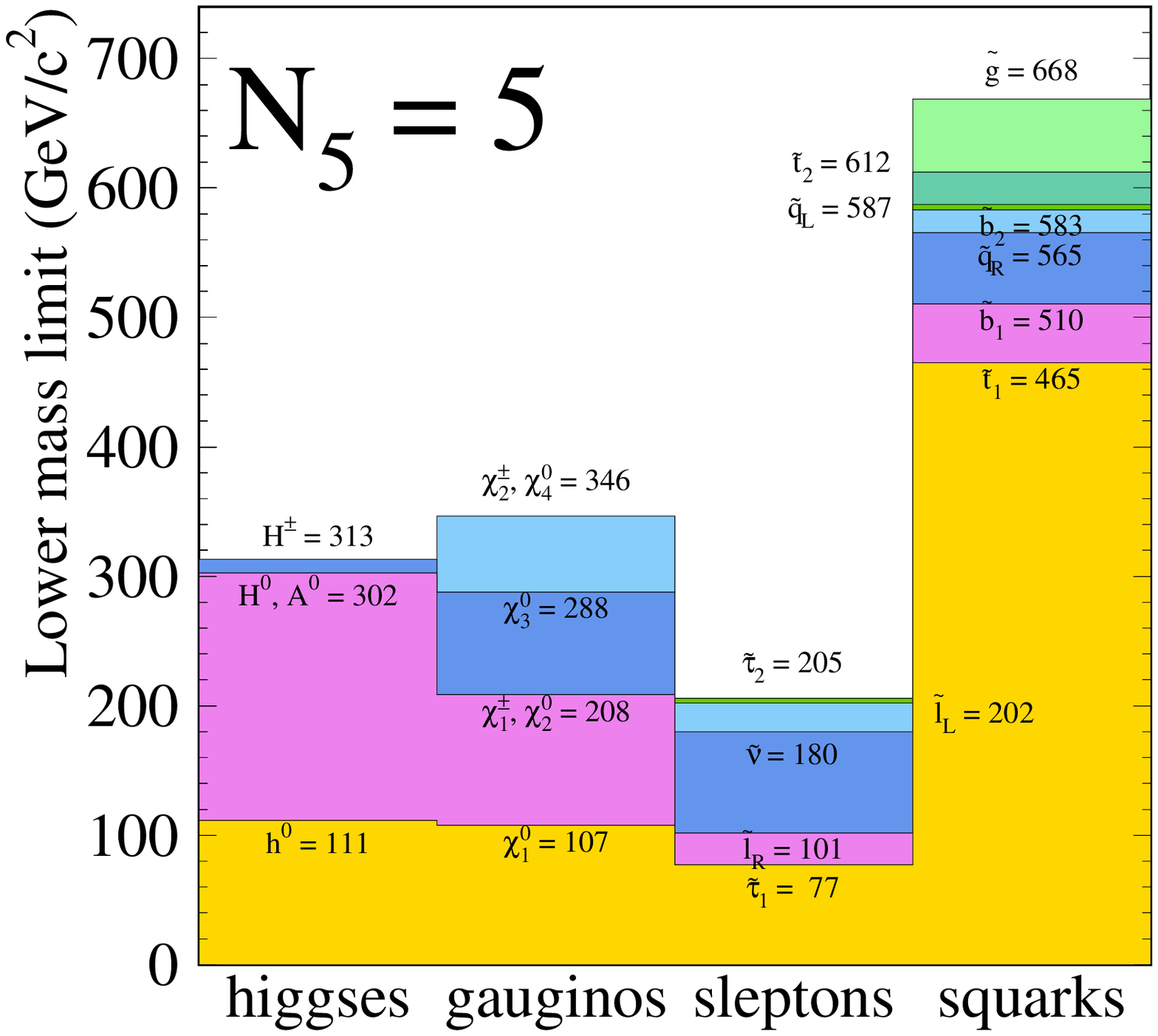}
\caption[Lower mass limit for all supersymmetric
particles as a function of the number of messenger families]
{\label{fig:allmasses}{\small Lower mass limit for all supersymmetric
particles as a function of the number of messenger families. The limits are
derived from the scan after applying GMSB and neutral Higgs boson searches
results. Higher masses may be excluded for certain regions of parameter
space. Thus these limits define the minimum allowed values for the masses
for all possible GMSB models inside the ranges described in the scan
(Tab.~\ref{tab:ranges}).}}
\end{center}
\end{figure}

To conclude, Fig.~\ref{fig:allmasses} gives the status of the full MSSM
spectrum in gauge mediated models after the LEP era. 
With the scan described in Chapter~\ref{scan} and using only 
the ALEPH exclusions as described here, the final excluded masses at 95\%
confidence level have been computed for each messenger index
$\N$.
The ongoing LEP wide combination~\cite{lepsusy} of results will improve
these limits. 

\section{Prospects}
\begin{figure}[ht]
\begin{center}
\includegraphics[width=0.45\linewidth]{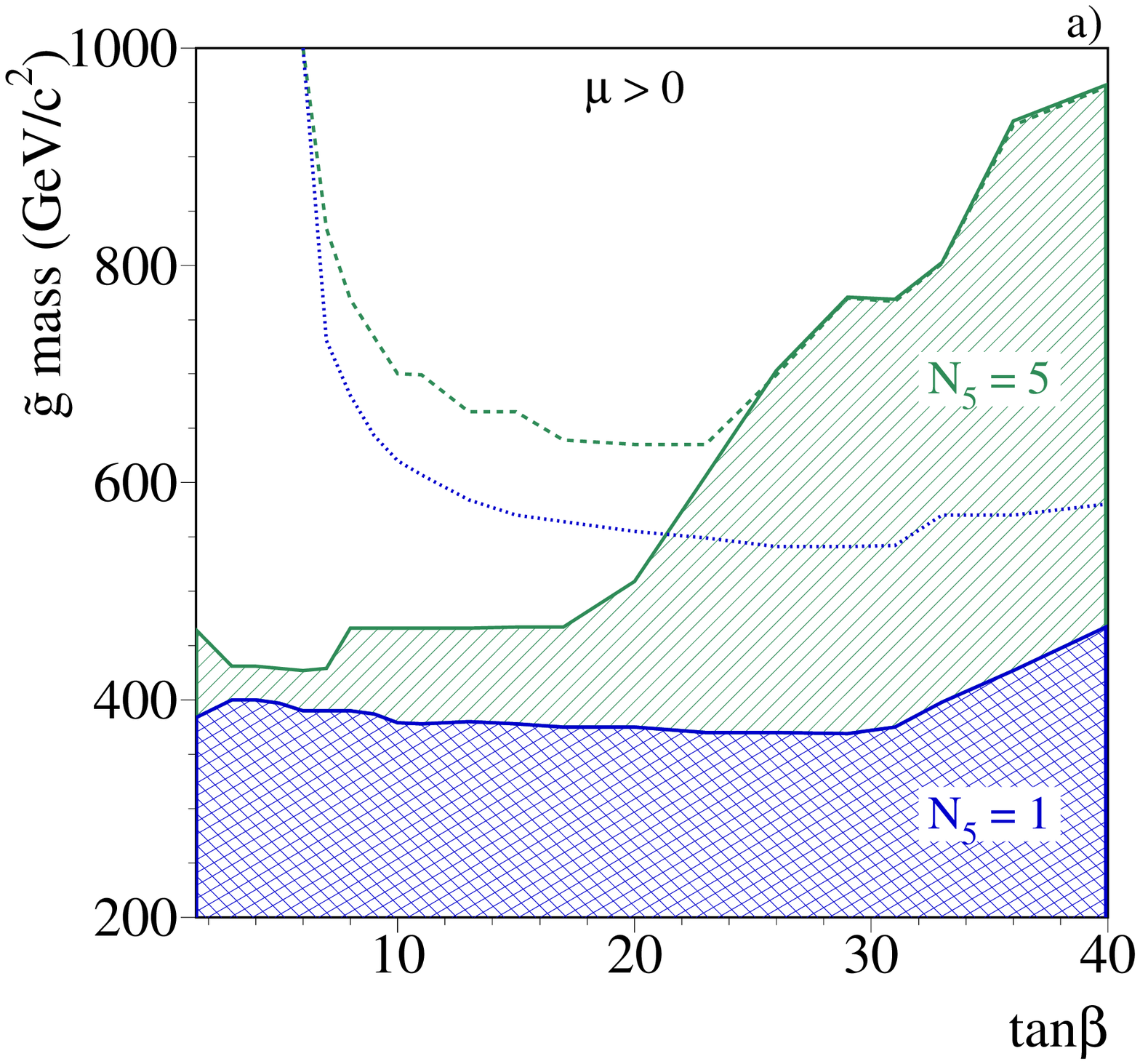}~~
\includegraphics[width=0.45\linewidth]{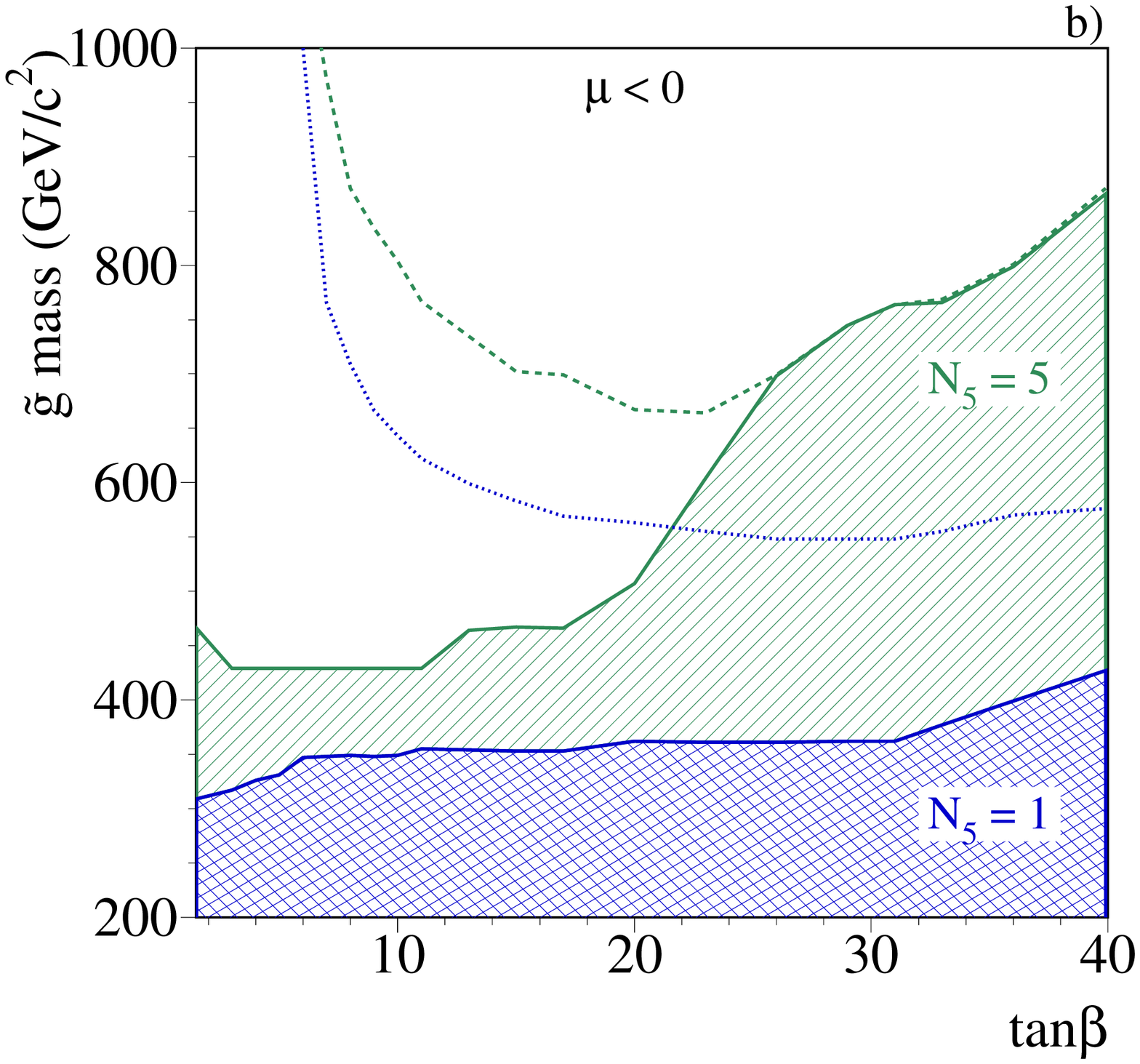}\\
\vspace{0.3cm}		 
\includegraphics[width=0.45\linewidth]{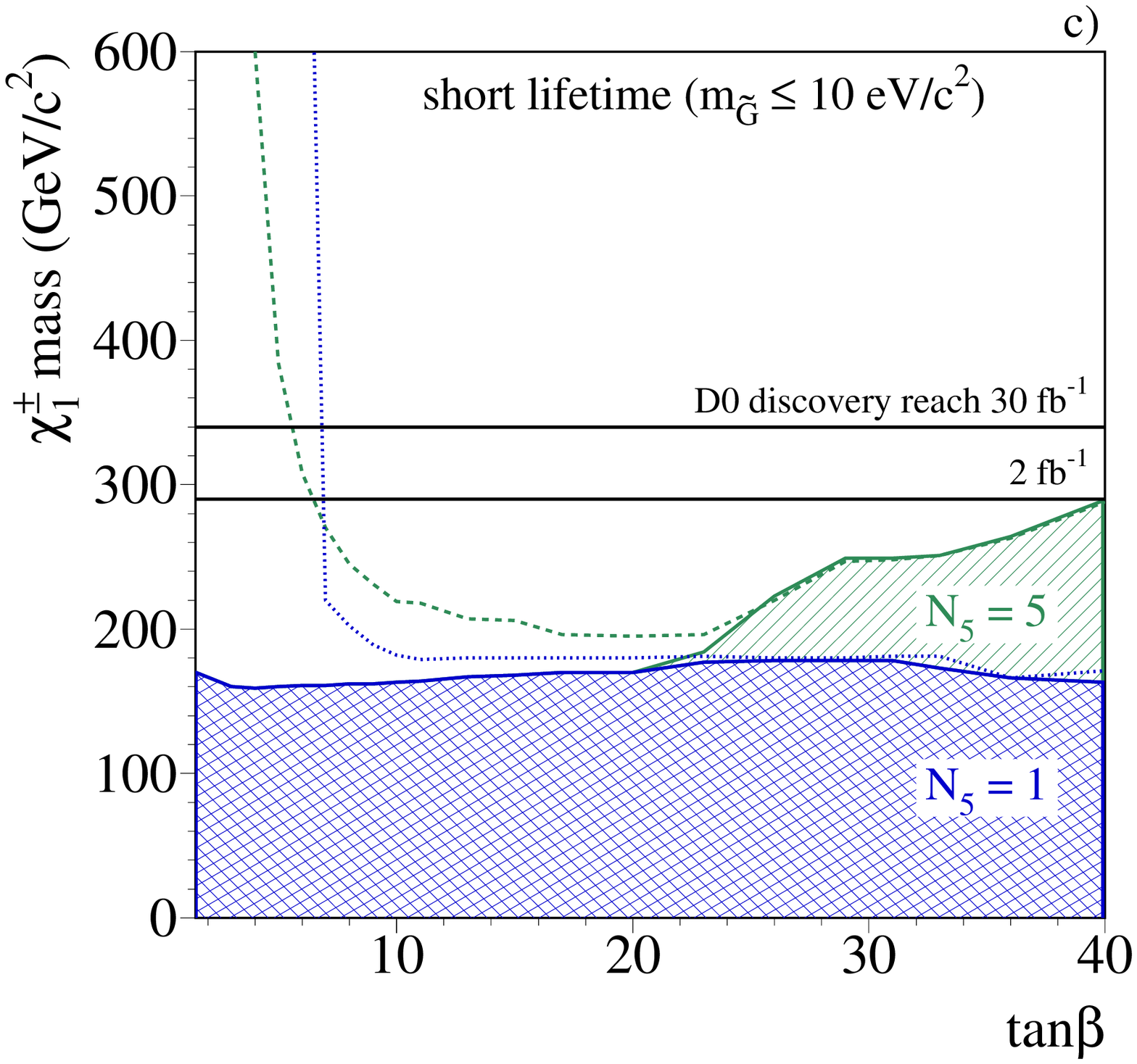}~~
\includegraphics[width=0.45\linewidth]{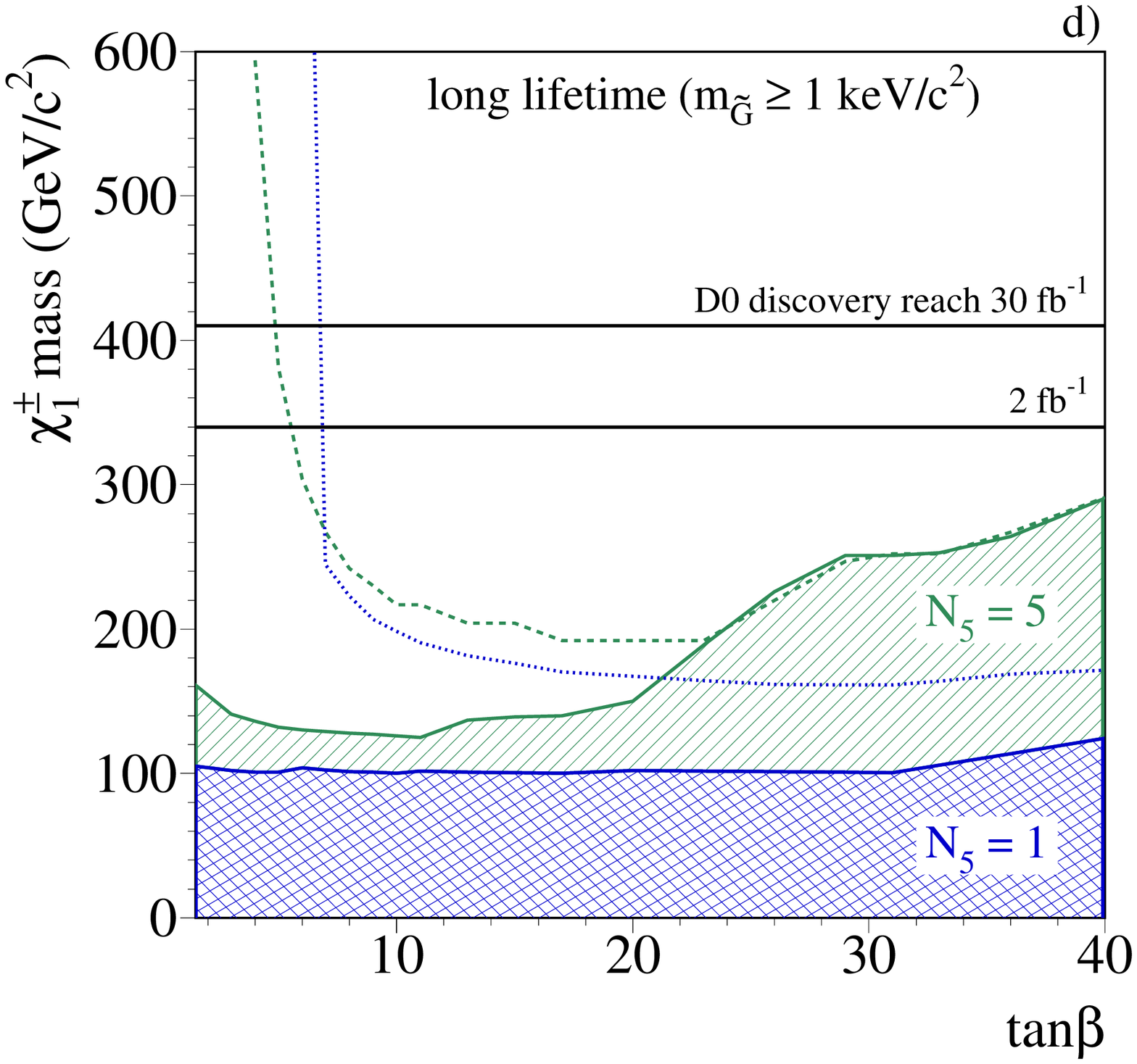}
\caption[Gluino and chargino excluded masses as a function of $\tanb$]
{\label{fig:gluchar}{\small 95\% C.L. exclusion on the gluino mass for (a)
positive and (b) negative $\mu$, for models with one or five messenger
families, as derived from the GMSB searches.
The lower limit on the chargino mass at 95\% C.L. is
displayed for (c) short and (d) long NLSP lifetimes respectively. 
The dotted lines represent the lower limit from GMSB and Higgs boson
searches for $\N=1$ and the dashed line for $\N=5$.}}
\end{center}
\end{figure}

The allowed parameter space in GMSB for searches where the stau is the NLSP and
selectrons and smuons are lighter than the neutralino is extremely large, and
will surely be tested in the near future with new data. 
At the incipient Run II of the Tevatron in Fermilab with $\roots=2\tev$, 
the dominant SUSY production is $p\bar{p}\to\chi_1^+\chi_1^-$ or
$\chi_1^{\pm}\chi_2^0$ through off-shell W and Z. Charginos will tend to
decay through their higgsino components by $\chi_1^{\pm}\to\chi_1^0\W^{\pm}$
and likewise for the neutralino $\chi_2^0\to\chi_1^0\Z^0$. Thus for a
neutralino NLSP scenario, final
states consisting of $\W^{\pm}\Z^0+\gaga+\Emiss_T$ and
$\W^+\W^-+\gaga+\Emiss_T$ will potentially lead to a discovery. 
The D$\emptyset$ collaboration~\cite{Qian:1998zs}, expects to be able to
discover charginos of masses up to 290 (340)$\gevcc$ in the $\gaga+\Emiss_T$
signature in the prompt $\chi$ NLSP decays for a collected luminosity of 
2\,fb$^{-1}$ (30\,fb$^{-1}$) by the end of the year 2004 (2007+).  
In the case of a stau NLSP, stau masses up to 160 (200)$\gevcc$ and chargino
masses up to 340 (410)$\gevcc$, will be discovered if the stau is
long-lived for the same luminosities. The LEP limits (2$\sigma$ exclusion),
as derived from the scan in this work, for the gluino and chargino masses, 
as a function of $\tanb$ and $\N$ can be seen in
Fig.~\ref{fig:gluchar}. Overlaid in the chargino mass plot are the D$\emptyset$
discovery reaches (5$\sigma$) in the $\neu$ NLSP case for short lifetimes
and the stau NLSP case for long lifetimes. 

The task for the Large Hadron Collider at CERN ($\sim$2007) would then be to
unravel the nature of SUSY breaking and measure the parameters of the
theory with $\sim$10\% accuracy. 
Present and future colliders are opening a new energy range in physics,
thus at the beginning of this century we are about to test new theories
that may reshape our understanding of spacetime\dots or even better, be
surprised by something completely unexpected. 
Whether it is supersymmetry or something else, it is surely going to be
exciting.  

\newpage
\addcontentsline{toc}{chapter}{List of Figures}
\listoffigures
\newpage
\addcontentsline{toc}{chapter}{List of Tables}
\listoftables
\newpage
\addcontentsline{toc}{chapter}{Bibliography}

%
\end{document}